\documentclass[a4paper, twoside, 11pt]{book}

\usepackage{bm, fancyhdr, color}
\usepackage{epsfig, float}
\usepackage[newcentury]{quotchap}
\usepackage[centering]{geometry}
\usepackage{booktabs}
\usepackage{setspace}
\usepackage{amsmath,amssymb,mathrsfs, url,cite,ifpdf,slashed,multirow,tabularx,type1cm}

\usepackage[colorlinks=true, linkcolor=blue, citecolor=blue,urlcolor=magenta]{hyperref} 

\makeatletter
\def\ps@fancy{
\def\chaptermark##1{\markboth{\ifnum \c@secnumdepth>\z@ \small\em Chapter \thechapter.\hskip 0.5em\relax \fi ##1}{}}
\def\sectionmark##1{\markright {\ifnum \c@secnumdepth >\@ne \thesection\hskip 0.5em\relax \fi ##1}}
\ps@@fancy
\gdef\ps@fancy{\@fancyplainfalse\ps@@fancy}
\ifdim\headwidth<0sp
\global\advance\headwidth123456789sp\global\advance\headwidth\textwidth
\fi}
\makeatother
\makeatletter
\def\headrule{\iffloatpage{}{{%
 \if@fancyplain\let\headrulewidth\plainheadrulewidth\fi
 \hrule\@height\headrulewidth height .2pt\@width\headwidth\vskip2pt%
 \hrule\@height\headrulewidth height 1.2pt\@width\headwidth\vskip-\headrulewidth\vskip-4pt
}}}
\makeatother
\renewcommand{\headrulewidth}{0.2pt}  

\makeatletter
\renewcommand{\@makechapterhead}[1]{
{\size@chapter{\sectfont\raggedleft
{\chapnumfont
\ifnum \c@secnumdepth >\m@ne
\if@mainmatter {\Large Chapter~} \\[-40pt] \thechapter
\fi\fi
\par\nobreak}
{\raggedleft\advance\interlinepenalty\@M #1\par}}
\nobreak\chapterheadendvskip}}
\makeatother

\def\thefootnote{\ifnum\c@footnote>\z@\textasteriskcentered\@arabic\c@footnote\fi}
\makeatletter
\renewcommand{\footnoterule}{%
\kern-3\p@
\hrule width 0.4\columnwidth
\kern 2.6\p@}
\def\thefootnote{\ifnum\c@footnote>\z@\@arabic\c@footnote\fi}
\makeatother

\allowdisplaybreaks

\newcommand{\TeV}{\,{\rm TeV}}
\newcommand{\GeV}{\,{\rm GeV}}
\newcommand{\MeV}{\,{\rm MeV}}

\newcommand{\hc}{\rm H.c.}
\newcommand{\susy}{\,{M_{\rm SUSY}}}

\newcommand{\non}{\nonumber \\ }
\newcommand{\matl}{\left( \begin{array}}
\newcommand{\matr}{\end{array} \right)}
\renewcommand{\sb}{\sin\beta}
\newcommand{\cb}{\cos\beta}
\newcommand{\tb}{\tan\beta}
\newcommand{\eq}[1]{Eq.~(\ref{#1})}

\newcommand{\bhline}[1]{\noalign{\hrule height #1}}   

\newcommand{\Slash}[1]{{\ooalign{\hfil \hspace*{-5pt}~#1\hfil\crcr\raise.167ex\hbox{/}}}}

\newcommand{\mtrem}[1]{{\color{green} }}

\newcommand{\coll}[1]{#1}

\newcommand{\abstchapter}[1]{\begin{center}\vspace*{-0.8cm}\parbox[c]{12cm}{{\fontsize{10pt}{3mm}\selectfont\hspace*{0.5cm}#1}}\end{center}}

\newcommand{\abstabst}[1]{\begin{center}\vspace*{-0.0cm}\parbox[c]{14cm}{{\fontsize{11pt}{3mm}\selectfont\hspace*{0.0cm}#1}}\end{center}}

\def\be{\begin{equation}}
\def\ee{\end{equation}}
\def\beq{\begin{eqnarray}}
\def\eeq{\end{eqnarray}}
\def\fr{\frac}
\def\({\left(}
\def\){\right)}
\def\<{\langle}
\def\>{\rangle}

\makeatletter
\newcommand{\@authornote}[2]{{\def\thefootnote{\fnsymbol{footnote}}\setcounter{footnote}{#1}#2\setcounter{footnote}{0}}}
\newcommand{\authornotemark}[1]{\@authornote#1{\addtocounter{footnote}{-1}\footnotemark}}
\newcommand{\authornotetext}[2]{\@authornote#1{\footnotetext{#2}}}
\makeatother

\begin{document}
\setlength{\baselineskip}{17pt}
\pagenumbering{roman}
\begin{titlepage}
\begin{flushright}
\hfill KEK--TH--1856\\
\hfill August, 2015\\
\end{flushright}

\vskip 1.5 cm
\begin{center}

{\huge \bf 
Aspects~of~High-Scale~Supersymmetry\\ \vspace{0.4cm}
in a Singlet-Extended Model
}

\vskip 1.55in

{{\LARGE \textbf{Teppei Kitahara}}\footnote[0]{${}${\it E-mail:} \textcolor{magenta}{kteppei@post.kek.jp}}
}
\vskip 0.4in
{\Large 
{\it KEK Theory Center, IPNS, KEK, Tsukuba, Ibaraki 305-0801, Japan}}
\vskip 1.55 in

\textit{{\Large Ph.D thesis submitted to\\
Department of Physics, University of Tokyo\\
December 2014}}

\end{center}
\vskip .25in

\end{titlepage}
\newpage
\thispagestyle{empty}
 ~ 
\newpage
\thispagestyle{empty}
\begin{center}
\begin{flushleft}

\textrm{{\fontsize{43pt}{5mm}\selectfont \\
\vspace{40pt}
\bf{Aspects~of~High-Scale\\
\vspace{3pt}
 Supersymmetry}} \\
\vspace{6pt}
{\fontsize{32pt}{5mm}\selectfont \bf{in a Singlet-Extended Model}}}\\
\vspace{14pt}
\hrule width 15.8cm height 0.1pt
\vspace{2pt}
\hrule width 9cm height 0.1pt
\vspace{14pt}

\hspace*{1cm}\textrm{{\Huge 
\bf{Teppei Kitahara}}}
\end{flushleft}

\vspace{50pt}

\begin{flushright}
\begin{figure}[h]
\begin{center}
\vspace*{ - 60pt}
\hspace*{9.5cm}
  \includegraphics[width=6.0cm]{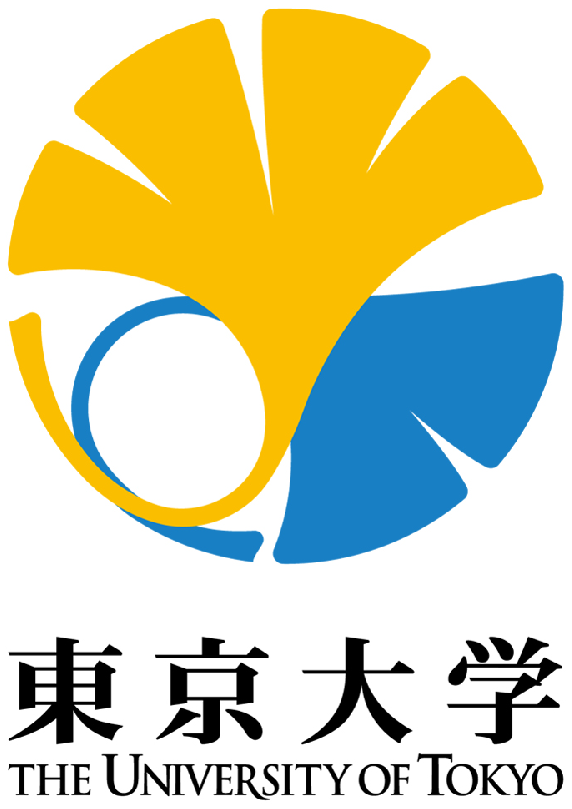}
\end{center}
\end{figure}

\textsc{{\fontsize{19pt}{5mm}\selectfont Department~of~Physics,~Faculty~of~Science\\ 
\vspace{5pt}
The University of Tokyo, Japan}}

\vspace{28pt}
\begin{flushright}
\rule{0.5\textwidth}{.1pt}
\end{flushright}
\vspace{6pt}

\textit{{\large  A thesis submitted for the degree of
Doctor of Philosophy in 2014}}\\
\vspace{4pt}

{\large Submitted : December 18, 2014}\\
{\large Revised : February 3, 2015}\\
{\large Accepted : February, 2015}
\end{flushright}

\end{center}
\vspace*{1.5cm}
{\LARGE{\vspace{80pt} \bf{
~~~~~~~~~~~~~~~~~~~~~~~~~~Abstract}}}
\abstabst{
\begin{spacing}{1.2}

~~~~The nearly Minimal Supersymmetric Standard Model (nMSSM) is one of the promising models of the new physics, since this model can avoid hierarchy problem, $\mu$ problem, cosmological domain wall problem, and  tadpole problem simultaneously.
In this thesis, we consider the phenomenology of the nMSSM. 
Especially, we focus on the phenomenology of the dark matter and  the baryon asymmetry in the universe generated by the electroweak baryogenesis mechanism.
We find that with high-scale supersymmetry breaking the singlino can obtain a sizable radiative correction to the singlino mass, which opens a window for the singlet dark matter scenario with resonant annihilation via the exchange of the Higgs boson. 
We also propose a new electroweak baryogenesis scenario in the nMSSM with additional vector-like multiplets.
 If the soft supersymmetry breaking scale is $\mathcal{O}(10)$ TeV, these scenarios are compatible with each other and an observed mass of the Higgs boson, constraints by the electric dipole moments measurements and the flavor experiments.
  As a result of these two studies, we  conclude that the nMSSM with a high-scale supersymmetry breaking is valid and can be  probed by the direct direction of the  singlino dark matter.

\end{spacing}}
\begin{center}
\newpage
\tableofcontents 
\thispagestyle{empty}
\newpage
\listoffigures
\thispagestyle{empty}
\newpage
\end{center}
\pagenumbering{arabic}
\renewcommand{\thefootnote}{\#\arabic{footnote}}
\chapter{Introduction}
\thispagestyle{empty}

\section{Overview}

The standard model (SM) of the particle physics has worked very well for a long time.
Amazingly, ten observations and eight predictions or theories, which are related with the SM,  had received the Novel Prize.
In 2012 the Higgs boson, which is a last missing piece of the SM, was observed by the Large Hadron Collider 
(LHC) experiments at CERN \cite{Aad:2012tfa, Chatrchyan:2012ufa}, and F. Englert and P. Higgs have won the Nobel Prize for the discovery of the Higgs mechanism.
This is a triumph of the SM, and the observation of the SM Higgs boson has given an important step towards understanding the electroweak symmetry breaking.

However, there are many unsolved problems within the SM, for example, the observed dark matter particles and baryon asymmetry of the universe. 
From theoretical viewpoint, the gauge hierarchy problem is still in question. 
Hence, there have been many attempts to solve such problems in framework beyond the SM.

\paragraph*{}
The supersymmetric (SUSY) models are good candidates as the physics beyond the standard model \cite{Wess:1974tw, Wess:1973kz, Wess:1974jb, Iliopoulos:1974zv,Girardello:1981wz}.
It is  because that they can solve the hierarchy problem naturally and ensure the unification of the gauge couplings. 
In addition, the lightest SUSY particle can be a natural candidate of the WIMP dark matter if the R parity is conserved.

The minimal SUSY extension of the SM (MSSM) contains a supersymmetric dimensional parameter $\mu$, which  is the mass term of the superpartner of the Higgs boson.
 However, this parameter  causes ``$\mu$ problem'', which is also one of the hierarchy problem \cite{Kim:1983dt}.
Although $\mu$  has to be a size of the SUSY breaking scale to realize the electroweak symmetry breaking properly, there is no reason for $\mu$ to be small compared to the Planck scale.

\paragraph*{}
One of the simplest ways to solve the $\mu$ problem is introducing a gauge-singlet superfield \cite{Fayet:1974pd}.
There are several models of singlet extension of the MSSM depending on the imposed additional symmetry. 
However, the additional singlet superfield causes a cosmological domain wall problem  \cite{Zeldovich:1974uw,Abel:1995wk}  and tadpole problem \cite{Ellwanger:1983mg}. 
The nearly Minimal (or new Minimal) Supersymmetric Standard Model (nMSSM)~\cite{ Panagiotakopoulos:1999ah, Panagiotakopoulos:2000wp, Dedes:2000jp} is based on a discrete $\mathbb{Z}^R_5$ R-symmetry.
Actually this model can avoid naturally the cosmological domain wall problem and tadpole problem, unlike $\mathbb{Z}_3$ symmetric models~\cite{Abel:1995wk, Panagiotakopoulos:1998yw}.
Therefore, the nMSSM is one of the promising models of the new physics: this model can avoid the $\mu$ problem, the domain wall problem, and the tadpole problem simultaneously. 
In addition this model has natural candidate of the dark matter and can generate the baryon asymmetry  of the universe. 

\paragraph*{}
In this thesis, we consider the phenomenology of the nMSSM, which are the based on the works by the author  \cite{Ishikawa:2014owa, Ishikawa:2014tfa}.
Especially, we focus on a phenomenology of the dark matter \cite{Ishikawa:2014owa} in Chapter~\ref{singlinochap} and  the baryon asymmetry in the universe generated by the electroweak baryogenesis mechanism  \cite{Ishikawa:2014tfa} in Chapter~\ref{EWBGchap}.

In Chapter~\ref{singlinochap}, we consider a singlino dark matter scenario in the nMSSM. 
The singlino is a fermion component of the additional singlet superfield.  
We find that with high-scale SUSY breaking the singlino can obtain a sizable radiative correction to the mass, which opens a window for the dark matter scenario with resonant annihilation via the exchange of the Higgs boson. 
We show that the current dark matter relic abundance and the Higgs boson mass can be explained simultaneously. 
This scenario can be probed by the search of the Higgs invisible decay and the direct direction of the dark matter.

In Chapter~\ref{EWBGchap}, we propose a new electroweak baryogenesis scenario in high-scale SUSY models, and consider 
the nMSSM introducing additional vector-like multiplets.
We show that the strongly first-order phase transition can occur at a high temperature comparable to the soft SUSY breaking scale.
In addition, the proper amount of the baryon asymmetry of the universe can be generated via the lepton number violating process in the vector-like multiplet sector.
The typical scale of our scenario, the soft SUSY breaking scale, can be any value. 
Thus our new electroweak baryogenesis scenario can be realized at arbitrary scales and we call this scenario as a scale free electroweak baryogenesis.
This soft SUSY breaking scale is determined by other requirements.
If the soft SUSY breaking scale is $\mathcal{O}(10) \TeV$, our scenario is compatible with the observed mass of the Higgs boson and the constraints by the electric dipole moments measurements and the flavor experiments.
Furthermore, the singlino can be a good candidate of the dark matter.
 
 As a result of these two studies, we will conclude  that the nMSSM with a high-scale SUSY breaking is valid and can be  probed by the direct direction of the  singlino dark matter.

\section{Organization of this thesis}

This thesis is organized as following.

{ 
In Chapter~\ref{chap2}, we review the SM, the supersymmetry and  current status of the supersymmetric minimal model.
In Section~\ref{2no1} and \ref{2no2}, 
in order to solve the hierarchy problem of the standard model, we first introduce the supersymmetry and the MSSM.
In Section~\ref{2no3}, we review the current situation of the MSSM.
In fact, an observed mass of the SM Higgs boson is $125\GeV$, and it gives a meaningful constraint on the parameter space of supersymmetric models.
Thus, we focus on the one-loop, two-loop and higher-loop radiative corrections to the Higgs boson mass. 
These calculations for the Higgs boson mass are reused in the study of the nMSSM (Chapter~\ref{singlinochap}).
Furthermore, we also discuss the constraints from the flavor violation and CP violation process in supersymmetric model.
In Section~\ref{2no4}, we summarize the current status of the MSSM.

{
In Chapter~\ref{chap3}, we review singlet extension models of the MSSM and the nMSSM.
In Section~\ref{3no1}, we first explain the  $\mu$ problem in the MSSM, and in order to solve the $\mu$ problem 
we introduce the additional gauge singlet superfield.
Next, we show that when one imposes extra symmetries to forbid unwanted terms of singlet superfield,
these symmetries lead to the domain wall problem and the tadpole problem in Section~\ref{3no2}.
The nMSSM is the one of the models which can solve the $\mu$ problem, the domain wall problem and the tadpole problem, and so we review the nMSSM in Section~\ref{3no3} and \ref{3no4}.
Anyhow, we need an extra symmetry to solve these problems.
}

In Chapter~\ref{singlinochap}, we study the phenomenology of the singlino dark matter in the nMSSM.
In Section~\ref{4no1}, we  briefly review a situation of the singlino dark matter in the nMSSM, and we also explain why we have considered it. 
In Section~\ref{DMS}, 
 using the low energy effective Lagrangian we calculate thermal relic abundance of the singlino dark matter which annihilate via the SM Higgs boson.
We point out that one-loop corrections to the singlino mass can raise its mass with relatively high-scale SUSY breaking, in Section~\ref{singlinomass}.
In Section~\ref{NumRes}, we numerically investigate the singlino resonant dark matter  scenario with high-scale SUSY breaking, and show this scenario is compatible with the observed SM Higgs boson mass.
Section~\ref{4nodis} and \ref{CON} are devoted to the conclusion and discussions in this chapter.

In Chapter~\ref{EWBGchap}, we study the baryon asymmetry in the universe generated by the electroweak baryogenesis mechanism  in the nMSSM, 
and we propose a new electroweak baryogenesis scenario in the nMSSM with high-scale SUSY breaking.
In Section~\ref{5no1}, \ref{sec_the_model} and \ref{sec:the_scenario},
we introduce the model: the nMSSM with vector-like multiplets,  and also present the overview of our scenario.
In Section~\ref{sec:the_FOPT}, 
we discuss about the strongly first-order phase transition and this section is divided into three parts.
In subsection~\ref{subsec:the_potential}, we show the full thermal potential at high temperatures.
In subsection~\ref{subsec:tree}, we provide an intuitive understanding for the behavior of the potential at high temperatures.
In subsection~\ref{subsec:numerical}, we analyze the full potential and show that the strongly first-order phase transition actually occurs at a temperature comparable to $M_{\rm SUSY}$.
We also show that the region with low $\tan\beta$ and a light charged Higgs boson is favored in our scenario.
In Section~\ref{sec_BAU}, we demonstrate the generation of the BAU with the lepton number violating process.
In Section~\ref{sec_DM}, we discuss the singlino dark matter scenario paying particular attention to the lifetime.
Section~\ref{5no7} and \ref{sec_conclusion} are devoted to the conclusion and discussions in this chapter.

Chapter~\ref{chap6} is devoted to the conclusion of this thesis.


\chapter{An Introduction to 
Supersymmetry}
\label{chap2}
\thispagestyle{empty}
\abstchapter{ 
In this chapter, in order to solve the hierarchy problem of the standard model, we first introduce the supersymmetry and the minimal supersymmetric standard model.
In fact, an observed mass of the SM Higgs boson is $125\GeV$, and it gives a meaningful constraint on the parameter space of supersymmetric models.
Therefore, we first review the one-loop, two-loop and higher-loop radiative corrections to the Higgs boson mass.
Furthermore, we also discuss the constraints from the flavor violation and CP violation process in supersymmetric model.
These facts imply that the na\"{\i}ve low scale minimal supersymmetric standard model, which is in spite of being favored in terms of the naturalness, is disfavored.
One of the solutions of these problems is the high-scale supersymmetry.}

\section{The Standard Model}
\label{2no1}
In nature, there are four fundamental forces: the electromagnetic, weak, and strong nuclear and gravitational interactions.
These interactions can be understood by interactions of elementary particles.
The standard model (SM) can describe the electromagnetic, weak, and strong nuclear interactions as a quantum field dynamics. 

The standard model is one of the gauge theories.
In the standard model, the imposed gauge symmetry is $G_{\rm SM} = $U$(1)_Y \times $ SU$(2)_L \times$ SU$(3)_c$ 
\cite{Glashow:1961tr, Weinberg:1967tq,Salam:1968rm,tHooft:1972fi}.
Gauge bosons are introduced at every gauge symmetry: a $B$ boson for U$(1)_Y$ gauge, isospin triplet $W$ bosons for SU$(2)_L $ gauge and color gluons octet for SU$(3)_c$  gauge.
However, in order to describe the electromagnetic and weak nuclear interactions, 
U$(1)_Y \times $ SU$(2)_L$ gauge symmetry (electroweak symmetry) should be spontaneous  broken to U$(1)_{EM}$ gauge symmetry (electromagnetic symmetry) by the Higgs mechanism \cite{Higgs:1964ia, Englert:1964et,Higgs:1964pj, Guralnik:1964eu}.
This mechanism predicts an existence of the Higgs boson $h$ (SM Higgs boson). 
The unitarity requirement for the high-energy scattering of the longitudinal $W$ boson leads to the upper bound on the mass of the  SM Higgs boson $m_h \lesssim 700 \GeV $ \cite{Lee:1977eg}.

\paragraph*{}
The standard model has been worked very well for a long time, and its last missing piece, the
Higgs boson, was finally discovered by the LHC experiment at CERN \cite{Aad:2012tfa,Chatrchyan:2012ufa}.
This is a triumph of the SM and a great step to understand physics at the electroweak scale.

The Lagrangian of the standard model is given as follows,
\beq
\mathcal{L}=\mathcal{L}_{\rm gauge} + \mathcal{L}_{\rm fermion} + \mathcal{L}_{\rm Yukawa} + \mathcal{L}_{\rm scalar},
\eeq
with
\beq
\mathcal{L}_{\rm gauge} &=& - \frac{1}{4} B^{\mu\nu} B_{\mu \nu} -\frac{1}{4} \sum^{3}_{a = 1} W^a_{\mu \nu} W^{a \mu \nu} - \frac{1}{4} \sum^{8}_{a= 1}G^{a}_{\mu \nu} G^{a \mu \nu}, \\
 \mathcal{L}_{\rm fermion} &= & i \bar {Q}_i \gamma^{\mu} \left( \partial_{\mu} - i \frac{g'}{6} B_{\mu}  - i \frac{g}{2} \sigma^a W^{a}_{\mu} - i \frac{g_s}{2}{\lambda^{a}} G^{a}_{\mu} \right)Q_i\non
 && + i \bar {U}_i \gamma^{\mu} \left( \partial_{\mu} - i \frac{2 g'}{3} B_{\mu} - i \frac{g_s}{2}{\lambda^{a}} G^{a}_{\mu} \right)U_i + 
  i \bar {D}_i \gamma^{\mu} \left( \partial_{\mu} + i \frac{g'}{3} B_{\mu} - i \frac{g_s}{2}{\lambda^{a}} G^{a}_{\mu} \right)D_i \non 
  && +  i \bar {L}_i \gamma^{\mu} \left( \partial_{\mu}  + i \frac{g'}{2} B_{\mu}  - i \frac{g}{2} \sigma^a W^{a}_{\mu} \right)L_i  +  i \bar {E}_i \gamma^{\mu} \left( \partial_{\mu}  + i g' B_{\mu} \right)E_i, 
\\
 \mathcal{L}_{\rm Yukawa} & = &   \bar{U}_i (y_u)_{i j}  H Q_j - \bar{D}_i (y_d)_{i j } H^{\dag} Q_j - \bar{E}_i (y_e)_{i j} H^{\dag} L_j +\hc, \\
 \mathcal{L}_{\rm scalar} &=& \left| \left( \partial_{\mu} - i \frac{g'}{2} B_{mu} - i \frac{g}{2}\sigma^a W^{a}_{\mu} \right) H \right|^2 - V(H), \\ 
\eeq
where the field strength $F^a_{\mu \nu}$ is defined as $F^a_{\mu \nu}= \partial_{\mu} A^a_{\nu} - \partial_{\nu} A_{\mu}^a + g_{A} f^{a b c} A^b_{\mu} A^{c}_{\nu} $, $\sigma^a ~(a =1,2,3 )$ is the Pauli matrix, $\lambda^a ~(a = 1,2,\dots,8)$ is the Gell-Mann matrix, index $i$ represents the generation $(i = 1,2,3)$, $y$ is the Yukawa couplings, the gauge couplings for U$(1)_Y$, SU$(2)_L$ and SU$(3)_c$ are denoted as $g'$, $g$ and $g_s$ respectively.
The contraction of the two SU(2) doublet is $AB = A^{T} i \sigma^2 B$.
A more detail definition is written in Appendix~\ref{AppA}.
$H$ is the Higgs doublet,
\beq
H = 
\left( \begin{array}{cc} H^{+}  \\
 H^{0} \end{array} \right).
\eeq
The Higgs potential can be written as follows,
\beq
V= - \mu^2 |H^{\dag} H| + \frac{\lambda_{\rm quartic}}{2} |H^{\dag} H|^2.
\label{SMH}
\eeq
Then, the Higgs doublet obtains the vacuum expectation value $v_{EW}$ (VEV), 
\beq
v_{EW}^2 = \frac{\mu^2} { \lambda_{\rm quartic}}.
\eeq
 It breaks the electroweak symmetry to the electromagnetic symmetry, SU$(2)_L \times $ U$(1)_Y \to $ U$(1)_{EM}$.
Using the freedom of  SU$(2)$ 
rotations, one can always align the VEV with the neutral Higgs direction, 
\beq
H = \left( \begin{array}{cc} 0  \\
 v_{EW} \end{array} \right) + 
 \left( \begin{array}{cc} G^{+}  \\
 \frac{1}{\sqrt{2}} \left( h + i G^0 \right) \end{array} \right),
\eeq
where $G$ is the Numbu-Goldstone bosons and $h$ is the SM Higgs boson.
Then, the mass of the SM Higgs boson is
\beq
m_h^2 = 2 \mu^2 = 2 \lambda_{\rm quartic} v_{EW}^2.
\label{SMmass}
\eeq
The Higgs VEV $v_{EW}$ generates the Dirac mass to the all fermion via the Yukawa interaction, $m_f = y_f v_{EW}$. The gauge bosons also obtain the mass,
\beq
M_Z^2 = \frac{g'^2 + g^2}{2} v_{EW}^2, ~~~~~~~M_W^2 = \frac{g^2}{2} v_{EW}^2,\non
m_{\gamma}  = m_{\rm gluon} = 0,~~~~~~~~~~~~
\eeq
where $A (\gamma) = \cos \theta_W B + \sin \theta_W W^3, ~Z = -\sin \theta_W B + \cos \theta_W W^3,~W^{\pm} = (W^1 \mp i W^2)/\sqrt{2} $, and $\theta_W$ is the Weinberg angle.

Therefore, the Higgs vev can determine the electroweak scale.
In fact, $v_{EW} = 174.10363 \pm 0.00004 \GeV$, which has been given by the measurement of the Fermi constant from the muon decay.
In the other words, the parameter $\mu^2$ in the Higgs potential determines the electroweak scale and the  SM Higgs boson mass.

\paragraph*{Hierarchy problem}
~

However, the radiative corrections to the parameter $\mu^2$ contain the quadratic divergence.
For example, the top quark loop gives the following quadratic divergence,
\beq
\Delta\mu^2 = -\frac{1}{8 \pi^2} y_t^2\left(  \Lambda^2 + \dots \right),
\label{topquaddiv}
\eeq
 where $\Lambda$ is the ultraviolet momentum cutoff.
 All diagrams which contain the quadratic divergence are shown in Figure~\ref{Figurequad}.
 If one assume that the standard model is valid up to the Planck scale, the ultraviolet cutoff is na\"{\i}vely the Planck scale, $\Lambda \sim 10^{19} \GeV$.
Although these radiative corrections can be renormalized by the bare parameter $\mu^2_0$,
it requires incredible fine-tuning cancellation between the bare mass $\mu^2_0$ and the quadratic radiative corrections $\Delta\mu^2$. 
In fact, in order for the $\mathcal{O}(100)\GeV$ electroweak scale to be realized, the  $10^{-34}$
 fine-tuning is  required.
This difficulty in the standard model is called the hierarchy problem.
 
\section{Supersymmetry and Minimal Model}
\label{2no2}

\begin{figure}[t]
\begin{center}
\begin{tabular}{cc|cc} 
\bhline{1.4pt}
\multicolumn{4}{c}{ } \\
\cline{1-2}
\multicolumn{2}{c|}{Standard model} &
\multicolumn{2}{c}{Supersymmetric minimal model} \\ 
\midrule
 \includegraphics[width=37mm]{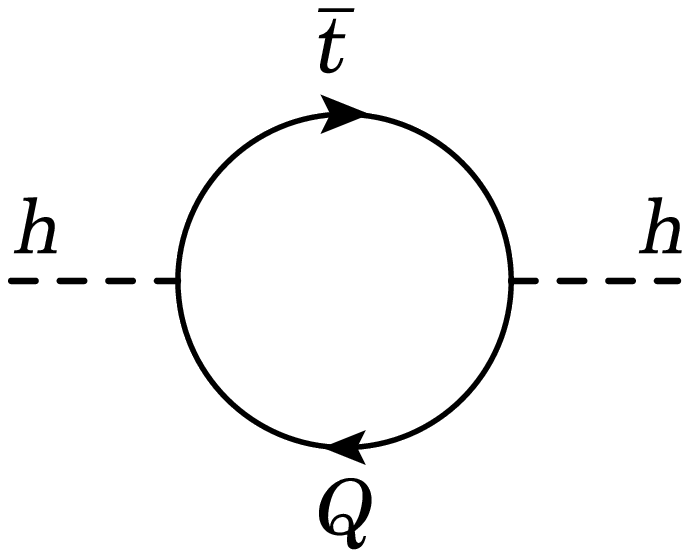}
  &
  \shortstack{
  $- \frac{1}{8 \pi^2} y^2 \Lambda^2$ \\ {\textcolor{white}{:}}\\ {\textcolor{white}{:}}\\  {\textcolor{white}{:}}\\  {\textcolor{white}{:}}}
  &
  \includegraphics[width=35mm]{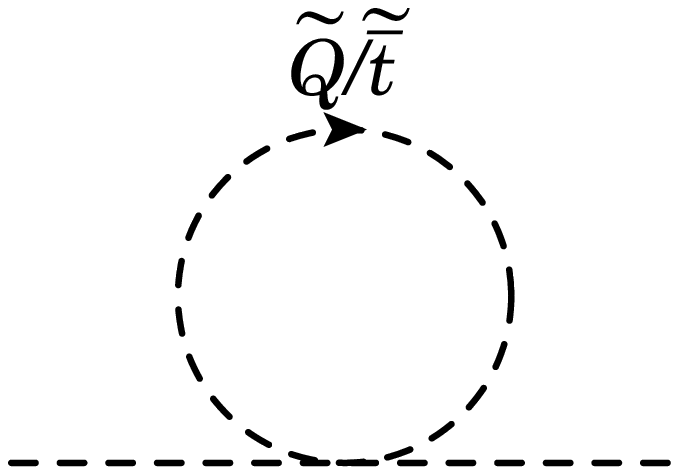}
  &
  \shortstack{
  $\frac{1}{16 \pi^2} y^2 \Lambda^2 \times 2$\\ {\textcolor{white}{:}}\\ {\textcolor{white}{:}}\\  {\textcolor{white}{:}}}
   \\ 
  \midrule
   \includegraphics[width=35mm]{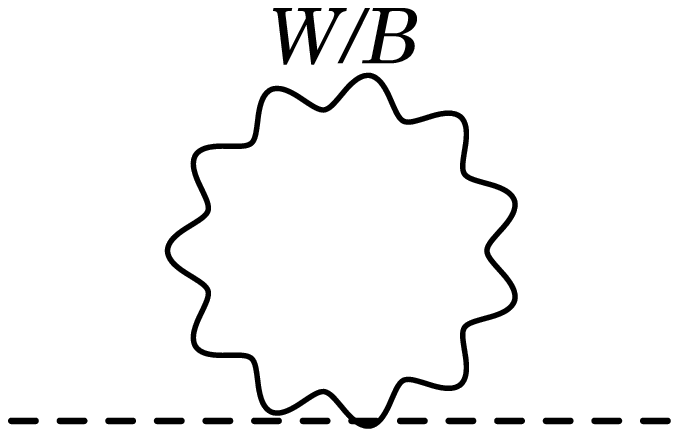}
  &
  \shortstack{
  $\frac{1}{4 \pi^2} g^2(t^a)^2 \Lambda^2$ \\ {\textcolor{white}{:}}\\ {\textcolor{white}{:}}\\  {\textcolor{white}{:}}}
  & & \\
  \includegraphics[width=35mm]{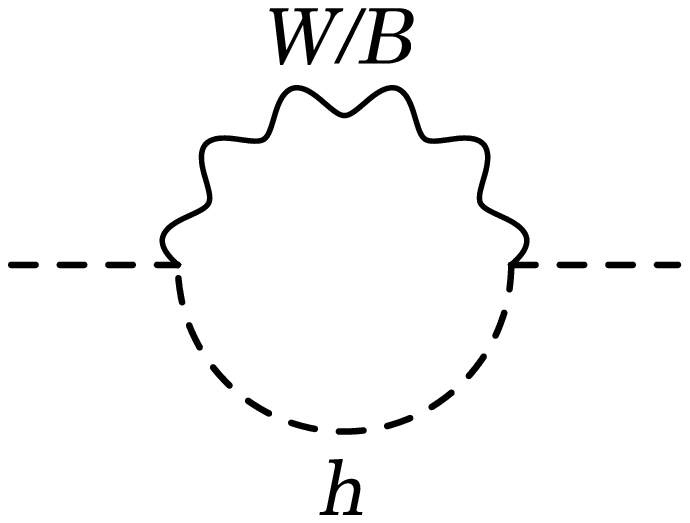}
  &
 \shortstack{
  $- \frac{1}{16 \pi^2} g^2 (t^a)^2 \Lambda^2$ \\ {\textcolor{white}{:}}\\ {\textcolor{white}{:}}\\  {\textcolor{white}{:}}\\  {\textcolor{white}{:}}}
  &
  \includegraphics[width=35mm]{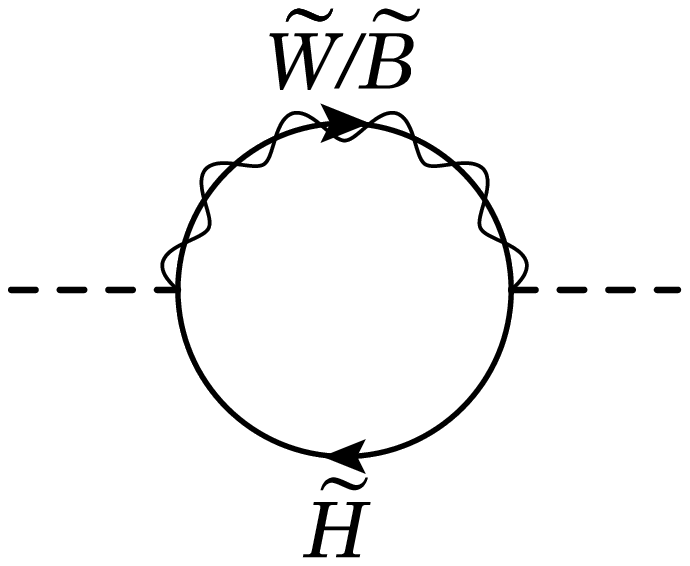}
  &
 \shortstack{
  $- \frac{1}{4 \pi^2} g^2 (t^a)^2 \Lambda^2$ \\ {\textcolor{white}{:}}\\ {\textcolor{white}{:}}\\  {\textcolor{white}{:}}\\  {\textcolor{white}{:}}}
   \\ 
   \includegraphics[width=35mm]{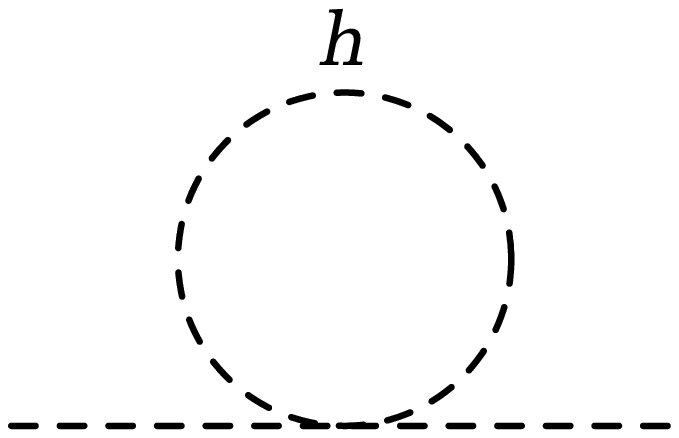}
  &
   \shortstack{
  $ \frac{1}{16 \pi^2} \lambda_{\rm quartic} \Lambda^2$\\ {\textcolor{white}{:}}\\ {\textcolor{white}{:}}\\  {\textcolor{white}{:}}}
  & & \\
\bhline{1.4pt}
\end{tabular}
\caption[The quadratic divergence to the mass of the Higgs boson in the standard model and the supersymmetric model. ]{The quadratic divergence to the mass of the Higgs boson in the standard model and the supersymmetric model. Since the supersymmetry assures that the Higgs quartic coupling is related to the gauge boson coupling like \eq{naive}, all quadratic divergence are always canceled out in the supersymmetric model.}
\label{Figurequad}
\end{center}
\end{figure}

The supersymmetry (SUSY) is the one of the symmetry which can solve the hierarchy problem.
In this section, we first introduce the supersymmetry and its minimal model, and show that the hierarchy problem is solved actually.
Next, we briefly review the Lagrangian and the mass of the Higgs boson in the supersymmetric minimal model.

\subsection{Motivations of Supersymmetry}
First, let us briefly introduce the supersymmetry, which can solve the hierarchy problem \cite{Wess:1974tw, Wess:1973kz, Wess:1974jb, Iliopoulos:1974zv,Girardello:1981wz}.
The supersymmetry is the extension symmetry of the Poincar{\'e} group through the introduction of anticommuting spinor generators $Q_{\alpha}, \bar{Q}_{\dot{\alpha}}$, where  $\alpha,~\dot{\alpha}$ are spinor index \cite{Golfand:1971iw}.
It is equal to the symmetry between   the boson and fermion,

\beq
\bar{Q}|{\rm boson}\rangle &=& |{\rm fermion}\rangle \\
Q|{\rm fermion}\rangle & = &|{\rm boson}\rangle.
\eeq
Then, ordinary space-time $x^{\mu}$ is extended to the superspace $(x^{\mu},\theta_{\alpha},\bar{\theta}_{\dot{\alpha}})$, where $\theta_{\alpha},\bar{\theta}_{\dot{\alpha}}$ are the anticommuting Grassman coordinates.
The supersymmetric theories are described by  chiral superfield, vector superfield and spinor chiral superfield.
The chiral superfield $\hat{\Phi}$ contains the scalar  and fermion field on the superspace,
 \beq
\hat{\Phi}(x,\theta,\bar{\theta}) = \phi (y)+ \sqrt{2} \theta \psi(y) + \theta \theta F(y),\label{chiral}
\eeq
with $y^{\mu}=x^{\mu} +  i \theta \sigma^{\mu} \bar{\theta}$, and the hat ($\hat{}$) represents the superfield. 
$\phi$ is a complex scalar field (sfermion), $\psi$ is two-components Weyl spinor field and $F$ is  an auxiliary field. 
The vector superfield $\hat{V}^a$ contains the vector boson and fermion field on the superspace,
\beq
\hat{V}^a(x,\theta,\bar{\theta})= - \theta \sigma^{\mu} \bar{\theta} V^a_{\mu}(y) + i \theta \theta \bar{\theta } \bar{\lambda}^a(y) - i \bar{\theta } \bar{\theta}\theta \lambda^a(y) + \frac{1}{2}\theta \theta \bar{\theta } \bar{\theta} \left( D^a(y) - i \partial^{\mu} V^a_{\mu} (y) \right),\label{vector}
\eeq
where $a$ is the index of the generator of the gauge group, $V^a_{\mu}$ is the gauge boson,  $\lambda_a$ is two-components Weyl spinor field (gaugino) and $D_a$ is  an auxiliary field. 
 The spinor chiral superfield $\hat{W}^a_{ \alpha}$  also contains the vector boson and fermion field on the superspace,
 \beq
\hat{W}^a_{ \alpha}(x,\theta,\bar{\theta}) = -i \lambda^a_{ \alpha} (y) + \left( \delta _{\alpha}^{\beta} D^a(y) -\frac{i}{2} (\sigma ^{\mu} \bar{\sigma}^{\nu})_{\alpha}^{\beta} V^a_{ \mu \nu}(y)\right) \theta_{\beta} 
+ \theta \theta \sigma_{\alpha \dot{\alpha}}^{\mu} \partial _{\mu} \bar{\lambda}^{a \dot{\alpha}}(y),\label{spinorchiral}
\eeq
where $V^a_{ \mu \nu} = \partial_{\mu} V^a_{\nu}-\partial_{\nu} V^a_{\mu}$.

The supersymmetric minimal model is called Minimal Supersymmetric Standard Model (MSSM). Table~\ref{MSSMtable} shows all chiral and vector superfields with their components for spin $0$, $1/2$ and $1$, and their representations for SU$(3)_c \times$ SU$(2)_L \times$ U$(1)_Y$  gauge group in the MSSM. 
Here the tilde ($\tilde{}$) represents the SUSY partner of the SM particle.
Note that, although the number of Higgs doublet is one in the SM, we must introduce two Higgs doublets in the SUSY model, 
\begin{eqnarray}
\hat{H_1} = \left( \begin{array}{cc}\hat{H_1^{0}} \\ \hat{H_1^{-}}\end{array} \right) , 
~~~~~~~~\hat{H_2} = \left( \begin{array}{cc}\hat{H_2^{+}}\\ \hat{H_2^{0}}\end{array} \right),
\end{eqnarray}
where $\tilde{H}$ is called Higgsino.
It is because that an existence of the Yukawa interaction with quark/lepton  and  the gauge anomaly cancelation  require two kinds of the Higgs doublets that the hypercharge is opposite.

\begin{table}[t]
 \caption{All chiral and vector superfields with their components for spin $0$, $1/2$ and $1$, and their representations for SU$(3)_c \times$ SU$(2)_L \times$ U$(1)_Y$  gauge group in the MSSM. }
\begin{center}
\begin{tabular}{r|cc|ccc} 
\bhline{1.4pt}
Chiral Supermultiplet & Spin 0 & Spin $\frac{1}{2}$ & $SU(3)_{C}$ & $SU(2)_L $&$ U(1)_Y$ \\ \midrule
Quark-Squark \ \ \ \ \ \  $\hat{Q}$ &$\tilde{Q} = ( \tilde{u}_L, \tilde{d}_L) ^T$ & $Q=(u_L,d_L)^T$ & $\bm{3}$ & $\bm{2}$ & $ \frac{1}{6}$ \\
$\hat{\bar{U}}$ & $\tilde{\bar{u}}_R$ & $u_R^{\dagger} $ & $\bm{3}$ & $\bm{1}$ & $ -\frac{2}{3}$ \\
$\hat{\bar{D}}$ & $\tilde{\bar{d}}_R$ & $d_R^{\dagger} $ & $\bm{3}$ & $\bm{1}$ & $ \frac{1}{3}$ \\ 
Lepton-Slepton \ \ \ \ \   $\hat{L}$ &$\tilde{L} = ( \tilde{\nu}_{e }, \tilde{e}_L) ^T$ & $L=(\nu_e,e_L)^T$ & $\bm{1}$ & $\bm{2}$ & $- \frac{1}{2}$ \\
$\hat{\bar{E}}$ & $\tilde{\bar{e}}_R$ & $e_R^{\dagger} $ & $\bm{1}$ & $\bm{1}$ & $ 1$ \\ 
Higgs-Higgsino \ \ \ \   $\hat{H}_1$ &$H_1 = ( H_1^0, H_1^{-}) ^T$ & $\tilde{H}_1=(\tilde{H}_1^0,\tilde{H}_1^{-})^T$ & $\bm{1}$ & $\bm{2}$ & $ -\frac{1}{2}$ \\
 $\hat{H}_2$ &$H_2 = ( H_2^{+}, H_2^{0}) ^T$ & $\tilde{H}_2=(\tilde{H}_2^{+},\tilde{H}_2^{0})^T$ & $\bm{1}$ & $\bm{2}$ & $ \frac{1}{2}$ \\ \toprule
 \midrule
Vector Supermultiplet & Spin $\frac{1}{2}$ & Spin 1 & $SU(3)_{C}$ & $SU(2)_L $&$ U(1)_Y$ \\ \midrule
Gluon-Gluimo  \ \ \   \ \ \ $V_G$  & $\tilde{g}$ & $G_{\mu}$ & $\bm{8} $ & $\bm{1} $ & 0 \\
W boson-Wino  \ \    \ \ \  $V_W$  & $\tilde{W}^{\pm}, \ \tilde{W}^0$ & $W^{\pm}_{\mu}, \ W^0_{\mu} $& $\bm{1} $ & $\bm{3} $ & 0 \\
B boson-Bino \ \ \    \ \ \  $V_B$ & $\tilde{B}^0$ & $B^0_{\mu}$ & $\bm{1} $ & $\bm{1} $ & 0 \\ \bhline{1.4pt}
\end{tabular}
\end{center}
\label{MSSMtable}
\end{table}

As we will discuss in detail later, 
the Higgs quartic coupling \eq{SMH} and the gauge coupling are related.
In the MSSM, this relation is 
 given as
\beq
\lambda_{\rm quartic} = g^2 (t^a)^2.
\label{naive}
\eeq
Figure~\ref{Figurequad} shows 
the quadratic divergence to the mass of the Higgs boson in the MSSM.
Obviously, the quadratic divergence  by the top quark loop \eq{topquaddiv} is canceled out by the stop loop which are SUSY partners of the top quark. 
Furthermore, 
the relationship \eq{naive} can cancel out the quadratic divergence by the gauge boson, Higgs boson, gaugino and Higgsino.
Actually, all quadratic divergence are always canceled out in the supersymmetric model \cite{Iliopoulos:1974zv,Girardello:1981wz}.
Therefore, the supersymmetry can solve the hierarchy problem.

\paragraph*{}

Other motivation of the supersymmetry is dark matter.
The SM does not include dark matter, which is stable and does not interact with the electromagnetic force. 
In the supersymmetric model, in order to forbid all harmful terms, which break baryon number ($B$) and lepton number ($L$) and thus cause the  proton decay, one should add a new symmetry.
This symmetry is called the $R$ parity \cite{Farrar:1978xj}.
It is defined as 
\beq
P_R = (-1)^{3 (B - L)+ 2s}, \label{Rparity}
\eeq
 where $s $ is the spin and $B = 1/3$ for $\hat{Q}$, $B = -1/3$ for $\hat{\bar{U}},~\hat{\bar{D}}$, $B = 0$ for all others, $L = 1$ for $\hat{L}$, $L = -1 $ for $\hat{\bar{E}}$ and $L = 0$ for all others. 
 Thus, the SM fermions, Higgs bosons and gauge bosons have even $R$ parity ($P_R = +1$), while the squarks, sleptons, Higgsinos and gauginos have odd $R$ parity ($P_R = -1$).
 This symmetry is equal to discrete $\mathbb{Z}_2$ R-symmetry.
 If the $R$ parity is exact symmetry, the lightest supersymmetric particle (LSP) becomes stable.
 Here the supersymmetric particles are defined as the $R$ parity odd ones.
Therefore, when the LSP does not have  the electromagnetic charge, it 
can be a natural candidate of the dark matter.

Another motivation of the supersymmetry is the gauge unification.
In fact, the electromagnetic and  weak nuclear interactions are unified by the electroweak theory above the unification energy $\mathcal{O}(100) \GeV$, it is so-called Glashow-Weinberg-Salam theory \cite{Glashow:1961tr, Weinberg:1967tq}.
In this sense, the electroweak interaction and strong interaction may be unified by grand unification theory (GUT).
In fact, although the gauge couplings can not unify in the SM, in the SUSY model the unification of the gauge couplings can occur \cite{Ellis:1990wk}.
Let us briefly observe this fact.
The one-loop level renormalization group equations (RGEs) for the gauge couplings in the SM are
\beq
\frac{d g'}{d {\rm ln}Q} = \frac{1}{(4 \pi)^2} \frac{41}{6} g'^3,~~~~~
\frac{d g}{d {\rm ln}Q}  =  - \frac{1}{(4 \pi)^2} \frac{19}{6} g^3,~~~~~
 \frac{d g_s}{d  {\rm ln}Q}  =  -\frac{1}{(4 \pi)^2} 7 g_s ^3,
 \label{kri}
\eeq
where $Q$ is the renormalization scale. 
While, the corresponding equations in the MSSM are
\beq
\frac{d g'}{d {\rm ln}Q}  = \frac{1}{(4 \pi)^2}  11 g'^3,~~~~~
\frac{d g}{d {\rm ln}Q} = \frac{1}{(4 \pi)^2}  g^3,~~~~~
\frac{d g_s}{d {\rm ln}Q}  =  -  \frac{1}{(4 \pi)^2}  3 g_s^3.
\label{kri2}
\eeq
Two-loop level RGEs for all couplings are summarized in Appendix~\ref{RGEqpp}.

\begin{figure}[tp]
\begin{center}
\includegraphics[width =7.5cm]{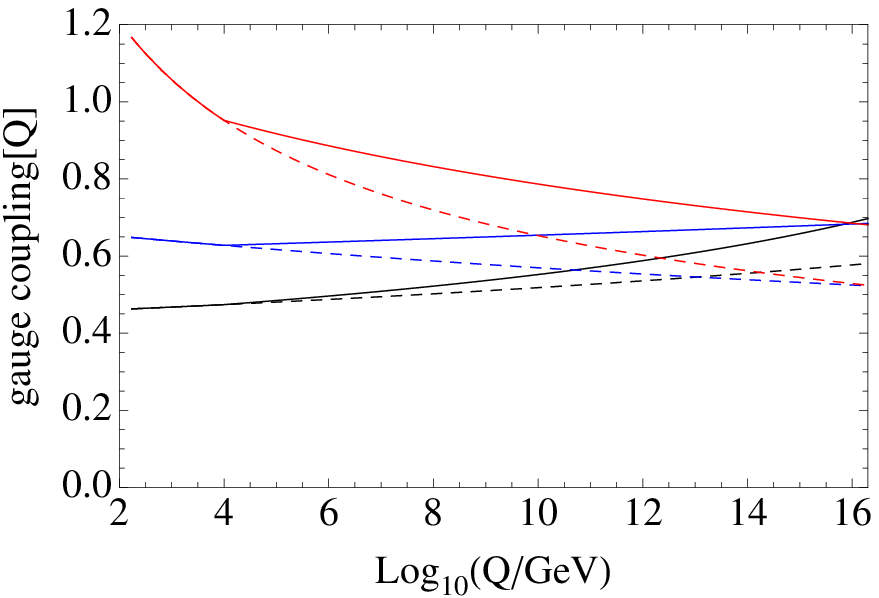}
~~
\includegraphics[width =7.5cm]{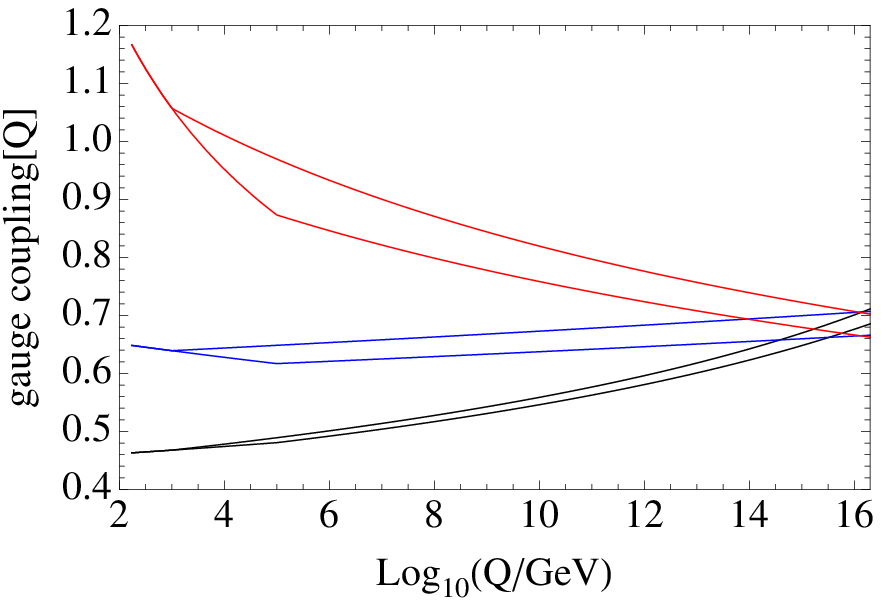}
\caption[Two-loop level running gauge couplings.]{Two-loop level running gauge couplings as a function of the  renormalization scale $Q$. 
The black lines represent $\sqrt{5/3} g'$, the blue lines represent $g$ and the red lines represent $g_s$. 
{\it Left} : The dashed lines correspond the running gauge couplings in the SM, while the solid lines correspond the running gauge couplings in the MSSM with $\susy  = 10\TeV$.
{\it Right} : The running gauge couplings in the MSSM with  $\susy  = 1\TeV$ (upper) and $\susy  = 100\TeV$ (lower). }
  \label{hoge1}
\end{center}
\end{figure}
\begin{figure}
\begin{center}
\includegraphics[width =7.6cm]{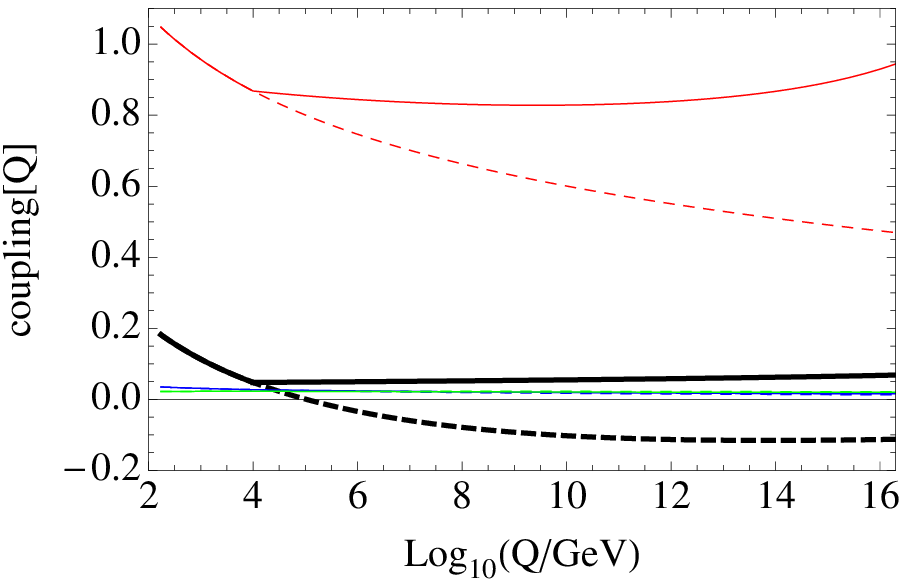}
~~
\includegraphics[width =7.6cm]{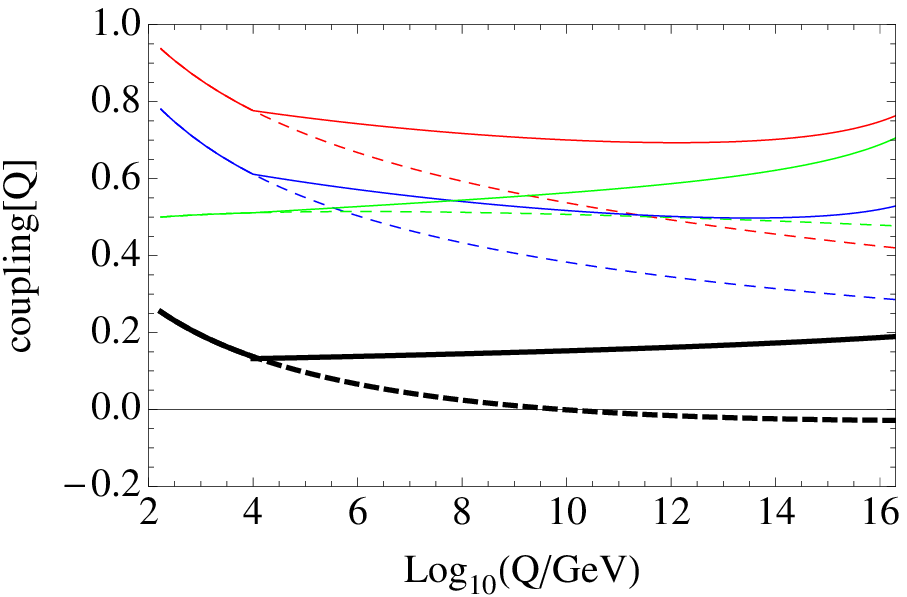}
\caption[Two-loop level running Yukawa couplings and the Higgs quartic coupling.]{Two-loop level running Yukawa couplings ($y_t$: red, $y_b$: blue, $y_{\tau}$: green) and the Higgs quartic coupling ($\lambda_{\rm quartic}$: black) as a function of the  renormalization scale $Q$.
The dashed lines correspond the running couplings in the SM, while the solid lines correspond the running couplings in the MSSM with $\susy  = 10\TeV$.
We take $\tb = 2$ in the left panel, and $\tb = 50$ in the right panel.
}
   \label{hoge2}
\end{center}
\end{figure}

Figure~\ref{hoge1} and \ref{hoge2} show the  two-loop level running couplings as a function of the  renormalization scale $Q$. In Figure~\ref{hoge1}, 
the black lines represent $\sqrt{5/3} g'(Q)$\footnote{The factor $\sqrt{5/3}$ is  a normalization factor for the GUT.}, the blue lines represent $g(Q)$ and the red lines represent $g_s(Q)$. 
In the {\it left} panel, the dashed lines correspond the running gauge couplings in the SM, while the solid lines correspond the running gauge couplings in the MSSM with $\susy  = 10\TeV$, here $\susy$ represents typical SUSY particles mass scale.
As one can see, in the MSSM three gauge couplings are actually unified at GUT sale: $M_{\rm GUT} \simeq 2.0 \times 10^{16} \GeV$, while in the SM it can not achieve. 
This fact gives one of the main motivations 
for the SUSY GUT models \cite{Dimopoulos:1981zb,Sakai:1981gr}.
In the {\it right} panel, we show the running gauge couplings in the MSSM with  $\susy  = 1\TeV$ (upper) and $\susy  = 100\TeV$ (lower). 
We find that when  $\susy  \gtrsim \mathcal{O}(100)\TeV$, the  gauge coupling unification is no longer achieved if one takes universal SUSY particle mass in the MSSM.
Thus, this figure implies that when mass scales of all SUSY particles are the same order, 
the GUT suggests that typical SUSY particles mass scale is $\susy \lesssim \mathcal{O}(100)\TeV$.

\paragraph*{}
 In Figure~\ref{hoge2},  the red lines represent the running top Yukawa coupling $y_t(Q)$, the blue lines represent the running bottom Yukawa coupling  $y_b(Q)$, the green lines represent the running tau Yukawa coupling  $y_{\tau}(Q)$ and the black lines represent the running Higgs quartic coupling $\lambda_{\rm quartic} (Q)$.
The dashed lines correspond the running couplings in the SM, while the solid lines correspond the running couplings in the MSSM with $\susy  = 10\TeV$.
We take $\tb = 2$ in the left panel, and $\tb = 50$ in the right panel. The definition of $\tb$ is given in next section.
The important point in these figures is that the Higgs quartic coupling becomes negative at the high energy scale in the SM. 
If one take $m_h \simeq 125 \GeV$, then  the Higgs quartic coupling becomes negative at $Q \gtrsim 10^{\rm 10-11} \GeV$  \cite{Degrassi:2012ry}.
This situation is avoided in the supersymmetric theory.
It is because that the relation between the quartic coupling and the gauge couplings \eq{naive} assures the stability of the $\lambda_{\rm quartic}$.

\subsection{Lagrangian and the Higgs Boson Mass in the MSSM}
\label{LagMSSM}

Next, we brief review the Lagrangian and the Higgs boson mass of the MSSM  \cite{Martin:1997ns}.

Using the chiral superfield \eq{chiral}, the vector superfield \eq{vector} and the spinor chiral superfield \eq{spinorchiral}, the SUSY invariant Lagrangian of the MSSM is given as follows,
\beq
\mathcal{L}=\int d \theta ^2 d \bar{\theta} ^2 K\left[ \hat{\Phi},\hat{\Phi}^{\dagger},\hat{V}\right]    +\left(\int d \theta ^2 W \left[ \hat{\Phi} \right] + \hc \right) + \left( \frac{1}{4}\int d \theta ^2 \hat{W}^{\alpha}\hat{W}_{\alpha}+ \hc \right), \label{lagMSSM}
\eeq
where $K$ is the following K\"{a}hler potential  
\beq
K\left[ \hat{\Phi},\hat{\Phi}^{\dagger},\hat{V} \right] = \sum_{\hat{\Phi}} \hat{\Phi}^{\dagger} {\rm Exp}(g t^{a} \hat{V}^a) \hat{\Phi},
\eeq
and $W$ is the following superpotential
\beq
W&=&\mu \hat{H}_2 \hat{H}_1+ W_{\text{Yukawa}},\label{25} \\
W_{\text{Yukawa}}&=&  \hat{\bar{U}}\mathbf{y_u} \hat{Q} \hat{H_2} - \hat{\bar{D}}\mathbf{y_d}\hat{Q} \hat{H_1} - \hat{\bar{E}}\mathbf{y_{e}}  \hat{L} \hat{H_1},
\eeq
where $\mu$ is the supersymmetric mass of the Higgs multiplets.
When the supersymmetry is broken, $\mu$ becomes the mass of the Higgsinos. 
While the superpotential $W_{\text{Yukawa}}$  gives the supersymmetric Yukawa couplings of the standard model.
In fact, the Yukawa matrices $\mathbf{y_u}, \mathbf{y_d} $ and $\mathbf{y_e}$ are
given approximately as follows
\beq
\mathbf{y_u} \simeq \begin{pmatrix} 0&0&0\\0&0&0\\0&0&y_t \end{pmatrix},~~~
\mathbf{y_d} \simeq \begin{pmatrix} 0&0&0\\0&0&0\\0&0&y_b \end{pmatrix},~~~ 
\mathbf{y_e} \simeq \begin{pmatrix} 0&0&0\\0&0&0\\0&0&y_{\tau} \end{pmatrix}.
\eeq
Here and the following, we use these approximations of Yukawa matrices.
Then, the superpotential becomes 
\beq
W_{\text{Yukawa}}& =  &  y_t \hat{\bar{t}}_R \hat{Q}  \hat{H_2} - y_b \hat{\bar{b}}_R \hat{Q}   \hat{H_1} - y_{\tau} \hat{\bar{\tau}}_R \hat{L} \hat{H_1} \non
&=&y_t(\hat{\bar{t}}_R  \hat{t}_L  \hat{H}^0_2 -  \hat{\bar{t}}_R \hat{b}_L \hat{H}^{+}_2) + y_b( \hat{\bar{b}}_R  \hat{b}_L \hat{H}^{0}_1  -  \hat{\bar{b}}_R \hat{t}_L \hat{H}^{-}_1 ) \non
&&  + y_{\tau}(  \hat{\bar{\tau}}_R \hat{\tau}_L  \hat{H}^{0}_1 -  \hat{\bar{\tau}}_R \hat{\nu}_{\tau} \hat{H}^{-}_1 ).
\label{yukawasp}
\eeq
Finally, as the spinor chiral superfields $\hat{W}^{\alpha}$ we take the U(1)$_Y$ gauge multiplets (the B boson and the bino),  the SU(2)$_L$ gauge multiplets (the W bosons and the winos)  and the SU(3)$_c$ gauge multiplets (the gluons and the gluinos).

Because the supersymmetry is broken in nature, one has to also introduce SUSY breaking terms  in Lagrangian. 
The supersymmetry must be broken {\it softly} in order to ensure the cancelation of the  quadratic divergences.
Generally, the soft SUSY breaking terms are gaugino masses, scalar masse, and trilinear coupling terms for scalars.
In the MSSM, the soft SUSY breaking Lagrangian is given as
\beq
\mathcal{L}_{\text{soft}} =  - V_{\text{soft gaugino}}  - V_{\text{soft Yukawa}} - V_{\text{soft Higgs}},
\label{soft}
\eeq
with
\beq
V_{\text{soft gaugino}}&=& \frac{1}{2} \left(  M_1 \tilde{B}\tilde{B}+ M_2 \tilde{W} \tilde{W}+  M_3 \tilde{g} \tilde{g} \right)+ \hc, \label{gauginomass}
\\
V_{\text{soft Yukawa}}&=&m^2_{{Q}} \tilde{Q}^{\dagger}\tilde{Q} + m^2_{{{U}}}\tilde{\bar{U}}^{\ast}\tilde{\bar{U}}+ m^2_{{{D}}}\tilde{\bar{D}}^{\ast}\tilde{\bar{D}}+m^2_{{L}} \tilde{L}^{\dagger}\tilde{L} + m^2_{{{E}}}\tilde{\bar{E}}^{\ast}\tilde{\bar{E}} \non
& &+ \left( \tilde{\bar{U}}  \mathbf{y_u}  \mathbf{A_u}\tilde{Q} \cdot H_2 - \tilde{\bar{D}}\mathbf{y_d} \mathbf{A_d} \tilde{Q}\cdot H_1 - \tilde{\bar{E}}\mathbf{y_{e}} \mathbf{A_{e}}  \tilde{L} \cdot H_1  + \hc  \right),\label{softyukawa}
\\
V_{\text{soft Higgs}}&=& m_{1}^2|H_1|^2+m_{2}^2 |H_2|^2+ \frac{1}{2}(  B \mu H_2  H_1 + \hc ),\label{Higgspo} 
\eeq
where $M_i$ are the gaugino masses, $m^2_{Q/U/D/L/E}$ are the squark/slepton masses, $m^2_{1/2}$ are soft SUSY breaking Higgs mass, $A_i$ are called $A$ terms and $B$ is called $B$ terms.
Note that $m^2_{\tilde{i}}$ are  flavor generation mixing $3\times 3 $ matrix generally. 
Without additional assumptions, these off-diagonal  squark/slepton masses are not suppressed.
Adding the soft SUSY breaking Lagrangian \eq{soft} into the SUSY invariant Lagrangian \eq{lagMSSM}, one can obtain the full Lagrangian of the MSSM.

\paragraph*{}

~

Now, let us see the  Higgs sector and the mass of the  neutral Higgs boron of the MSSM.
The tree-level Higgs scalar potential is given directly from Eqs.~(\ref{soft},~\ref{lagMSSM}) as  

\beq
V_{\rm Higgs} &=& \frac{g^2}{2}\left| H_1^{\dagger} H_2 \right|^2+ \frac{g^2 + g^{\prime 2}}{8}\left( |H_1|^2 - |H_2|^2 \right)^2 +\left( |\mu|^2 + m_{1}^2\right)|H_1|^2  + \left(|\mu|^2 + m_{2}^2\right) |H_2|^2  \nonumber
\\ & &  + \frac{1}{2}( B \mu H_2  H_1 + \hc ). \label{MSSMp}
\eeq
Using the freedom of SU$(2)$ rotations, one can always choose $\langle H_2^+ \rangle = 0$. Then, a potential minimization condition $\partial V/\partial H_2^+ =0$ leads to  $\langle H_1^- \rangle = 0$. 
Thus, scalar potential of the neutral Higgs boson is given as
\beq
V_{\rm Higgs}&=&  \frac{g^2 + g^{\prime 2}}{8}\left( | H_1^0|^2 - |H_2^0|^2\right)^2 +\left(|\mu|^2 + m_{1}^2 \right)|H_1^0|^2  + \left( |\mu|^2 + m_{2}^2\right)|H_2^0|^2  \nonumber
\\ & &  - \frac{1}{2}\left(  B \mu H_2^0 H_1^0 + \hc \right). \label{VN2}
\eeq

The potential minimization conditions $\partial V/\partial H_1^0 =0$ and $\partial V/\partial H_2^0 =0$ leads to
\beq
(m_1^2 + |\mu|^2 )v_1^{\ast} + \frac{g^2 + g^{\prime 2}}{4}(|v_1|^2 - |v_2|^2)v_1^{\ast}-\frac{1}{2}B \mu v_2 &=&0,\label{224} \\
(m_2^2 + |\mu|^2 )v_2^{\ast} - \frac{g^2 + g^{\prime 2}}{4}(|v_1|^2 - |v_2|^2)v_2^{\ast}-\frac{1}{2}B \mu v_1 &=&0,\label{225}
\eeq
where $v_{1/2}$ is the dev of $H_{1/2}^0$.
If the determinant of $\partial^2 V/(\partial H_i^0 \partial H_j^0)$ is negative, one linear combination of $H_1^0$ and $H_2^0$ has a negative square mass, then $H_1^0$ and $H_2^0$ obtain nonzero vevs.
This condition at the vicinity of $H_1^0 = H_2^0 =0$ is
\beq
B^2 \mu^2 > 4 (m_1^2 + |\mu|^2 )(m_2^2 + |\mu|^2 ).
\eeq
In this case, using a redefinition of the phase of $H_1^0$ and $H_2^0$, one can always choose that $B\mu$, $v_1$ and $v_2$  are real and positive.
Let us defined the ratio of the Higgs vev as
\beq
\tan \beta &\equiv &\frac{v_2}{v_1},\label{tan} \\
v_{EW} =\sqrt{ v_1^2+v_2^2} &  =  &174.1\text{ GeV}. \label{174} 
\eeq
Then, the potential minimization conditions Eqs.~(\ref{224},~\ref{225}) become
\beq
m_1^2 &=& - |\mu|^2 + \frac{1}{2}M_A^2 -\frac{1}{2}\left( M_A^2+M_Z^2 \right) \cos 2 \beta,
\label{minimalcon1}
\\
m_2^2 &=&- |\mu|^2 + \frac{1}{2}M_A^2 +\frac{1}{2}\left( M_A^2+M_Z^2\right) \cos 2 \beta,
\label{minimalcon2}
\eeq
where
\beq
M_A^2 =  2 |\mu|^2 + m_1^2 +m_2^2.    \label{MAA}
\eeq
The vacuum fluctuations of the Higgs field are defined as
\beq
H_1^0 &=& v_1 + \frac{1}{\sqrt{2}} ( H_{1R}+i H_{1I} ), \non
H_2^0 &=& v_2 + \frac{1}{\sqrt{2}} ( H_{2R}+i H_{2I} ).
\eeq

Now, one can obtain the following mass matrix for the CP even Higgs bosons, 
\beq
-\mathcal{L}_{\text{CP-even}}= 
\frac{1}{2}\begin{pmatrix} H_{1 R} H_{2 R}\end{pmatrix} \begin{pmatrix}M_A^2 \sin^2 \beta + M_Z^2 \cos^2 \beta  &-\frac{1}{2}(M_A^2+M_Z^2)\sin 2 \beta \\
-\frac{1}{2}(M_A^2+M_Z^2)\sin 2 \beta  & M_A^2 \cos^2 \beta + M_Z^2 \sin^2 \beta  \end{pmatrix}
\begin{pmatrix}H_{1 R}\\ H_{2 R}\end{pmatrix}.\label{222}
\eeq
Diagonalizing this mass matrix, the following mass eigenvalues are given\footnote{Similarly to the CP even Higgs boson mass, the tree-level mass of the CP-odd Higgs boson $A$  and charged Higgs boson $ H^{\pm}$  are given as 
\beq
M_A^2 &= & 2 |\mu|^2 + m_1^2 +m_2^2 = \frac{B\mu}{\sin  2 \beta},\\
m_{H^{\pm}}^2 & = & M_A^2 + M_W^2.\label{MSSMcharged}
\eeq } 
\beq
m_H^2&=& \frac{1}{2}\left( M_A^2 + M_Z^2 + \sqrt{ (M_A^2 -M_Z^2)^2 + 4 M_A^2 M_Z^2 \sin ^2 2 \beta} \right),\\
m_h^2&=& \frac{1}{2}\left( M_A^2 + M_Z^2 - \sqrt{ (M_A^2 -M_Z^2)^2 + 4 M_A^2 M_Z^2 \sin ^2 2 \beta} \right), \label{243}
\eeq
here we call a eigenstate, which has the heaviest/lightest mass eigenvalue, $H$/$h$.
In the decoupling limit $M_A^2 \gg M_Z^2$, the eigenstate $h $ becomes the SM Higgs boson perfectly. It is because that the couplings of $h$ with SM particles are equal to the one of the SM Higgs boson with SM particles in this limit. 
Here and the following, we call this  eigenstate $h $ SM-like Higgs boson.
Thus, one can recognize \eq{243} as the tree-level mass of the SM-like Higgs boson in the MSSM.

In fact,  the tree-level mass of the SM-like Higgs boson \eq{243} has an upper bound,
\beq
m_h^2 \leq  M_Z^2  \cos ^2 2 \beta.\label{mhh}
\eeq
In the SUSY model, in order to cancel out the quadratic divergence in the radiative corrections, the Higgs quartic couplings must be related with the gauge couplings (see Figure~\ref{Figurequad} and \eq{naive}).
In addition, the Higgs quartic couplings decide the mass of the SM-like Higgs boson (see \eq{SMmass}).
These are the reason why there are the upper bound on the Higgs boson mass in the SUSY model.
Thus, the SM-like Higgs boson mass must be lighter than the Z boson mass  at tree-level in the MSSM. 
However, this mass bound is too severe to explain an observed SM Higgs boson mass, $m_h \simeq 125 \GeV$ \cite{Aad:2012tfa,Chatrchyan:2012ufa}.

\paragraph*{Radiative corrections to the Higgs boson mass}

~ 

On the contrary, in 1990, several groups found that radiative corrections can contribute  to the mass of the Higgs boson significantly  \cite{Okada:1990vk,Ellis:1990nz,  Haber:1990aw, Okada:1990gg,Ellis:1991zd,Drees:1991mx} \footnote{At that time, the Large Electron-Positron Collider (LEP) had not  excluded the Higgs boson mass  up to the Z boson mass region \cite{Drees:1990fs}.
}.
The dominant radiative correction comes from the top/stop loop diagrams, which are depicted in Figure~\ref{Higgsmasstopstop}.
\begin{figure}[t]
\begin{center} 
\includegraphics[width=16 cm]{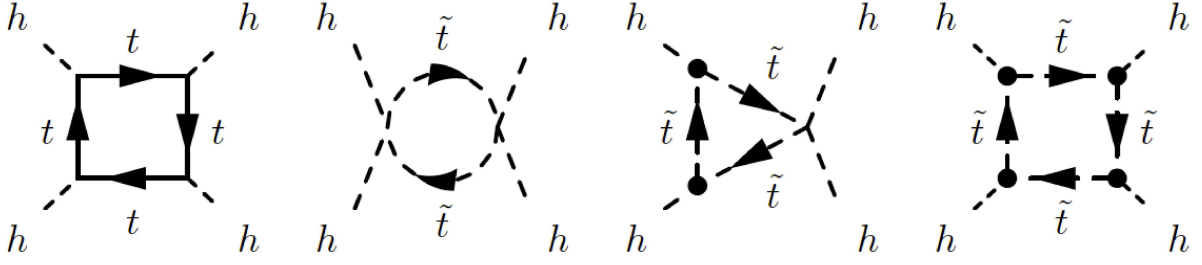}
 \caption[The Feynman diagrams for the dominant one-loop radiative corrections to the Higgs boson mass.]{The Feynman diagrams for the dominant one-loop radiative corrections to the Higgs boson mass. Here, the bullet $\bullet$ represents $A_{t}$ term interaction.}
\label{Higgsmasstopstop}
 \end{center} 
\end{figure}

The top/stop one-loop effective potential (the Coleman-Weinberg potential \cite{Coleman:1973jx}) $V_{\rm CW}^{\rm t}$  is given as,
\beq
V_{\rm CW}^{\rm t} 
&=& \frac{3}{32 \pi^2} 
\biggl[ m_{\tilde{t}_1}^4 \left( \ln \left(\frac{m_{\tilde{t}_1}^2}{Q^2}\right) -\frac{3}{2}\right)+m_{\tilde{t}_2}^4 \left( \ln \left(\frac{m_{\tilde{t}_2}^2}{Q^2}\right) -\frac{3}{2}\right)\non
&&
~~~~~~- 2  m_t^4 \left( \ln \left(\frac{m_t^2}{Q^2}\right) -\frac{3}{2}\right) 
\biggl]\,,
\label{topstopCW}
\eeq
where the top quark mass is $m^2_t = y_t^2 |H_2^0|^2$ and $Q$ is the renormalization scale. 
 The stop masses $m_{\tilde{t}_1}$ (lighter) and $m_{\tilde{t}_2}$ (heavier) are given as the eigenvalues of the following stop mass matrix,
\beq
-\mathcal{L}_{\text{stop~mass}}= 
\begin{pmatrix} \tilde{t}_{L}^{\ast} \tilde{t}_{R}^{\ast} \end{pmatrix} \begin{pmatrix}M_{\tilde{t}_{11}}^2  &M_{\tilde{t}_{12}}^2 \\
M_{\tilde{t}_{21}}^2& M_{\tilde{t}_{22}}^2  \end{pmatrix}
\begin{pmatrix} \tilde{t}_{L}\\ \tilde{t}_{R}\end{pmatrix}, 
\eeq
with
\beq
M_{\tilde{t}_{11}}^2 & =& m_{Q}^2 +  y_t^2 |H_2^0|^2 + \left( \fr{2}{3} \fr{M_W^2}{v_{EW}^2} - \fr{1}{6} \fr{M_Z^2}{v_{EW}^2} \right) \left( |H_1^0|^2 - |H_2^0|^2 \right),\non
M_{\tilde{t}_{22}}^2  &=& m_{U}^2 +  y_t^2 |H_2^0|^2 + \left(- \fr{2}{3} \fr{M_W^2}{v_{EW}^2} + \fr{2}{3} \fr{M_Z^2}{v_{EW}^2} \right) \left( |H_1^0|^2 - |H_2^0|^2 \right),\non
M_{\tilde{t}_{12}}^2  &=& ( M_{\tilde{t}_{12}}^2)^{\ast} \non
& = & y_t  \left(A_t^{\ast} (H_2^0)^{\ast} - \mu H_1^0 \right).
\eeq

This effective potential gives the following one-loop corrections to the CP even Higgs mass matrix \eq{222},

\begin{eqnarray}
\Delta \mathcal{M}^{2}_{11}&=&\frac{3}{8 \pi ^2}y_t^4 v_2^2 \mu ^2 R_t^2 (m^2_{\tilde{t}_1} - m^2_{\tilde{t}_2} ) ^2 g(m^2_{\tilde{t}_1},m^2_{\tilde{t}_2} ) \nonumber \\
&+&\frac{3}{16 \pi ^2}y_t^2A_t \mu \tan \beta f(Q^2, m^2_{\tilde{t}_1},m^2_{\tilde{t}_2} ) ,\label{quark4} \\
\Delta \mathcal{M}^{2}_{22}&=&\frac{3}{8 \pi ^2}y_t^4 v_2^2 \biggl(  \ln{\frac{ m^2_{\tilde{t}_1 } m^2_{\tilde{t}_2}}  {m_t^4 } } +2 A_t R_t \ln{\frac{m^2_{\tilde{t}_1}} {m^2_{\tilde{t}_2} }  } +A_t^2 R_t^2 ( m^2_{\tilde{t}_1} - m^2_{\tilde{t}_2} ) ^2 g(m^2_{\tilde{t}_1},m^2_{\tilde{t}_2} ) \biggl) \nonumber \\
&+&\frac{3}{16 \pi ^2}y_t^2A_t \mu \cot \beta  f(Q^2, m^2_{\tilde{t}_1},m^2_{\tilde{t}_2} ) ,\\
\Delta \mathcal{M}^{2}_{12}&=& - \frac{3}{8 \pi ^2}y_t^4 v_2^2 \mu  R_t \biggl(  \ln{  \frac{ m^2_{\tilde{t}_1}} {m^2_{\tilde{t}_2}}  } +A_t R_t (m^2_{\tilde{t}_1} - m^2_{\tilde{t}_2} ) ^2 g(m^2_{\tilde{t}_1},m^2_{\tilde{t}_2} )  \biggl) \nonumber \\
&-&\frac{3}{16 \pi ^2}y_t^2A_t \mu  f(Q^2, m^2_{\tilde{t}_1},m^2_{\tilde{t}_2} ) \label{quark5},
\end{eqnarray}
with
$R_t  = (A_t -\mu \cot \beta )/(m^2_{\tilde{t}_1} - m^2_{\tilde{t}_2})$, here we assume $A_t$ and $\mu$ to be real for simplisity. 
The loop functions $f$ and $g$ are given in Appendix~\ref{appfunctions}.
These corrections give the following upper bound on the SM-like Higgs boson mass,
\beq
m_h^2&\leq & M_Z^2 \cos^2 2 \beta + \frac{3}{8 \pi^2} y_t^4 v^2 \sin^2 \beta  \biggl(\ln{\frac{m^2_{\tilde{t}_1} m^2_{\tilde{t}_2}}{m_t^4}}  + 2  \frac{(A_t - \mu \cot \beta)^2} {m^2_{\tilde{t}_1}- m^2_{\tilde{t}_2}} \ln{\frac{m^2_{\tilde{t}_1}}{m^2_{\tilde{t}_2}}} \non
& & \ \ \ \ \ \ \  \ \ \ \ \  \ \ \ \ \  \ \ \ \ \  \ + \frac{(A_t - \mu \cot \beta)^4}{(m^2_{\tilde{t}_1}- m^2_{\tilde{t}_2})^2 } \left( 2- \frac{m^2_{\tilde{t}_1}+ m^2_{\tilde{t}_2}}{m^2_{\tilde{t}_1}- m^2_{\tilde{t}_2}} \ln{\frac{m^2_{\tilde{t}_1}}{m^2_{\tilde{t}_2}}}   \right) \biggl).\label{259}
\eeq
If we take the universal soft SUSY breaking stop masses, 
\beq
 m^2_{{Q}}=m^2_{{{U}}}=m^2_{\tilde{q}},
\eeq
the stop masses are written down easily as follows,
\beq
m^2_{\tilde{t}_{1/2}}\sim   m^2_{\tilde{q}} + m_t^2 \mp m_t (A_t - \mu \cot \beta).\label{mq}
\eeq
Then, the upper bound \eq{259} becomes 
\beq
m_h^2 \lesssim M_Z^2 \cos^2 2 \beta +  \frac{3}{4 \pi^2} y_t^4 v^2 \sin^2 \beta \biggl(  \ln{\frac{m^2_{\tilde{q}} }{m_t^2}} + \frac{(A_t-\mu \cot \beta)^2}{m^2_{\tilde{q}}}- \frac{1}{12}\frac{(A_t - \mu \cot \beta)^4}{m^4_{\tilde{q}} } \biggl).\label{yanagida}
\eeq
Therefore, the radiative corrections can give significant contributions to the mass of the SM-like Higgs boson.
Note that the logarithmic term is given by the first and second diagram from the left in the  Figure~\ref{Higgsmasstopstop}. 
As one can see, when the stop mass is heavy, this term generates large correction.
While, the last two terms are given by the  first and second diagram from the right in the  Figure~\ref{Higgsmasstopstop}, where the bullet represents the $A_t$ term interaction. 
A notable point is that these contributions are maximized by
\beq
A_t - \mu \cot \beta = \pm \sqrt{6} m_{\tilde{q}}.
\label{sqrt6}
\eeq
Then these contributions can be comparable to the logarithmic one.

\section{Current status of the MSSM}
\label{2no3}
In this section, we will review the current situation of the MSSM in terms of an observed $125 \GeV$ Higgs boson and the constraints from the flavor violation and CP violation process.

\subsection{125~GeV}
\label{125gev}

On Wednesday, July 4 2012, the ATLAS and CMS collaborations at the LHC experiment declared  
astonishing announcements that they had observed a new particle which is consistent with the SM Higgs boson \cite{Aad:2012tfa,Chatrchyan:2012ufa}.
A mass of the new particle had been around $126\GeV$.
The latest measured value of the Higgs boson mass is
\beq
m_h &=& 125.36 \pm 0.37 ~({\rm stat.}) \pm  0.18 ~({\rm syst.}) \GeV~~~~~({\rm ATLAS})~ \textrm{\cite{Aad:2014aba}},  \\
m_h &=& 125.03  ~{}^{+0.26}_{-0.27} ~({\rm stat.})   ~{}^{+0.13}_{-0.15} ~({\rm syst.}) \GeV~~~~~~~~~({\rm CMS})~ \textrm{\cite{CMS:2014ega}},
\eeq
where  correspond to integrated luminosities are 
$ {\rm ~4.5 ~fb}^{-1} {\rm ~at~} 7 {\rm ~TeV ~and  ~20.3 ~fb}^{-1} {\rm  ~at ~8 ~TeV}$ (ATLAS) and
 $ {\rm ~5.1 ~fb}^{-1} {\rm ~at~} 7 {\rm ~TeV ~and  ~19.7 ~fb}^{-1} {\rm  ~at ~8 ~TeV}$ (CMS).
 A na\"{\i}ve average of the ATLAS and CMS results is $125.15 \pm 0.25 \GeV$ \cite{Bagnaschi:2014rsa}.
 
 As we discussed in the previous section, such a Higgs boson mass can {\it not} be realized in the tree-level estimation of the MSSM.
However, the one-loop radiative corrections  can actually raise the mass of the  SM-like Higgs boson, and so it can be realized.
Therefore, considering the radiative corrections to the Higgs boson mass is important and essential in SUSY models.

Although the one-loop order radiative corrections can contribute significantly to the Higgs boson mass, in fact two-loop order  radiative corrections give a negative contribution to the one and it is not a negligible contribution \cite{Hempfling:1993qq,Heinemeyer:1998kz,Zhang:1998bm,Espinosa:2000df,Degrassi:2001yf, Martin:2002wn}.
It is because that the QCD corrections first appear in two-loop order diagrams, and they give opposite contributions to the SM-like Higgs boson (cf. see the last term of second line of the two-loop RGE for $\lambda_{\rm quartic}$ \eq{lq2loop}).
Therefore, in order to predict a reliable Higgs boson mass in SUSY models, 
one should take the two-loop order radiative corrections to the Higgs boson mass into account.

It is known that there are three ways to achieve the $125 \GeV$ SM-like Higgs boson  considering the two-loop order radiative corrections.
First way is the heavy stop scenario. 
When stop masses are about $10\TeV$, the logarithmic corrections which are generated by the stop loop become an appropriate magnitude.
Second way is the large stop mixing scenario, namely the large $A_t$ term scenario \eq{sqrt6}. 
In this way, even when the stop masses are about $1\TeV$ (and $A_t \sim 2.5 \TeV$), the radiative corrections become an appropriate magnitude \cite{Arbey:2011ab}.
Third way is an extended models of the MSSM \cite{Fayet:1974pd}. 
The appropriate extension models which can explain the observed Higgs boson mass are 
singlet extension models, vector-like matter extended models \cite{Moroi:1991mg,Moroi:1992zk}, U$(1)$ gauge extended models \cite{Han:2004yd}, etc. 
In the singlet extension models, an additional $F$-terms can raise the tree-level mass of the Higgs boson\footnote{If the additional CP-even singlet scalar is lighter than the Higgs boson,  the singlet-doublet mixing can raise the
mass of the Higgs boson \cite{Jeong:2014xaa}.}.
In the vector-like matter extended models, similarly to the stop loop, the vector-like particle loop gives the sizable radiative corrections to the Higgs boson mass. 
In the U$(1)$ gauge extended models, an additional $D$-terms can raise the tree-level mass of the Higgs boson.
Especially,  we will focus on a singlet extension model in this thesis.

In the following two sub sections, we will review the two-loop and higher-loop order analysis of   the Higgs boson mass in the MSSM.
It is because that we have applied this two-loop order analysis of the Higgs boson mass to a singlet extension model.

\subsubsection{Two-loop level analysis of the Higgs boson mass}
\label{HiggsmasscalcRGE}

Inclusion of  the two-loop level radiative corrections is important and indispensable in the calculation of the mass of the Higgs boson in supersymmetric models.  
In this section, 
we first review the two-loop level calculation of the mass of the Higgs boson using the RGE \cite{Giudice:2011cg, Draper:2013oza,Bagnaschi:2014rsa}. 
Then, we will show a behavior of the mass of the Higgs boson as a function of SUSY breaking scale and $\tb$ in the MSSM.

Let us assume $v_{\rm EW} \ll M_{\rm SUSY}$ and $M_{\rm gaugino} \sim \mu \sim \sqrt{m^2_0} =\mathcal{O}(M_{\rm SUSY})$ for simplicity, where $m^2_0$ represents dimension two soft SUSY breaking mass term of Higgs and sfermion, namely 
all the sfermions, heavy Higgs doublet $A$, Higgsinos and gauginos are integrated out at the scale $\susy$. 
This assumption of the mass spectrum represents the effective theory below SUSY breaking scale $M_{\rm SUSY}$ to be the SM.
In this section, the SM-like (surviving) Higgs doublet $\Phi_h$ and heavy Higgs doublet $\Phi_H$ are defined as 
\beq
\left( \begin{array}{cc}\Phi_h \\  \Phi_H \end{array} \right) = 
\left( \begin{array}{cc} \cos\beta &\sin\beta \\ - \sin\beta & \cos\beta \end{array} \right)
\matl{cc} -\epsilon H_d^{\ast} \\ H_u \matr,
\eeq
where $\epsilon$ is the antisymmetric tensor $\epsilon_{12}=1$.
In the component representation, this equation is the same as follows, 
\beq
\Phi_h &=& \cb \matl{cc} - (H_1^{-})^{\ast} \\ (H_1^0)^{\ast} \matr + \sb \matl{cc} H_2^+ \\ H_2^0\matr, \\
\Phi_H &=& -\sb \matl{cc} - (H_1^{-})^{\ast} \\ (H_1^0)^{\ast} \matr + \cb \matl{cc} H_2^+ \\ H_2^0\matr.
\eeq
The potential of the SM-like Higgs $\Phi_h$ below SUSY breaking scale can be given by 
\beq
V(\Phi_h)  =  \frac{\lambda_{\rm quartic}}{2} (\Phi_h^{\dag} \Phi_h - v_{EW}^2)^2.
\eeq
This potential becomes as follows when it is  expanded by vacuum fluctuation of the Higgs boson,
$\Phi_h^{0} = v_{EW} + \frac{1}{\sqrt{2}} ( h  + i G^0)$, 
\beq
V(h) = 2 v_{EW}^2 \lambda_{\rm quartic} \left( \frac{h}{\sqrt{2}} \right)^2 + 2 v_{EW} \lambda_{\rm quartic} \left( \frac{h}{\sqrt{2}} \right)^3 + \frac{\lambda_{\rm quartic} }{2} \left( \frac{h}{\sqrt{2}} \right)^4.
\eeq
Namely the  tree-level mass of the SM Higgs scalar $h$ is 
\beq
m_h^2 = 2 v_{EW}^2 \lambda_{\rm quartic}.
\label{treeHiggs}
\eeq

In order to derive the physical mass of the SM Higgs scalar $h$ from SUSY breaking scale, 
we connect the two scale, that is SUSY breaking and electroweak scale, 
using the two-loop RGE for the Higgs quartic coupling.
The full set of two-loop RGEs for the coupling constants of the SM  using $\overline{MS}$ regularization are presented in appendix \ref{RGESM}.
Then, we impose  mating conditions of the couplings for the RGE at the SUSY breaking (high) and electroweak (weak) scale.
Note that we estimate the mass of the Higgs boson at two-loop level (next-to-leading order), we need to include one-loop threshold corrections in these matching conditions.

\paragraph{Matching at high scale}
~

Because supersymmetry ensures the relationship among the dimensionless coupling constants,  
the Higgs quartic coupling must satisfy the following matching condition at SUSY breaking scale,  
\beq
\lambda_{\rm quartic} (\susy) = \lambda_{\rm LO}(\susy) \equiv  \frac{1}{4}\left( g^2(\susy)  + g'^2(\susy)\right) \cos^2 2 \beta.
\label{treematch}
\eeq
Now, in order to accurately calculate the mass of the Higgs boson, we need the matching condition including  next-to-leading order corrections.
The matching condition including 
one-loop level  threshold corrections is given as follows \cite{Giudice:2011cg, Bagnaschi:2014rsa}
\beq
\lambda_{\rm quartic} (\susy) = \lambda_{\rm LO}(\susy) + \frac{1}{(4\pi)^2}  \lambda_{\rm NLO}(\susy),
\label{matchhigh}
\eeq
where
\beq
 \lambda_{\rm NLO}=   \lambda_{\rm NLO}^{\rm reg} + \lambda_{\rm NLO}^{\phi} + \lambda_{\rm NLO}^{\chi^1} + \lambda_{\rm NLO}^{\chi^2}.
 \label{NLOmatch}
\eeq

The First term in the right hand $\lambda_{\rm NLO}^{\rm reg}$ is a convention factor from $\overline{MS}$ to $\overline{DR}$ regularization scheme, which gives a correction to the tree-level relation of  Eq.~(\ref{treematch}) even in SUSY limit,
\beq
\lambda_{\rm NLO}^{\rm reg}
 &=& -\left[ \frac{1}{4}g'^4 + \frac{1}{2}g'^2 g^2 +     (\frac{3}{4} - \frac{\cos^2 2\beta}{6}) g^4 \right].
\eeq
The other terms, $ \lambda_{\rm NLO}^{\phi}, \lambda_{\rm NLO}^{\chi^1}, \lambda_{\rm NLO}^{\chi^2}$ are computed using the $\overline{DR}$ regularization scheme.

The second term $\lambda_{\rm NLO}^{\phi} $ is obtained when one integrate out the heavy Higgs multiplet and sfermion at the matching scale.
Neglecting all Yukawa coupling except the top quark $y_t$, this term is given as follows, 
\beq
 \lambda_{\rm NLO}^{\phi}  &=&  3 y_t^2 \left[  y_t^2  + \frac{1}{2}(g^2 - \frac{1}{3}g'^2 ) \cos 2 \beta \right] \ln \frac{m_Q^2}{\susy^2}+3 y_t^2 \left[ y_t^2 + \frac{2}{3} g'^2 \cos 2 \beta \right] \ln\frac{m_U^2}{\susy^2} \non
 & &+ 3 y_t^4 \left[ 2 X_t^2 \tilde{F}(\frac{m_Q}{m_U})-\frac{X_t^4}{6} \tilde{G}(\frac{m_Q}{m_U})\right]
 + \frac{3}{4} y_t^2 X_t^2 \left[g'^2 \tilde{H}_1(\frac{m_Q}{m_U} +  g^2 \tilde{H}_2(\frac{m_Q}{m_U})  )\right] \cos 2 \beta\non
 & &- \frac{y_t^2}{4} X_t^2 \cos^2 2 \beta (g'^2 + g^2) \tilde{H}(\frac{m_Q}{m_U}) - \frac{3}{16}(g'^2 + g^2)\sin^2 4 \beta\non
 & &+ \frac{1}{192}\left[29 g'^4 +42 g'^2 g^2 + 53 g^4 - 4 \cos 4 \beta( g'^4 + 6 g'^2 g^2 + 7g^4) \right.\non
 & & \left. - 9 \cos 8 \beta (g'^2 + g^2)^2 \right]\ln\frac{m_A^2}{\susy^2} + \frac{\cos^2 2 \beta}{4}\left[ 2 g'^4 \ln\frac{m_E^2}{\susy^2} \right.\non
 & & +  \frac{8}{3}g'^4 \ln\frac{m_U^2}{\susy^2} + \frac{2}{3} g'^4 \ln\frac{m_D^2}{\susy^2}\non
 & & \left.+ \frac{1}{3}(g'^4 + 9 g^4)\ln\frac{m_Q^2}{\susy^2} + (g'^4 + g^4)\ln\frac{m_L^2}{\susy^2}\right],
 \label{lambdaNLOphi}
\eeq
where loop functions $\tilde{F}, \tilde{G}, \tilde{H}, \tilde{H}_1, \tilde{H}_2$ are defined in appendix \ref{appfunctions}, and they are normalized such that $\tilde{F}(1) = \tilde{G}(1) =\tilde{H}(1) = \tilde{H}_1(1)= \tilde{H}_2(1)=1$.
The stop mixing parameter $X_t$ is defined by
\beq
X_t = \frac{A_t - \mu \cot \beta}{\sqrt{m_Q m_U}}.
\eeq
When the all masses of the heavy Higgs multiplet and sfermons are the same as SUSY breaking scale, $\lambda_{\rm NLO}^{\phi} $ becomes as follows 
\beq
\lambda_{\rm NLO}^{\phi}   &=& 6 y_t^4 \left[ X_t^2 - \frac{X_t^4}{12} \right]
 + \frac{y_t^2}{4} X_t^2 \left( g'^2 + g^2 \right) (3 - \cos 2\beta ) \cos 2 \beta \non
& &  -  \frac{3}{16}(g'^2 + g^2)\sin^2 4 \beta.
\eeq
Here dominant term is the first one, and it reproduces the last two terms of the one-loop corrections Eq.~(\ref{yanagida}).
As we discussed before,  this $y_t^4$ correction is maximized when $X_t \simeq \sqrt{6}$ \eq{sqrt6}.
Therefore, the largest threshold correction comes from the mixing of the stops, 
\beq
\lambda_{\rm NLO,~max}^{\phi} \simeq 18 y_t^4.
\label{18}
\eeq
Note that the contribution to the mass of the Higgs boson from this threshold correction (\ref{18})  becomes smaller when SUSY breaking scale is higher.
There are two reasons why the contribution becomes small.
First, the value of the top Yukawa coupling becomes smaller  at high scale by the RGE corrections.
Second, the renormalization flow of $\lambda_{\rm quartic}$ has a focusing effect.
Namely, in order to obtain the 125 GeV Higgs boson, $\lambda_{\rm quartic}$ of the matching condition should be small when SUSY breaking scale is high.
Then the dominant contribution comes from not threshold corrections but RGE corrections from SUSY breaking scale to weak scale.

The third one $\lambda_{\rm NLO}^{\chi^1}  $ is the modification term to the tree-level relation of  Eq.~(\ref{treematch}) through the Higgsino-gaugino loop, 
\beq
 \lambda_{\rm NLO}^{\chi^1}  =  - \frac{1}{6}\cos^2 2 \beta \left[ 2 g^4 \ln\frac{M_2^2}{\susy^2} + (g'^4 + g^4)\ln\frac{\mu^2}{\susy^2}\right].
\eeq

The last term $\lambda_{\rm NLO}^{\chi^2}$  is obtained as follows when one integrate out the Higgsinos and gauginos at the matching scale, 
\beq
 \lambda_{\rm NLO}^{\chi^2} & = & \frac{1}{2} \tilde{\beta}_{\lambda} \ln\frac{\mu^2}{\susy^2} + \left[ - \frac{7}{12} \tilde{f}_1 (r_1) (\tilde{g}_{1d}^4 + \tilde{g}_{1u}^4) - \frac{9}{4} \tilde{f}_2(r_2) (\tilde{g}_{2d}^4 + \tilde{g}_{2u}^4) \right. \non
 & & -\frac{3}{2} \tilde{f}_3 (r_1) \tilde{g}_{1d}^2 \tilde{g}_{1u}^2 -\frac{7}{2} \tilde{f}_4(r_2) \tilde{g}_{2d}^2 \tilde{g}_{2u}^2 -\frac{8}{3} \tilde{f}_5 ( r_1, r_2) \tilde{g}_{1d}    \tilde{g}_{1u} \tilde{g}_{2d}  \tilde{g}_{2u} \non
 & &         - \frac{7}{6} \tilde{f}_6(r_1,r_2) ( \tilde{g}_{1d}^2  \tilde{g}_{2d}^2 +  \tilde{g}_{1u}^2  \tilde{g}_{2u}^2     ) -\frac{1}{6} \tilde{f}_7 (r_1, r_2) ( \tilde{g}_{1d}^2  \tilde{g}_{2u}^2 +  \tilde{g}_{1u}^2  \tilde{g}_{2d}^2   ) \non
 & & - \frac{4}{3} \tilde{f}_8 (r_1, r_2) ( \tilde{g}_{1d} \tilde{g}_{2u} + \tilde{g}_{1u}\tilde{g}_{2d}    )(\tilde{g}_{1d} \tilde{g}_{2d} + \tilde{g}_{1u} \tilde{g}_{2u}) \non
 & &+ \frac{2}{3} \tilde{f}(r_1) \tilde{g}_{1d} \tilde{g}_{1u} \left[ \lambda_{\rm LO} - 2 (\tilde{g}_{1d}^2 + \tilde{g}_{1u}^2) \right] + 2 \tilde{f}(r_2) \tilde{g}_{2d} \tilde{g}_{2u} \left[ \lambda_{\rm LO} - 2 (\tilde{g}_{2d}^2 + \tilde{g}_{2u}^2) \right] \non
 & & \left. + \frac{1}{3} \tilde{g}(r_1) \lambda_{\rm LO} (\tilde{g}_{1d}^2 + \tilde{g}_{1u}^2) + \tilde{g}(r_2) \lambda_{\rm LO} (\tilde{g}_{2d}^2 + \tilde{g}_{2u}^2) \right],
\eeq
with
\beq
\tilde{g}_{1d} &=& g' \sb,~~~~~~\tilde{g}_{2d} = g \sb, \non
\tilde{g}_{1u} & =& g' \cb, ~~~~~~\tilde{g}_{2 u } = g \cb, 
\label{gauginoyukawatree}
\eeq
\beq
\tilde{\beta}_{\lambda}  &=&2 \lambda_{\rm LO} (\tilde{g}_{1d}^2 + \tilde{g}_{1u}^2 + 3 \tilde{g}_{2d}^2 + 3 \tilde{g}_{2u}^2) - \tilde{g}_{1d}^4 - \tilde{g}_{1u}^4 - 5 \tilde{g}_{2d}^4 - 5 \tilde{g}_{2u}^4 \non
 & &- 4 \tilde{g}_{1d}\tilde{g}_{1u} \tilde{g}_{2d} \tilde{g}_{2u} - 2 (\tilde{g}_{1d}^2 + \tilde{g}_{2u}^2) (\tilde{g}_{1u}^2 + \tilde{g}_{2d}^2),
\eeq
and $r_1 = M_1/\mu$, $r_2 = M_2/\mu$\footnote{In the {\it split } case, $v_{\rm EW} \sim M_{\rm gaugino} \sim \mu \ll M_{\rm SUSY}$, one must include this threshold correction $ \lambda_{\rm NLO}^{\chi^2} $ in matching condition  at not the high scale but weak scale, as we discuss later.}.
The loop functions $\tilde{f}_i, \tilde{f}, \tilde{g}$ ($i=1,2,\dots,8 $) are defined in appendix \ref{appfunctions}, and they are normalized such that $\tilde{f}(1) =\tilde{g}(1) =\tilde{f}_{1/2/3/4}(1)=\tilde{f}_{5/6/7/8}(1,1)=1$.
When the all masses of the Higgsino and gauginos are the same as SUSY breaking scale, $\lambda_{\rm NLO}^{\chi^2} $ becomes as follows 
\beq
 \lambda_{\rm NLO}^{\chi^2} & = & \frac{1}{8}(1 + \sin 2 \beta) \left[ -\frac{13}{3} g'^4 - 8 g'^2 g^2 -  17 g^4 \right. \non
 & & \left. + (\frac{1}{3} g'^2 +  g^2) \left(( g'^2 +  g^2) \cos 4 \beta + 2 (- g'^2 +  g^2)\sin 2 \beta\right) \right].
\eeq

\paragraph{Matching at weak scale}
~

In order to calculate the mass of the Higgs boson at the next-to-leading order, we should include the one-loop corrections to the tree-level mass (\ref{treeHiggs}).
The pole mass of the Higgs boson and the top quark are related to $\lambda_{\rm quartic}(Q)$ and $y_t(Q)$ at the $\overline{MS}$ scale $Q$ as
\beq
m^2_{h,\textrm{pole}} &=& 2 v_{EW}^2 (\lambda_{\rm quartic}(Q) + \delta_{\lambda}(Q) ), \label{Higgsmatch}\\
m_{t,\textrm{pole}} &=& \frac{y_t(Q) v_{EW}}{1 + \delta_t(Q)}, 
\eeq
Here
$\delta_{\lambda}$ is the full one-loop radiative corrections via SM particle loop derived by Sirlin and Zucchini \cite{Sirlin:1985ux}\footnote{Recently, the full two-loop corrections to the Higgs quartic coupling are evaluated \cite{Buttazzo:2013uya}.}. 
$ \delta_{t}$ is the three-loop corrections $\mathcal{O}(\alpha_s^3)$  and the full one-loop radiative corrections via the SM particle loop \cite{Hempfling:1994ar, Chetyrkin:1999qi}.
\beq
\delta_{\lambda} (Q)&=& - \frac{\lambda_{\rm quartic} G_F M_Z^2}{8 \pi^2 \sqrt{2}} \left(\xi F_1(Q) + F_0(Q) + \frac{F_{-1}(Q)}{\xi}\right),\label{theweak1}\\
\delta_{t}(Q) &=& \delta_t^{\rm QCD}(Q) + \delta_{t}^{\rm EW}(Q),\label{theweak2}
\eeq
where $G_F $ is the Fermi constant from the muon decay, $(\sqrt{2} G_F)^{-1/2}=246.21971\pm0.00006\GeV$, and $\xi = m_h^2 / M_Z^2$, $M_Z = 91.1876\pm0.0021\GeV$. The loop function $F_0, F_1$ and $F_{-1}$ are defined in appendix \ref{appfunctions}.

We have used the matching condition for the Higgs quartic coupling Eq.~(\ref{Higgsmatch}) at $Q = m_t$ with 
\beq
m_t &=& (174.34 \pm 0.37({\rm stat}) \pm 0.52({\rm syst}) ) \GeV \textrm{\cite{Tevatron:2014cka}},\\
\alpha_s (M_Z) &=& 0.1184 \pm 0.0007 \textrm{\cite{Bethke:2009jm}}. 
\eeq
The values of the SM couplings at $Q = m_t$ are computed by Ref.~\cite{Draper:2013oza}, using two-loop 5-flavor $\overline{MS}$ RGE (as initial values they use the $\overline{MS}$  mass of the bottom quark $m_b(m_b) = 4.18 \GeV$ and the pole mass of the tau lepton $m_{\tau} = 1.777\GeV$.). 
We use these values in matching condition at the weak scale, 
\beq
g_s (Q = m_t) &=&1.1666 +0.00314 \frac{\alpha_s (M_Z) - 0.1184}{0.0007} - 0.00046(\frac{m_t}{\GeV} - 173.35),\non
\label{sono1}
\eeq
\beq
g(Q=m_t) &=& 0.6483,\\
g'(Q= m_t) &=& 0.3587,\\
y_b(Q = m_t) &=& 0.0156,\\
y_{\tau} (Q = m_t)& =& 0.0100.
\eeq
Then, the numerical values of the next-to-leading order threshold correction Eqs.~(\ref{theweak1},~\ref{theweak2}) are 
\beq
\delta_{\lambda} (m_t )&\simeq& 0.0075 \lambda_{\rm quartic}(m_t),\\
\delta_t^{\rm QCD} (m_t) &=& -\frac{4}{3 \pi} \alpha_s (m_t) - 0.92 \alpha_s^2(m_t) - 2.64 \alpha_3^3 (m_t),\\
\delta_{t}(m_t) & =& \delta_{t}^{\rm QCD}(m_t) +\delta_{t}^{\rm EW}(m_t)    \simeq - 0.0600 + 0.0013.
\eeq

\paragraph*{}
We have numerically calculated the mass of the Higgs boson
 at two-loop level, using two-loop SM RGEs (appendix \ref{RGEqpp}) from SUSY breaking scale to top quark mass scale and including  one-loop
threshold corrections (Eqs.~(\ref{matchhigh},~\ref{Higgsmatch})).
In the Figure \ref{hiscaleHiggs}, 
we show the predicted mass of the Higgs boson as a function of SUSY breaking scale and $\tb$. The green regions represent the appropriate Higgs mass $125 \GeV <m_h<126\GeV$.
We take $M_{\rm gaugino}  = \mu  =\sqrt{m_0^2} =  M_{\rm SUSY} $ for simplicity.
The stop mixing parameter is fixed at $A_t (\susy) =0$ ({\it upper figure}), $A_t (\susy) =\susy$ ({\it middle figure}) and $X_t(\susy) =\sqrt{6} $ ({\it lower figure}).
Our results are consistent with the Figure~2 in Ref \cite{Bagnaschi:2014rsa}.

We find that at large $\tb$ region,  an appropriate Higgs mass is achieved when SUSY breaking scale is $\mathcal{O} (10) \TeV$. 
On the other hand, at small $\tb$ region, 
it is achieved when SUSY breaking scale is $\mathcal{O}(10^{ \textrm{5-10}}) \GeV$.
We also show that  the stop mixing contribution Eq.~(\ref{18}), that is the dominant threshold correction, is effective at large $\tan\beta$ region.
For example, even when the soft SUSY breaking scale is  $\mathcal{O}(1)\TeV$, the maximal stop mixing can raise the Higgs boson mass to be observed value (see the lower figure).
However, we discussed before, this contribution is not effective at small $\tb$ and high-scale SUSY breaking region.

\begin{figure}[hp]
\begin{center} 
\vspace*{-2.2cm}
\includegraphics[width=110mm]{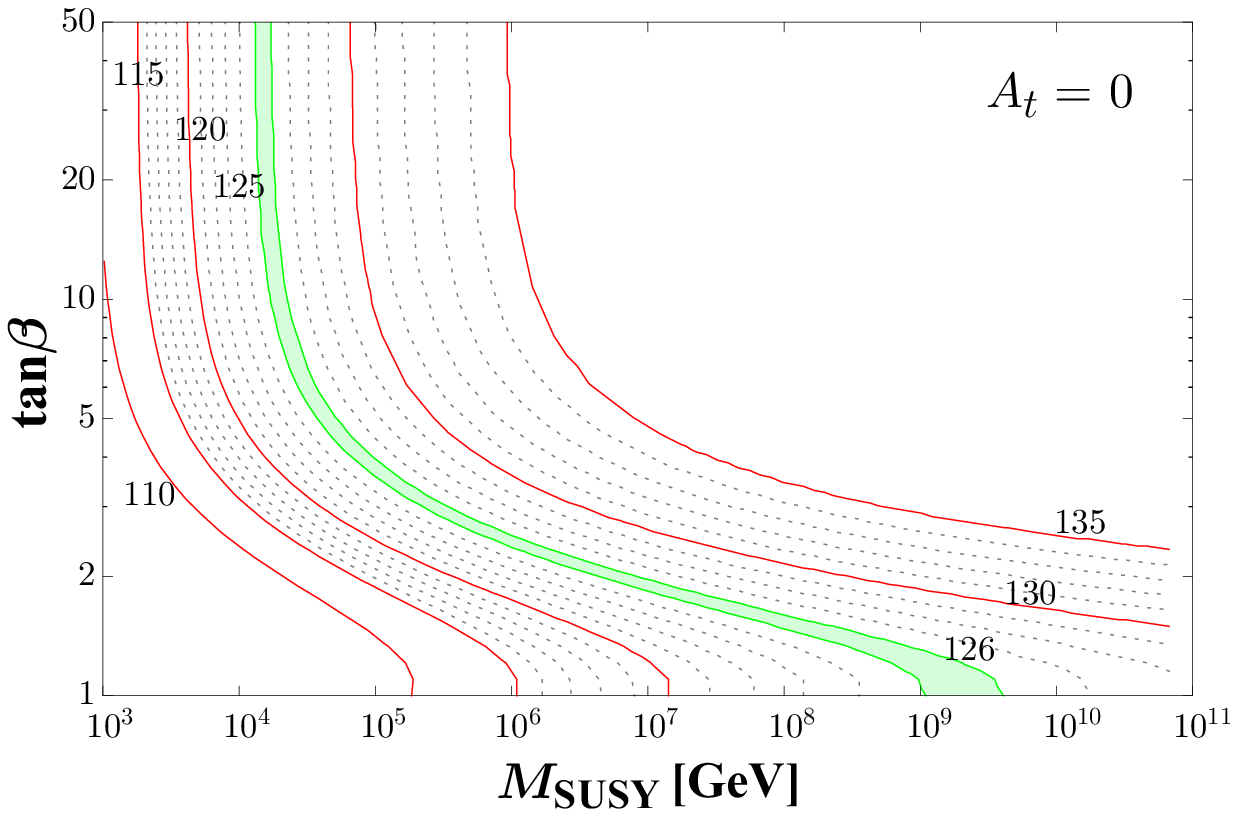} \\
\vspace{0.2cm}
\includegraphics[width=110mm]{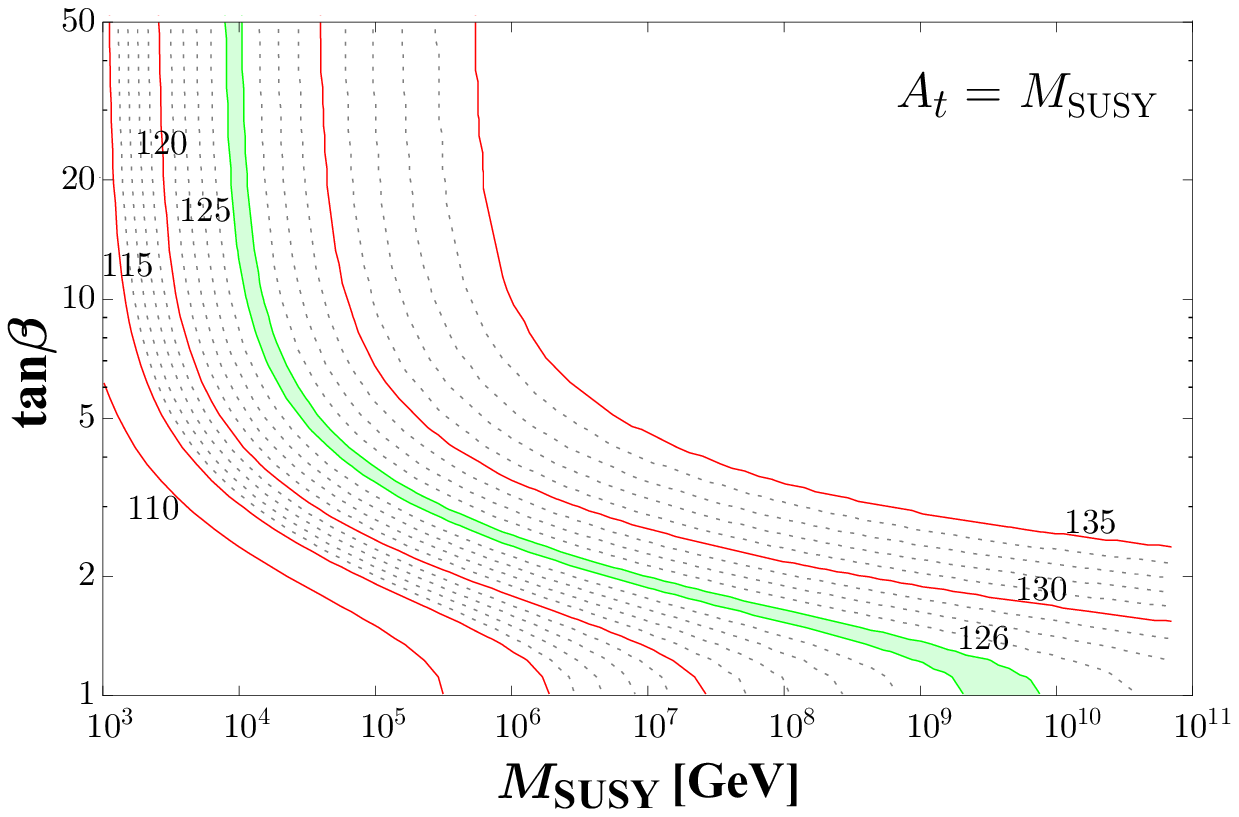} \\
\vspace{0.2cm}
\includegraphics[width=110mm]{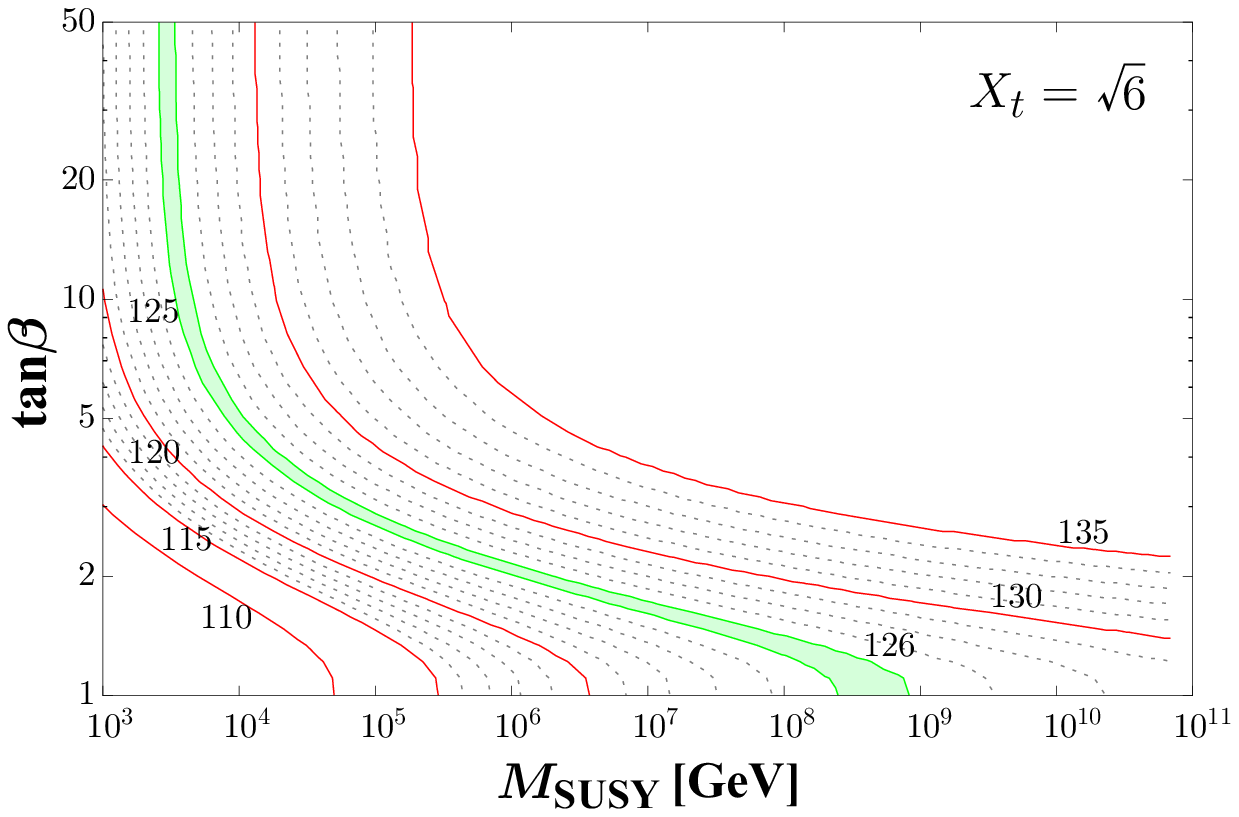}
\end{center} 
\caption[Two-loop level mass of the Higgs boson as a function of SUSY breaking scale and $\tb$.]{Two-loop level mass of the Higgs boson as a function of SUSY breaking scale and $\tb$. The result is evaluated by using two-loop SM RGEs from SUSY breaking scale to top quark mass scale and including  one-loop threshold corrections.
The green regions represent the appropriate Higgs mass $125 \GeV <m_h<126 \GeV$.
We take $M_{\rm gaugino}  = \mu  =\sqrt{m_0^2} =  M_{\rm SUSY} $.
The stop mixing parameter is fixed at $A_t (\susy) =0$ ({\it upper figure}), $A_t (\susy) =\susy$ ({\it middle figure}) and $X_t(\susy) =\sqrt{6} $ ({\it lower figure}).}
\label{hiscaleHiggs}
\end{figure}

\paragraph*{}

Note that in {\it split} case, that is  $v_{\rm EW} \sim M_{\rm gaugino} \sim \mu \ll M_{\rm SUSY} \sim \sqrt{m_0^2}$, 
 Higgsino and gaugino also affect the RGEs of the quartic coupling of the Higgs boson  at $v_{\rm EW}  < Q < M_{\rm SUSY} $ in addition to the SM particles.   
In other words, the following terms is still active at $v_{\rm EW}  < Q < M_{\rm SUSY} $,
\beq
\mathcal{L} = - \frac{1}{\sqrt{2}}\Phi_h^{T} i \sigma_2 (- \tilde{g}_{2 d} \sigma^a \tilde{W}^a + \tilde{g}_{1d} \tilde{B}  )\tilde{H}_d- 
\frac{1}{\sqrt{2}}\Phi_h^{\dag}  ( \tilde{g}_{2 u} \sigma^a \tilde{W}^a + \tilde{g}_{1u} \tilde{B}  )\tilde{H}_u   +\hc,
\label{yukawagaugino}
\eeq
 where $\tilde{g}_{1d/u}$ and $\tilde{g}_{2d/u}$  are Yukawa-like gaugino couplings.
Then, since supersymmetry is no longer ensued at $Q < M_{\rm SUSY}$, the Yukawa-like gaugino couplings are different from the corresponding gauge couplings\footnote{
We have also considered the corrections to the gaugino couplings from the corresponding gauge couplings \cite{Endo:2013lva, Endo:2013xka} in the split mass spectrum case, in which sleptons are also light.  We found  that this corrections can be as large as $\sim 10 \%$ in the parameter region in which the muon $g-2$ anomaly can be solved, and that the gaugino couplings can be  measured from the production cross section of the right-handed selectrons at $1 \%$ accuracy at ILC with $\sqrt{s} = 500\GeV$.}.
Namely, the relations Eq.~(\ref{gauginoyukawatree}) are not satisfied  at $Q < M_{\rm SUSY}$.
Therefore, in the split case,
we should take into account the RGE of not only the SM couplings but also the Yukawa-like gaugino couplings at $v_{EW} < Q < M_{\rm SUSY}$.
The study of the split mass spectrum case is written in Refs.~\cite{Giudice:2011cg, Bagnaschi:2014rsa} in detail.

\subsubsection{Higher-loop radiative corrections to the Higgs boson mass}

Recently, the contributions of higher-loop correction to the mass of the Higgs mass have been studied.  
As a result, one find that these new contributions are important when SUSY breaking scale is not low.
In this section, we briefly review the higher-loop corrections.

Figure \ref{HiggsmassFig} shows the mass of the Higgs boson as a function of SUSY breaking scale  including the  higher-loop radiative corrections.
We have numerically analyzed the mass of the Higgs boson using the public code {\tt FeynHiggs2.10.0} \cite{Heinemeyer:2000nz,Frank:2002qf,Degrassi:2002fi,Frank:2006yh,Heinemeyer:1998yj,Hahn:2009zz,Heinemeyer:1998np,Hahn:2013ria}.
We take $m_{Q} = m_{U} = M_3 = \mu = M_A = M_{\rm SUSY}$, $\tan \beta = 50$ and $A_t = 0$.
In this analysis, the mass of the top quark is $174.34 \GeV$ \cite{Tevatron:2014cka}, which is the latest result.
The green region is $125 \GeV < m_h < 126 \GeV$.
The dotted line represents the full one-loop result and the dashed line represents leading $\mathcal{O}(y_t^2 g_s^2)$ plus subleading $\mathcal{O}(y_t^4)$ two-loop result.
Note that full one-loop, leading two-loop and subleading two-loop corrections are calculated by Feynman-diagrammatic approach, and these contributions include not only the logarithmic corrections  but also finite corrections.
The thick line represents the result of two-loop plus higher-loop corrections, which is evaluated  by a resummation of the leading and subleasing logarithmic corrections from the scalar top sector.
The resummation have been obtained from an analysis of the RGE at two-loop level.
The blue and  red regions represent one sigma bands from the theoretical uncertainty.
The theoretical uncertainty is dominated by two contribution, the experimental error on the mass of the top quark and unknown higher-order corrections.
The theoretical uncertainty from the experimental error on  the mass of top quark is numerically estimated \cite{Heinemeyer:2003ud}, and it is obtained as follows, 
\beq
\delta m_h^{m_t} \sim \delta m_t^{\rm exp} \sim 1\GeV.
\eeq
On the other hand, the theoretical uncertainty from the unknown higher-order corrections is estimated at $\delta m_h^{\rm higher}\sim 3$ - $ 5 \GeV$ in two-loop level calculations.
However, as you can see Figure \ref{HiggsmassFig},  including the resummation  of the leading and subleading  logarithmic corrections, this uncertainty dramatically decrease \cite{Hahn:2013ria,Binger:2004nn } at  
\beq \delta m_h^{\rm higher}\lesssim 1 \GeV. \eeq 
\begin{figure}[t]
\begin{center} 
\includegraphics[width=130mm]{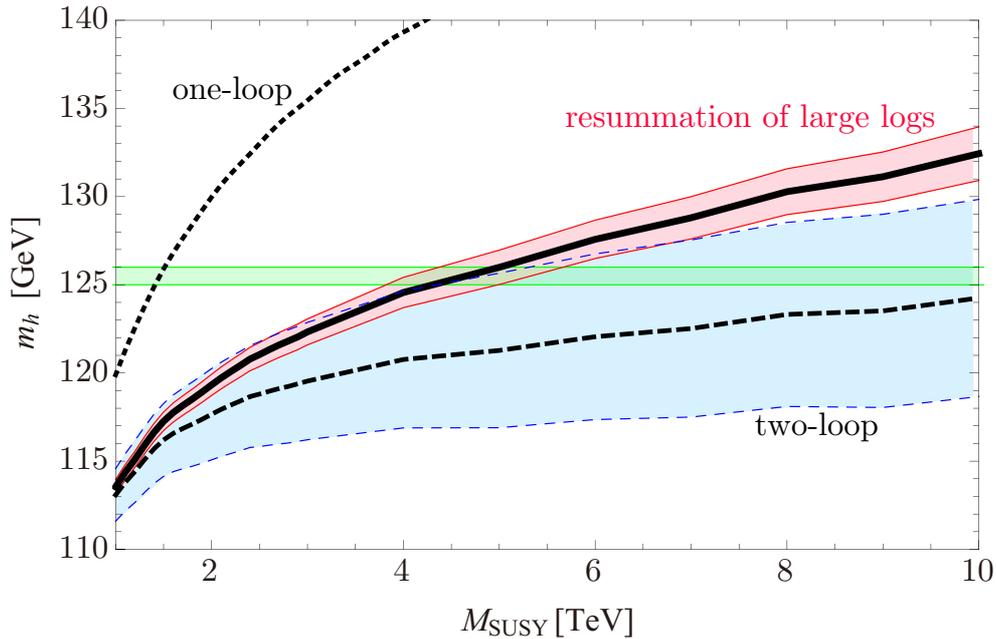}
\end{center} 
\caption[The mass of the Higgs boson as a function of SUSY breaking scale including the higher-loop radiative corrections.]{The mass of the Higgs boson as a function of SUSY breaking scale including the higher-loop radiative corrections.
We take $m_{Q} = m_{U} = M_3 = \mu = M_A = M_{\rm SUSY}$, $\tan \beta = 50$, all $A = 0$, and  $m_t = 174.34 \GeV$ \cite{Tevatron:2014cka}.
The dotted (dashed) line represents the full one-loop (leading plus subleading two-loop) result.
The thick line represents the result of two-loop plus higher-loop corrections, which is evaluated  by a resummation of the leading and subleasing logarithmic corrections from the scalar top sector.
The blue and  red regions represent one sigma bands from the theoretical uncertainty.
The green region is $125 \GeV < m_h < 126 \GeV$.}
\label{HiggsmassFig}
\end{figure}

In this Figure, we show that 
the higher-order corrections  by the resummation of the leading and subleading logarithmic corrections can raise the mass of the Higgs boson.
We also find that
when there is no help of $A_t$ term, SUSY breaking scale is predicted $M_{\rm SUSY} \sim 5 \TeV$ in order to obtain  
an appropriate mass of the Higgs boson ($\tb = 50$ case).

\subsection{SUSY FCNC~/~CP Problem}
\label{FCNC}

In general, the supersymmetry introduces the new flavor and CP violating sources through the SUSY breaking sector.
These new contributions seem to cause the flavor changing neutral currents (FCNC) and electric dipole mordents (EDM) of quarks and leptons.
 On the other hand, in the SM, the FCNC is suppressed by the GIM mechanism \cite{Glashow:1970gm}, and the EDM is suppressed by the  three (four) loop suppression and by the smallness of component of the CP violating phase in the CKM matrix.
Hence, the FCNC and EDM which are generated from the SUSY breaking sector can give dominant contributions, and they can be probed by the low energy flavor experiments and  the EDM measurements.
Nevertheless, the corresponding signals have not been observed yet.
The current experimental results set the constraint to the  new flavor and CP violating sources in the supersymmetric models.

In the MSSM, the new flavor violations are generated by the off-diagonal sfermion soft SUSY breaking terms $(m^2_{\tilde{f}})_{ij} ~(i\neq j)$ and the off-diagonal $A$ terms  $A_{ij} ~(i\neq j)$.
If the $A$ terms are proportional to the Yukawa matrix, the main sources of the flavor violation are   given by the sfermion masses.
Besides these off-diagonal components are expected the same order as the diagonal components if  there are no additional symmetries\footnote{The large off-diagonal components cause the negative eigenvalue of sfermion mass square, which leads to the charged/colored breaking vacuum. Thus, actually the off-diagonal components are expected $\mathcal{O}(0.1\times (m^2_{\tilde{f}})_{ii})$.}.
Thus, the supersymmetry would cause relatively large FCNC process.

One of the severe constraints comes from the branching ratio of $\mu \to e \gamma$.
In the MSSM, when we assume $(m^2_{\tilde{\ell}_L})_{ij} = (m^2_{\tilde{\ell}_R})_{ij}=(m^2_{\tilde{\nu}})_{ij}$ and neglect the $A$ term contributions, the branching ratio of $\mu \to e \gamma$ at the one-loop order is given as \cite{Endo:2013lva,Moroi:1995yh},
\beq
\textrm{Br} (\mu \to e \gamma )  &\simeq& \fr{1}{\Gamma_{\rm tot}} \fr{\alpha m_{\mu}}{16} \biggl|  \fr{(m^2_{\tilde{\ell}})_{23}(m^2_{\tilde{\ell}})_{31} }{m^4_{\tilde{\mu}}}\fr{m_{\tau}}{m_{\mu}}  a_{\mu,~\textrm{bino(neutralino)~loop}  } \non
&&~~~~~~~~~~~~~ +  \fr{(m^2_{\tilde{\ell}})_{21}} {m^2_{\tilde{\mu}}} \biggl(  a_{\mu,~\textrm{chargino~loop} }   +   a_{\mu,~\textrm{wino-Higgsino(neutralino)~loop}  } \non
&& ~~~~~~~~~~~~~~~~~~~~~~~~~~~~ + a_{\mu,~\textrm{bino-Higgsino(neutralino)~loop}  }  \biggl)  \biggl|^2,
\eeq
with
\beq
 a_{\mu,~\textrm{chargino~loop} }
 &=& \frac{g^2}{16\pi^2} \frac{m_\mu^2}{M_2 \mu} \tan\beta
f_C
 \left( \frac{M_2 ^2}{m_{\tilde{\nu}_2}^2}, \frac{\mu ^2}{m_{\tilde{\nu}_2}^2}  \right), \label{g-21}
 \\
a_{\mu,~\textrm{wino-Higgsino(neutralino)~loop}  }
 &=& - \frac{g^2 }{32\pi^2} \frac{m_\mu^2}{M_2 \mu} \tan\beta
 f_N
 \left( \frac{M_2 ^2}{m_{\tilde{\mu }}^2}, \frac{\mu ^2}{m_{\tilde{\mu }}^2} \right),  
  \\ 
a_{\mu,~\textrm{bino-Higgsino(neutralino)~loop}  }
  &=& - \frac{g'^2 }{32\pi^2} \frac{m_\mu^2}{M_1 \mu} \tan\beta
 f_N 
 \left( \frac{M_1 ^2}{m^2_{\tilde{\mu }}}, \frac{\mu ^2}{m^2_{\tilde{\mu }}} \right),   
 \\ 
a_{\mu,~\textrm{bino(neutralino)~loop}  }
 &=& \frac{g'^2}{16\pi^2} \frac{m_{\mu }^2 M_1 \mu}{m_{\tilde{\mu }}^4 }  \tan \beta
 f_N \left( \frac{m_{\tilde{\mu }}^2}{M_1^2}, \frac{m_{\tilde{\mu }}^2}{M_1^2}\right), \label{g-22}
\eeq
 where $(m^2_{\tilde{\ell}_L})_{22} \equiv m^2_{\tilde{\mu}} $, the total decay width of muon $\Gamma_{\rm tot} = 2.99 \times10^{-19}\GeV$,
 and the loop functions $f_C$ and $f_N$ are defined in Appendix~\ref{appfunctions}.
 Note that  the SUSY contributions are given by the chargino-sneutrino loop and the neutralino-smuon loop.
 As one can see, these effects are proportional to $\tan^2 \beta$.
 Naturally, these SUSY contributions are decoupled in the $\susy \to \infty$.
 
The current bound on the $\mu \to e \gamma$ had been given by MEG Collaboration, and the result is Br($\mu \to e \gamma)  < 5.7 \times 10^{-13} ~(90 ~\%$ CL) \cite{Adam:2013mnn}.
The Ref.~\cite{Moroi:2013sfa} showed that when   $\sqrt{m^2_{\tilde{\mu}} }=\mu =\susy$, $3 M_1/(5g'^2) = M_2/g^2 = \susy/g_s^2$, $\tb = 50$, $(m^2_{\tilde{\ell}})_{21}/m^2_{\tilde{\mu}}=0.1$ and  $(m^2_{\tilde{\ell}})_{23}/m^2_{\tilde{\mu}}=(m^2_{\tilde{\ell}})_{31}/m^2_{\tilde{\mu}} = 0$ are taken,  the current bound implies that the soft SUSY breaking scale should be heavier than $30\TeV$, $\susy \gtrsim 30 \TeV$.
Thus, this bound severely constrains  the low scale SUSY models if there are no additional symmetries or mechanisms.
Note that this bound can be relaxed if we take small $\tan \beta$.

\paragraph*{}
One the other hand, in the MSSM the new CP violation sources are the  off-diagonal sfermion soft SUSY breaking masses Im($(m^2_{\tilde{f}})_{ij}$) $ ~(i\neq j)$, $A$ terms  Im($A_{ij}$), the gaugino mass Im($M_{\rm gaugino}$) and the Higgsino mass Im($\mu$).
The physical CP violation source is relative phases of those.
Although the quark (lepton)  EDM is generated  by three (four) loop diagram in the SM, the EDM is actually generated by one loop diagram in the SUSY models.
Therefore, the supersymmetry also would cause relatively large EDM.

One of the severe constraints to the CP violation phenomena comes from the electron EDM.
In the MSSM, when we take the same assumption as the above, the electron EDM at the one-loop order is given as \cite{Endo:2013lva},
\beq
\fr{d_e}{e} &\simeq& - \fr{1}{2 m_e} \biggl[  \textrm{arg}[M_2 \mu] \left( a_{e,~\textrm{chargino~loop} }+a_{e,~\textrm{wino-Higgsino(neutralino)~loop}  } \right) \non
&& ~~~~~~~ + \textrm{arg}[M_1 \mu] \left(a_{e,~\textrm{bino-Higgsino(neutralino)~loop}  } + a_{e,~\textrm{bino(neutralino)~loop}  } \right) \biggl],
\eeq
where we have replaced subscript $\mu$ with $e$ in the Eqs.~(\ref{g-21}-\ref{g-22}).
Similarly to the $\mu\to e \gamma $ process,  the SUSY contributions are given by
the chargino-sneutrino loop and the neutralino-smuon loop.
These contributions are proportional to the CP violating relative phase of the Higgsino and the gaugino masses and proportional to $\tb$.
When the all dimensional parameter are the same values $M_1 = M_2 = \mu = \sqrt{m^2_{\tilde{\nu}_1}} = \sqrt{ m^2_{\tilde{e}}}  = \susy$, and if we take the universal CP violating phase of gaugino mass $\textrm{arg}(M_1 \mu) = \textrm{arg}(M_2 \mu) = \sin \phi $,  the one loop order electron EDM is given as,
\beq
|d_e| &\simeq& \fr{5 g^2 + g'^2}{2\cdot 48\pi \cdot 4 \pi} \fr{m_e}{\susy^2}\tb \sin\phi \times 197\times 10^{-16} ~~[e \textrm{~cm}],\non
&=& 2.9\times 10^{-27} \left(\fr{10\TeV}{\susy}\right)^2 \left( \fr{\tb}{50}\right) \sin\phi~~[e \textrm{~cm}].
\label{MSSMEDM}
\eeq

The current bound on the electron EDM had been given by  ACME Collaboration,
and the result is $|d_e |  < 8.7 \times 10^{-29}$ [$e$~cm] ~$(90 ~\%$ CL) \cite{Baron:2013eja}.
The Ref.~\cite{Moroi:2013sfa} showed that when  $\sqrt{m^2_{\tilde{e}} }=\mu =\susy$, $3 M_1/(5g'^2) = M_2/g^2 = \susy/g_s^2$, $\tb = 50$,  $(m^2_{\tilde{\ell}})_{ij}/m^2_{\tilde{e}}= 0 ~(i\neq j)$ and the maximal CP violating phase are taken,  the current bound implies that the soft SUSY breaking scale should be heavier than $100\TeV$, $\susy \gtrsim 100 \TeV$.
Although this bound can be relaxed if one takes small $\tan \beta$, 
it also severely constrains  the low scale SUSY models.

\coll{Another severe constraint comes from measurement of  the CP violation in the kaon decay \cite{Gabbiani:1996hi}.
Under the CP transformation a neutral kaon $K^0 ~(d \bar{s})$ becomes an anti-neutral kaon $\bar{K}^0 ~(s \bar{d})$,
and it is represented  by convention, 
\beq
\bm{{\rm CP}} | K^0 \rangle = + | \bar{K}^0 \rangle, \non
\bm{{\rm CP}} | \bar{K}^0 \rangle = + | K^0 \rangle.
\eeq
Hence, $( | K^0 \rangle + | \bar{K}^0 \rangle )/\sqrt{2}  $ and $  (  | K^0 \rangle  -|  \bar{K}^0 \rangle )/\sqrt{2} $ states are the CP even and odd eigenstates. 
Since the CP symmetry is slightly broken in nature, 
the mass eigenstates of the neutral kaons become the following forms,
\beq
|K^0_S \rangle &=& \fr{1}{\sqrt{2}} \left[ e^{ - i \phi_K } (1 + \epsilon_K) |K^0 \rangle + e^{ i \phi_K } (1 - \epsilon_K) |\bar{K}^0 \rangle \right], \non
|K^0_L \rangle &=& \fr{1}{\sqrt{2}} \left[ e^{ - i \phi_K } (1 + \epsilon_K) |K^0 \rangle -  e^{ i \phi_K } (1 - \epsilon_K) |\bar{K}^0 \rangle \right], 
\eeq
where $\epsilon_K $ is an indirect CP violation parameter (CP violation in $K^0$-$\bar{K}^0$ mixing) and $\phi_K$ is a direct CP violation parameter (CP violation in the neutral kaon decay), and actually $\epsilon_K \ll 1, ~\phi_K \ll 1$.
The indirect CP violation parameter $\epsilon_K$ is determined by an imaginary part of the  $K^0$-$\bar{K}^0$ mixing \cite{Bigi:2000yz, Buras:2008nn},
\beq
\epsilon_K = e^{i \phi_{\epsilon}}\sin \phi_{\epsilon} \left( - \fr{\textrm{Im} (\mathcal{M}_{K^0\textrm{-}\bar{K}^0})}{\Delta M_K} + \phi_{K,0} \right),
\label{eK}
\eeq
with
\beq
\mathcal{M}_{K^0\textrm{-}\bar{K}^0} &=&  \langle K^0 |  \mathcal{H}_{\rm eff}^{4fermi} (\bar{d} s \bar{d} s ) | \bar{K}^0 \rangle, \non
\tan \phi_{\epsilon} &=& \fr{2 \Delta M_K}{\Delta \Gamma_K} = (43.52 \pm 0.05)^{\circ},
\eeq
where $\phi_{K,0}$ is the phase of the 0-isospin amplitude in kaon decay $K^0 \to 2 \pi$, and 
\beq
\Delta M_K &=& M_L - M_S = (3.484 \pm 0.006) \times 10^{-12} ~\MeV, \non
\Delta \Gamma_K &=& \Gamma_S - \Gamma_L = (7.2823 \pm 0.0098) \times 10^{-12} ~\MeV.
\eeq
The second term of the \eq{eK} contributes an $\mathcal{O}$(5\%) correction to $\epsilon_K$ \cite{Buras:2008nn}.

In the SM, the leading contribution of the four-fermion Hamiltonian $  \mathcal{H}_{\rm eff}^{4fermi} (\bar{d} s \bar{d} s ) $ comes from one-loop box diagrams with the weak interactions and the imaginary component of the CKM matrix.
While in the MSSM, the leading SUSY contribution comes from one-loop gluino-squark box  diagrams with the strong interaction, and these contributions are easy to affect estimation of $\epsilon_K$.   
The main CP-violating source of this diagram is the off-diagonal sfermion soft SUSY breaking masses Im$[(m^2_{\tilde{d}_L})_{12} (m^2_{\tilde{d}_R})_{12} ]$.
The detailed formulae of  the SUSY contributions to $\epsilon_K$ are given in Ref.~\cite{Baek:2001kh}.

The latest experimental value is $ |\epsilon_K^{\rm (exp)} | = (2.228 \pm 0.011) \times 10^{-3}$ \cite{Beringer:1900zz} and the SM  prediction (including next-to-next-to-leading-order) is $|\epsilon_K^{\rm (SM)}| = (1.81 \pm 0.28) \times 10^{-3}$ \cite{Brod:2011ty}.
These numerical values set a conservative upper bound of the SUSY contributions, $|\epsilon_K^{\rm (SUSY)}| < 0.98 \times 10^{-3}$ \cite{Moroi:2013sfa}.   
The Ref.~\cite{Moroi:2013sfa} showed that when  $\sqrt{m^2_{\tilde{d}} }=M_3 =\susy$,  $ |(m^2_{\tilde{d}_L})_{ij}/m^2_{\tilde{d}} | =|(m^2_{\tilde{d}_R})_{ij}/m^2_{\tilde{d}} | = 0.1 ~(i\neq j)$ and the maximal CP violating phase are taken,  the current bound implies that the soft SUSY breaking scale should be heavier than $500\TeV$, $\susy \gtrsim 500 \TeV$.
If one imposes the SO(10) relation at the GUT scale; $(m^2_{\tilde{d}_R})_{ij} =(m^2_{\tilde{d}_L})^{\ast}_{ij}$, the imaginary components of the soft SUSY breaking squark masses vanish at the GUT scale and they are generated only through the RG effects, and thus $\epsilon_K^{\rm (SUSY)}$ is suppressed.
  Then, the current bound implies $\susy \gtrsim 40 \TeV$.
}

\paragraph*{}

Therefore, when the soft SUSY breaking scale is low scale which is favored in terms of the naturalness, 
too large FCNC and too large CP-violation are induced by the SUSY particles one-loop radiative corrections. These difficulties are called ``the SUSY FCNC problem" and ``the SUSY CP problem".


\section{Discussions}
\label{2no4}

In previous section, we have reviewed the current situation of the MSSM in terms of an observed 125 GeV Higgs boson and the constraints from the flavor violation and CP violation process.
Actually, we have shown the following two points in the MSSM,
\begin{itemize}
\item
 In order to realize the observed Higgs boson mass, the masses of the stops have to be heavier than 5 TeV or the stop mixing has to be maximized (or their compromised parameter region).  

\item In order to avoid the severe constraints from the flavor violation and CP violation measurements, the masses of the sleptons should be heavy and the parameter $\tan\beta$ should be not so large  if there are no additional symmetries or mechanisms.
\end{itemize}
Therefore, in this chapter we na\"{\i}vely conclude that one of the suitable solutions is the {\it high-scale supersymmetry}, which has $\mathcal{O}$(10-100) TeV  soft SUSY breaking terms.

Once we give up thought on the naturalness, such a high-scale supersymmetric model is very attractive.
First, as we discussed, the observed 125 GeV Higgs boson and the explanation of the current bound on the flavor violation and CP violation process are simultaneous realized naturally.
Second, all the SUSY particles are out of the reach of the current LHC, thus it can explain that there are no signals of the SUSY particles at the LHC experiments. 
Third, it can naturally solve the cosmological gravitino problem \cite{Kawasaki:2008qe}: although an unstable  gravitino spoils the big-bang nucleosynthesis, the heavy gravitino (typically heavier than $5$-$8$ TeV) can avoid the disaster\footnote{Here, we implicitly  assumed that the soft SUSY breaking scale and the gravitino mass are the same scale.}. 
Fourth, this heavy SUSY breaking scale remains the gauge couplings unification at the GUT scale 
(see Figure~\ref{hoge1}).

Furthermore, the future prospects of the flavor violation and the CP violation measurements have
the potential to probe the supersymmetry beyond the reach of the LHC by orders of magnitude \cite{Baldini:2013ke,Blondel:2013ia,Kuno:2013mha,Abrams:2012er,Kuno:2012pt, Sakemi:2011zz, Kawall:2011zz, Kara:2012ay}.
Therefore,
we hope that these precision measurements can indirectly probe the high-scale supersymmetry.


\chapter{Singlet Extension}
\label{chap3}
\thispagestyle{empty}

\abstchapter{
In this chapter, in order to solve the $\mu$ problem of the supersymmetric minimal model, 
we first introduce a gauge singlet superfield and singlet extension supersymmetric minimal model.
Then, we should impose extra symmetries to forbid unwanted terms of singlet superfield, which spoil the solution of the $\mu$ problem.
In general, these symmetries lead to the domain wall problem and the tadpole problem.
The nearly minimal supersymmetric standard model is the one of the models which can solve the $\mu$ problem, the domain wall problem and the tadpole problem simultaneously.
Finally, we review the Lagrangian and the Higgs sector of the nearly minimal supersymmetric standard model, and we show that there is a sizable tree-level contribution to the Higgs boson mass    due to an extra F-term contribution to the Higgs quartic coupling. }

\section{$\mu$~Problem}
\label{3no1}

In the MSSM, the potential minimization conditions Eqs.~(\ref{minimalcon1},~\ref{minimalcon2})
lead to the following relationship,
\beq
M_Z^2 &=& \frac{m_2^2 - m_1^2}{\cos 2 \beta} - M_A^2  = \frac{m_2^2 - m_1^2}{\cos 2 \beta} - m_1^2 - m_2^2 - 2|\mu|^2,
\label{muproblemeq}
\eeq
where $\mu$ is the supersymmetric mass of the Higgs multiplets.
Especially, when $\tb \gg 1$, this equation can be expanded by $\tb$ and becomes 
\beq
M_Z^2 = -2 (m_2^2 + |\mu|^2) + \frac{2}{\tan^2 \beta}(m_1^2 - m_2^2) + \mathcal{O}(\fr{1}{\tan^4\beta}). 
\eeq
These relationships imply that since actually $M_Z \sim \mathcal{O}(100)\GeV$ (electroweak scale),
$\sqrt{m_1^2},~\sqrt{m_2^2}$ and $\mu$ should be na\"{\i}vely at the electroweak scale, or
$\sqrt{m_1^2},~\sqrt{m_2^2}$ and $\mu$ should be the same scale to able to cancel out.
In other words,  the magnitude  of $\mu$ have to be the soft SUSY breaking scale $\susy$ or be  
less than the scale.

One the other hand, the $\mu$ term 
is stable under the all orders in perturbation theory of the effective Lagrangian  due to the non-renormalization theorem \cite{Seiberg:1993vc},
and 
the supersymmetry  provides a valid description that  the scale of  the $\mu$  parameter is  as large as GUT scale or Planck scale\footnote{
Another elegant possibility is $\mu = 0 $, which is respected the some symmetries (e.g. Peccei-Quinn symmetry). 
However, nature have not chosen this values.
It is because that the lightest chargino search by the LEP set lower bound on chargino mass,  
$|\mu| > 103.5 \GeV$ \cite{Abdallah:2003xe}.
}.
If $\mu$ is at the Planck scale, it leads to $M_Z \sim M_{\rm Pl}$ (see \eq{muproblemeq}), and the  appropriate  electroweak symmetry breaking can not occur.
Therefore, $\mu$ has to  know the soft SUSY breaking scale
in order to realize the nature, namely the $Z$ boson mass is the electroweak scale.
This problem is called ``$\mu$~problem" \cite{Kim:1983dt}: 
\beq
\textrm{Why }~\mu \ll M_{\rm GUT}, ~M_{\rm Pl}~\textrm{?}\non
\textrm{Why does}~\mu~\textrm{know}~\susy~\textrm{?}
\eeq

\begin{table}[t]
 \caption{The singlet superfield with their components for spin $0$ and $1/2$, and their representations for SU$(3)_c \times$ SU$(2)_L \times$ U$(1)_Y$  gauge group. }
\begin{center}
\begin{tabular}{r|cc|ccc} 
\bhline{1.4pt}
Chiral Supermultiplet~~ & Spin 0 & Spin $\frac{1}{2}$ & $SU(3)_{C}$ & $SU(2)_L $&$ U(1)_Y$ \\ \midrule
Singlet scalars-Singlino   $\hat{S}$ &$S = (S_{R} + i S_{I})/\sqrt{2} $ & $\tilde{S}$ & $\bm{1}$ & $\bm{1}$ & $  0 $ \\
 \bhline{1.4pt}
\end{tabular}
\end{center}
\label{nMSSMtable}
\end{table}

\paragraph*{}

One of the natural solutions of the $\mu$ problem is the singlet extension models of the MSSM \cite{Fayet:1974pd}.
These models have  the following superpotential,
\beq
W  = \lambda \hat{S} \hat{H}_2 \hat{H}_1 + f[\hat{S}] + W_{\rm Yukawa},
\eeq
where $\hat{S}$ is an additional gauge singlet superfield, $\lambda$ is a dimensionless coupling constant, and $f[\hat{S}]$ is the superpotential which does not depend on  superfields of the MSSM at the renormalizable level.
The singlet superfield $\hat{S}$ with their components for spin $0$ and $1/2$ are shown in Table~\ref{nMSSMtable}.
When supersymmetry is broken, singlet superfiled also receives the soft SUSY breaking mass or $A$ term. At this time, \coll{singlet scalar boson} can naturally obtain a vev which is the order of the soft SUSY breaking scale,
$\langle S \rangle \sim \mathcal{O}(\susy)$, and so its vev gives the effective $\mu$ term for Higgs multiplets,
\beq
\mu_{\rm eff} = \lambda \langle S \rangle.
\eeq
Therefore, the $\mu$ problem can be solved by the singlet extension models of the MSSM \footnote{
One of the other solutions of the  $\mu$ problem is provided by the Giudice and Masiero \cite{Giudice:1988yz}.
Let us consider the following K\"{a}hler potential,
\beq
K   = c_1 \frac{1}{M_{\rm Pl}} \hat{X}^{\dag} \hat{H}_2 \hat{H}_1 + c_2 \hat{H}_2 \hat{H}_1 +\hc,
\eeq
where $ c_1$ and $c_2$ are $\mathcal{O}(1)$ couplings and $\hat{X}$ is an additional gauge singlet superfield. 
Note that the second term is permitted in the supergravity theory.
This  K\"{a}hler potential can be rewritten to the following superpotential,
\beq
W = \left(c_1 \frac{F_X^{\dag}}{M_{\rm Pl}} + \left(c_2 + c_1 \fr{\langle X^{\ast} \rangle }{M_{\rm Pl}}\right) m_{3/2} \right) \hat{H}_2 \hat{H}_1,
\eeq
where $m_{3/2}$ is the gravitino mass which is equivalent to a vev of the superpotential. Therefore one can obtain the appropriate effective $\mu$ term,
\beq
\mu_{\rm eff} = c_1 \frac{F_X^{\dag}}{M_{\rm Pl}} + \left(c_2 + c_1 \fr{\langle X^{\ast} \rangle }{M_{\rm Pl}}\right) m_{3/2}  \sim \mathcal{O}(\susy).
\eeq
 }.

\paragraph*{}
In the case that the additional superfiled for the MSSM is only one singlet superfield, 
the superpotential $f[\hat{S}]$ can be written as
\beq
f[\hat{S}] = \xi_F \hat{S} + \frac{1}{2} \mu ' \hat{S}^2 + \fr{1}{3}\kappa \hat{S}^3.
\label{genS1}
\eeq
Then, the soft SUSY breaking terms are 
\beq
V_{\rm soft} &=& m_1^2 |H_1|^2 + m_2 |H_2|^2 + m_S^2 |S|^2  \non
&& + \left( \lambda A_{\lambda} S H_2 H_1 + \xi_S S + \frac{1}{2} m'^2_S S^2 + \fr{1}{3}\kappa A_{\kappa }S^3 + \hc \right),
\label{genS2}
\eeq
here $ \sqrt{m_1^2},~\sqrt{m_2^2},~\sqrt{m_S^2},~A_{\lambda},~(\xi_S)^{1/3},~\sqrt{m'^2_S} $ and $A_{\kappa }$ are the soft SUSY breaking terms, and their magnitudes are typically $\susy$.
The minimization equations of the singlet scalar potential\footnote{
Strictly speaking, we should also solve a condition for an absolute minimum of the scalar potential.}  lead to the vev of the singlet scalar boson.
In other words, the vev of singlet scalar $\langle S \rangle$ is a solution the following equation,
\beq
\langle S \rangle& =& - \fr{\xi_S + \xi_F \mu ' - \lambda v_1 v_2(A_{\lambda} + \mu ')}{m_S^2 + m'^2_S
 + \mu '^2 + 2 \kappa \xi_F + \kappa A_{\kappa} \langle S \rangle   + 2 \kappa^2 \langle S \rangle^2 + 3 \kappa \mu ' \langle S \rangle  + \lambda^2 (v_1^2 + v_2^2) - 2 \lambda \kappa v_1 v_2}\non
 &\sim&
  - \fr{\susy^3 + \xi_F \mu ' }{2 \susy^2 
 + \mu '^2 + 2 \kappa \xi_F + \kappa \susy \langle S \rangle   + 2 \kappa^2 \langle S \rangle^2 + 3 \kappa \mu ' \langle S \rangle },
 \label{vevexact}
\eeq
here we have neglected the terms which includes the Higgs vev since $v_1,v_2 \ll \susy$.
Because the solution of this equation can not be estimated intuitively,  
let us consider the case of $\kappa \ll 1$.
The equation becomes
\beq
\langle S \rangle&\sim&
  - \fr{\susy^3 + \xi_F \mu ' }{2 \susy^2 
 + \mu '^2 }.
\eeq
When $\xi_F \lesssim \susy^2$ and $\mu ' \lesssim \susy$, then the singlet scalar can obtain an appropriate vev, $\langle S \rangle \sim \mathcal{O}(\susy)$.
However, since $ \xi_F $ and $\mu ' $ are the dimensional supersymmetric tadpole and mass term, 
they have no reason that $\xi_F \lesssim \susy^2$ and $\mu ' \lesssim \susy$, and their magnitudes are typically the GUT scale or the Planck scale if there are no symmetry.
Then, the singlet scalar obtains an inadequate vev for the solution of the $\mu$ problem, $\langle S \rangle \sim \mathcal{O}(M_{\rm Pl})$.

These facts imply that in order to solve the $\mu$ problem we have to impose some symmetries, which suppress or forbid some terms in the superpotential $f[\hat{S}]$. 
For example, when some symmetries forbid the dimensional  supersymmetric tadpole $ \xi_F $  and mass term  $\mu ' $, the singlet scalar obtains the following vev \cite{Ellwanger:2009dp},
\beq
\langle S \rangle&\sim& \fr{1}{4 \kappa} \left( - A_{\kappa} + \sqrt{A_{\kappa}^2 - 8 m_S^2}\right).
\label{NMSSMvev}
\eeq
Thus,  the singlet scalar can obtain an appropriate vev, $\langle S \rangle \sim \mathcal{O}(\susy)$.
In this manner, there are various singlet extension models which are classified by the symmetries to control the superpotential $f[\hat{S}]$.
Note that,  when $f[\hat{S}] = 0$,  there are extra global U$(1)$ symmetries in the Lagrangian.
This symmetry leads to an unwanted visible Nambu-Goldstone boson or a visible axion, when the symmetry is spontaneously broken that is associated the electroweak symmetry breaking.

\paragraph*{}
What symmetry is useful for controlling the superpotential $f[\hat{S}]$?

\mtrem{gauge symmetryを課してコントロールする方法は？破れてもNG bosonは喰われて重く成ってdecouplingするはず。Proton decayとかとあわないのかな？？}

The imposition of extra {\it global} symmetries is unreasonable.
Because when the singlet scalar boson obtains vev, this symmetry is spontaneously broken and  gives an unwanted visible Nambu-Goldstone boson or a visible axion.
On the other hands, {\it discrete symmetries} are suitable in order to control the singlet superfield \cite{AlvarezGaume:1983gj,Ellis:1986mq,Ellis:1988er}.
It is because that even if this symmetry is   spontaneously broken, all extra scalar bosons can have a heavy mass which is   phenomenologically acceptable.
Furthermore, {\it discrete R-symmetries} could be  more suitable.
The R-symmetry can be imposed in the supersymmetric models,  and the  R-symmetry breaking is related  to the  supersymmetry breaking. 
Therefore,  we are able to  easily estimate and control  terms in the superpotential  which are generated via the R-symmetry breaking\footnote{
Moreover, such a desecrate R-symmetry should be embedded in a gauge symmetry.
In other wards, the discrete R-symmetry should be  anomaly-free.
Otherwise since the R-symmetry is anomalous broken,
 one can not estimate the terms  which break the classical R-symmetry.
Therefore, the anomaly-free discrete R-symmetry should be employed.
Actually, the discrete $\mathbb{Z}^R_5$ R-symmetry can be anomaly-free if one introduces the extra charged singlet superfields  to the nearly minimal supersymmetric standard model \cite{Paraskevas:2012kn}.
}.

\section{Domain Wall Problem~/~Tadpole Problem}
\label{3no2}

In order to control the singlet superfield, one can impose a discrete symmetry.
However,
the spontaneous broken of the discrete symmetry, which is caused by the electroweak symmetry breaking, leads to a disastrous cosmological domain walls.
Furthermore, if we introduce the explicit breaking terms of the  discrete symmetry to avoid the domain walls, there are cases where these terms lead to a disastrous tadpole.
In this section, we briefly review these two problems with the desecrate  $\mathbb{Z}_3$ symmetry as an example.

\begin{table}[tbp]
\caption{The charge assignments under the Abelian symmetries which are considered in he text. The index $i$ denotes the generations.}
\begin{center}
\label{tab_ch3}
\begin{tabular}{c|c c c c c c c c | c}
\hline \hline 
 &$\hat{H}_1$&$\hat{H}_2$&$\hat{S}$&$\hat{Q}_i$&$\hat{\bar{U}}_i$&$\hat{\bar{D}}_i$&$\hat{L}_i$
&$\hat{\bar{E}}_i$& W \\
 \hline
 $\textrm{U(1)}_Y\textrm{~gauge}$ & -1/2 & 1/2 & 0 & 1/6 & - 2/3 & 1/3 & -1/2 & 1 & 0 \\
 \hline
$\mathbb{Z}_3 \subset {\rm U(1)}_{PQ} $
& 1 & 1 & -2 & -1  & 0 & 0 & -1& 0 & 0 \\ \hline
 ${\rm U(1)}_R$ & 0 & 0 & 2 & 1 & 1 & 1 & 1 & 1 & 2 \\ \hline
$\mathbb{Z}_5^R  \subset {\rm U(1)}_{R'} $
&1&1& 4&2&3&3&2&3&6~(1~(mod~5))\\  \hline \hline
\end{tabular}
\end{center}
\end{table}

The Next-to-Minimal Supersymmetric Standard Model (NMSSM) \cite{Fayet:1974pd,Nilles:1982dy,Frere:1983ag,Derendinger:1983bz,Greene:1986th,Drees:1988fc,Ellwanger:1993hn,Ellwanger:1993xa} is  one of the singlet extension models, 
and its matter content is the MSSM matter and an additional gauge singlet superfield.
In order to control the singlet superfield, 
the NMSSM is imposed the desecrate $\mathbb{Z}_3$ symmetry.
In this symmetry the superfields are transformed  as
\beq
\hat{\Phi }_i \to e^{i \fr{2 \pi q_i }{3}  } \hat{\Phi}_i,
\eeq
where  the charge assignments $q_i $ is listed in Figure~\ref{tab_ch3}.
Then, the NMSSM has the following superpotential,
\beq
W = \lambda \hat{S} \hat{H}_2 \hat{H}_1 + \frac{1}{3} \kappa \hat{S}^3 + W_{\rm Yukawa}.
\label{NMSSMW}
\eeq
Thus, the superpotential of the NMSSM does not have the dimensional supersymmetric tadpole  and mass term.
The vev of the singlet scalar is given \eq{NMSSMvev}, and the $\mu$ problem is solved.
Note that without the singlet cubic term $\kappa \hat{S}^3$, this superpotential becomes invariant under an anomalous global U(1$)_{PQ}$ symmetry, which includes the desecrate $\mathbb{Z}_3$ symmetry as a subgroup.
When the scalar bosons obtain its vev, this symmetry is spontaneously broken and 
gives an unwanted visible axion \cite{Abel:1995uu}.
Thus, one needs this singlet cubic term as discussed previous section.

\paragraph*{Domain wall problem}

~

This desecrate $\mathbb{Z}_3$ symmetry has to be spontaneously broken 
in order to give the appropriate electroweak scale.
The discrete symmetry is broken in different ways in different domains, which are separated  in a larger distance  than the horizon size  or correlation length.
Eventually, these different domains are divided by the domain walls (domain boundaries).
If the discrete symmetry is exact symmetry, these domain wall configurations are topologically stable.
 
When the domain walls had not disappeared in  the early universe, they would produce  destruction of the observed homogeneity and isotropy of our universe.
In addition,   in the radiation dominated era, the energy density contributions to the universe by the domain walls can become comparable to the energy density of the universe, and so the existence of the domain walls would change  the evolution of the universe   significantly.
Therefore, one should avoid the disastrous cosmological domain walls. 
This difficulty is called ``domain wall problem".

One of the solutions of the domain wall problem is an addition of  tiny {\it explicit} discrete symmetry breaking terms.
These tiny explicit breaking terms  remove the  vacuum degeneracy of the different domains.
 It can be interpreted in the decay of the domain walls, and eventually the universe is covered by a unique vacuum.

 Let us consider the NMSSM case, that is desecrate $\mathbb{Z}_3$ symmetry and its explicit breaking terms.
 The additional dimension-5  operators, which are the Planck suppressed explicit  $\mathbb{Z}_3$ symmetry breaking terms, are given as,
 \beq
 \lambda' \fr{\hat{S}^4}{M_{\rm Pl}},~~~~~~ \lambda' \fr{\hat{S}^2 \hat{H}_1 \hat{H}_2}{M_{\rm Pl}},~~~~~~ \lambda' \fr{ (\hat{H}_1 \hat{H}_2)^2}{M_{\rm Pl}},
 \label{dim5}
 \eeq
in the superpotential \eq{NMSSMW}.
Note that the additional dimension-5  operators in the K\"{a}hler potential, $(\hat{S} + \hat{S}^{\dag})  (\hat{H}_i \hat{H}_i^{\dag}) / M_{\rm Pl}$  and $(\hat{S}^{\dag} \hat{H}_1 \hat{H}_2  +\hc)/M_{\rm Pl}$, can be absorb into the last two terms of the superpotential \eq{dim5} by the  redefinitions of the superfields. 
 These explicit breaking terms give the pressure to the domain walls.
 The strongest constrains come from the big-bang nucleosynthesis.
 It is because that the decay of the domain walls produce the entropy, and it destroys the great success of the  nucleosynthesis.
Thus,  we require that the domain walls have to decay before   the onset of the   nucleosynthesis, 
and it sets the following lower bound on the parameter $\lambda'$ \cite{Abel:1995wk}, 
\beq
\lambda' \gtrsim 10^{-7}.
\label{DW7}
\eeq

\paragraph*{Tadpole problem}

~

However, such a model, which is no longer protected by the exact discrete symmetry, permits the renormalizable explicit discrete symmetry breaking  terms in the superpotential at the loop level.
Especially, there is the quadratic divergence in the tadpole, which is generated via the loop diagram including the explicit symmetry breaking non-renormalizable terms.
The  cancelation of the quadratic divergences of the two-point functions remains even when  the supersymmetry is softly broken. 
However, in fact, when the supersymmetry is softly broken, the quadratic divergences of the tadpole are not canceled out \cite{Harada:1981ug}.
It is because  that the coefficients of the quadratic divergence have dimension one, and so they depend on the soft SUSY breaking masses and A terms. 
Thus, the supersymmetry can not keep the cancelation of the quadratic tadpole divergences due to the soft SUSY breaking terms.
It means reintroduction of the hierarchy problem.
These facts lead too large tadpole of the singlet superfield  at the loop level, and 
it gives  too large vev of the singlet scalar terms \eq{vevexact}.
Therefore, the singlet scalar no longer solve the $\mu $ problem.
This difficulty is called ``tadpole problem".

Let us consider the NMSSM which includes the explicit symmetry breaking terms \eq{dim5}.
The leading quadratic tadpole divergences appear from the two-loop diagrams.
These diagrams are shown in the Figure 4 in Ref.~\cite{Abel:1995wk}. 
These contributions to the Lagrangian are 
\beq
\mathcal{L}_{\rm tad} &\simeq &\fr{1}{(16 \pi^2)^2}\fr{\lambda'}{M_{\rm Pl}} \fr{\kappa}{3}  (S + S^{\ast}) \Lambda^2 m_S^2   + \fr{1}{(16 \pi^2)^2}\fr{\lambda'}{M_{\rm Pl}} \lambda (S + S^{\ast}) \Lambda^2 m_S^2 \non
&& + \fr{1}{(16 \pi^2)^2} \fr{\lambda'}{M_{\rm Pl}} \fr{\kappa}{3}(F_S + F_S^{\ast}) \Lambda^2 A_{\kappa}
+ \fr{1}{(16 \pi^2)^2} \fr{\lambda'}{M_{\rm Pl}} \lambda(F_S + F_S^{\ast}) \Lambda^2 A_{\lambda},\non
&\sim & \fr{\lambda' } {(16 \pi^2)^2} \left(\fr{\kappa}{3} + \lambda \right) M_{\rm Pl}\susy^2 S + \fr{\lambda'}{(16 \pi^2)^2}\left( \fr{\kappa}{3} + \lambda \right) M_{\rm Pl} \susy F_S\non
&& +\hc,
\eeq
where we have taken the UV cut off $\Lambda$ to be the Planck scale.
The first term is regarded as the tadpole in the soft SUSY breaking scalar potential, and the second term is regarded as the tadpole in the (effective) superpotential.
Thus the tadpole is generated by the loop diagram which has the soft SUSY breaking effect.

Including these tadpoles in the scalar potential, the vev of the singlet scalar becomes too large (see \eq{vevexact}).
The demand that the scale of the effective $\mu$ term should be the soft SUSY breaking scale in terms of the solution of the  $\mu$ problem sets the following upper bound on the parameter $\lambda'$ \cite{Abel:1995wk}, 
\beq
\lambda' \lesssim 10^{-11}.
\label{tad11}
\eeq
In  this estimation, $\lambda$ and $\kappa$ are assumed $\mathcal{O}(1)$.

\paragraph*{}
As one can see, two demands for the solution of the domain wall problem \eq{DW7} and for the solution of the tadpole problem \eq{tad11} are inconsistent obviously.
Therefore, the NMSSM with $\mathbb{Z}_3$ symmetry, which is explicitly broken by the Planck suppressed operators,  can {\it not} solve the domain wall problem and tadpole problem simultaneously \cite{Bagger:1993ji, Bagger:1995ay, Abel:1995wk,Abel:1996cr}\footnote{
Hamaguchi,  Nakayama and Yokozaki have pointed out that  the NMSSM in gauge mediation SUSY breaking  with vector-like exotic matters,  which are charged under the hidden QCD,  can solve the domain wall and tadpole problem \cite{Hamaguchi:2011nm,Hamaguchi:2011kt}.
In this model, $\mathbb{Z}_3$ symmetry is anomalous, so that the domain wall problem can solve \cite{Preskill:1991kd}.
}.

\section{Solution of $\mu$ Problem,  Domain Wall Problem and Tadpole Problem}
\label{3no3}

 Refs.~\cite{Panagiotakopoulos:1998yw,Panagiotakopoulos:1999ah,Panagiotakopoulos:2000wp,Dedes:2000jp} have shown that the desecrate $\mathbb{Z}_5 $ R-symmetry, which can control the superpotential $f[\hat{S}]$ and the singlet scalar vev, can avoid the the domain wall problem and tadpole problem simultaneously.
So, we review this symmetry in the following.

When the superpotential $f[\hat{S}] = 0$, apart from ordinary Lepton and Baryon number symmetries,  the Lagrangian has two additional global continuous symmetries.
That is the anomalous Peccei-Quinn symmetry U(1)$_{PQ} $ and a non-anomalous R-symmetry U(1)$_R$. The charge assignments for these symmetries is given in Table.~\ref{tab_ch3}.
Therefore, 
the Lagrangian is also invariant under global U(1)$_{R'}$ transformation, where charges of U(1)$_{R'}$  symmetry are defined as 
\beq
R' = 3 R + PQ.
\eeq
In order to avoid an unwanted  visible Nambu-Goldstone boson related with the spontaneous  global U(1)$_{R'}$ symmetry breaking, let us introduce the discrete symmetry.
Actually,  the maximal discrete sub-symmetries of U(1)$_{R'}$  is
discrete $\mathbb{Z}_5^R$  R-symmetry.
In this symmetry the superfields are transformed  as
\beq
\hat{\Phi }_i &\to& e^{i \fr{2 \pi q_i }{5}  } \hat{\Phi}_i,\non
W &\to& e^{i \fr{2 \pi }{5}  } W, 
\eeq
where  the charge assignments $q_i $ is listed in Figure~\ref{tab_ch3}.

When one imposes  this discrete R-symmetry on both the K\"{a}hler potential and the superpotential, the Lagrangian becomes the desired form up to a possible
singlet tadpole term, which is  generated from the non-renormalizable sector.
 The discrete symmetry is imposed up to the higher dimensional K\"{a}hler potential and superpotential completely.
 This is the different point to the previous section.
 Then,  
 the desecrate $\mathbb{Z}_5^R$ R-symmetry is spontaneously broken by \coll{R-symmetry breaking sector $W_{\not{R}}$, whose $\mathbb{Z}_5^R$ charge is one. This R-symmetry breaking sector generates the vev of the superpotential $W_0$, which is required for small cosmological constant. In addition, $W_0$ is related with the gravitino mass as follows, 
 \beq
 \langle W_{\not{R}} \rangle = W_0 = m_{3/2} M_{\textrm{Pl}}^2.
 \eeq} 

 \coll{The tadpole is generated by the following higher dimensional interaction with the R-symmetry breaking sector,
 \beq
 K &=& \frac{c_1}{M_{\rm Pl}^2}W_{\not{R}} \hat{S}+\hc, \non
 W &=& \frac{c_2}{M_{\rm Pl}^4} W_{\not{R}}^2 \hat{S},
 \eeq
 where $c_1$ and $c_2$ are dimensionless $\mathcal{O}(1) $ coupling constants.
 Because of an imposition of the discrete R-symmetry,  a harmful quadratic tadpole divergences does not appear from loop diagrams. }
\coll{
 Once the discrete R-symmetry is broken by $W_0$, the tadpoles of the singlet fields are induced, 
  \beq
  \mathcal{L}_{\rm tad} \sim \susy^3 S +   \susy^2  F_S + \hc,
  \eeq
  where we assume $m_{3/2} \sim \susy$.}
As we will show next section explicitly, the singlet scalar obtains the following appropriate vev (see \eq{vevexact}),
\beq
\langle S \rangle \sim - \fr{t_S}{m_S^2} \sim \mathcal{O}(\susy),
\eeq
with
\beq
t_S \sim \susy^3.
\eeq
Therefore, the generated tadpole and effective $\mu$  term do not destabilize the hierarchy, and this symmetry can naturally solve the $\mu $ problem and tadpole problem. 

Furthermore, the existence of the  tadpole term, which is effectively generated by the R-symmetry breaking, breaks the discrete symmetry explicitly.
Thus, this symmetry can naturally avoid the domain wall problem \cite{Panagiotakopoulos:2000wp}. 

Hence, the singlet extension model imposed the desecrate $\mathbb{Z}_5^R$  R-symmetry can  
naturally solve three problems simultaneously: the $\mu$ problem, the domain wall problem and the tadpole problem.
This singlet extension model is called 
``{\it  nearly minimal supersymmetric standard model (nMSSM)} \cite{Panagiotakopoulos:1999ah, Panagiotakopoulos:2000wp, Dedes:2000jp}.


In addition, the desecrate $\mathbb{Z}_5^R$  R-symmetry prohibits  the dangerous $D \leq 5 $ Baryon or Lepton violating operators like $\hat{Q}\hat{Q}\hat{Q}\hat{L}$ and $\hat{\bar{U}}\hat{\bar{U}}\hat{\bar{D}}\hat{\bar{E}}$ in the nMSSM matter contents \cite{Panagiotakopoulos:2000wp}. Thus, the constraint of the proton decay is satisfied.

 Next section, we review the Lagrangian and the Higgs sector of the nMSSM. 

\section{Nearly MSSM}
\label{3no4}

In this section, we review the Lagrangian, the Higgs sector and the Landau pole constraint of the nearly minimal supersymmetric standard model \cite{Panagiotakopoulos:1999ah, Panagiotakopoulos:2000wp, Dedes:2000jp}.

 \subsection{Lagrangian}
 \label{nMSSMLag}

In the nMSSM, the superpotential is given as
\beq
	W &= \lambda \hat{S} \hat{H}_{2}  \hat{H}_1 + \frac{m_{12}^2}{\lambda} \hat{S} + W_{\textrm{Yukawa}}\,,
	\label{nMSSMsp}
\eeq
where $W_{\textrm{Yukawa}}$ is defined as \eq{yukawasp}.
The soft SUSY breaking terms are given as 
\beq
V_{\rm soft}&=&m_1^2 |H_1|^2   + m_2^2  |H_2|^2 +   m_S^2|S|^2+
\left( \lambda A_\lambda H_2 H_1 S+t_S S+\hc  \right)  \non
 & &+V_\textrm{soft~gaugino}+V_\textrm{soft~Yukawa}\,,
 \label{nMSSMsb}
\eeq
where   $V_\textrm{soft~gaugino}$ and $V_\textrm{soft~Yukawa}$ are defined as Eqs.~(\ref{gauginomass},~\ref{softyukawa}).
As discussed previous section, although the terms $m_{12}^2$ and $t_S$ are forbidden by the discrete $\mathbb{Z}^R_5$ R-symmetry, when the R-symmetry is broken 
they are generated.
Let us parameterize these tadpole terms as follows,
\beq
m_{12}^2 & = & \lambda c_F \susy^2\,, \\
t_S & = &c_S \susy^3\,, 
\eeq
where $\susy$ denotes the SUSY breaking scale. 
Here, $c_F$ and $c_S$ are $\mathcal{O}(1)$ complex constants and then $m_{12}^2$ and $t_S$ become $O(\susy^2)$ and $O(\susy^3)$ respectively\footnote{
Although the trilinear $\kappa S^3$ term is also generated, it is highly suppressed by
 Planck scale.}.
With these values, $S$ has a vacuum expectation value $\langle S \rangle \sim -t_S/m_S^2\sim O(\susy)$ as we will discuss next subsection.
 Thus the generated effective $\mu$ term is $O(\susy)$ and the $\mu$ problem can be solved.

\paragraph*{}
Note that this model is imposed the desecrate  $\mathbb{Z}^R_5$  R-symmetry, which is broken by the R-symmetry breaking terms in the hidden sector.
Then, as a low-scale (TeV scale) effective theory, the nMSSM Lagrangian Eqs.~(\ref{nMSSMsp},~\ref{nMSSMsb}) are generated.
It is known that, however, there are other models which generates  nMSSM Lagrangian Eqs.~(\ref{nMSSMsp},~\ref{nMSSMsb}) as a low-scale (TeV scale) effective theory.
An example is Peccei-Quinn invariant NMSSM \cite{Jeong:2011jk,Choi:2013lda}.
The Lagrangian of this model is given as,
\beq
\mathcal{L} = \int d^2 \theta \lambda \hat{S} \hat{H}_2 \hat{H}_1 + \int d^4 \theta \fr{\kappa }{M_{\rm Pl}} ( \hat{X}^{\dag 2} \hat{S} + \hc),
\eeq 
where $\hat{X}$ is an axion superfield.
This Lagrangian is controlled by U(1) Peccei-Quinn symmetry for ($\hat{H}_1,~\hat{H}_2, ~\hat{S},~\hat{X}$) carrying the charges as (1, 1, -2, -1).
The vev of the scalar component of $\hat{X}$ becomes the axion decay constant $f_a \sim 10^{10}$-$10^{12}$ GeV, and the vev of the auxiliary field $F_X$ is zero.
Then, similarly to the Giudice Masiero mechanism, by the supergravity interaction  the holomorphic term in the K\"{a}hler potential can be rewritten to the superpotential.
As a result, the (TeV scale) effective superpotential is given as
\beq
W_{\rm eff} = \lambda \hat{S} \hat{H}_2 \hat{H}_1  + \kappa m_{3/2} \fr{f_a^2}{M_{\rm Pl}} \hat{S}.
\eeq 
As one can see, the tadpole is generated and its dimensional coupling is actuary $\mathcal{O}(\susy^2)$. 
Therefore, this model can solve not only $\mu$ problem, the tadpole problem but also the strong CP problem, and effective Lagrangian can be regarded as the nMSSM.

Other example is secluded U(1$)'$-extended minimal supersymmetric standard model (sMSSM) \cite{Barger:2006dh}.
The sMSSM contains a U(1$)'$ gauge symmetry, $Z'$ gauge boson, and  four  additional singlets.
The superpotential is 
\beq
W&=&  \lambda \hat{S} \hat{H}_2 \hat{H}_1 + \lambda_s \hat{S}_1 \hat{S}_2 \hat{S}_3  + \mu_1 \hat{S} \hat{S}_1 + \mu_2 \hat{S} \hat{S}_2,
\eeq 
and this additional gauge symmetry is motivated by the GUT.
The U(1$)'$ charges satisfy $q_{H_1} + q_{H_2} + q_{S} = 0$ and $- q_{S} = q_{S_1} = q_{S_2} = - 1/ 2 q_{S_3}$.
If the dimensional couplings $\mu_i$ is controlled by some mechanisms, and three singlet $\hat{S}_i$ masses  are  heavier than soft SUSY breaking scale, 
the effective Lagrangian can be regarded as the nMSSM,
\beq
W_{\rm eff} = \lambda \hat{S} \hat{H}_2 \hat{H}_1  + (\mu_1 \langle S_1 \rangle + \mu_2 \langle S_2 \rangle )\hat{S}.
\eeq

\coll{Other example is Fat Higgs model \cite{Harnik:2003rs}.
In the Fat Higgs model, the Higgs doublet fields  are  composite bound states of fundamental  fields, which 
couple to a extra supersymmetric strong SU(2) gauge.  This gauge theory is UV complete and calculable.
Below a scale $\Lambda_H$, although the strong SU(2) gauge theory becomes non-perturbative, the theory can be described by the composite fields and their perturbative couplings.
As a low energy effective Lagrangian, the following superpotential is dynamically generated, 
\beq
W_{\rm eff} = \lambda \hat{N} (\hat{H}_1 \hat{H}_2 - v_0^2),
\eeq
where $\hat{N}$ is a SU(2$)_L$ singlet composite field and $\hat{H}_1,~ \hat{H}_2$ are SU(2$)_L$ doublet  composite fields.
The coupling $\lambda$ is generated at $\lambda(\Lambda_H) \sim 4 \pi$, and it decreases towards the low energy scale due to the RGEs.
$v_0^2$ is given by $v_0^2 \sim m \Lambda_H /(4 \pi)^2 $, where $m$ is a fundamental parameter in the UV theory. 
The tadpole of $\hat{N}$ is generated and $N$ obtains a its suitable vev. 
Therefore, Fat Higgs model also can solve $\mu$ problem, and the effective Lagrangian can be regarded as the nMSSM.
One of the features of this model is that the coupling $\lambda$ is typically $\mathcal{O}(1)$. 
}

The important thing is that the phenomenology of the nMSSM  is almost independent  of the symmetry of a 
UV theory (i.e. the desecrate  $\mathbb{Z}^R_5$ symmetry, the U(1) Peccei-Quinn symmetry, the secluded U(1$)'$ gauge symmetry,  \coll{the strong SU(2) gauge symmetry,} etc.).
Therefore, we  consider only the nMSSM Lagrangian Eqs.~(\ref{nMSSMsp},~\ref{nMSSMsb}) as a low-scale (TeV scale) effective theory  in the next two chapters.
We will discuss the phenomenology of the nMSSM, especially the dark matter and the baryon asymmetry of the universe.

 \subsection{Higgs Sector}

In the nMSSM, a Higgs potential is  given as,
\beq
V_0 &=& m_1^2 |H_1|^2 + m_2^2 |H_2|^2 + m_S^2 |S|^2 + \lambda^2 |H_2 H_1|^2  + \lambda^2 |S|^2 (|H_1|^2 + |H_2|^2) \non
 && + \frac{\bar{g}^2}{8} (|H_2|^2 - |H_1|^2)^2 + \frac{g^2}{2}|H_1^{\dag}H_2|^2\non
 &&
 + (\lambda A_{\lambda} S H_2  H_1 + t_S S + m_{12}^2 H_2  H_1 + H.c. ),
 \label{HiggsV}
\eeq 
where $\bar{g}^2$ is defined as $\bar{g}^2=g'^2+g^2$ where $g' ~(g)$ is the U(1$)_Y$ (SU(2)) gauge coupling constant.
In this potential, there are seven independent parameters,
\beq
\lambda,~m_1^2,~m_2^2,~m_S^2,~A_{\lambda},~m_{12}^2,~t_S.
\label{nMSSMpara}
\eeq
Thanks to SU(2) rotation, we can take $\langle H_1^{-} \rangle = 0$ at elsewhere.
Thus, when  one takes $H_1^{-}  = 0$, the Higgs potential can be expanded as follows
\beq
V_0 &=& m_1^2 |H_1^0|^2 + m_2^2 (|H_2^0|^2+ |H_2^{+}|^2) + m_S^2 |S|^2 + \lambda^2  |H_1^0|^2 |H_2^0|^2    + \lambda^2 |S|^2 (|H_1^0|^2 + |H_2^0|^2 +  |H_2^{+}|^2) \non
 &+& \frac{\bar{g}^2}{8} (  | H_1^0|^4 +  |H_2^0|^4 +  |H_2^{+}|^4 - 2 |H_1^0|^2 |H_2^0|^2 -2 |H_1^0|^2   |H_2^{+}|^2     +2  |H_2^0|^2 |H_2^{+}|^2 )\non
 & +& \frac{g^2}{2} |H_1^0|^2   |H_2^{+}|^2    + ( - \lambda A_{\lambda} S H_1^{0} H_2^0  + t_S S - m_{12}^2 H_1^0 H_2^0 + H.c. ).
 \label{Higgspotekuwashi}
\eeq  

Next, the vacuum fluctuations of the scalar field are defined as
\beq
H_1^0 &=& v_1 + \frac{1}{\sqrt{2}} ( H_{1R}+i H_{1I} ), \non
H_2^0 &=& v_2 + \frac{1}{\sqrt{2}} ( H_{2R}+i H_{2I} ),\non
S&=& s + \frac{1}{\sqrt{2}} ( S_{R}+i S_{I} ),
\eeq
here we can choose  $\langle H_2^{0} \rangle = v_2$ to be real and positive by  phase redefinition of $H_2$.
The minimization conditions for electroweak symmetry breaking give the following conditions,
\beq
m_1^2 &=& (m_{12}^{2} + \lambda A_{\lambda} s )^{\ast}\fr{v_2}{v_1} - \fr{\bar{g}^2}{4} (|v_1|^2 - v_2^2) -|\lambda|^2 (v_2^2  + |s|^2),\\
m_2^2 &=& (m_{12}^{2} + \lambda A_{\lambda} s)^{\ast}\fr{v_1^{\ast}}{v_2} + \fr{\bar{g}^2}{4} (|v_1|^2 - v_2^2) -|\lambda|^2 (|v_1|^2  + |s|^2),\label{mannaka} \\
m_S^2 &=& \lambda^{\ast } A_{\lambda}^{\ast} \frac{v_1^{\ast} v_2 }{s}  - \frac{t_S^{\ast}}{s} - |\lambda|^2 (|v_1|^2 + v_2^2).
\label{saigono}
\eeq
By the phase redefinition of $H_2 H_1$, one can choose $(m_{12}^{2} + \lambda A_{\lambda} s)$ to be real and positive, and so it leads $v_1$ is real form \eq{mannaka}.
Similarly to the MSSM, let us define the parameter $\tb$,
\beq
\tan\beta = \fr{v_2}{v_1}, ~~~~~~v_{EW}  = \sqrt{v_1^2 + v_2^2 } = 174.1\GeV.
\eeq
In addition, the \eq{saigono} leads to the vev of the singlet scalar, 
\beq
s = - \frac{t_S^{\ast} - \lambda^{\ast} A_{\lambda}^{\ast} v_1 v_2}{m_S^2 + |\lambda|^2 v_{EW}^2}.
\label{singletvev}
\eeq
Here, we can choose  $s$  and $({t_S^{\ast} - \lambda^{\ast} A_{\lambda}^{\ast} v_1 v_2})$ to be real  by  phase redefinition of $S$.
At the this time, the Higgs potential \eq{HiggsV} can obtain an e effective $\mu$ term and a effective $B\mu$ term (see \eq{VN2}),
\beq
\mu_{\rm eff} &=& \lambda s,\label{mueff}\\
(B\mu)_{\rm eff} &=& 2 (m_{12}^{2} + \lambda A_{\lambda} s).
\eeq
Since $m_{12}^2 \simeq \mathcal{O}(\susy^2)$ and $t_S \simeq \mathcal{O}(\susy^3)$, \eq{singletvev} leads to $s \simeq \mathcal{O}(\susy) $.
Therefore, $\mu_{\rm eff}  \simeq \mathcal{O}(\susy)$ and $(B\mu)_{\rm eff} \simeq \mathcal{O}(\susy^2)$ are realized.

\paragraph*{}
Note that since this model has four complex parameter \eq{nMSSMpara} and one can carry out three phase  redefinition,  {\it one} physical CP-violating phase remains in the Higgs sector.
If one take $A_{\lambda} $ to be zero, there is no CP-violating phase in the Higgs sector.
But relative phase with gaugino masses and other A terms, arg$(m_{12}^2 t_S^{\ast} M_{\rm gaugino})$ and arg$(m_{12}^2 t_S^{\ast} A_i)$, keep being the physical CP-violating phase \cite{Balazs2007}.

\paragraph{Mass of Scalar bosons}

~

We assume that $\lambda, ~t_S, ~m_{12}^2$ and $ A_{\lambda}$ are real in the following for simplicity. Then, a classification according to the CP-even or CP-odd for the scalar boson is justified.
The singlet superfiled consists of singlet CP-even and CP-odd scalar components and singlet Majorana spinor component.
Therefore, the nMSSM has $3 \times 3 $ CP-even mass matrix, $3 \times 3 $ CP-odd mass matrix and $5 \times 5 $ neutralino mass matrix, which are points of difference in the MSSM.


Now, one can obtain the following mass matrix for the CP-even and CP-odd Higgs bosons from the Higgs potential \eq{Higgspotekuwashi},
\begin{eqnarray}
-\mathcal{L}_{\rm CP-even/odd} &=&\frac{1}{2} \begin{pmatrix} H_{1R}  H_{2R}  S_{R} \end{pmatrix} \begin{pmatrix}  \mathcal{M}^2_{R 11} &\mathcal{M}^2_{R 12} &\mathcal{M}^2_{R 13}\ \\ 
\mathcal{M}^2_{R 12} &\mathcal{M}^2_{R 22} &\mathcal{M}^2_{R 13} \\
\mathcal{M}^2_{R 13} &\mathcal{M}^2_{R 23} &\mathcal{M}^2_{R 33}   \end{pmatrix}
 \left( \begin{array}{ccc} H_{1R} \\ H_{2R} \\ S_{R}\end{array} \right)  \nonumber \\ 
& & + \frac{1}{2} \begin{pmatrix} H_{1I}  H_{2I}  S_{I} \end{pmatrix} \begin{pmatrix}  \mathcal{M}^2_{I 11} &\mathcal{M}^2_{I 12} &\mathcal{M}^2_{I 13}\ \\ 
\mathcal{M}^2_{I 12} &\mathcal{M}^2_{I 22} &\mathcal{M}^2_{I 13} \\
\mathcal{M}^2_{I 13} &\mathcal{M}^2_{I 23} &\mathcal{M}^2_{I 33}   \end{pmatrix}
 \left( \begin{array}{ccc} H_{1I} \\ H_{2I} \\ S_{I} \end{array}\right),
 \label{Hmasstree}
 \end{eqnarray}
with
\beq
\mathcal{M}^2_{R 11}&=&M_A^2 \sin ^2 \beta + M_{Z}^2 \cos ^2 \beta ,\\
\mathcal{M}^2_{R 22}&=&M_A^2\cos ^2 \beta + M_{Z}^2 \sin ^2 \beta ,\\
\mathcal{M}^2_{R 33}&=& - \frac{1}{s}( t_S - \lambda  A_{\lambda} v_{EW}^2 \sb \cb ) = m_S^2 + \lambda^2 v_{EW}^2,\\
\mathcal{M}^2_{R 12}&=& -\frac{1}{2}(M_{A}^2+M_{Z}^2 - 2 \lambda ^2 v_{EW}^2)\sin 2\beta ,\\
\mathcal{M}^2_{R 13}&=& v_{EW} (2  \lambda^2 s \cb - \lambda A_{\lambda}\sb),\\
\mathcal{M}^2_{R 23}&=& v_{EW} (2  \lambda^2 s \sb - \lambda A_{\lambda}\cb),
\eeq
\beq
\mathcal{M}^2_{I 11}&=&M_A^2\sin ^2 \beta, \\
\mathcal{M}^2_{I 22}&=& M_A^2\cos ^2 \beta ,\\
\mathcal{M}^2_{I 33}&=& - \frac{1}{s}( t_S - \lambda  A_{\lambda} v_{EW}^2 \sb \cb ) = m_S^2 + \lambda^2 v_{EW}^2,\\
\mathcal{M}^2_{I 12}&=& \frac{1}{2}M_{A}^2  \sin 2\beta,\\
\mathcal{M}^2_{I 13}&=&  \lambda A_{\lambda }v_{EW} \sin \beta,\\
\mathcal{M}^2_{I 23}&=&  \lambda A_{\lambda }v_{EW} \cos \beta ,\label{Hmasstree2}
\eeq
where we have defined the  following $M_A$,
\begin{eqnarray}
M_{A}^2 & =  & m_1^2 + m_2^2 + 2 \mu_{\rm eff}^2  + \lambda^2 v_{EW}^2 
       =\frac{(B \mu)_{\textrm{eff}} }{\sin 2 \beta}.
\end{eqnarray}
The determinant of the  CP-odd mass matrix is zero, and so the Nambu-Goldstone boson apperas.

At the large $\susy$ case, these mass matrices give
\beq
m_H^2 \sim M_A^2, ~~~
m_{HS}^2 \sim m_S^2,~~~~~~~~~~~\non
m_h^2 \sim M_Z^2 \cos^2 2\beta + \lambda^2 v_{EW}^2 \left( 1 - \frac{A_{\lambda}^2}{m_S^2}\right) \sin^2 2 \beta, \non
m_A^2 \sim M_A^2, ~~~m_{AS}^2 \sim m_S^2,~~~~~~~~~~~\label{nMSSMHiggsmass}
\eeq
where ${HS}$ (${AS}$) is the heavy CP-even (-odd) scalar boson, whose component is mainly $S$, and $h$ is the SM-like Higgs boson.
Note that the mass of the SM-like Higgs boson can become large at a low $\tb$ region in comparison with the MSSM.
It is given by an additional F-terms as we have discussed in Section~\ref{125gev}.

However, the radiative contributions to the Higgs boson mass are still important.
In order to estimate the mass of the SM-like Higgs boson including two-loop radiative corrections, 
we have extended the two-loop level calculation using the RGE of the MSSM, which  is given in  Section~\ref{HiggsmasscalcRGE}.
In the nMSSM, since the SM-like Higgs boson receives an extra F-term contribution to the Higgs  quartic coupling $\lambda_\textrm{quartic}$, there is a sizable tree-level contribution to the Higgs boson mass.
When integrating out heavy SUSY particles, the matching condition at high scale from the SUSY to the SM is shifted by~\cite{Giudice:2011cg}
\beq
\lambda_{\rm quartic} (\susy) = \lambda_{\rm LO}(\susy) +\delta \lambda_{\textrm{quartic}}(\susy)  + \frac{1}{(4\pi)^2}  \lambda_{\rm NLO}(\susy),
\label{matchhigh}
\eeq
with
\beq
\label{masscor}
\delta \lambda_{\textrm{quartic}} \simeq \frac{\lambda^2 }{2} \frac{m_S^2 - A_{\lambda}^2}{m_S^2} \sin^2 2 \beta\,,
\eeq
where $ \lambda_{\rm LO}$ is given by \eq{treematch}, and $\lambda_{\rm NLO} $ is given by \eq{NLOmatch}.
Therefore, at large $\lambda $ and small $\tan \beta$ can give an additional  sizable contribution to the Higgs boson mass.
Note that this extra contribution can be controlled by $A_{\lambda}$.
\coll{We have not considered extra loop corrections in $  \lambda_{\rm NLO}$ term, which depend on the coupling $\lambda$. 
When $\lambda$ is larger than $1$, these extra radiative corrections become significantly contributions to the SM-like Higgs boson mass \cite{Nakayama:2011iv}. 
In this thesis, we have taken $\lambda \lesssim 0.8$ due to consideration to a Landau pole constraint below the GUT scale\footnote{The detail of the Landau pole  is written in the next subsection.}, and thus these extra radiative corrections do not give the significantly contributions.}

\paragraph*{}

Next, one can obtain the following mass matrix for the charged Higgs boson from the Higgs potential \eq{Higgspotekuwashi},
\beq
-\mathcal{L}_{\text{charged}}&=& \left( \frac{(B \mu)_{\rm eff} }{2} +(M_W^2 - \lambda^2 v_{EW}^2 )\frac{ \sin 2 \beta}{2} \right) \begin{pmatrix} H_1^{-} H_2^{+ \ast}\end{pmatrix} \begin{pmatrix}\tan \beta &1\\
1 & \cot \beta \end{pmatrix}
\begin{pmatrix}H_1^{- \ast}\\ H_2^{+}\end{pmatrix}, \non
&=&  \left( \frac{(B \mu)_{\rm eff}}{\sin 2 \beta}  +M_W^2 - \lambda^2 v_{EW}^2  \right)\begin{pmatrix} H^{+ \ast} NG^{+ \ast}\end{pmatrix} \begin{pmatrix} 1 &0 \\
0 & 0 \end{pmatrix}
\begin{pmatrix}H^{+}\\ NG^{+}\end{pmatrix},
\eeq
where 
\beq
H^{+} &=& - \sb H_1^{-\ast} - \cb H_2^+,\non
 NG^{+} &=& -\cb H_1^{-\ast} + \sb H_2^+. \label{chargedcoupling}
 \eeq
Thus, the mass of the charged Higgs boson is given as,
\beq
m^2_{H \pm }&=&  M_A^2+M_W^2 - \lambda^2 v_{EW}^2.\label{ch1}
\eeq
Note that the mass of the charged Higgs boson is lighter than the one of the MSSM by $\lambda^2 v_{EW}^2$ term (see \eq{MSSMcharged}).
Thus,  the theoretical condition $M_A^2 + M_W^2 > \lambda^2 v_{EW}^2$ is needed  to avoid a vev of the charged Higgs scalar field  in the nMSSM.

 \subsection{Landau Pole Constraint}
 \label{Landausec}

A large $\lambda$ coupling can raise the mass of the Higgs boson at the tree level.
However, when the  $\lambda$ coupling is too large at low energy, it causes a Landau pole (or Landau singularity) at a higher energy scale.
The Landau pole means that some dimensionless running coupling constants become non-perturbative couplings at the finite energy scale, and so  a perturbative approximation  is broken down by the strong (non-perturbative) couplings.
In the nMSSM, a one-loop RGE for $\lambda$ is given as
\beq
\fr{d \lambda}{d \textrm{ln}Q} = \fr{1}{(4 \pi)^2} \lambda \left( 4 {\lambda}^2    +3 y_t^2   +3 y_b^2  + y_{\tau}^2 - g'^2  -3 g^2 \right). 
\label{kurikomiii}
\eeq
This RGE implies that an absolute value of $\lambda$ monotonically increases with energy scale $Q$, and so $\lambda$ may eventually  become non-perturbative. 
Let us demand that the theory is perturbative ($\lambda $ and other dimensionless couplings do not blow up) below the GUT scale, $M_{\rm GUT} \simeq 2 \times 10^{16}\GeV$.
Then, one can obtain the upper bound on $\lambda$ at the soft SUSY breaking scale.

Figure \ref{Landau} shows the  upper bound on the coupling $\lambda ~ (\lambda_{\rm max})$ under the condition of the no Landau pole up to  the GUT scale as a function of $\tb$.
Here, we have demanded $g_i^2(M_{\rm GUT})/4 \pi < 1$, where $g_i$ denotes all dimensionless couplings. 
Here we have used two-loop RGEs for calculations of the running couplings.
The two-loop level RGEs for all couplings are summarized in Appendix \ref{RGESUSY}.
$\lambda_{\rm max}$  is the values at $Q = \susy$, where we have used the SM RGEs below the soft SUSY breaking scale $\susy$ and the matching condition at electroweak scale (see Section~\ref{HiggsmasscalcRGE}).
We take $\susy = 1\TeV$ (dotted line), $10\TeV$ (dashed line) and $100\TeV$ (solid line).

\begin{figure}[t]
\begin{center}
\includegraphics[width =13cm]{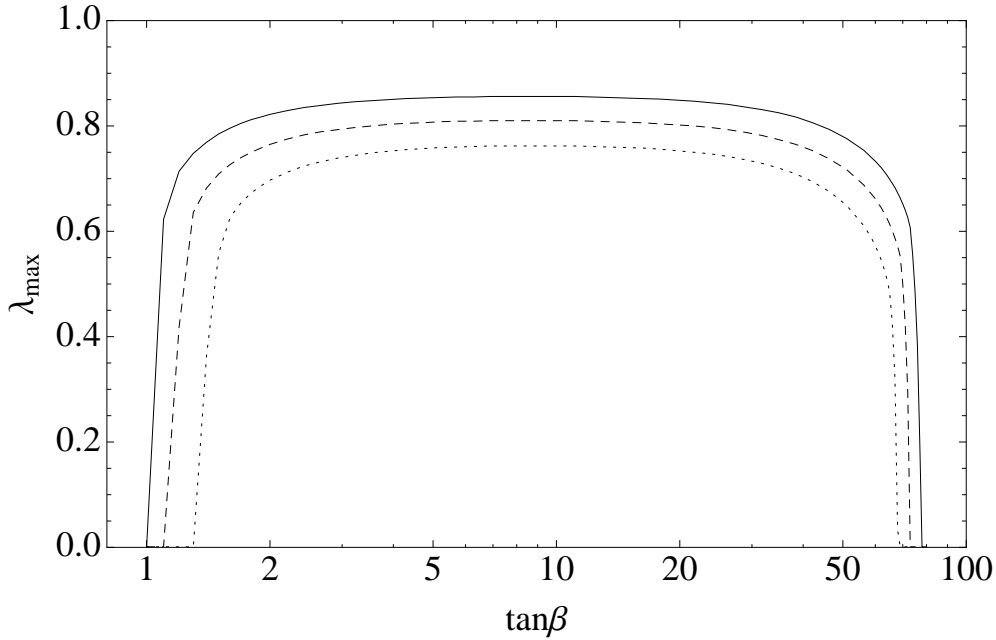}
\caption[The upper bound on the coupling $\lambda~ (\lambda_{\rm max})$ under the condition of the no Landau pole up to  the GUT scale.]{The upper bound on the coupling $\lambda ~ (\lambda_{\rm max})$ under the condition of the no Landau pole up to  the GUT scale using two-loop RGEs.
 The horizontal axis is $\tb$. 
$\lambda_{\rm max}$  is the values at $Q = \susy$. 
We take $\susy = 1\TeV$ (dotted line), $10\TeV$ (dashed line) and $100\TeV$ (solid line).}
\label{Landau}
\end{center}
\end{figure}

We find that the  $\lambda$ coupling should be smaller than 0.75-0.85, which depends on the soft SUSY breaking scale.
Note that this bound can be alleviated by introducing an additional gauge symmetry or extra particles \cite{Kyae:2012df}. It is because that the contributions of additional gauge coupling to the RGE for $\lambda$ is negative or  the ordinary gauge couplings $g',~g$ become large at the high energy scale due to the extra matters, and so its contributions to the RGE for $\lambda$ is negative (see \eq{kurikomiii}).

At low $\tb$ region, $\tb \lesssim 1.5$, the upper bound on  $\lambda$ decreases drastically.
It is because that the top Yukawa coupling becomes large values at the low energy scale, $y_{t}  = m_t/(v_{EW} \sb)$, then it causes the Landau pole of top Yukawa coupling at the high energy scale.
While, at large $\tb$ region, $\tb \gtrsim 65$, the upper bound on  $\lambda$ also decreases drastically. It is caused by the bottom/tau Yukawa couplings, which become large values  at the low energy scale, $y_{b/\tau}  = m_{b/\tau}/(v_{EW} \cb)$.


\chapter{Resonant Singlino Dark Matter}
\label{singlinochap}
\thispagestyle{empty}

\abstchapter{
This chapter is based on the work by the author \cite{Ishikawa:2014owa}.
We consider a singlino dark matter scenario in the nearly minimal supersymmetric standard model. We find that with high-scale supersymmetry breaking the singlino can obtain a sizable radiative correction to the mass, which opens a window for the dark matter scenario with resonant annihilation via the exchange of the Higgs boson. 
We show that the current dark matter relic abundance and the Higgs boson mass can be explained simultaneously. 
This scenario can be probed by the search of the Higgs invisible decay and the direct direction of the dark matter.}

We have shown that the nMSSM can solve the $\mu$ problem, the domain wall problem and the tadpole problem simultaneously in the previous chapter.
Next two chapter, we will consider the phenomenology of the nMSSM, which are the based on the works by the author  \cite{Ishikawa:2014owa, Ishikawa:2014tfa}.
As a results, both the two studies conclude  that the nMSSM with a high-scale SUSY breaking is valid.

\section{Dark matter in the nMSSM}
\label{4no1}

In this chapter, we focus on the dark matter phenomenology in the nMSSM.
This chapter is based on the work by the author \cite{Ishikawa:2014owa}.
First, we briefly review a situation of the dark matter in the nMSSM, and we also explain why we have considered it. 

Recent various cosmological observations have established the $\Lambda$CDM cosmological model and the relic abundance of the cold dark matter is measured accurately by WMAP and Planck~\cite{Hinshaw:2012aka,Ade:2013zuv}. 
In the nMSSM,  as discuss later, the singlino, which is the fermionic component of the extra gauge singlet superfield, can be a candidate of the dark matter~\cite{Dedes:2000jp,Menon:2004wv,Barger:2005hb,Balazs:2007pf,Cao:2009ad,Wang:2012ry}.
In fact,  the singlino mass and its couplings with SM particles are  suppressed by the soft SUSY breaking scale in the nMSSM, and such a dark matter  leads to the overabundance in the universe.
Therefore, 
the singlino dark matter scenario seems to be incompatible with relatively high-scale (TeV scale) supersymmetry breaking, which is inferred from the measured SM Higgs boson mass~\cite{ATLAS:2013mma, CMS:yva} and the null signals of the sparticle searches at the LHC~\cite{TheATLAScollaboration:2013fha, CMS:2013cfa}.

However, if one-loop corrections to the singlino mass are taken into account, this situation will change.
As will be shown later, similar to the mass of the SM Higgs boson in the SUSY models, the singlino can obtain a sizable mass, which opens a window for a resonant dark matter scenario via the s-channel annihilation with the exchange of the SM Higgs boson.

Let us consider the helicity of the annihilation process of the neutralino dark matter to fermions $\chi \chi\to \bar{f}f$.
Since  neutralinos are Majorana fermions, the helicity of the two neutralinos are opposite to each other in the center-of-momentum frame when the case that neutralino pair annihilates via the CP-even scalar boson s-channel exchange.
While the helicity of the final state two fermions are   
facing in the same direction.
Therefore this process needs either the $p$-wave ($L=1$) suppression or chirality suppression. 
Since the annihilation rate of the singlino via the SM Higgs boson s-channel exchange is also $p$-wave suppressed, one needs a relatively large value of SM Higgs boson-singlino coupling compared with the scalar dark matter in the Higgs portal model~\cite{Kanemura:2010sh}.
This fact implies that the singlino dark matter could be probed more easily than the scalar one.

\paragraph*{}
Now, let us focus on the phenomenology of the singlino in the nMSSM.
The singlino  participates in a member of usual neutralino as a new gauge eigenstate.
At the tree level, the $5\times 5$ neutralino mass matrix in the basis 
$X^{\rm T} = (\tilde{B}, \tilde{W}^0, \tilde{H}_1^0, \tilde{H}_2^0, \tilde{S})$ is given by
\begin{align}
\label{NMASS}
	\mathcal{L}&= - \frac{1}{2}X^{\rm T} \mathcal{M}_\text{tree}^{\chi^0} X +\hc,\\
	\mathcal{M}_\text{tree}^{\chi^0}&=
       \begin{pmatrix}	
	M_1 & 0 & -\frac{g' v_1}{\sqrt{2}}& \frac{g' v_2}{\sqrt{2}} &0\\
	 0& M_2 &  \frac{g v_1}{\sqrt{2}}& -\frac{g v_2}{\sqrt{2}} &0\\
	  -\frac{g' v_1}{\sqrt{2}}& \frac{g v_1}{\sqrt{2}}&0&-\mu_{\rm eff}&-\lambda v_2\\
	 \frac{g' v_2}{\sqrt{2}} & -\frac{g v_2}{\sqrt{2}} &-\mu_{\rm eff}&0&-\lambda v_1\\
	 0&0&-\lambda v_2&-\lambda v_1&0
	 \end{pmatrix}\,,
	 \label{Nmassmatrix}
\end{align}
where 
$\tilde{S}$ is the fermionic component of $\hat{S}$.
Since a  determinant of this mass matrix is nonzero, all (five) neutralinos obtain a nonzero mass. 
If one impose  additional 
 matter parity $P_M$, 
 the lightest neutralino becomes stable and can be a  candidate for the dark matter in the universe\footnote{ The desecrate $\mathbb{Z}_5^R$ 
   R-symmetry does not contain the R parity \eq{Rparity}.  
 In fact, although the R parity is conserved accidentally in the renormalizable teams, it is broken in the non-renormalizable terms: $\hat{S}\hat{S}\hat{L}\hat{H}_2$, $\hat{S}\hat{L}\hat{L}\hat{\bar{E}}$ and $\hat{S}\hat{L}\hat{Q}\hat{\bar{D}}$, 
 which  can not lead to an observable proton decay. 
Considering the effects of such as R parity breaking non-renormalizable (Planck suppressed) operators, 
the life time of the lightest neutralino is longer than the age of the universe \cite{Panagiotakopoulos:2000wp}.
However,   there are experimental bound on the lifetime of the dark matter from the cosmic ray searches, which is longer enough than the age of the universe. 
It implies that the life time of the dark matter which is comparable to the age of universe have been excluded. 
Therefore, we should impose the additional matter parity or some symmetries which do not forbid the tadpole.}.
\coll{The one of the  origin of the matter parity $P_M$ is  a remnant discrete subgroup of the local U(1$)_{B-L}$ symmetry. This U(1$)_{B-L}$ symmetry is broken above the electroweak scale.}
The neutralino mass matrix can be diagonalized by a unitary matrix $N$, 
\beq
N^{\ast } \mathcal{M}_\text{tree}^{\chi^0} \mathcal{M}_\text{tree}^{\chi^0,\dag} N^{{\rm T}}  = \textrm{diag}(m^2_{\chi^0_1},m^2_{\chi^0_2},m^2_{\chi^0_3},m^2_{\chi^0_4},m^2_{\chi^0_5}),
\eeq
here we call $\chi^0_1$ the lightest neutralino.

From the tree-level calculations, in a very well approximation the mass of the lightest neutralino is given as \cite{Hesselbach:2008vt},
\beq
|m_{\chi^0_1}^\textrm{tree}| = {\rm Min} \left[\frac{1}{2} \left| B - \sqrt{B^2 - 4 C}\right|, \frac{1}{2} \left| B + \sqrt{B^2  - 4 C} \right| \right], 
\eeq
where
\begin{align}
B = \frac{M_1 M_2}{M_1 + M_2} + \left(\frac{\nu^2}{\mu_{\rm eff}^2 + \nu^2 } - \frac{M_Z^2}{\mu_{\rm eff}^2 + \nu^2} \frac{\tilde{M}}{M_1 + M_2} \right) \mu_{\rm eff} \sin 2 \beta - \frac{M_Z^2 \nu^2}{(M_1 + M_2) (\mu_{\rm eff}^2 + \nu^2)},
\end{align}\begin{align}
C = \frac{\nu^2}{\mu_{\rm eff}^2 + \nu^2 } \left( \frac{M_1 M_2}{M_1 + M_2} \mu_{\rm eff} \sin 2 \beta - \frac{\tilde{M}}{M_1 + M_2} M_Z^2 \right),
\end{align}
with $\nu^2 =\lambda^2 v^2_{EW}$ and $\tilde{M} = M_1 \cos^2 \theta_{W} + M_2 \sin^2 \theta_W$.
When the case $M_Z \ll \mu_{\rm eff}$ and $M_Z \ll  M_{\rm gaugino}$, 
a mass-eigenstate neutralino whose component is mainly $\tilde{S}$ becomes  the lightest neutralino.
We denote $\tilde{s}$ as the mass-eigenstate neutralino whose component is mainly $\tilde{S}$. 
Note that, here and in the following, we call $\tilde{s}$ as a ``singlino" in order to make understanding easy.
Then, the mass of the singlino  $m_{\tilde{s}}$ is  evaluated by expansions of $M_Z/\mu_{\rm eff}$ and $M_Z/M_{\rm gaugino}$,
\begin{align}
|m_{\chi^0_1}^\textrm{tree}|=m_{\tilde{s}}^\textrm{tree}&  \simeq \frac{\mu_{\rm eff} \nu^2 }{ \mu_{\rm eff}^2 + \nu^2} \sin 2 \beta\\ 
& \sim \lambda^2 \frac{v_{EW}^2}{\susy} \sin{2\beta}\,, \label{treecoup} 
\end{align}
where we denotes the typical soft SUSY breaking scale by $\susy$ and we use the fact that a value of $\mu_{\rm eff}$ become $\mathcal{O}(\susy)$ \eq{mueff}.
As you can see, the mass of the singlino has suppressions by soft SUSY breaking scale and by $\sin 2 \beta$.
 Therefore, when the soft SUSY breaking scale is relatively high ($\susy \gtrsim 1\TeV$) as suggested by the LHC experiments~\cite{ATLAS:2013mma, CMS:yva, TheATLAScollaboration:2013fha, CMS:2013cfa}, the singlino becomes the LSP and it can be a candidate of the  dark matter.
Furthermore, by the tree-level analysis one can see $m_{\tilde{s}} \lesssim 50 \GeV$.

Since the singlet superfield $\hat{S}$ interacts only the Higgs multiplets, 
the singlino can be coupled with the SM particles only through the mixing with Higgsinos. 
Thus, the singlino-SM  particle coupling are suppressed by the soft SUSY breaking (Higgsino mass) scale in the nMSSM.
Moreover, since the singlino is the LSP it can not decay.
Generally, an annihilation cross section of  such a stable particle is small and it would freeze-out at relatively early time in the universe.
In other words, the relic density of such a stable particle  would be overabundant.
Therefore, 
the singlino dark matter typically leads to the overabundance in the universe, and one needs to dilute the relic density of the singlino by some mechanisms. 

In the literature, it is known that there are two solutions for avoiding overabundance of the singlino dark matter. 
First, when the lightest CP-odd Higgs boson $A_1$ is dominantly singlet-like and its mass is $m_{A_1} \sim 2 m_{\tilde{s}}$, the singlino annihilation cross section can be resonant via the s-channel $A_1$ exchange \cite{Cao:2009ad,Cao:2010na,Cao:2011pg,Cao:2012yn,Cao:2013gba}. 
This resonant annihilation cross section gives much dilution of singlino dark matter, then the relic density is suppressed. 
After discovering the SM Higgs boson $h$, however, this scenario is severely constrainted from branching ratio of $h \to \tilde{s}\tilde{s} $ (invisible) and $h \to A_1 A_1 \to $ All \cite{Cao:2012yn}.
Strictly speaking, if $\mu_{\rm eff} $ is light,  the mass spectrum becomes $m_{\tilde{s}} < m_h /2 < m_{A_1} \sim 2 m_{\tilde{s}}  $, then the SM Higgs boson decay to the two singlino is kinematically allowed.
While, if $\mu_{\rm eff} $ is heavy, the singlino mass becomes suppressed and the mass spectrum becomes $m_{\tilde{s}} <  m_{A_1} \sim 2 m_{\tilde{s}} < m_h /2 $, then the SM Higgs boson decay to the two $A_1$ is also kinematically allowed.
Such an additional decay channel of the SM Higgs boson is currently constrained from the  measurements of the Higgs coupling.

Next, when $m_Z \sim 2 m_{\tilde{s}}$, the singlino annihilation cross section can be  resonant via the s-channel $Z$ boson exchange \cite{Menon2004, Balazs:2007pf}.
This scenario is constrained from $Z \to \tilde{s}\tilde{s} $ and  $h \to \tilde{s}\tilde{s}$.
Furthermore, this scenario leads to light $\mu_{\rm eff}$ in order to obtain the sizable singlino mass $m_{\tilde{s}} \sim 45 \GeV$ \eq{treecoup}.
Although such a light soft SUSY breaking scale is favored with naturalness,
it is disfavored from the point of view of the SUSY flavor/CP problem.

However, these arguments are incomplete.
It is because they are given by the tree-level analyses, and in fact, we find at first time that the one loop radiative corrections give significantly contributions to the mass of the singlino.
We will show the numerical analysis of the full one-loop level singlino mass in Section~\ref{singlinomass}.
Interestingly, we find that this radiative correction is roughly proportional to the soft SUSY breaking scale, $m_{\tilde{s}}^{\rm 1-loop} \propto \susy / (4 \pi)^2$.
Thus, this contrition can dominate the singlino mass in relatively large $\susy$.
Furthermore, thanks to the radiative corrections the singlino mass can reach $62\GeV$, which is half of the mass of the SM Higgs boson.
Therefore, the singlino annihilation cross section can be resonant via the s-channel SM Higgs boson exchange. This is a new scenario in the nMSSM.

In next section, using effective Lagrangian, we will calculate the thermal relic abundance  of the singlino dark matter and experimental constraints from Higgs invisible decay searches and from direct dark matter searches.
In order to compare with literature,
we will consider the resonant case with the Higgs boson exchange and with the Z boson exchange.
In Section~\ref{singlinomass}, we will evaluate the full one-loop singlino mass.

\section{Resonant singlino dark matter via SM Higgs boson}
\label{DMS}

In this section,  using the low energy effective Lagrangian we calculate thermal relic abundance of the dark matter which annihilate via the SM Higgs boson or the Z boson.


Let us consider the case where only the singlino $\tilde{s}$ is light and other SUSY particles are relatively heavy, $v_{EW}  \ll \susy$.
Note that the masses of the singlet scalars are also heavy in the nMSSM with $v_{EW}  \ll \susy$, see \eq{nMSSMHiggsmass}. 
In this case, the low energy effective Lagrangian includes singlino $\tilde{s}$, the SM Higgs boson $h$, Z boson, other fermions and other gauge bosons, and it can be written as
\begin{align}
\label{low}
	-\mathcal{L}_{\rm eff}\supset \frac{m_{\tilde{s}}}{2}\bar{\tilde{s}}\tilde{s}
	+\frac{\lambda_{\rm eff}}{2}h\bar{\tilde{s}}\tilde{s} +\frac{ g_{Z\tilde{s}\tilde{s}}}{2}Z^{\mu}\bar{\tilde{s}}\gamma_{\mu}\gamma_{5}\tilde{s}\,, 
\end{align}
where singlino $\tilde{s}$ is written by Dirac 4-component spinor,  $\lambda_{\rm eff}$ is the SM Higgs-singlino effective coupling and $ g_{Z\tilde{s}\tilde{s}}$ is the Z boson-singlino axial effective coupling\footnote{
Let us derive this neutralino-neutralino-Z boson axial vector coupling.
Generally speaking, the neutralino-neutralino-Z boson coupling is given as
\beq
\mathcal{L} &=& \frac{1}{2}\bar{\chi^0_i} O_{ij}^L \gamma_{\mu} P_L \chi^0_jZ^{\mu} - \frac{1}{2}\bar{\chi^0_i} O_{ij}^{L,\ast} \gamma_{\mu} P_R \chi^0_j Z^{\mu}\non
&=& \bar{\chi^0_i} \frac{ O_{ij}^L-O_{ij}^{L,\ast}}{4}  \gamma_{\mu}\chi^0_jZ^{\mu}  -\bar{\chi^0_i} \frac{ O_{ij}^L + O_{ij}^{L,\ast}}{4}  \gamma_{\mu}\gamma_5 \chi^0_jZ^{\mu},
\eeq
with
\beq
O_{ij}^L = \frac{g}{2 \cos \theta_W} ( -N_{i3} N_{j3}^{\ast} + N_{i4} N_{j4}^{\ast}).
\eeq
When the neutralino mass matrix includes no CP-violating phase, one gets $O_{ij}^L = O_{ij}^{L\ast} $.
Therefore, we obtain 
\beq
\mathcal{L} &=& - \bar{\chi^0_i} \frac{g_{Z\chi_i^0\chi_i^0} }{2}   \gamma_{\mu}\gamma_5 \chi^0_jZ^{\mu} ,
\eeq
with $g_{Z\chi_i^0\chi_i^0} = O_{ij}^L $.
}.
In this Lagrangian, the singlino dark matter can annihilate to  SM particles, 
 via the s-channel exchange of the SM Higgs boson or the Z boson.
In the following, we estimate the thermal relic abundance of singlino dark matter with this effective model regarding $\lambda_{\rm eff}$, $g_{Z\chi_i^0\chi_i^0} $ and $m_{\tilde{s}}$ as free parameters by solving Bolzmann equation~\cite{Gondolo:1990dk}.

\paragraph*{The resonant case with the Higgs boson exchange}
~

\begin{figure}[t]
\begin{center}
\includegraphics[width =8cm]{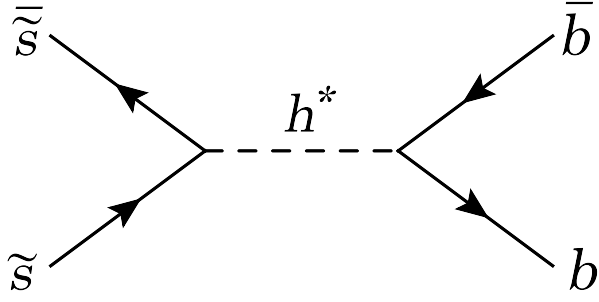}
\caption{One of the Feynman diagrams for the  resonant annihilation of the  singlino dark matter via the SM Higgs boson exchange. }
\label{singlinoresonance}
\end{center}
\end{figure}
First, we consider a dark matter  annihilation via the  SM Higgs boson exchange, $\bar{\tilde{s}}\tilde{s}\to h^{\ast}\to $ All.
Figure~\ref{singlinoresonance} represents a relevant diagram for the  resonant annihilation of the  singlino dark matter.
In order to obtain an annihilation cross section of the singlino, we use the optical theorem,
\beq
\sigma(\bar{\tilde{s}}\tilde{s}\to h^{\ast} \to  \textrm{All})=\frac{\textrm{Im} \mathcal{M}(\bar{\tilde{s}}\tilde{s}\to h^{\ast} \to  \bar{\tilde{s}}\tilde{s})}{2 E_{CM} p_{CM}},
\eeq
with
$E_{CM} = \sqrt{s}$, $p_{CM} = \sqrt{s(1 - 4 m^2_{\tilde{s}}/s)}/2$.
We assume the following  full propagator of Higgs boson,
\beq
G_h(s) = \frac{i}{s - m_h^2 + i \sqrt{s} \Gamma_h(s)},
\eeq
where $m_h = 125.5 \GeV$ and 
\beq
\Gamma_h (s) = \Gamma_{h,0}\frac{\sqrt{s}}{m_h},
\label{HiggsGamma}
\eeq
with $\Gamma_{h,0} = 4.07 \times 10^{-3}\GeV$ \cite{Denner:2011mq}\footnote{Although \eq{HiggsGamma} is our assumption, this is satisfied in leading order.
The dominant decay channel of the SM Higgs  is $\bar{b}b$ (Br($h\to\bar{b}b)\sim0.5$).
This transition rate is
\beq
\Gamma(s)= \frac{3}{16 \pi} y_b^2 \sqrt{s} \left( 1 - \frac{4 m_b^2}{s} \right)^{\frac{3}{2}}.
\eeq 
Therefore, the decay width is roughly proportional to $E_{CM}$ around $m_h =125.5\GeV$.
}.
Then the pair annihilation cross section of the singlino can be obtained as follows,
\beq
\sigma(\bar{\tilde{s}}\tilde{s}\to h^{\ast} \to \textrm{All}) = \frac{\lambda_{\rm eff}^2}{2} \sqrt{1 - \frac{4 m_{\tilde{s}}^2}{s}} \frac{\sqrt{s} \Gamma_{h}}{(s - m_h^2)^2 + s \Gamma_{h}^2}.
\eeq
Thus, one can see $\sigma(\bar{\tilde{s}}\tilde{s}\to h^{\ast} \to \textrm{All}) \propto \mbox{\boldmath $v$}_{\tilde{s}}$.
Since the freeze-out of the WIMP dark matter occurs when they are non relativistic (roughly estimation, $\mbox{\boldmath $v$} \sim 0.3$), this dependence of the velocity of the singlino gives suppression to the annihilation cross section.
In other words,
this annihilation rate is $p$-wave suppressed.
To distinguish whether the dark matter annihilation cross section is s-wave process (not suppresed)  from the interaction structure of the dark matter is possible, and it is summarized in Ref~\cite{Kumar:2013iva}.
For example, $\bar{\chi}\chi\bar{f}f$ is $p$-wave annihilation,  $\bar{\chi} \gamma_5 \chi\bar{f} \gamma_5 f$ (CP-odd Higgs exchange) and $\phi \phi \bar{f} f$ are $s$-wave annihilation, and so on.

The thermal average of the annihilation cross section  times the relative velocity of the annihilating particles $v_{\rm rel}$\footnote{Strictly speaking,
the relative velocity in the thermal average $\langle \dots \rangle$ is not the nonrelativistic relative velocity $v_r = |\mbox{\boldmath $v$}_1 - \mbox{\boldmath $v$}_2|$ but the so-called M$\o$ller velocity,
\beq
\bar{v} = \sqrt{(\mbox{\boldmath $v$}_1-\mbox{\boldmath $v$}_2)^2 - (\mbox{\boldmath $v$}_1 \times \mbox{\boldmath $v$}_2)^2}.
\eeq
 } can be  obtained as follows,
 \beq
 \langle \sigma v_{\rm rel} \rangle(T) = \frac{\int d^3 \mbox{\boldmath $p$}_1  d^3 \mbox{\boldmath $p$}_2 e^{-E_1/T} e^{-E_2/T } \sigma(s) v_{\rm rel}    }{\int d^3 \mbox{\boldmath $p$}_1  d^3 \mbox{\boldmath $p$}_2 e^{-E_1/T} e^{-E_2/T }}.
 \eeq
One can reduce this formula to the single-integration \cite{Gondolo:1990dk},
\beq
 \langle \sigma v_{\rm rel} \rangle(T)  = \frac{1}{8 m^4 T (K_2 [m/T])^2} \int_{4 m^2}^{\infty} d s \sqrt{s} (s - 4 m^2) K_1 \left[ \frac{\sqrt{s}}{T} \right] \sigma(s),
 \label{thermalaverage}
\eeq
where $K_i[x]$ are the modified Bessel functions of order $i$ \cite{Bessel}.

Using thermal average of the annihilation cross section, 
the relic density of the singlino in the expanding universe  can be  evaluated by solving the following Bolzmann equation,
\beq
\frac{d n_{\tilde{s}}}{dt} + 3 H n_{\tilde{s}} =  - \langle \sigma v_{\rm rel} \rangle(T)  (n_{\tilde{s}}^2 - n_{\tilde{s}, {\rm eq}}^2),
\label{bol}
\eeq
where $H$ is the Hubble parameter, and $n_{\tilde{s}}$ is the number density of the singlino.
The details of the calculation for the relic density of the dark matter are written in Ref.~\cite{Steigman:2012nb}.
The Hubble parameter is related to the total energy density during the radiation-dominated era,
\beq
H\equiv \fr{1}{a} \fr{d a}{d t} = \sqrt{\frac{8 \pi  G \rho}{3}},
\label{Hubbledef}
\eeq
where $a$ is  the cosmological scale factor, $G$ is the gravitational constant $G = 1/M_{\rm Pl}^2 =  6.708\times 10^{- 39} \GeV^{-2}$, and the total energy density $\rho = (\pi^2 / 30) g_{\ast}(T) T^4$.
$g_{\ast}(T)$ denotes the relativistic degrees of freedom ($m <T$).
We have used  the fitting formula of $g_{\ast}(T)$  of Ref.~\cite{Wantz:2009it}, where $g_{\ast,R}$ in the reference corresponds to  $g_{\ast}$\footnote{Although the value of $g_{\ast,R}$ is one of the SM, one can use it in the nMSSM. 
In our scenario the mass of the singlino dark matter is about 60 GeV, and so its freeze out temperature is $\mathcal{O}(1)\GeV$. Therefore, there are no sparticle contributions to  $g_{\ast}$.    }.

\begin{figure}[t]
\begin{center}
\includegraphics[width =13cm]{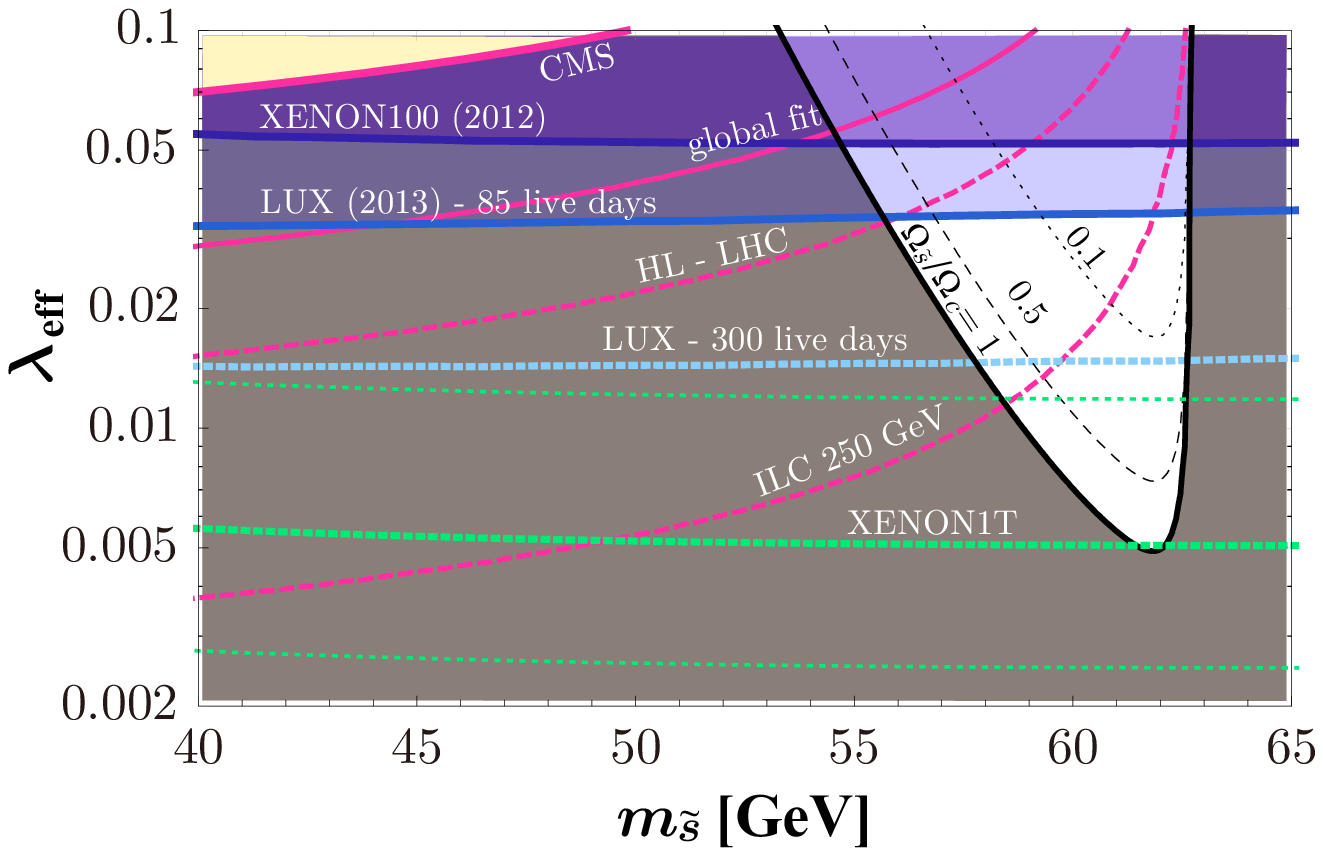}
\caption[The singlino thermal relic abundance and experimental constraints/future prospects in the case of the singlino resonant annihilation via the SM  Higgs boson s-channel exchange.]{The singlino thermal relic abundance and experimental constraints/future prospects  in the case of the singlino resonant annihilation via the SM  Higgs boson s-channel exchange.
The black contour  denotes the ratio of the thermal relic abundance $\Omega_{\tilde{s}} h^2$ to the current dark matter density $\Omega_c h^2 = 0.1199$~\cite{Ade:2013zuv}.
The singlino thermal relic density overclose the universe at the dark-shaded region.
The regions above the red solid lines are excluded by the Higgs invisible decay ($h \to \tilde{s} \tilde{s}$) searches of CMS (Br${}_h^{\textrm{inv.}}$ $\leq$ 58~\%) \cite{Chatrchyan:2014tja} for upper line (yellow-shaded region) and by the global fit of the Higgs couplings ($19~\%$)~\cite{Belanger:2013xza} for lower line.
The dashed red lines correspond to the future sensitivity of high luminosity LHC (6.2~\%)~\cite{Dawson:2013bba} and ILC with $\mathcal{L} = 1150 \textrm{fb}{}^{-1}$ at $\sqrt{s} = 250 \GeV$ (0.4~\%) \cite{Asner:2013psa}.
The blue-shaded regions are excluded by XENON100~\cite{Aprile:2012nq} and LUX~\cite{Akerib:2013tjd}.
The regions above the blue and the green dashed lines can be probed by the future direct dark matter searches of LUX~\cite{Akerib:2012ys} and XENON1T~\cite{Aprile:2012zx}.
\coll{As the future sensitivities of XENON1T, we have used $0.14$ (top), $0.326$ (middle) and $0.66$ (bottom) to the parameter $f_N$.}
}\label{abundance}
\end{center}
\end{figure}
In Figure~\ref{abundance}, the black lines show the ratio of the thermal relic abundance $\Omega_{\tilde{s}} h^2$ to the current dark matter density $\Omega_c h^2 = 0.1199$~\cite{Ade:2013zuv} where we take the Higgs boson mass as $m_h=125.5$ GeV.
The horizontal axis is the mass of the singlino $m_{\tilde{s}}$ and the vertical axis is the singlino-singlino-Higgs coupling $\lambda_{\rm eff}$. 
Note that we  have handled  $m_{\tilde{s}}$ and $\lambda_{\rm eff}$ as free parameters.
The thick black line represents the appropriate parameter region in which the dark matter relic abundance is consistent with the current dark matter abundance.
While, the singlino relic density overclose the universe at the dark-shaded region.
As discussed previous section, since the couplings of the singlino with SM particles are too small and the annihilation cross section is also too small, the singlino dark matter could not  be sufficiently diluted in the universe.
On the other hand, around $m_{\tilde{s}} \sim m_h/2$, thanks to the s-channel exchange of the SM Higgs boson  the annihilation cross section becomes large, and the singlino dark matter can  be appropriately diluted (white region). 

\paragraph*{}
This effective potential \eq{low} receives two kinds of experimental constraints:
constraint from the branching ratio of the Higgs to singlino pair (Higgs invisible decay searches) and the direct detection of the  dark matter. 
The decay width of the Higgs to singlino pair is given by
\beq
\Gamma(h\to \bar{\tilde{s}}\tilde{s}) = \frac{\lambda_{\rm eff}^2}{16 \pi} m_h \left( 1 - \frac{4 m_{\tilde{s}}^2}{m_h^2} \right)^{\frac{3}{2}}.
\eeq
Thus  the branching ratio of the Higgs to singlino pair is 
\beq
\textrm{Br}(h\to \bar{\tilde{s}}\tilde{s}) = \frac{\Gamma(h\to \bar{\tilde{s}}\tilde{s})}{\Gamma_{h,0} + \Gamma(h\to \bar{\tilde{s}}\tilde{s}) }.
\eeq
In collider, this search can be particularly performed on the Higgs boson produced via association with Z boson,
\beq
pp(e^+e^-)\to Z^{\ast} \to Z h \to \ell \ell +\textrm{invisible}.
\eeq
In the Figure~\ref{abundance}, 
the regions above the red solid lines are excluded by the Higgs invisible decay
searches of CMS, $\textrm{Br}(h\to \bar{\tilde{s}}\tilde{s})  <  0.58$ at $95~\%$ C.L. (upper line)~\cite{Chatrchyan:2014tja}, 
and by the global fit of the Higgs couplings, $\textrm{Br}(h\to \bar{\tilde{s}}\tilde{s})  <  0.19$ at $95~\%$ C.L.  (lower line)~\cite{Belanger:2013xza}.
The regions above the red dashed lines can be probed by the future Higgs invisible decay searches of high luminosity LHC, $\textrm{Br}(h\to \bar{\tilde{s}}\tilde{s})  <  0.062$ at $95~\%$ 
 (upper line)~\cite{Dawson:2013bba} and ILC at $\sqrt{s } = 250\GeV$,  $1~\textrm{at}^{-1}$, $\textrm{Br}(h\to \bar{\tilde{s}}\tilde{s})  <  0.004$ at $95~\%$ 
 (lower line)~\cite{Asner:2013psa}.
 
 \paragraph*{}
The direct dark matter searches can set limits on the spin-independent cross section of dark matter-nucleon elastic scattering, $\tilde{s} N \to\tilde{s} N $.
If one integrate out the SM Higgs boson, effective Lagrangian becomes
\beq
\mathcal{L}_{\rm eff} \supset \frac{\lambda_{\rm eff}}{2 \sqrt{2} m_h^2} \bar{\tilde{s}}\tilde{s} \left( \sum_i (y_i \bar{f_i} f_i)  - \frac{\alpha_s}{4 \pi v_{EW}} G_{\mu\nu}G^{\mu\nu}\right),
\eeq
where $f_i$ are the SM fermions, and $G^{\mu\nu}$ is a field strength of SU(3) gauge.
Note that, here we have not considered the Z boson exchange in the  spin-independent cross section of dark matter-nucleon elastic scattering. It is because the Z boson-singlino coupling $g_{Z\tilde{s}\tilde{s}}$ always has more suppression by $\susy$ than the Higgs-singlino coupling $\lambda_{\rm eff}$ (see Eqs.~(\ref{geff}),~(\ref{lambdaeff})).
We have also not considered the squark exchange since we have assumed sparticles are heavy enough.   
Using this effective Lagrangian, we can obtain the cross section of dark matter-nucleon spin-independent elastic scattering as follows \cite{Kanemura:2010sh,Djouadi:2011aa},
\beq
\sigma(\tilde{s} N \to \tilde{s}N) = \frac{\lambda_{\rm eff}^2}{2 \pi v_{EW}^2}f_N^2 \frac{ m_N^4  m_{\tilde{s}}^2}{m_h^4 (m_{\tilde{s}} + m_N)^2},
\eeq
with $m_N = 0.939 \GeV$ and
\beq
f_N  m_N \equiv \langle N | \sum_q  m_q \bar{q} q - \frac{\alpha_s}{4 \pi} G_{\mu\nu}G^{\mu\nu} |N  \rangle.
\eeq
We use $f_N = 0.326$, which is the lattice result \cite{Young:2009zb}.
Then,
\beq
\sigma(\tilde{s} N \to \tilde{s}N)& =& 8.051 \times 10^{-43} \times \lambda_{\rm eff}^2 \frac{m_{\tilde{s}}^2}{(m_{\tilde{s}} + 1\GeV)^2} ~~[\textrm{cm}^2]\\
&\simeq& 0.8 \times  10^{-46} \left( \frac{\lambda_{\rm eff}}{0.01} \right)^2  ~~[\textrm{cm}^2].
\eeq
Note that, there are various estimation of the parameter $f_N$ in the literature.
It is because a treatment of the heavy quarks, especially the  contribution  of the strange quark to the nucleon,  is difficult.
These results in the literature give a range   to the value of $f_N$ as follows \cite{Andreas:2008xy},
\beq
0.14 < f_N < 0.66.
\eeq 
Including this uncertainty,  the cross section of dark matter-nucleon spin-independent elastic scattering is gives as,
\beq
\sigma(\tilde{s} N \to \tilde{s}N) \simeq (\textrm{0.15~-~3.3})\times  10^{-46} \left( \frac{\lambda_{\rm eff}}{0.01} \right)^2  ~~[\textrm{cm}^2].
\eeq 

In the Figure~\ref{abundance}, 
the blue-shaded regions are excluded by the current direct dark matter searches of XENON100~\cite{Aprile:2012nq} and LUX~\cite{Akerib:2013tjd}.
The region above the blue dashed line can be probed by the future direct dark matter search of LUX~\cite{Akerib:2012ys}.
For applying these constraints and future prospects, we assume $\Omega_{\tilde{s}} h^2 = \Omega_c h^2$.
Here we have used $f_N = 0.326$.
\coll{The regions above the green dashed lines can be probed by the future direct dark matter searches of XENON1T~\cite{Aprile:2012zx}.
As the future sensitivities of XENON1T, we have used $0.14$ (top), $0.326$ (middle) and $0.66$ (bottom) to the parameter $f_N$.}
One can see that the region where $\tilde{s}$ is consistent with the current dark matter relic abundance lies around $\lambda_{\rm eff}\sim \mathcal{O}(0.01)$ and ${m_{\tilde{s}}}\sim 60$ GeV. 
In this region, resonant pair-annihilation of $\tilde{s}$ occurs via the Higgs boson with $m_{\tilde{s}} \sim m_h/2$.
This allowed region can be covered by the future Higgs invisible decay searches and direct dark matter searches, especially by XENON1T.

\paragraph*{The resonant case with the Z boson exchange}
~

Next, let us consider a dark matter  annihilation via the  Z boson exchange, $\bar{\tilde{s}}\tilde{s}\to Z^{\ast}\to\bar{f}f$.
Then the pair annihilation cross section can be obtained as follows \cite{Nihei:2002ij},
\beq
\sigma(\bar{\tilde{s}} \tilde{s}\to Z^{\ast} \to \bar{f}f ) = \frac{2}{\sqrt{s (s - 4 m_{\tilde{s}}^2)}} \omega^Z(s).
\label{sigmaZ}
\eeq
The Lorentz-invariant function $\omega^Z(s)$ is defined as
\beq
\omega^Z(s) = \frac{1}{32\pi}\sum_{f}\left( N_{c_f} \theta(s - 4 m_f^2) \beta_f(s,m_f) \tilde{\omega}^Z_f(s)\right),
\eeq
where $N_{c_f}$ is the color factor ($N_{c_f} = 3$ for quarks and $N_{c_f} = 1 $ for leptons), and a kinematic factor $\beta_f$ is given as
\beq
\beta_f(s, m_f) = \sqrt{1 - \frac{4 m_f^2}{s}},
\eeq 
with
\beq
\tilde{\omega}^Z_f(s) &=& \frac{4}{3} \left|\frac{g_{Z\tilde{s}\tilde{s}}}{s -M_Z^2 + i \Gamma_Z M_Z} \right|^2 
\biggl[ 12 |C_A^{ffZ}|^2 \frac{m_{\tilde{s}}^2 m_f^2}{M_Z^4} (s - M_Z^2)^2 \non
& &  ~~~~~~~  + \left( |C_V^{ffZ}|^2 (s + 2 m_f^2) + |C_A^{ffZ}|^2 (s - 4 m_f^2) \right) (s - 4 m_{\tilde{s}}^2)\biggl].
\eeq
Here total with of Z boson is $2.4952\pm0.0023\GeV$, and Z boson-fermion couplings are
\beq
C_V^{ffZ} &=&  - \frac{g}{2 \cos \theta_W} (T_{3,f} - 2 \sin^2 \theta_W Q_f ), \\
C_A^{ffZ} &= & - \frac{g}{2 \cos \theta_W} T_{3,f}.
\eeq
Substituting the pair annihilation cross section via the  Z boson exchange \eq{sigmaZ} into the thermal average of the annihilation cross section \eq{thermalaverage}, and solving the Bolzmann equation \eq{bol}, we have estimated 
the thermal relic abundance of the singlino dark matter.

\begin{figure}[t]
\begin{center}
\includegraphics[width =13cm]{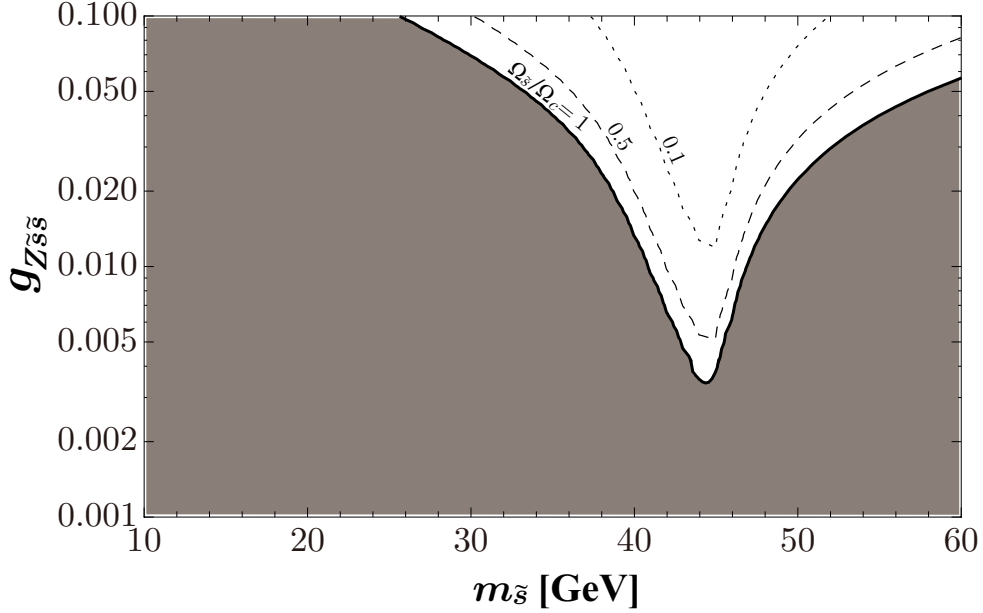}
\caption[The singlino dark matter thermal relic abundance  in the case of the singlino resonant annihilation via the Z boson s-channel exchange.]{
The singlino dark matter thermal relic abundance  in the case of the singlino resonant annihilation via the Z boson s-channel exchange.
The black contour  denotes the ratio of the thermal relic abundance $\Omega_{\tilde{s}} h^2$ to the current dark matter density $\Omega_c h^2 = 0.1199$~\cite{Ade:2013zuv}.
The singlino thermal relic density overclose the universe at the dark-shaded region.
}\label{Zreso}
\end{center}
\end{figure}

\paragraph*{}
In Figure~\ref{Zreso}, we show the singlino dark matter thermal relic abundance  in the case of the singlino resonant annihilation via the Z boson s-channel exchange\footnote{Similar to  the resonant case with the Higgs boson,
 although there is the upper bound on the invisible Z width, $\Gamma(Z \to \chi_1^0 \chi_1^0) < 1.76\times 10^{-3}\GeV$ \cite{ALEPH:2005ab} as a constraint from the experiment, we have not calculated this bound.}.
 The black contour  denotes the ratio of the thermal relic abundance $\Omega_{\tilde{s}} h^2$ to the current dark matter density $\Omega_c h^2 = 0.1199$~\cite{Ade:2013zuv}.
Thus, the singlino thermal relic density overclose the universe at the dark-shaded region.
Similar to the case of the resonance annihilation  via the Higgs boson s-channel exchange,
the greater part of the parameter region except $m_{\tilde{s}} \sim M_Z/2$ are suffering from  the overabundance of the universe. 
We find that $g_{Z\tilde{s}\tilde{s}} $ can become small  to $\mathcal{O}(10^{-3}$-$10^{-2})$ at the case when one assume the singlino annihilation cross section is resonant by the Z boson exchange. 
While, one can estimate $g_{Z\tilde{s}\tilde{s}} $ at
\beq
g_{Z\tilde{s}\tilde{s}} &=& \frac{g}{2 \cos \theta_W} ( -N_{53} N_{53}^{\ast} + N_{54} N_{54}^{\ast})\\
&
\sim & \frac{\lambda^2 g}{2 \cos\theta_W} \frac{v_{EW}^2} {\susy^2} \cos  2\beta.
\label{geff}
\eeq
This estimation implies $\susy \lesssim 1$-$2\TeV$ at the resonant case with the Z boson exchange.

However, as discuss next section, the Higgs-singlino coupling can also be estimated 
relating to $\susy$.
It implies  that when $\susy \lesssim 1$-$2\TeV$, $\lambda_{\rm eff} \gtrsim 0.1$ (see \eq{lambdaeff1}).
Such a large $\lambda_{\rm eff}$ and light singlino mass ($m_{\tilde{s}} \sim 45 \GeV$) may be excluded by current experimental bounds\footnote{Strictly speaking, 
they are not excluded at the case when   $\tb$ is large and there is a extra contribution to the singlino mass which is beyond the nMSSM.
It is because the Higgs-singlino coupling $\lambda_{\rm eff}$ is suppressed at large $\tb$  (see \eq{lambdaeff1}), but the Z boson-singlino coupling $g_{Z\tilde{s}\tilde{s}} $ is not suppressed.  } (see Figure~\ref{abundance}).
Therefore, in terms of the overabundance of the singlino dark matter in the universe, the resonant scenario with the SM Higgs boson is the last resort for the nMSSM.

\section{Radiative Singlino mass}
\label{singlinomass}

Now, we calculate the  singlino mass $m_{\tilde{s}}$ and the Higgs-singlino coupling $\lambda_{\rm eff}$ in the nMSSM.
From the tree-level calculations, these values are evaluated as 
\beq
	\lambda_\textrm{eff}^\textrm{tree} &= &
	\sqrt{2}\lambda   \left(    Z_{11}^H  N_{14}^{\ast}N_{15}^{\ast}+  Z_{12}^H N_{13}^{\ast} N_{15}^{\ast}  + Z_{13}^H N_{13}^{\ast} N_{14 }^{\ast} \right)\non
&&  + g' \left( Z_{11}^H N_{11}^{\ast } N_{13}^{\ast}  - Z_{12}^{H} N_{11}^{\ast} N_{14}^{\ast} \right)
	 - g \left( Z^{H}_{11}N_{12}^{\ast} N_{13}^{\ast}  + Z_{12}^{H} N_{12}^{\ast} N_{14}^{\ast} \right)  	\label{lambdaeffex}\\
&\sim&	 \sqrt{2} \lambda \frac{v_{EW}}{\susy}\sin{2\beta}\,, 
	\label{lambdaeff1}
\eeq
where $Z_{ij}^H$ is a unitary matrix which can diagonalize the CP-even mass matrix as follows
\beq
Z^H  \begin{pmatrix}  \mathcal{M}^2_{R 11} &\mathcal{M}^2_{R 12} &\mathcal{M}^2_{R 13}\ \\ 
\mathcal{M}^2_{R 12} &\mathcal{M}^2_{R 22} &\mathcal{M}^2_{R 13} \\
\mathcal{M}^2_{R 13} &\mathcal{M}^2_{R 23} &\mathcal{M}^2_{R 33}   \end{pmatrix} Z^{H,\dag} = \left( (m^H_1)^2, (m^H_2)^2, (m_3^H)^2\right)_{\rm diag},
\eeq
where $m^H_1$ is the lightest CP-even Higgs mass, that is the SM-like Higgs boson.
On the other hand, the tree-level singlino mass is evaluated as Eq.~(\ref{treecoup}).
Obviously $\lambda_{\rm eff}\sim \mathcal{O}(0.01)$ and $m_{\tilde{s}}\sim 60$ GeV can not be satisfied at the same time.

However, one-loop corrections to the neutralino mass can raise the singlino mass with relatively large $\susy$.
The typical diagram, Higgs-Higgsino loop diagram,  which contributes to the singlino mass is given in Figure~\ref{diagram}.
\begin{figure}[t]
\begin{center}
\includegraphics[width =8cm]{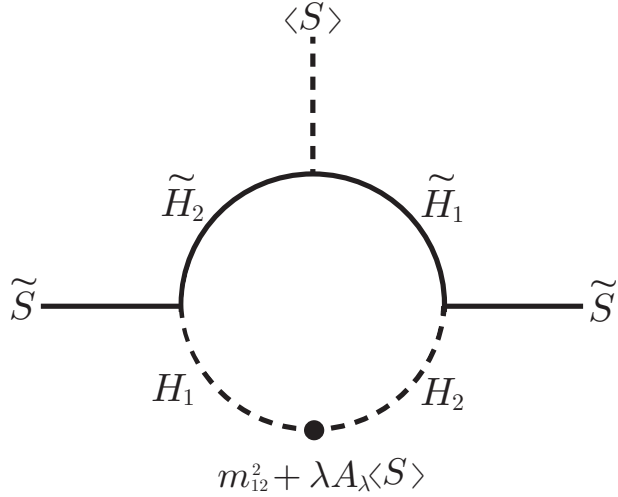}
\caption{Typical one-loop diagram which contributes to the mass of the  singlino. 
}\label{diagram}
\end{center}
\end{figure}
Note that a sum of the contributions of the  neutral Higgs-Higgsino vanishes at the leading order.
The vertex of the CP-even Higgs-Higgsino-singlino is different from the one of the CP-odd Higgs-Higgsino-singlino in only ``i".
Thus, since the loop contribution is proportional to the square of the vertex,  the CP-even Higgs loop contribution cancel out  the CP-odd Higgs one  up to their mass dependence.
So, one can easily estimate the radiative corrections to the singlino mass by calculating the charged Higgs-Higgsino loop.
This loop gives the following contribution to the singlino mass,
\begin{align} \label{oneloopmass}
	m_{\tilde{s}}^\textrm{1-loop}&=
	\frac{\lambda^2}{(4\pi)^2}\mu_{\textrm{eff}}\sin2\beta \cdot F\left( \frac{2 (m_{12}^2 + A_{\lambda} \mu_{\textrm{eff}})}{\mu_{\textrm{eff}}^2 \sin 2 \beta}\right) \nonumber \\
	&\sim
	\frac{\lambda^2}{(4\pi)^2}\susy\sin 2\beta \,, 
\end{align}
where the loop function $F\left( x\right)$ is defined as 
\beq
F\left(x\right) \equiv \frac{x \log x}{x-1},
\eeq
and it satisfies $F\left(1\right) = 1$.
We find that radiative corrections is proportional to the soft SUSY breaking scale.
It is because this loop diagram should have a chirality flip on the Higgsino propagator, and this flip gives $\susy$ to a numerator of the loop contribution.

Next,  we numerically calculate the neutralino $5\times5$ mass matrix including {\it full one-loop} corrections~\cite{Staub:2010ty,private}\footnote{ 
In the limit of $\kappa=0$, one-loop corrections in the NMSSM reduce to the one in the nMSSM.}.
\coll{We assume that the all complex parameters to be real in neutralino mass matrix \eq{Nmassmatrix}, then the neutralino mass term becomes
\beq
\mathcal{L} = -\frac{1}{2}  \bar{\psi}_{\chi} \mathcal{M}^{\chi^0}_{\rm tree} \psi_{\chi},
\eeq
where $\psi_{\chi}$ is the Dirac 4-component spinor $\psi_{\chi}^{\rm T} = ( ( ^{\tilde{B}}_{\tilde{B}^{\ast}}),~( ^{\tilde{W}^0}_{\tilde{W}^{0 \ast}}),~( ^{\tilde{H}_1^0}_{\tilde{H}_1^{0 \ast}}),~( ^{\tilde{H}_2^0}_{\tilde{H}_2^{0 \ast}}),~( ^{\tilde{S}}_{\tilde{S}^{\ast}}) )$.
At one-loop level, the radiative corrections to the neutralino sector are given by \cite{Staub:2010ty, Eberl:2001eu} 
\beq
\mathcal{M} = \frac{1}{2} \biggl[\bar{\psi}_{\chi} \left( \Slash{p} -\mathcal{M}^{\chi^0}_{\rm tree}\right) \psi_{\chi}+  \frac{1}{(4 \pi)^2}\left( \bar{\psi}_{\chi} \Slash{p} \Sigma^L P_L \psi_{\chi} + \bar{\psi}_{\chi} \Slash{p} \Sigma^R P_R \psi_{\chi} +   \bar{\psi}_{\chi} \Sigma^S P_L \psi_{\chi} + \hc \right)\biggl],
\eeq
where the correction $ \Sigma^S(p^2)$ comes from a self-energy for neutralinos, and the corrections $ \Sigma^{L/R}(p^2)$ come from the wave function constants for the  neutralinos.
The momentum of the external line is represented by $p$. 
A pole mass of the neutralino at one-loop level is obtain by the following equation, }
\beq
\mathcal{M}_{\rm 1~loop}^{\chi^0} (p^2) &=& \mathcal{M}_{\rm tree}^{\chi^0} - \frac{1}{2} \frac{1}{(4 \pi)^2} \left[ \Sigma^S(p^2) +   \left(\Sigma^{S}(p^2)\right)^{\rm T} + \left(   \left( \Sigma^{L}(p^2)\right)^{\rm T} +  \Sigma^{R}(p^2) \right)   \mathcal{M}_{\rm tree}^{\chi^0} \right.\non
& & ~~~~~~~~~~~~~~~~~~~ + \left. \mathcal{M}_{\rm tree}^{\chi^0}\left( \Sigma^{L}(p^2) +  \left( \Sigma^{R}(p^2)\right)^{\rm T} \right) \right].
\label{singlinomassexact}
\eeq
The explicit formulae of the one-loop corrections to the neutralino $ \Sigma^S(p^2)$  and  $ \Sigma^{L/R}(p^2)$  are given in appendix \ref{staubloop}.
Here, we have performed $\overline{DR}$ renormalization in which we have subtracted the $1/\bar{\epsilon}$ poles (see Eqs.~(\ref{B0app},~\ref{B1app})). 
Note that 55 component of the neutralino mass matrix does not diverge,  which is consistent with the fact that the quadratic term of the singlet superfield is not included in the superpotential.
The cancelation of the divergence of the neutral Higgs-Higgsino loop is trivial, since the vertices are different in only ``i"  as discussed before.  
While, the cancelation of the divergence of the charged Higgs-Higgsino loop is rather non-trivial.
The charged Higgs-Higgsino-singlino coupling is proportional to $\sb\cb$, on the other hand the charged NG boson-Higgsino-singlino coupling is proportional to  $-\sb\cb$ (see \eq{chargedcoupling}). Since the divergence is independent of the mass of the virtual particle,
thus, all divergences are  canceled  out.
\coll{This cancellation can be understood  in terms of the supergraph. One-loop correction to the singlet superfiled is roughly given as
\beq
\Gamma \simeq \int d \theta^4 \frac{\lambda^4}{(4 \pi)^2} \frac{1}{\susy^2}(\hat{S}^{\dag} \hat{S})^2,
\eeq
and thus  the $55$ component of the neutralino mass matrix is finite.}

\begin{figure}[t]
\begin{center}
\includegraphics[width =13cm]{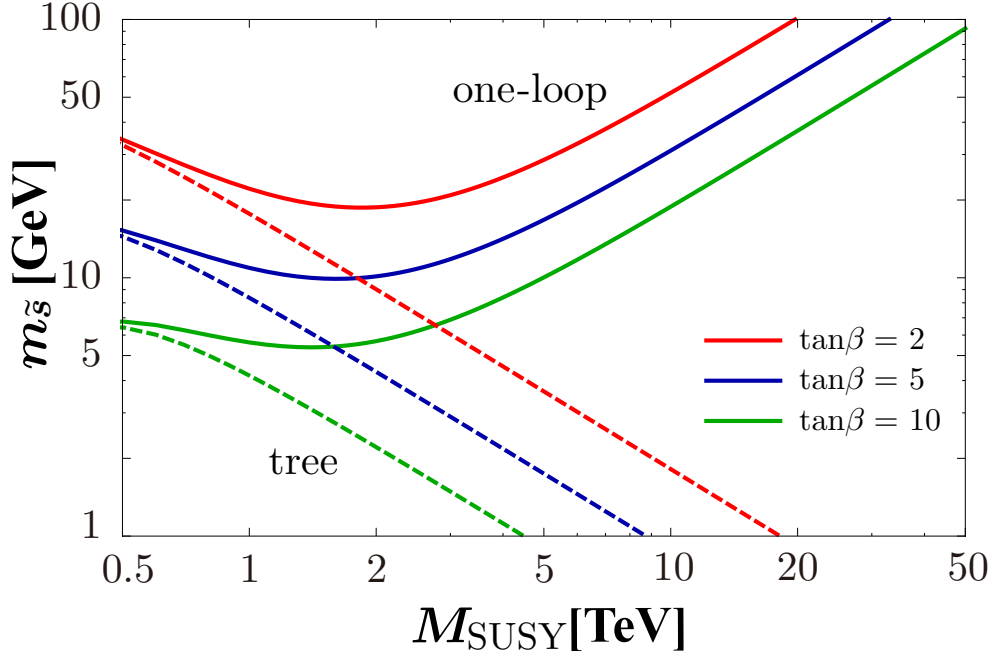}
\caption[The singlino mass at the tree level  and the full one-loop level as a function of $\susy$.]{
 The singlino mass, which is the lightest mass eigenvalues of the neutralino mass matrix, at the tree level (dashed) and the full one-loop level (solid) as a function of $\susy$.
We take $\lambda = 0.75$, all dimensional parameters equal to $\susy$ and $\tb = 2$ (red), 5 (blue) and 10 (green).
}\label{singlinomass}
\end{center}
\end{figure}
Figure~\ref{singlinomass} shows the dependence of the singlino mass to $\susy$, which is the lightest mass eigenvalues of the neutralino mass matrix, at the tree level (dashed) and the full one-loop level (solid). 
In this figure, we take $\lambda = 0.75$, all dimensional parameters equal to $\susy$ and $\tb = 2$ (red), 5 (blue) and 10 (green).
We find that the singlino obtains sizable one-loop corrections to the mass when $\susy \gtrsim 1 \TeV$.
Since this feature is due to the suppression of the singlino mass at the tree level, the two-loop level corrections to the singlino mass is estimated to be smaller than the one-loop one.
We also find that both the tree level and the one-loop level mass have a $\tb$ suppression. 

We have checked the validity of the our full one-loop calculation code by three  ways.
First, we have compared the full one-loop result with the estimation by Eq.~(\ref{oneloopmass}), then we find that these results agree with very well.
Second, we have checked that  55 component of  the full one-loop correction matrix vanishs in the SUSY limit, $\susy \to 0$.
Third, we have chosen the  $A_{\lambda}$ to vanish the combination $m_{12}^2 + \lambda A_{\lambda}$ varying the soft SUSY breaking scale. Then we find that even if  $\susy$ is much higher scale, 55 component of the full one-loop correction matrix  vanishs.

Note that with $\susy\sim O(10)$ TeV, $\tan\beta  \sim O(1)$ and $\lambda \sim O(1)$, one can simply obtain $\lambda_{\rm eff}\sim \mathcal{O}(0.01)$ and $m_{\tilde{s}}\sim 60$ GeV\footnote{The one-loop $\lambda_{\textrm{eff}}$ can be roughly estimated as $\lambda_\textrm{eff}^\textrm{1-loop} \sim \frac{\lambda^4}{(4 \pi)^2} \frac{v_{EW}}{\susy} \sin 2 \beta$, which is negligible in comparison with $\lambda_\textrm{eff}^\textrm{tree}$.}.
Moreover, the Higgs boson mass becomes around $125$ GeV in such parameter sets with the help of the additional quartic coupling $\lambda$.
We will show these validity by using the numerical calculations in the next section.

\section{Numerical Results}
\label{NumRes}
\mtrem{もっと細かく書く}
In this section, we numerically investigate the singlino resonant dark matter scenario and the Higgs boson mass in the nMSSM.
We calculate the mass of the  Higgs boson  using the two-loop renormalization group equation including the matching condition Eq.~(\ref{masscor})~\cite{Giudice:2011cg}.
The detail of the calculation for the Higgs boson mass is given in Section~\ref{HiggsmasscalcRGE}.

\begin{figure}[t]
\begin{center}
\includegraphics[width =13cm]{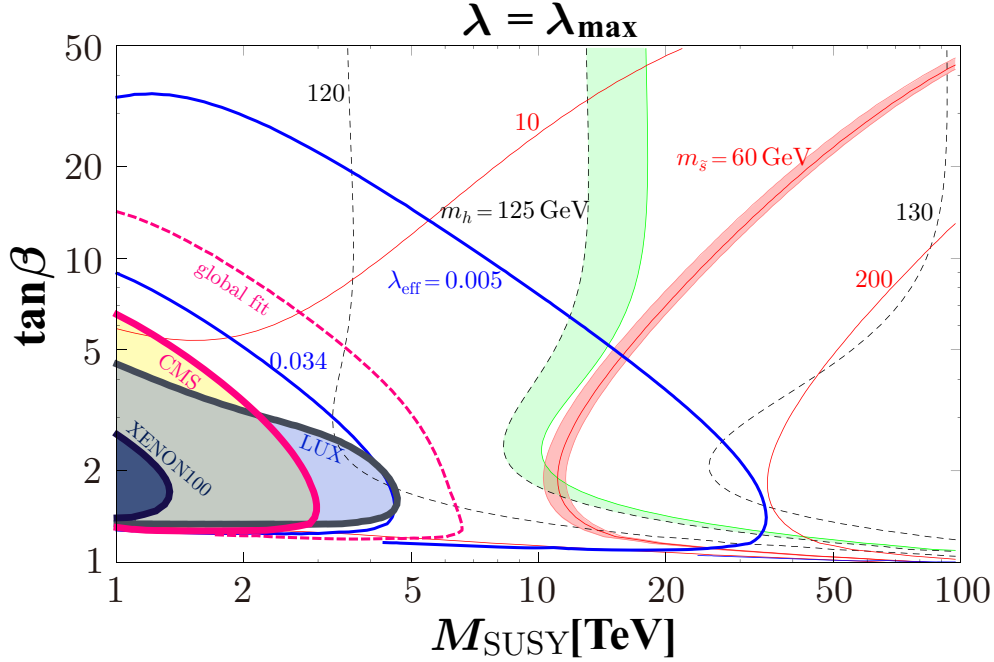}
\caption[Contours of $m_{\tilde{s}}$, $\lambda_{\textrm{eff}}$ and $m_h$ in $\susy$-$\tan\beta$ plane assuming $\lambda = \lambda_{\textrm{max}}$ at each point.]{Contours of $m_{\tilde{s}}$ (red lines), $\lambda_{\textrm{eff}}$ (blue lines) and $m_h$ (black dashed lines) in $\susy$-$\tan\beta$ plane assuming $\lambda = \lambda_{\textrm{max}}$ at each point.
On the red-shaded region ( $55.5 \GeV < m_{\tilde{s}} < 62.7 \GeV$ ), the resonant annihilation via the Higgs boson can occur.
The green-shaded region satisfies $125 \GeV < m_h < 126 \GeV$.
The blue (dark blue)-shaded region is excluded by the current limits from LUX~\cite{Akerib:2013tjd} (XENON~\cite{Aprile:2012nq}).
The yellow-shaded region is excluded by the Higgs invisible decay search at the CMS~\cite{Chatrchyan:2014tja} and the magenta dashed line is the current bound by the global fit of the Higgs coulings~\cite{Belanger:2013xza}.
}\label{lambdamax}
\end{center}
\end{figure}

In Figure~\ref{lambdamax}, we show the singlino mass $m_{\tilde{s}}$ (red lines), the effective Higgs-singlino dark matter coupling $\lambda_{\textrm{eff}}$ (blue lines) and the Higgs boson mass $m_h$ (black dashed lines) in $\susy$-$\tan\beta$ plane.
Here, the singlino mass $m_{\tilde{s}}$  is obtained as  the lightest mass eigenvalue of the one-loop full neutralino mass matrix \eq{singlinomassexact}, and $\lambda_{\rm eff} $ is given by the tree-level calculation \eq{lambdaeffex}.
For simplicity, all parameters are chosen to be real.
The trilinear coupling $\lambda$ is taken to be $\lambda_{\textrm{max}}$ which is a maximal value avoiding Landau singularities up to  the GUT scale, $2\times10^{16}\GeV$ (see Section~\ref{Landausec}).
All soft SUSY breaking parameters except $A_\lambda$ are set to $\susy$ ($\lambda  c_F = c_S = 1$).
In order to obtain a sizable contribution to the Higgs boson mass, we choose $A^2_\lambda = \frac{2}{5} \susy^2$.
As one can see from Figure~\ref{abundance}, the viable regions for the singlino dark matter are $55.5 \GeV < m_{\tilde{s}} < 62.7 \GeV$ and  $0.005 < \lambda_{\textrm{eff}} < 0.034 $.
In Figure~\ref{lambdamax}, these regions correspond to the red-shaded band and the region between the two blue lines respectively. 
One can see that the singlino resonant dark matter scenario is successful with $\tan\beta \sim \mathcal{O}(1)$ and $\susy \sim \mathcal{O}(10) \TeV$. 
On the other hand, the green band represents $125 \GeV < m_h < 126 \GeV$.
We also find that the current dark matter relic abundance and the Higgs boson mass can be explained simultaneously  with $\tan\beta \sim \mathcal{O}(1)$ and $\susy \sim \mathcal{O}(10) \TeV$. 
The blue (dark blue)-shaded region is excluded by the current limits from LUX~\cite{Akerib:2013tjd} (XENON~\cite{Aprile:2012nq}).
The yellow-shaded region is excluded by the Higgs invisible decay search at the CMS~\cite{Chatrchyan:2014tja} and the magenta dashed line is the current bound by the global fit of the Higgs couplings~\cite{Belanger:2013xza}.
Note that, in the calculation of  these experimental bounds,
we have assumed $m_h = 125.5\GeV$ at each point in the plane.
It means that these experimental abounds exclude the low-scale and $\tb\sim \mathcal{O}(1)$ nMSSM.

\begin{figure}[t]
\begin{center}
\includegraphics[width =13cm]{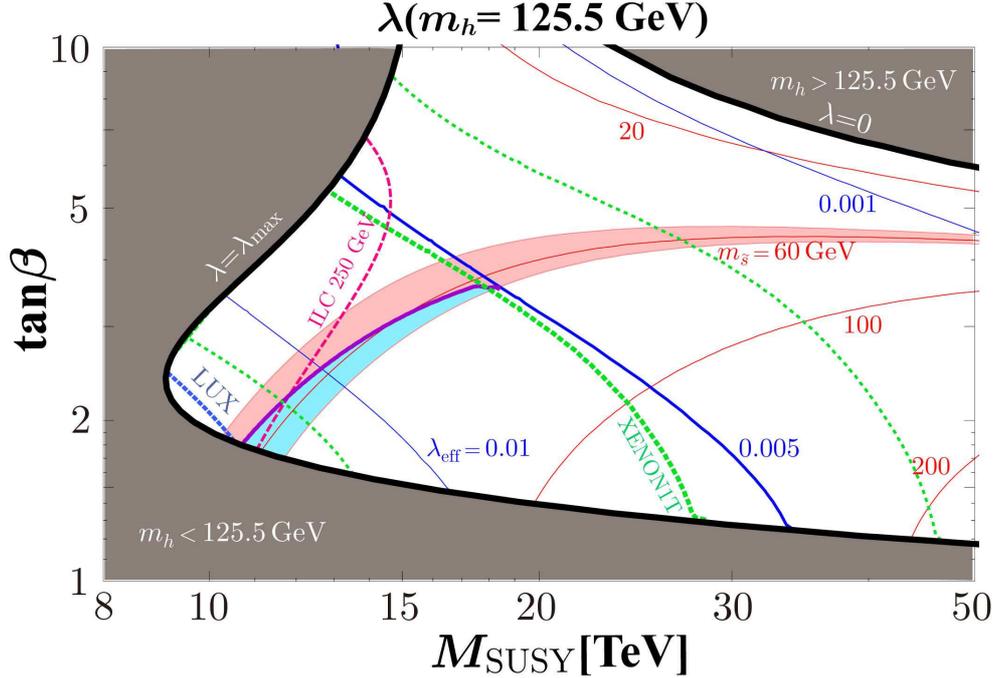}
\caption[Contours of $m_{\tilde{s}}$, $\lambda_{\textrm{eff}}$ in $\susy$-$\tan\beta$ plane under $m_h = 125.5 \GeV$ by changing $\lambda$, $0 \leq \lambda \leq \lambda_\textrm{max}$.]{Contours of $m_{\tilde{s}}$ (red lines), $\lambda_{\textrm{eff}}$ (blue lines) in $\susy$-$\tan\beta$ plane under $m_h = 125.5 \GeV$ by changing $\lambda$, $0 \leq \lambda \leq \lambda_\textrm{max}$.
On the purple line, the singlino relic abundance $\Omega_{\tilde{s}} h^2$ is consistent with the current value, $\Omega_c h^2 = 0.1199$~\cite{Ade:2013zuv}. 
In the light blue region, $\Omega_{\tilde{s}}h^2 \leq \Omega_c h^2$. 
The left side of the blue (green) dashed line can be probed by the future dark matter search LUX~\cite{Akerib:2012ys} (XENON1T~\cite{Aprile:2012zx}).
\coll{As the future sensitivities of XENON1T, we have used $0.14$ (left), $0.326$ (middle) and $0.66$ (right) to the parameter $f_N$.}
ILC~\cite{Asner:2013psa} can cover the left side of the magenta dashed line.
Other lines are the same in Figure~\ref{lambdamax}.
}\label{higgs1255}
\end{center}
\end{figure}

If we choose the lower value of $A^2_\lambda$, the green line moves to left because the Higgs boson mass obtains more contribution from the quartic coupling (see Eq.~(\ref{masscor})). 
On the other hand, with smaller value of $m_{12}^2+\lambda A_\lambda \langle S\rangle$ the singlino mass becomes lighter and the red-shaded region moves to right (see \eq{oneloopmass}).
The blue lines are not sensitive to the choice of $m_{12}^2$ and $A_\lambda$, because $\lambda_{\textrm{eff}}$ is determined by the soft SUSY breaking scale and $\tan\beta$.
The important point is that in any case with $\susy\sim \mathcal{O}(10)$ TeV and low $\tan \beta$ the current dark matter abundance and the measured Higgs boson mass can be realized simultaneously.
This opens a window for the singlino dark matter in high-scale supersymmetry.

\paragraph*{}
Finally, in Figure~\ref{higgs1255} we show these regions in detail.
The input parameters are the same as Figure~\ref{lambdamax} except $\lambda$.
All soft SUSY breaking parameters except $A_\lambda$ are set to $\susy$ ($\lambda  c_F = c_S = 1$), and we take $A^2_\lambda = \frac{2}{5} \susy^2$.
The $\lambda$ is varied at each point to fix the Higgs mass boson to be $125.5$ GeV.
The range of varying $\lambda$ is $0 \leq \lambda \leq \lambda_\textrm{max}$.
Therefore, in this figure the Higgs boson mass is fixed to be $125.5$ GeV
except  the dark-shaded regions, where one can not explain $m_h = 125.5\GeV$. 
The singlino relic abundance $\Omega_{\tilde{s}} h^2$ is consistent with the current value on the purple line, $\Omega_c h^2 = 0.1199$~\cite{Ade:2013zuv}. 
In the light blue region, $\Omega_{\tilde{s}}h^2 \leq \Omega_c h^2$. 
The left side of  the dashed lines can be covered by LUX (blue) \cite{Akerib:2012ys}, XENON1T (green) \cite{Aprile:2012zx} and  ILC (magenta) \cite{Asner:2013psa}.
\coll{As the future sensitivities of XENON1T, we have used $0.14$ (left), $0.326$ (middle) and $0.66$ (right) to the parameter $f_N$.}

Again, we show that the current dark matter relic abundance (purple line and light blue region) and the observed Higgs boson mass (not dark-shaded regions) can be explained simultaneously with $\tan\beta \sim \mathcal{O}(1)$ and $\susy \sim \mathcal{O}(10) \TeV$. 
In addition, we find that the future experiments, especially the direct dark matter search by the XENON1T, can probe a sign of the singlino dark matter.

\paragraph*{}
\coll{There are other proposals of the direct dark matter search whose future sensitivities are comparable to or stronger than prospect of XENON1T.
One of them is XMASS experiment, which is located at Kamioka Observatory.
The XMASS experiment also use an ultra-pure liquid xenon and its scattering with dark matter.
An expected sensitivity of the XMASS1.5 (XMASSII) project is that spin independent nucleon-dark matter cross section is $10^{-46}$ ~($10^{-47}$) [cm$^2$] at $m_{\rm DM} \sim 60 \GeV$ \cite{XMASStalk, private2} (cf. $10^{-45}$  [cm$^2$]  (XENON100),  $10^{-47}$  [cm$^2$]  (XENON1T), $10^{-45}$  [cm$^2$]  (LUX85days) and $10^{-46}$  [cm$^2$]  (LUX300days)).
Other ones are DarkSide-G2 experiment and LUX-ZEPLIN (LZ) experiment.
Their  expected sensitivities of the spin independent nucleon-dark matter cross section are  $10^{-47}$  [cm$^2$]  (DarkSide-G2 \cite{DStalk}) and   $10^{-48}$  [cm$^2$]  (LZ \cite{LZtalk}) at $m_{\rm DM} \sim 60 \GeV$.
}

\section{Discussions}
\label{4nodis}


The NMSSM is another model of the singlet extension of MSSM~\cite{Ellwanger:2009dp}.
As we have discussed previous chapter, this model is imposed the discrete $\mathbb{Z}_3$ symmetry and the superpotential is given as
\beq
W_{\textrm{NMSSM}} = \lambda \hat{S} \hat{H}_2 \hat{H}_1 + \frac{\kappa}{3} \hat{S}^3 + W_{\textrm{Yukawa}}.
\eeq
In the NMSSM, the singlino can obtain a radiative correction to the mass in addition to the tree-level mass $m_{\tilde{s}}^\textrm{tree} \sim 2 \kappa \langle S \rangle$.
The singlino resonant dark matter scenario may be successful with small $\tan\beta$  and small $\kappa$ in high-scale SUSY scenario.
In the small $\kappa$ limit, a singlet-like CP-odd scalar boson $A_1$ becomes a pseudo Nambu-Goldstone boson because of the existence of the global U(1) Peccei-Quinn symmetry.
Therefore, one may be able to make a distinction between the singlino resonant scenario in the nMSSM and NMSSM by the search for $h \to A_1 A_1 $~\cite{Cao:2013gba}.
 
 \paragraph*{}
Since there are some new sources of CP violating phases in the nMSSM, the EDM are generally generated through relative phase between $\mu_{\textrm{eff}}$ and $M_{\textrm{gaugino}}$ at the one-loop level.
We have estimate the one-loop electron EDM, which is generated from chargino-sneutrino loop and neutralino-selectron loop.
The expected electron EDM is (see \eq{MSSMEDM})
\beq
\left|\frac{d_e}{e} \right| &=& \frac{5 g^2 + g'^2 }{384 \pi^2} \frac{m_e}{\susy^2} \sin\phi \tan \beta  \hspace{0.6em}[\GeV^{-1}] \nonumber \\
&\sim& 6 \times 10^{-29} \left( \frac{10\TeV}{\susy} \right)^2 \sin\phi \tan\beta \hspace{0.6em}[\textrm{cm}],
\eeq
where $\phi = \textrm{arg}\left(\mu_{\textrm{eff}} M_{\textrm{gaugino}}\right)$.
One can obtain  $|d_e| \sim \mathcal{O}(10^{-29})$ $e~\textrm{cm}$ with $\tan\beta \sim \mathcal{O}(1)$, $\susy \sim \mathcal{O}(10) \TeV$ and $\sin \phi \sim \mathcal{O}(1)$.
Interestingly, the electron EDM of this size does not conflict with the current bound~\cite{Baron:2013eja} and can be probed by some future experiments~\cite{Sakemi:2011zz, Kawall:2011zz, Kara:2012ay}.

\section{Conclusion of the Resonant Singlino Dark Matter}
\label{CON}
In this chapter, we have studied the phenomenology of the singlino resonant dark matter scenario. 
We find that  including one-loop corrections to the neutralino masses, 
the singlino can explain the current dark matter relic abundance through the resonant annihilation via the Higgs boson, if the soft SUSY breaking scale is high scale.
We have shown that with high-scale SUSY breaking $\sim 10$ TeV and low $\tan\beta$, the  dark matter relic abundance and the SM Higgs boson mass can be explained simultaneously in this scenario.

Even for the high-scale SUSY, we have also shown that the parameter region where the singlino dark matter is consistent with the current dark matter relic abundance can be probed by the future experiments (see Figure~\ref{abundance},~\ref{higgs1255}).
Therefore, the singlino dark matter signal can be {\it ``a first sign"} of  the high-scale supersymmetry.


\chapter{Towards a Scale Free Electroweak Baryogenesis}
\label{EWBGchap}
\thispagestyle{empty}


\abstchapter{
This chapter is based on the work by the author \cite{Ishikawa:2014tfa}.
We propose a new electroweak baryogenesis scenario in high-scale SUSY models.
We consider a singlet extension of the minimal SUSY standard model introducing additional vector-like multiplets.
We show that the strongly first-order phase transition can occur at a high temperature comparable to the soft SUSY breaking scale.
In addition, the proper amount of the baryon asymmetry of the universe can be generated via the lepton number violating process in the vector-like multiplet sector.
The typical scale of our scenario, the soft SUSY breaking scale, can be any value. 
Thus our new electroweak baryogenesis scenario can be realized at arbitrary scales and we call this scenario as a scale free electroweak baryogenesis.
This soft SUSY breaking scale is determined by other requirements.
If the soft SUSY breaking scale is $\mathcal{O}(10) \TeV$, our scenario is compatible with the observed mass of the Higgs boson and the constraints by the electric dipole moments measurements and the flavor experiments.
Furthermore, the singlino can be a good candidate of the dark matter.
 }

\section{Electroweak Baryogenesis in the nMSSM}
\label{5no1}

In this chapter, we focus on the baryon asymmetry in the universe and the electroweak baryogenesis mechanism in the nMSSM.
This chapter is based on the work by the author \cite{Ishikawa:2014tfa}.
We will propose a new electroweak baryogenesis scenario in the nMSSM with high-scale SUSY breaking.
First, we briefly summarize the our electroweak baryogenesis scenario in the nMSSM.

The Electroweak baryogenesis (EWBG)~\cite{Kuzmin:1985mm,Shaposhnikov:1986jp,Shaposhnikov:1987tw} is one of the most promising mechanisms to generate the baryon asymmetry of the universe (BAU) \coll{$\eta\equiv n_B/s = (0.86 \pm 0.01 ) \times 10^{-10} \sim 10^{-10}$~\cite{Ade:2013zuv}.}
In this mechanism, the first-order phase transition of the Higgs field occurs and the bubbles are nucleated initially.
Then the CP asymmetric distributions of the particles are generated around the bubble walls if there is a source of CP asymmetry.
Finally, these CP asymmetric distributions turn into the BAU due to the decoupling of the sphaleron process.
This phase transition which associates with this sphaleron decoupling effect is called as the {\it strongly} first-order phase transition.

Within the standard model, this EWBG mechanism can not be realized by two reasons.
First, the strongly first-order phase transition can not occur while maintaining  the Higgs boson mass 125 GeV~\cite{Bochkarev:1987wf,Kajantie:1995kf}. 
Second, there is no CP-violating source enough to generate the proper amount of the baryon asymmetry~\cite{Gavela:1993ts,Huet:1994jb,Gavela:1994dt}.
Thus, this mechanism requires new physics which can cause the strongly first-order phase transition with new CP-violating sources.
The typical scale of this new physics seems to be comparable to the electroweak scale since this mechanism is supposed to occur around the electroweak scale.
Now, the new physics models with such a relatively low scale suffer from severe constraints from the collider searches, the EDM measurements and the flavor experiments.

In this chapter, we propose a new EWBG scenario in which EWBG occurs at arbitrary scales.
As a new physics model, we consider SUSY models which have a new physical scale, the soft SUSY breaking scale, $M_{\rm SUSY}$.
In this new scenario, the particles with the masses of $\mathcal{O}(M_{\rm SUSY})$ play important roles.
When the temperature of the universe drops across $\mathcal{O}(M_{\rm SUSY})$, the appearance of the universe changes drastically.
First, the dominant terms of the potential for the scalar fields change from the thermal terms to the soft SUSY breaking terms.
Second, the particles with masses $\mathcal{O}(M_{\rm SUSY})$ disappear due to the Boltzmann suppression.
These changes deform the shape of the potential for the Higgs fields and they may cause  the strongly first-order phase transition at the temperature $\mathcal{O}(M_{\rm SUSY})$.
In this mechanism, the value of $M_{\rm SUSY}$ is not constrained.
Thus, EWBG can be realized at arbitrary scale $M_{\rm SUSY}$ {if there is a proper amount of the CP-violating sources.

We consider the nMSSM~\cite{Panagiotakopoulos:1999ah,Panagiotakopoulos:2000wp,Dedes:2000jp} specially.
The potential of the nMSSM is suitable for the first-order phase transition. 
The ordinary EWBG scenarios in the nMSSM have been well studied in the literature~\cite{Menon:2004wv,Huber:2006wf}.
In our new scenario, we add extra vector-like multiplets to the nMSSM which are coupled to the singlet superfield.
In addition, we introduce a lepton number violating term in the vector-like multiplet sector.

Here, let us see the outline of our scenario.
In this scenario, the singlet scalar field obtains sizable thermal potential from the vector-like multiplets only at high temperatures.
Then, the absolute field value of the singlet scalar field becomes smaller at high temperatures than at the zero temperature.
As a result, the potential for the Higgs field gets deformed.
Furthermore, the global minimum of the potential for the Higgs field is generated far from the origin when the temperature is around $M_{\rm SUSY}$.
At this time, the strongly first-order phase transition occurs from the origin (symmetric vacuum) to this minimum (breaking vacuum).
Subsequently, the baryon$(B)$+lepton$(L)$ number is generated
\footnote{The concrete estimation of the $B+L$ number generated by the first-order phase transition is beyond the scope of this chapter and it is devoted to future work.}.
After the strongly first-order phase transition, the Higgs field is trapped at the breaking vacuum.
As the temperature decreases below $M_{\rm SUSY}$, the breaking vacuum is lifted up and disappears.
Then, the Higgs field returns to the symmetric vacuum.
In this interval, non zero $B-L$ number is generated from the $B+L$ number by the lepton number violating term.
As a result, the BAU is not washed out by the sphaleron process at the symmetric vacuum.
The lepton number violating process is active only when $T\gtrsim M_{\rm SUSY}$ since the number densities of the vector-like multiplets get Boltzmann-suppressed when $T\lesssim M_{\rm SUSY}$.
Thus, the BAU is generated and fixed at the temperatures smaller than $M_{\rm SUSY}$.
Finally, the Higgs field goes to the electroweak symmetry breaking vacuum when the temperature becomes the electroweak scale.

In this scenario, the whole processes occur at $T\sim M_{\rm SUSY}$.
Surprisingly, the scale $M_{\rm SUSY}$ becomes a free parameter up to the small electroweak scale corrections which are needed to realize the electroweak symmetry breaking vacuum.
Thus we call this scenario as a scale free electroweak baryogenesis.
On the other hand, the favored value of the scale $M_{\rm SUSY}$ can be determined by other experiments.
Considering the Higgs mass $125$ GeV~\cite{ATLAS:2013mma,CMS:yva} and SUSY flavor/CP problem, $M_{\rm SUSY}\sim \mathcal{O}(10)\text{ TeV}$ seems to be favored.
Moreover, the singlino, the fermionic component of the singlet superfield, can be a good candidate of the dark matter.
With $M_{\rm SUSY}\sim \mathcal{O}(10)\text{ TeV}$, the proper amount of the singlino dark matter can be obtained by resonant annihilation via the exchange of the standard model Higgs boson~\cite{Ishikawa:2014owa}.
We show that the lifetime of the singlino dark matter is long enough even though there is the lepton number violating term which induces its decay.
Therefore, this scenario can realize the proper Higgs boson mass, the right amount of the dark matter and the BAU without SUSY flavor/CP problem if $M_{\rm SUSY} \sim \mathcal{O}(10)$ TeV.
 
\section{The nMSSM with Vector-like Matters}
\label{sec_the_model}

In this section, we briefly introduce our model, the nMSSM~\cite{Panagiotakopoulos:1999ah,Panagiotakopoulos:2000wp,Dedes:2000jp} with vector-like multiplets.
We show the matter contents,  the symmetries and the interactions in our model.

The superpotential and the soft SUSY breaking terms of the nMSSM are given in Section~\ref{nMSSMLag}.
In addition, we add extra vector-like multiplets to the nMSSM.
These vector-like multiplets play important roles.
First, they give the sizable thermal corrections for $S$ to cause the first-order phase transition.
Second, they give the lepton number violation at high temperatures.
As the vector-like multiplets, we add ${\bf 16}~(\hat{Q}',\hat{\bar{U}}',\hat{\bar{D}}',\hat{L}',\hat{\bar{E}}',\hat{N}')+{\bf \overline{16}}~(\hat{\bar{Q}}',\hat{U}',\hat{D}',\hat{\bar{L}}',\hat{E}',\hat{\bar{N}}')$ multiplets (with SO(10) notation).
We express the MSSM multiplets as $\hat{Q}_i,\hat{\bar{U}}_i,\hat{\bar{D}}_i,\hat{L}_i,\hat{\bar{E}}_i$ with $i=1,2,3$ denoting the generation. 
In order to forbid unwanted terms of the vector-like multiplets, we impose additional $\mathbb{Z}_3$ and $\mathbb{Z}_2$ discrete symmetries (see Table~\ref{tab_ch}).
\coll{$\mathbb{Z}_3$ symmetry forbids the terms like $\hat{S}^2 \hat{L} \hat{H}_2$ which cause a rapid decay of the singlino, the dark matter candidate in our model (see Sec.~\ref{sec_DM} for details).}
$\mathbb{Z}_2$ symmetry is the vector-like multiplet parity where all vector-like multiplets are assigned as odd while the other multiples are assigned as even.
We consider the situation where this vector-like multiplet parity $\mathbb{Z}_2$ is slightly broken and the small mixings between the vector-like multiplets and the MSSM multiplets exist. 
\begin{table}[tbp]
\caption{The charge assignment.}
\begin{center}
\label{tab_ch}
\begin{tabular}{c|c c c c c c c c c c c c c c c c c}
\hline \hline 
{}$\mathbb{Z}_2${{\small -even}}&$\hat{H}_1$&$\hat{H}_2$&$\hat{S}$&$\hat{Q}_i$&$\hat{\bar{U}}_i$&$\hat{\bar{D}}_i$&$\hat{L}_i$
&$\hat{\bar{E}}_i$& & & & & & \\
{}$\mathbb{Z}_2${{\small -odd}}&& & &$\hat{Q}'$&$\hat{\bar{U}}'$&$\hat{\bar{D}}'$&$\hat{L}'$
&$\hat{\bar{E}}'$&$\hat{\bar{Q}}'$&$\hat{U}'$&$\hat{D}'$&$\hat{\bar{L}}'$&$\hat{E}'$&
$\hat{N}'$&$\hat{\bar{N}}'$ \\ \hline
$\mathbb{Z}_5^{R}$&1&1& 4&2&3&3&2&3&0&4&4&0&4&0&2  \\ \hline
$\mathbb{Z}_3$&0&0&0&2&1&1&2&1&1&2&2&1&2&2&1 \\  \hline \hline
\end{tabular}
\end{center}
\end{table}

The allowed superpotential by the symmetries $\mathbb{Z}_5^{R}$, $\mathbb{Z}_3$ and $\mathbb{Z}_2$ in the vector-multiplet sector is
\begin{align}
\label{w_sym}
	W_{\rm sym} &=\lambda_1 \hat{S} \left(\hat{\bar{Q}}' \hat{{Q}}'
+\hat{\bar{U}}'\hat{U}'+\hat{\bar{D}}'\hat{D}'+\hat{\bar{L}}'\hat{L}'
+\hat{\bar{E}}'\hat{E}'+\hat{\bar{N}}'\hat{N}'
\right)\nonumber \\
&+k_1\hat{L}'\hat{H}_1\hat{\bar{E}}'+k_2\hat{\bar{L}}'\hat{H}_1\hat{N}' 
+k_3\hat{Q}'\hat{H}_1\hat{\bar{D}}'+k_4\hat{Q}'\hat{H}_2\hat{\bar{U}}'\,,
\end{align}
where we take a universal coupling $\lambda_1$ for $\hat{S}\hat{X}'\hat{\bar{X}}'$ type terms for simplicity.
There are corresponding soft SUSY breaking terms like $A$-terms $A_{\lambda_1}SX'\bar{X}', A_{k_1}L'H_1\bar{E}$ and soft mass terms $m^2_{X'}|X'|^2,m^2_{\bar{X}'}|\bar{X}'|^2$ .
As mentioned above, we assume that the vector-like multiplet parity $\mathbb{Z}_2$ is slightly broken
\footnote{The $R$-symmetry $\mathbb{Z}_5^R$ is also broken softly.
Though, we assume that the terms introduced by the broken of $\mathbb{Z}_5^R$ are negligible except the tadpole terms of $\hat{S}$.
In addition, we assume that the size of these tadpole terms are still $\mathcal{O}({M_{\rm SUSY}})$ with our setup.}.
The terms which appear after the broken of $\mathbb{Z}_2$ are
\begin{align}
\label{sp:z2b}
	W_{\not{\mathbb{Z}}_2}&=\epsilon_S^i \hat{S} \left(\hat{\bar{Q}}' \hat{{Q}}_i
+\hat{\bar{U}}_i\hat{U}'+\hat{\bar{D}}_i\hat{D}'+\hat{\bar{L}}'\hat{L}_i
+\hat{\bar{E}}_i\hat{E}'
\right)\nonumber \\
&+\epsilon^i  
\left(
\hat{Q}_i\hat{H}_1\hat{\bar{D}}'+\hat{Q}' \hat{H}_1\hat{\bar{D}}_i
+\hat{Q}_i\hat{H}_2\hat{\bar{U}}'+\hat{Q}'\hat{H}_2\hat{\bar{U}}_i
+\hat{L}_i\hat{H}_1\hat{\bar{E}}'+\hat{L}'\hat{H}_1\hat{\bar{E}}_i
\right) \nonumber\\
&+\epsilon_N \hat{\bar{N}}'^3\,.
\end{align}
We set partially universal couplings $\epsilon_S^i$, $\epsilon^i$ and $\epsilon_N$ for simplicity.
In this chapter, we consider the superpotential
\beq
W=W_{\rm Yukawa}+W_{\rm nMSSM}+W_{\rm sym}+W_{\not{\mathbb{Z}}_2},
\eeq
where $W_{\rm Yukawa}$ is the ordinary Yukawa terms in the MSSM superpotential.
There are also the soft SUSY breaking terms for the MSSM multiplets like the soft masses for the stops $m^2_{\tilde{t}}$.

The lepton number ($L$) and the baryon number ($B$) of the vector-like multiplets are set as follows.
$\hat{Q}'$, $\hat{\bar{U}}'$, $\hat{\bar{D}}'$, $\hat{L}'$, and $\hat{\bar{E}}'$ have the same quantum numbers as the corresponding MSSM multiplets.
$\hat{\bar{X}}'$ has the opposite charge of $\hat{X}'$. The lepton number of the $\hat{N}'$ is decided by the term $k_2 \hat{\bar{L}}'\hat{H}_1\hat{N}'$ to conserve the lepton number:~$\hat{\bar{N}}'$ has the same quantum number with $\hat{\bar{E}}$.
Note that the term only $\epsilon_N\hat{\bar{N}}'^3$ violates the lepton number explicitly.

In this model, a singlino which is the fermionic component of the singlet superfield can be a good candidate of the dark matter~\cite{Ishikawa:2014owa}.
However, the singlino has a finite lifetime in this model since the $R$-parity is slightly broken due to the $\epsilon_N \hat{\bar{N}}'^3$ term.
In Sec.~\ref{sec_DM}, we show that our electroweak baryogenesis scenario is compatible with the singlino dark matter scenario.

\section{Overview of our scenario}
\label{sec:the_scenario}

In this section, we present the overview of our scenario.
Since there are several steps in this scenario, we briefly outline the series of the thermal history below.
The details of each step are given in the subsequent sections.

\begin{figure}[tbp]
\begin{center}
\includegraphics[width =16cm]{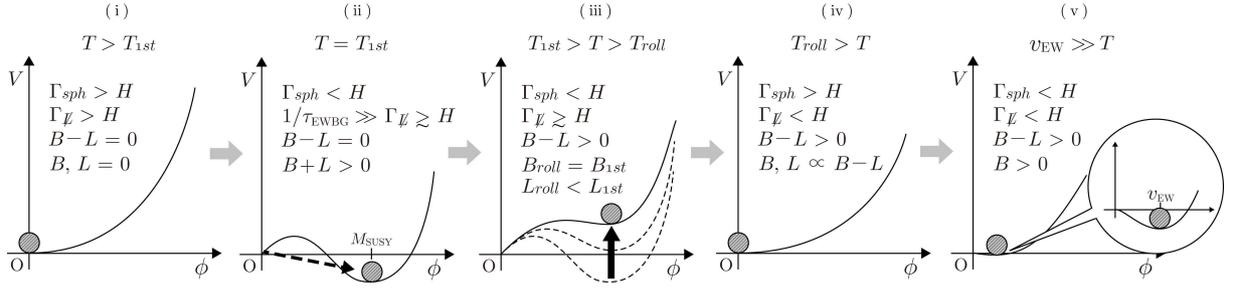}
 \vspace{-.2cm}
\caption{The outline of the thermal history of our scenario.
The details are given in the text.}
\label{fig:ponchi}
\vspace{-.2cm}
\end{center}
\end{figure} 

Figure~\ref{fig:ponchi} shows the rough sketch of the thermal history in our scenario.
Each graph shows the potential for the Higgs field and the graphs are aligned from left (i) to right (v) as time goes.
The shaded circle indicates the field value of the Higgs field.
$T$ denotes the temperature of the universe and $B~(L)$ denotes the baryon (lepton) number in the universe.
$H$ is the Hubble parameter at each time point.
The Hubble parameter during the radiation dominated era is given as \eq{Hubbledef}.
$\Gamma_{sph}$ is the effective sphaleron rate where the sphaleron process changes the $B+L$ number with conserving the $B-L$ number only if $\Gamma_{sph}>H$.
The situation $\Gamma_{sph}>H$ is realized when the field value of the Higgs field is smaller than the temperature (see Eq.~(\ref{sph_rate})).
$\Gamma_{\not{L}}$ is the effective lepton number decreasing rate coming from $\epsilon_N \hat{\bar{N}}'^3$ term. 
The lepton number violating process which changes the $L$ number is active only if $\Gamma_{\not{L}}>H$.
This condition $\Gamma_{\not{L}}>H$ corresponds to $T\gtrsim  M_{\rm SUSY}$.
If $T< M_{\rm SUSY}$, the number densities of the vector-like multiplets are suppressed exponentially since their masses are $\mathcal{O}(M_{\rm SUSY})$.
As a result, this lepton number violating process would be decoupled since this process is caused by the scattering (or decay) processes of the vector-like multiplets (see Eq.~(\ref{notL})). 

Here, we briefly outline the thermal history (see Figure~\ref{fig:ponchi}).
\begin{itemize}
\item[(i)]
At enough high temperatures compared to $\mathcal{O}(M_{\rm SUSY})$, the potential for the Higgs field is lifted and the Higgs field exists at the origin of the potential (symmetric vacuum).
Both $\Gamma_{sph}$ and $\Gamma_{\not{L}}$ are larger than $H$.
At this time, $B=L=0$ holds since there is no conserved number in the thermal equilibrium.

\item[(ii)]
As the temperature decreases, the global minimum(breaking vacuum) of the potential for the Higgs field appears far away from the origin.
The first-order phase transition of the Higgs field occurs at $T=T_{1st}$.
Note that both the temperature $T_{1st}$ and the field value of the Higgs field at the breaking vacuum are $\mathcal{O}(M_{\rm SUSY})$.
At this time, EWGB occurs and the $B+L$ number is generated in the interval of $\tau_{EWBG}$~\cite{Kuzmin:1985mm,Shaposhnikov:1986jp,Shaposhnikov:1987tw}.
In the interval of $\tau_{EWBG}$, $\Gamma_{\not{L}}$ does not work $(1/\tau_{EWBG}\gg \Gamma_{\not{L}})$ and the $B-L$ number is not generated.
On the other hand, the field value of the Higgs field at the breaking vacuum is larger than the temperature in this scenario.
It makes the sphaleron rate smaller $\Gamma_{sph}<H$ at the breaking vacuum.
Thus the sphaleron process is decoupled and generated $B+L$ number is not changed at the breaking vacuum.

\item[(iii)]
After EWBG, the Higgs field is trapped at the breaking vacuum.
During this time, the sphaleron process is decoupled $(\Gamma_{sph}<H)$.
On the other hand, the lepton number violating process is active $(\Gamma_{\not{L}}\gtrsim H)$ and the $L$ number decreases gradually.
Thus, the $B$ number is conserved and the generated $B+L$ number is converted to the $B-L$ number.

\item[(iv)]
At $T=T_{roll}\lesssim M_{\rm SUSY}$, the breaking vacuum (the local minimum of the potential for the Higgs field) disappears.  
Then the Higgs field returns to the symmetric vacuum again through the second-order phase transition.
The sphaleron process becomes active again $(\Gamma_{sph}>H)$ since the Higgs field exists at the symmetric vacuum.
On the other hand, the lepton number violating process becomes decoupled due to the Boltzmann suppression of the vector-like multiplets at this time $(\Gamma_{\not{L}}\lesssim H)$.
As a result, the generated  $B-L$ number is conserved.
Thus the $B$ number and $L$ number are fixed in the thermal equilibrium.

\item[(v)]
After the temperature becomes lower than the electroweak scale $\mathcal{O}(v_{EW})$, the Higgs field settles down at the electroweak symmetry breaking vacuum.
At this time, both the sphaleron process and the lepton number violating process are decoupled.
Thus, the generated $B-L$ number is conserved and the BAU exists until today.
\end{itemize}

In this scenario, there are two nontrivial points.
\begin{itemize}
\item
The strongly first-order phase transition of the Higgs field occurs at $T_{1st}\sim \mathcal{O}(M_{\rm SUSY})$.
\item
The lepton number violating process is active only when $T\gtrsim \mathcal{O}(M_{\rm SUSY})$
\end{itemize}
The first point is discussed in Sec.~\ref{sec:the_FOPT}.
The second point is discussed in Sec.~\ref{sec_BAU}.
In these sections, we show that these conditions are satisfied actually.
The essential point is that the typical scales of the system such as the potential and the masses of the relevant particles are all $\mathcal{O}(M_{\rm SUSY})$.
On the other hand, the scale $M_{\rm SUSY}$ is not constrained by this scenario.
Thus, we call this scenario as a scale free electroweak baryogenesis
\footnote{We do not consider the CP-violation sources explicitly.
The estimation including them is devoted to future work.}.

In addition, the singlino dark matter scenario~\cite{Ishikawa:2014owa} can be compatible with this scenario.
This fact is nontrivial since the $R$-parity is explicitly broken due to the lepton number violating term in our model.
Fortunately, the lifetime of the singlino is long enough and the singlino can be a good candidate of the dark matter, as we show in Sec.~\ref{sec_DM}.

\section{Strongly First-Order Phase Transition}
\label{sec:the_FOPT}

In this section, we show that the strongly first-order phase transition of the Higgs field occurs at $T\sim \mathcal{O}(M_{\rm SUSY})$.
In Sec.~\ref{subsec:the_potential},  we introduce the relevant potentials. 
Sec.~\ref{subsec:tree} is devoted to the intuitive understanding of its behavior.
In Sec.~\ref{subsec:numerical}, we analyze the full potential defined in Sec.~\ref{subsec:the_potential}.

\subsection{Full Scalar Potential}
\label{subsec:the_potential}

In this chapter, we consider the following potential
\beq
V(\phi_i, T) = V_0(\phi_i) + V_{\rm CW}(\phi_i) + V_T(\phi_i, T)\,,
\eeq
where $\phi_i$ ($i = 1, 2, s$) are the field values of $H_1^0, H_2^0, S$.
$V_0$, $V_{\rm CW}$ and $V_T$ are the tree-level, the Coleman-Weinberg and the thermal potential respectively.

Here we assume some conditions to make the potential simpler since the complete one-loop potential is highly complicated.
First, only $\mathcal{O}(1)$ couplings are taken into account.
Thus, we neglect the MSSM Yukawa couplings except the top Yukawa coupling $y_t$.
We also do not consider $\epsilon$ couplings which are introduced by the broken of the vector-like multiplet parity $\mathbb{Z}_2$ (see Eq.~(\ref{sp:z2b})).
The couplings of the Higgs field with the vector-like multiplets are assumed as $k\equiv k_1=k_2=\mathcal{O}(1)$ and $k_3,k_4\ll 1$ to make the potential simple (see Eq.~(\ref{w_sym})).
Then, the superpotential becomes
\begin{align}
	W_{\rm pot}&=y_t \hat{Q}_3\hat{H}_2\hat{\bar{U}}_3 +\lambda \hat{S} \hat{H}_2 \hat{H}_1 + \frac{m_{12}^2}{\lambda} \hat{S}
	\nonumber \\&+
	\lambda_1 \hat{S} \left(\hat{\bar{Q}}' \hat{{Q}}'
+\hat{\bar{U}}'\hat{U}'+\hat{\bar{D}}'\hat{D}'+\hat{\bar{L}}'\hat{L}'
+\hat{\bar{E}}'\hat{E}'+\hat{\bar{N}}'\hat{N}'
\right)\nonumber \\
&+k\hat{L}'\hat{H}_1\hat{\bar{E}}'+k\hat{\bar{L}}'\hat{H}_1\hat{N}' \,.
\label{eq:W_pot}
\end{align}
Second, we partially neglect the $H_2$ and $S$ dependences of the one-loop potential.
As we will see later, the strongly first-order phase transition occurs in $\tan\beta\sim 0$ direction and these dependences are irrelevant.
Third, we set all $A$-terms to be zero
\footnote{The CP-violating sources can enter in $A$-terms.
However, we do not consider them since we show the possibility of the strongly first-order phase transition at high temperatures in this chapter.
The study with explicit CP-violating sources can be found elsewhere.}
 and some soft SUSY breaking masses to be the same values for simplicity.
Fourth, we assume that the scalar components of the vector-like multiplets are heavy enough and their effects to the thermal self energy can be neglected.

Here we show the each potential $V_0$, $V_{\rm CW}$ and $V_T$.
\begin{itemize}
\item[$V_0$:]
We can write the tree-level potential from the superpotential Eq.~(\ref{eq:W_pot}) and the soft terms Eq.~(\ref{nMSSMsb}) as
\beq
V_0(\phi_i) = - M^2 \phi^2 + m^2_{s,0} \phi_s^2 + 2 t_S\phi_s + \lambda^2 \phi^2 \phi_s^2 + \bar{\lambda}^2\phi^4\,,\label{treeV}
\eeq
where
\beq
M^2 &\equiv& - m_1^2 \cos^2 \beta - m_2^2 \sin^2\beta + m_{12}^2 \sin 2 \beta\,,\label{def:m}\\
\bar{\lambda}^2 &\equiv& \frac{\lambda^2}{4} \sin^2 2 \beta + \frac{\bar{g}^2}{8} \cos^2 2 \beta\,,
\label{treeV2}
\eeq
and $\phi^2 = \phi_1^2 +\phi_2^2$, $\tan \beta = \phi_2/\phi_1$.

\item[$V_{\rm CW}$:]
For the Coleman-Weinberg potential, we consider the terms from the top/stops $V_{\rm CW}^{\rm t}$ and from the vector-like multiplets $V_{\rm CW}^{\rm vec}$
\begin{align}
	V_{\rm CM}=V_{\rm CM}^{\rm t}+V_{\rm CW}^{\rm vec}\,.
\end{align}
Each term has the form as
\begin{align}
\frac{N_C}{32 \pi^2} 
\left[ \sum_{i=\rm{scalars}} M_{i}^4 \left( \ln \left(\frac{M_{i}^2}{Q^2}\right) -\frac{3}{2}\right)
-  \sum_{i=\rm{fermions}}M_i^4 \left( \ln \left(\frac{M_i^2}{Q^2}\right) -\frac{3}{2}\right) 
\right]\,,
\end{align}
where $N_C$ is the color factor.
$M_i$'s are the masses of the corresponding particles.

$V_{\rm CW}^{\rm t}$ can be written as
\begin{align}
V_{\rm CW}^{\rm t} 
&= \frac{3}{32 \pi^2} 
\left[ \sum_{\pm} M_{\tilde{t},\pm}^4 \left( \ln \left(\frac{M_{\tilde{t},\pm}^2}{Q^2}\right) -\frac{3}{2}\right)
-  2 M_t^4 \left( \ln \left(\frac{M_t^2}{Q^2}\right) -\frac{3}{2}\right) 
\right]\,,\\
M_t &= y_t \phi_2\,, \\
M_{\tilde{t},\pm}^2& =m_{\tilde{t}}^2 + (y_t \phi_2)^2\pm y_t\lambda |\phi_s|\phi_1\,,
\end{align}
where $M_t$ is the mass of the top quark and $M_{\tilde{t},\pm}$ are the diagonalized masses of the stops with given $\phi,\phi_s$. 
Here, we assume the universal soft mass $m^2_{\tilde{t}}$ for the left- and the right-handed stops. 
Note that in this potential  we have neglected the stop $A_t$ term and $D$ terms, and we have taken account of the filed dependence on the singlet scalar. 
These are different points from \eq{topstopCW} in Section  \ref{LagMSSM}.

For the vector-like multiplets, we can diagonalize the mass matrix analytically with the assumption written in Sec.~\ref{subsec:the_potential}. 
Thus $V_{\rm CW}^{\rm vec}$ can be written as
\begin{align}
         V_{\rm CW}^{\rm vec}&=
         \frac{1}{32\pi^2}\left[
         2\sum_{{\pm},i=1,2}
         M_{si\pm}^4 \left( \ln \left(\frac{M_{si\pm}^2}{Q^2}\right) -\frac{3}{2}\right)
         -4\sum_{\pm}
         M_{f\pm}^4 \left( \ln \left(\frac{M_{f\pm}^2}{Q^2}\right) -\frac{3}{2}\right)
         \right]\,.
\end{align}
where $M_{s1\pm}, M_{s2\pm}$ and $M_{f\pm}$ are the diagonalized masses of the vector-like particles.
The mass matrices of the vector-like matters are given in Appendix~\ref{vecMASS}.
The diagonalized massed of the vector-like particles are,
\begin{align}
         \label{s1pm}
         M_{s1\pm}^2&=\frac{1}{2}\left(
         m_{L'}^2+m_{N'}^2+2\lambda_1^2\phi_s^2+k^2\phi_1^2\pm
         \sqrt{(m_{L'}^2-
         m_{N'}^2 + k^2\phi_1^2
         )^2+4\lambda_1^2k^2\phi_s^2\phi_1^2}
         \right)\,,\\
         \label{s2pm}
         M_{s2\pm}^2&=\frac{1}{2}\left(
         m_{L'}^2+m_{N'}^2+2\lambda_1^2\phi_s^2+k^2\phi_1^2\pm
         \sqrt{(m_{L'}^2-
         m_{N'}^2-k^2\phi_1^2
         )^2+4\lambda_1^2k^2\phi_s^2\phi_1^2}
         \right)\,,\\
         \label{fpm}
         M_{f\pm}^2&=\frac{1}{2}\left(
         2\lambda_1^2\phi_s^2+k^2\phi_1^2\pm
         \sqrt{k^4\phi_1^4+4\lambda_1^2k^2\phi_s^2\phi_1^2}
         \right)\,.
\end{align}
Here, we assume $m^2_{L'}=m^2_{\bar{L}'}$, $m^2_{N'}=m^2_{\bar{N}'}=m^2_{E'}=m^2_{\bar{E}'}$.

\item[$V_T$:]
For the thermal potential, we consider the improved one-loop thermal potential.
It means that the thermal self energy for all scalars and the longitudinal components of the gauge bosons are taken into account.
Thus we consider the following set of the thermal potential
\beq
V_T(\phi_i,T) = V_T^H(\phi, T) +V_T^{A}(\phi,\phi_s,T)+ V_T^S(\phi_s, T)+V_T^{\rm mix}
(\phi,\phi_s,T)\,.
\eeq
Each term have the form as $\sum_{i=\rm{particles}}C_iV_{\textrm{th}}^{B/F}(M_i/T,T)$ where $C_i$'s are the numerical constants and $V_{\textrm{th}}^{B/F}$ is defined as~\cite{Dolan:1973qd}
\beq
\label{eq:thermal}
V_{\textrm{th}}^{B/F}(x,T)=\pm\frac{T^4}{\pi^2}\int_0^{+\infty}dz ~z^2\ln\left( 1\mp e^{-\sqrt{z^2+x^2}}
         \right)\,, \\
\frac{V_{\textrm{th}}^{B/F}(x,T)}{T^4} \sim \begin{cases}
         -\frac{\pi^2}{45}+\frac{x^2}{12},~~(x \ll 1)~~\text{for boson(B)}\,,\\
         -\frac{7\pi^2}{360}+\frac{x^2}{24},~~(x \ll 1)~~\text{for fermion}(F)\,.
         \end{cases}
\eeq
$V_T^H$ is the improved one-loop thermal potential for the Higgs field coming from the Z-boson, the W-boson and the top-quark.
\begin{align}
         V_T^H(\phi, T)& = 
   6V^F_{\textrm{th}}\left(M_t/T,T\right)+
\frac{2}{3}
         \left[3V_{\rm th}^B\left(M_W/T,T \right)+\frac{3}{2}V_{\textrm{th}}^B\left(M_Z/T,T\right)\right]
         \nonumber \\
         &+\frac{1}{3} \left[3V_{\rm th}^B\left(\tilde{M}_W/T,T \right)
         +\frac{3}{2}V_{\textrm{th}}^B\left(\tilde{M}_Z/T,T\right)\right]\,,\\
\end{align}
where $M_W^2=g^2\phi^2/2$, $M_Z^2=\bar{g}^2\phi^2/2$, $\tilde{M}_W^2=M_W^2+19g^2T^2/6$ and $\tilde{M}_Z^2=M_Z^2+19g^2T^2/6+59g'^2T^2/18$.
$V_{\textrm{th}}^{B/F}$ is defined as~\cite{Dolan:1973qd}.
Note that if $\phi\lesssim T$ holds, the Higgs field $\phi$ obtains thermal mass terms:
\begin{align}
\label{thm_h}
	V_T^{H}(\phi,T)\simeq
	\left( \frac{y_t^2}{4}\sin^2\beta T^2+\frac{3}{4}\left(2\bar{g}^2+g^2\right)T^2\right)\phi^2\,.
\end{align}

$V_T^{A}$ comes from the thermal loops of the charged Higgs boson and the CP-odd Higgs boson.
We have to take this effect into account since a relatively light charged/CP-odd Higgs boson is favored to induce the first-order phase transition.
This term can be written as
\begin{align}
         V_T^{A}(\phi,\phi_s,T)&=V_{\rm th}^B
         \left(\tilde{M}_{\rm charged}/T,T \right)
         +\frac{1}{2}V_{\rm th}^B\left(\tilde{M}_{\rm odd}/T,T \right)\,,\\
         \tilde{M}_{\rm charged}^2&=m_1^2+m_2^2+2\lambda^2\phi_s^2+\frac{{g}^2}{2}\phi^2+\Pi_A\,,\\
          \tilde{M}_{\rm odd}^2&=m_1^2+m_2^2+2\lambda^2\phi_s^2+\lambda^2\phi^2+\Pi_A\,,\\
          \Pi_A&= \frac{\bar{g}^2}{4}T^2+ \frac{g^2}{2}T^2+
          \frac{y_t^2}{4}T^2+ \frac{\lambda^2}{3}T^2+\frac{k^2}{6}T^2\,.
\end{align}
where $\tilde{M}_{\rm charged}$ is the mass of the charged Higgs boson and $\tilde{M}_{\rm odd}$ is the mass of the CP-odd Higgs boson.

$V_T^S$ is the one-loop thermal potential for $\phi_s$ coming from the colored vector-like fermions and the Higgsinos as
\beq
 V_T^S(\phi_s,T) = 24 V^{F}_{\textrm{th}}\left({\lambda_1 \phi_s}/{T},T\right) + 4 V^{F}_{\textrm{th}}\left({\lambda \phi_s}/{T},T\right)\,.
\eeq
The second term comes from the Higgsinos and we neglect their small mixing to the singlino and the gauginos.
Note that if $\phi_s \lesssim T$ holds, $\phi_s$ obtains the thermal mass terms:
\begin{align}
\label{thm_s}
	V_T^{S}(\phi_s,T)\simeq \left(\lambda^2_1T^2+\frac{\lambda^2}{6}T^2\right) \phi_s^2\,.
\end{align}

$V_T^{\rm mix}$ is the one-loop thermal potential coming from the vector-like multiplets $\bar{L}',$ $L', $ $\bar{E}',$ $E',\bar{N}',N'$. 
This potential can be written as
\begin{align}
         V_{T}^{\rm mix}&=
         2\sum_{i\pm,i=1,2}V^{B}_{\rm th}(\tilde{M}_{si\pm}/T,T)
         +4\sum_{\pm} V^{F}_{\rm th}(M_{f\pm}/T,T)\,.
\end{align}
$\tilde{M}_{si\pm}$ can be obtained by the replacement of $m_{L'}^2\rightarrow m_{L'}^2+3g^2T^2/8+k^2 T^2/6$ and $m_{N'}^2\rightarrow m_{N'}^2+k^2T^2/3$ in $M_{si\pm}^2$ (see Eq.~(\ref{s1pm}, \ref{s2pm})).
Here, we neglect the corrections of order $\mathcal{O}( g'^2T^2)$ in the thermal self energy.
\end{itemize}

\subsection{Tree-Level Analysis including Thermal Mass Terms}
\label{subsec:tree}

In this section, we give the intuitive understanding of the potential.
We consider the simplified potential which has only the tree-level terms and the thermal mass terms. 
As the thermal mass terms, we include the terms $T^2\phi_i^2$.
Analysis including the full terms is written in the next subsection.
Here, we show that the potential is deformed due to the thermal mass terms for the singlet field $\phi_s$. 
We also show that the global minimum of the potential for the Higgs field appears far away from the origin only at high temperatures.

The potential with only the tree-level terms and the thermal mass terms $V_{\rm tr+th}$ can be written as
\begin{align}
	V_{\rm tr+th}(\phi,\beta,\phi_s,T)
	&=(y_\phi^2T^2- M^2) \phi^2 + (y_S^2T^2+m_{s,0}^2) \phi_s^2 + 2 t_S \phi_s + \lambda^2 \phi^2 \phi_s^2 + \bar{\lambda}^2\phi^4\,,
\end{align}
where $y_{\phi}^2=\frac{y_t^2}{4}\sin^2\beta+\frac{3}{4}(2\bar{g}^2+g^2)$ and $y_{S}^2=\lambda^2_1+\frac{\lambda^2}{6}$ 
(see Eq.~(\ref{thm_h},\ref{thm_s})).
The field value of the singlet scalar field can be driven from the minimization condition $\partial V_{\rm tr+th}(\phi_i,T) /\partial \phi_s = 0$. 
It is derived as 
\beq
\phi_{s} = -\frac{t_S }{m^2_{s,0} + \lambda^2 \phi^2+y_S^2T^2} \sim \mathcal{O}(M_{\rm SUSY})\,.
\label{vevs}
\eeq
Since $t_S \sim \mathcal{O}(M^3_{\rm SUSY})$~\cite{Panagiotakopoulos:1999ah,Panagiotakopoulos:2000wp,Dedes:2000jp}, the absolute field value of the singlet scalar field becomes $\mathcal{O}(M_{\rm SUSY})$.
Note that it decreases when the field value of the Higgs field $\phi$ or the temperature $T$ increases.
This is one of the key features of our model.
After substituting the field value of the singlet scalar field, the potential becomes 
\begin{align}
	V_{\rm tr+th}(\phi,\beta, T)=-M^2 \phi^2  + y_{\phi}^2 T^2 \phi^2 + \bar{\lambda}^2 \phi^4 -\frac{t_S^2}{m_{s,0}^2 + \lambda^2 \phi^2 + y_S^2 T^2}\,.
\end{align}
For convenience, we rewrite the potential as the following form
\begin{align}
	v(X,\beta, T) &\equiv V_{\rm tr+th}(\phi,\beta, T) \frac{f(T)m_{s,0}^2}{t_S^2}
	\nonumber \\
	&=a(\beta,T)^2 X^2 + \left(- b(\beta,T)^2 + c(\beta,T)^2 \right) X-\frac{1}{1+X }\,,
\end{align}
where
\begin{align}
f(T)\equiv 1+{y_S^2}\frac{T^2}{m_{s,0}^2}\,,\hspace{0.8cm}
	X\equiv \frac{1}{f(T)}\frac{\lambda^2{\phi}^2}{m_{s,0}^2}\,,
\end{align}
\begin{align}
	a(\beta,T)^2\equiv [f(T)]^3\frac{\bar{\lambda}^2m_{s,0}^6 }{\lambda^4 {t}_{S}^2}\,,\hspace{0.2cm}
	\label{def:b}
	b(\beta,T)^2\equiv [f(T)]^2\frac{{M}^2 m_{s,0}^4}{\lambda^2 {t}_S^2}\,,\hspace{0.2cm}
	c(\beta,T)^2 \equiv [f(T)]^2\frac{y_{\phi}^2 {T}^2 m_{s,0}^4}{\lambda^2 {t}_S^2}\,.
\end{align}
Note that  $f(T)\geq1$ holds.
In addition, $a(\beta,T)$, $b(\beta ,T)$ and $c(\beta ,T)$ are increasing functions with respect to $T$. 

From here, we consider the following conditions:
\begin{itemize}
\item[(i)]
Only the electroweak symmetry breaking vacuum is realized at the zero temperature.
\item[(ii)]
The global minimum of the potential for the Higgs field appears far away from the origin at high temperatures.
\end{itemize}
For simplicity, we mainly consider two directions.
One is the direction with $\beta_{\rm vac}$ being the angle of the vacuum at the zero temperature.
The other is the direction with $\beta_{\rm tr}$ being the typical angle of the first-order phase transition.
As we will see later, $\beta_{\rm tr}\sim 0$ is favored to realize the first-order phase transition.

\subsubsection*{Zero temperature conditions}
First, let us consider the conditions to have only the electroweak symmetry breaking vacuum at the zero temperature.

For the $\beta_{\rm vac}$ direction, in order to realize the electroweak symmetry breaking vacuum properly, we need 
\begin{align}
	\frac{\partial V_{\rm tr+th}(\phi,\beta_{\rm vac},T=0)}{\partial \phi}\big|_{\phi\sim0}&<0\,,\\
	\frac{\partial V_{\rm tr+th}(\phi,\beta_{\rm vac},T=0)}{\partial \phi}\big|_{\phi=v_{EW}}&=0\,,
\end{align}
where $v_{\rm EW}\simeq 174.1$ GeV is the vacuum expectation value of the Higgs field at the zero temperature.
$\phi\sim0$ indicates that $\phi$ is at the vicinity of the origin.
These conditions can be rewritten as 
\begin{align}
\label{cnd:b1}
         &b(\beta_{\rm vac},0)>1\,,\\
	b(\beta_{\textrm{vac}},0)^2&=\frac{1}{(1+X_{EW})^2}+2a(\beta_{\textrm{vac}},0)^2X_{EW}\nonumber\\
	&\simeq 1+2X_{EW}(a(\beta_{\textrm{vac}},0)^2-1)\,,
\end{align}
where $X_{EW}\equiv \lambda^2v_{EW}^2/m_S^2$.
Note that $X_{EW}\ll1$ holds since we assume the soft SUSY breaking scale is much larger than the electroweak scale.
In order to satisfy these conditions, we need
\begin{align}
	\label{vacvac}
	a(\beta_{\textrm{vac}},0)&>1\,,\\
	\label{vacvacvac}
	b(\beta_{\textrm{vac}},0)&=1+\mathcal{O}\left(\frac{v_{EW}^2}{m_S^2}\right)\,.
\end{align}

In addition, we impose the condition not to generate the minimum at $\beta~\Slash{{$\simeq$}}~\beta_{\rm vac}$
\begin{align}
\frac{\partial V_{\rm tr+th}(\phi,\beta,T=0)}{\partial \phi}\big|_{\phi\sim0}>0\hspace{1cm}{\rm for\ } \beta~\Slash{{$\simeq$}}~\beta_{\rm vac}\,.
\end{align}
This condition can be rewritten as
\begin{align}
b(\beta,0)<1\hspace{1cm}{\rm for\ } \beta~\Slash{{$\simeq$}}~\beta_{\rm vac}\,.\label{b_cond}
\end{align}

For the $\beta_{\rm tr}$ direction, there should be no global minimum at the zero temperature.
Thus, the condition $ V_{\rm tr+th}({}^{\forall}\phi,\beta_{\rm tr},0) -V_{\rm tr+th}(0, \beta_{\rm tr},0)>0$ is imposed and can be rewritten as
\begin{align}
\label{tr0}
       \left(a(\beta_{\rm tr},0)-1\right)^2+b(\beta_{\rm tr},0)^2<1\,.
\end{align}

\subsubsection*{High temperatures conditions}
Next, let us consider the conditions to have the global minimum far away from the origin at high temperatures.

Suppose that at the critical temperature $T_C$, two minima of the potential appear at the origin and at $X=X_C,\beta=\beta_{\rm tr}$.
The condition becomes
\begin{align}
         v(X_C,\beta_{\rm tr},T_C)&=v(0,\beta_{\rm tr},T_C)\,,\\
         v'(X_C,\beta_{\rm tr},T_C)&=0\,,
\end{align}
where the prime means the partial derivative by $X$.
To have the positive solutions of $X_C$ and $T_C$, the necessary and sufficient conditions are
\beq
\label{tr1}
         a(\beta_{\rm tr},T_C)  < 1\,,\hspace{2cm}\\     
\label{tr2}
         \left(a(\beta_{\rm tr},T_C)-1\right)^2 +b(\beta_{\rm tr},T_C)^2  > 1\,.
\eeq

\subsubsection*{Solutions}
Let us see that the conditions Eqs.~(\ref{vacvac}),~(\ref{vacvacvac}),~(\ref{b_cond}),~(\ref{tr0}),~(\ref{tr1}),~(\ref{tr2}) can be satisfied simultaneously.
We divide these conditions to the pairs of Eqs.~(\ref{vacvacvac},~\ref{b_cond}), Eqs.~(\ref{vacvac},~\ref{tr1}) and Eqs.~(\ref{tr0},~\ref{tr2}).

First, we see the conditions Eqs.~(\ref{vacvacvac},~\ref{b_cond}).
To satisfy these conditions simultaneously, let us parameterize $b(\beta,0)$ as the following form
\begin{align}
\label{bb_bb}
b(\beta,0)=b_1+b_2\cos (2\beta-2b_3)\,.
\end{align}
Note that $b_1$, $b_2$ and $b_3$ are the function of $m_1^2$, $m_2^2$, $m_{12}^2$, $\lambda^2$, $t_S$ and $m_{s,0}^2$
(see Eqs.~(\ref{def:m},~\ref{def:b})).
If we take  these values to satisfy $b_1+b_2\simeq 1$, $b_2>0$ and $b_3\simeq \beta_{\rm vac}$, these conditions can be satisfied easily.

Second, we consider the conditions Eqs.~(\ref{vacvac},~\ref{tr1}).
Note that the conditions Eqs.~(\ref{vacvac},~\ref{tr1}) are the opposite conditions.
In addition, $a(\beta,T)$ is an increasing function of $T$.
Thus, the two conditions Eqs.~(\ref{vacvac},~\ref{tr1}) can not be satisfied with only one direction.
However, these conditions can be satisfied with the  two directions $\beta_{\rm vac}$ and $\beta_{\rm tr}$ are needed.
Next, let us see that $\beta_{\rm tr}\sim 0$ is favored.
Note that if the ratio $a(\beta_{\rm vac},0)/a(\beta_{\rm tr},0)$ is larger, it is easier to satisfy the two conditions Eqs.~(\ref{vacvac},~\ref{tr1}) at the same time.
On the other hand, if $\lambda^2>\bar{g}^2/2$ holds, $a(\beta,0)$ can be parameterized as
\begin{align}
	a(\beta,0)=a_1-a_2\cos(4\beta)\,,
\end{align}
with $a_1,a_2>0$.
Thus, if $\beta_{\rm vac}$ is near $\pi/4$, $\beta_{\rm tr}\sim 0$ is favored to give the ratio $a(\beta_{\rm vac},0)/a(\beta_{\rm tr},0)$ larger and satisfy these two conditions.

Finally, let us consider the conditions Eqs.~(\ref{tr0},~\ref{tr2}).
The discrepancy between the conditions Eqs.~(\ref{tr0},~\ref{tr2}) can be reconciled by $f(T_C)$.
In other words, the thermal mass of $\phi_s$ can work to generate the global minimum of the potential for the Higgs field only at high temperatures.
Actually, if we find the values of $a(\beta_{\rm tr},0)$, $b({\beta_{\rm tr}},0)$ and $f(T_C)$ which satisfy
\begin{align}
\left(a(\beta_{\rm tr},0)-1\right)^2+b(\beta_{\rm tr},0)^2 &< 1\,,\\
\left(f(T_C)^{3/2}a(\beta_{\rm tr},0)-1\right)^2+f(T_C)^2b(\beta_{\rm tr},0)^2 &> 1\,,
\end{align}
the conditions Eq.~(\ref{tr0},~\ref{tr2}) can be satisfied.

The above solutions can be achieved simultaneously with appropriate parameters.
Thus the global minimum far away from the origin can be generated only at high temperatures due to the thermal mass for the singlet field $\phi_s$.
Note that small value of $a(\beta_{\rm tr},0)$ and large value of $b(\beta_{\rm tr},0)$ are favored in order to satisfy the above conditions.
Small $a(\beta_{\rm tr},0)$ is satisfied easily} with $\tan \beta_{\rm vac}\sim1$.
On the other hand, large value of $b(\beta_{\rm tr},0)$ corresponds to small $m_{12}^2$ compared to $|m_1^2+m_2^2|$.
This situation makes the charged Higgs boson light.
As we will see in the full potential analysis of the next subsection, the strongly first-order phase transition can actually occur at the high temperature. 
In addition, the region with $\tan\beta_{\rm vac}\sim \mathcal{O}(1)$ and the light charged Higgs boson is favored.

\subsection{Numerical Analysis with  Full Potential}
\label{subsec:numerical}

In this section, we analyze the full potential introduced in Sec.~\ref{subsec:the_potential}. 
We show that the strongly first-order phase transition can actually occur at the temperature comparable to $M_{SUSY}$.
At first, we show the thermal history at a benchmark point.
Next, the conditions for the strongly first-order phase transition are discussed.
Then, we present a scatter plot and show that the region with  low $\tan\beta_{\rm vac}$ and the light charged Higgs boson is favored in our scenario.

\begin{table}[tbp]
\caption{The parameters at the benchmark point.}
\begin{center}
\label{table}
\begin{tabular}{c c c c c c c c c c c}
\hline \hline
$\tan\beta_{\rm vac}$&$\lambda^2$&$\lambda_1^2$&
$k$& $t_s/m_{s,0}^3$&
$m_1^2/m_{s,0}^2$&$m_2^2/m_{s,0}^2$&$m_{12}^2/m_{s,0}^2$
&$Q/m_{s,0}$  \\ \hline
2.0  & 0.50   & 0.50   & 1.0&0.58&
-0.1657 &-0.1675 &0.001226 &1 \\ \hline \hline
\end{tabular}
\end{center}
\end{table}

\subsubsection*{Thermal history}
Now, let us see the typical thermal history of our scenario.
Table~\ref{table} shows the benchmark parameters.
\coll{We take $\tb_{\rm vac} = 2.0$,~$\lambda^2 = \lambda_1^2 = 0.50$ and $\kappa = 1.0$.}
The standard model coupling constants have scale dependence.
We take the values at the scale $10 {\rm ~TeV}$: $y_t^2=0.753$, $\bar{g}^2=0.528$ and $g^2=0.394$.
For simplicity, we assume that all of the soft SUSY breaking masses are same $m^2_{\tilde{t}}=m^2_{X'}=m^2_{\bar{X}'}=m^2_{s,0}$.
In order to realize the electroweak symmetry breaking vacuum at the zero temperature, $\mathcal{O}(v_{\rm EW}^2/m_{s,0}^2)$ corrections are needed.
However, such small corrections are negligible for the high temperature dynamics. 
Thus we do not consider the corrections
\footnote{We impose the zero temperature conditions as  $a(\beta_{\rm vac},0)>1$ and $b(\beta_{\rm vac})=1$ (see Eq.~(\ref{vacvac},~\ref{vacvacvac})).
In order to impose these conditions easily, we absorb the tadpole and quadratic terms of the Coleman-Weinberg potential into the tree parameter. 
The explicit formulae of the tadpole and quadratic terms are given in Appendix~\ref{CWApp}}.
Note that we have checked that at the benchmark point there is no Landau pole of couplings $k$ up to the GUT scale using one-loop RGEs \cite{Joglekar:2013zya}.

\begin{figure}[tbp]
\begin{center}
\includegraphics[width =12cm]{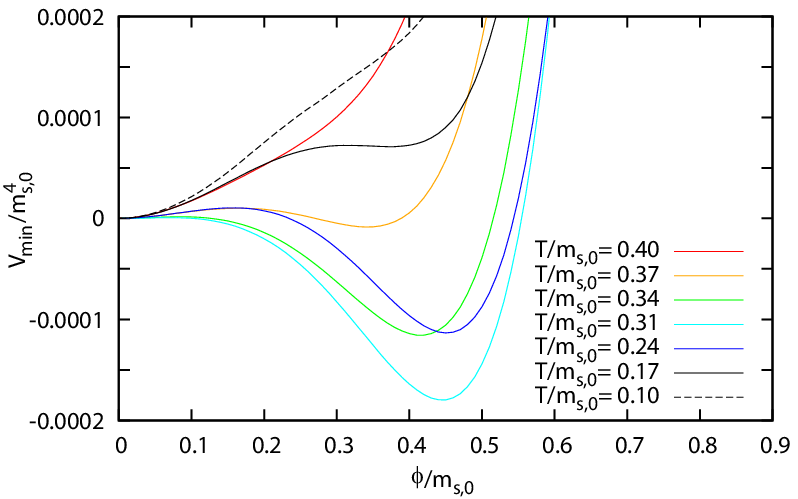}
~
\vspace*{1cm}

~
\includegraphics[width =12cm]{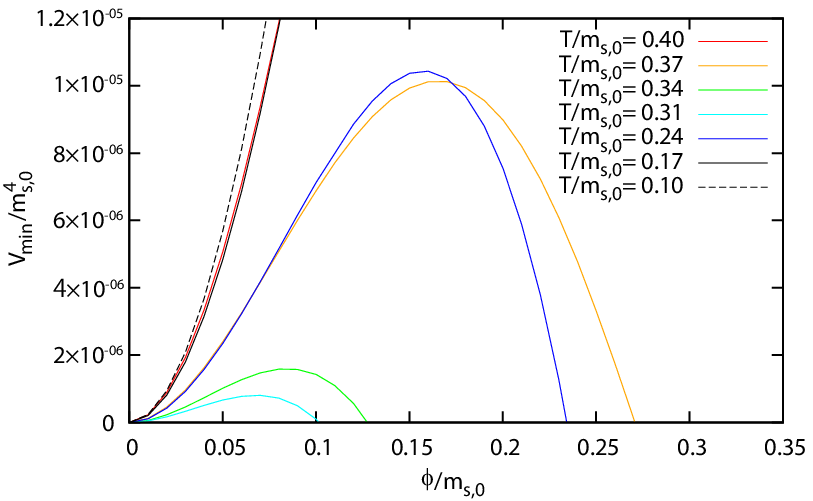}
 \vspace{-.2cm}
\caption[The potential for the Higgs field $V_{\rm min}(\phi,T)$ as a function of $\phi$ with varying temperatures $T$.]{The potential for the Higgs field $V_{\rm min}(\phi,T)$ as a function of $\phi$ with varying temperatures $T$.
$\phi_s$ and $\tan\beta$ are calculated to minimize the potential for each given $\phi$ and $T$.
Here, we subtract the constant term from the potential to set $V_{\rm min}(\phi=0,T)=0$.
The region with small $\phi$ is enlarged in the right figure.}
\label{fig:pot}
\vspace{-.7cm}
\end{center}
\end{figure} 

Figure~\ref{fig:pot} shows the potential for the Higgs field $V_{\rm min}(\phi,T)$ as a function of $\phi$ with varying temperatures $T$.
$\phi_s$ and $\tan\beta$ are calculated to minimize the potential for each given $\phi$ and $T$.
Typically, $\phi_s / m_{s,0} \sim -0.5 $ and $\tan \beta\sim 0.01 - 0.1$ hold.
At the high temperature (the red line $T/{m_{s,0}}=0.4$), the origin is the only minimum of the potential. 
As the temperature decreases, a global minimum appears at $\phi/m_{s,0}\sim 0.4$ (see the orange line $T/{m_{s,0}}=0.37$). 
Then, after $T/m_{s,0}=0.31$ (the cyan line), the potential is lifted up and the local minimum  disappears at $T/m_{s,0}=0.17$ (the black line).

Note that $m_{s,0}$ can be any value in this analysis.
If $m_{s,0}$ is varied, the size of the corrections $\mathcal{O}(v_{\rm EW}^2/m_{s,0}^2)$ changes.
In addition, the values of the standard model couplings change since their values depend on the scale to calculate.
However, up to these small corrections, the results do not depend on the value of $m_{s,0}$.
Thus we call this scenario as a scale free electroweak baryogenesis.

\subsubsection*{Strongly first-order phase transition}
Here, we show the conditions for the strongly first-order phase transition.

First, let us see the condition for the first-order phase transition to occur.
If the global minimum of the potential exists except the origin, the vacuum tunneling from the origin to the minimum can occur.
We call this global minimum as the breaking vacuum and the origin as the symmetric vacuum.
The finite temperature vacuum tunneling rate $\Gamma_{\rm tran.}$ per unit space-time volume $V$ is given as the following form:
\beq
\frac{\Gamma_{\rm tran.}}{V} \sim T^4 e^{-S(T)}\,,
\eeq
where $S(T) \equiv S_3/T$ and $S_3$ is the three-dimensional Euclidean action which is evaluated on the bounce solution~\cite{Linde:1980tt, Linde:1981zj}.
The condition for the first-order phase transition to occur is given by
\beq
\int dt \frac{1}{H^3}  T^4 e^{-S(T)} = 1\,.
\eeq
For $T\sim \mathcal{O}(\text{TeV})$, the first-order phase transition occurs at $S(T) \lesssim 130$~\cite{Quiros:1999jp}.
Here, we adopt the condition $S(T)=130$ for the first-order phase transition to occur.
Since this condition has only a logarithmic dependence on the temperature, we ignore this dependence. 
We show this condition of the first-order phase transition  explicitly in Appendix~\ref{finiteTVacuumdecay}.

Second, we show the condition for the {\it strongly} first-order phase transition.
After the vacuum tunneling occurs, the Higgs field is trapped at the breaking vacuum.
To cause EWBG, the sphaleron process have to be decoupled at the breaking vacuum since the $B+L$ number should not be washed out.
The sphaleron rate is evaluated as~\cite{Klinkhamer:1984di}
\beq
\label{sph_rate}
\Gamma_{sph} \propto T e^{-2 \frac{4 \sqrt{2} \pi}{g} \frac{ \Delta \phi }{T}}\,,
\eeq
with $\Delta \phi\equiv \sqrt{\phi_1^2+\phi_2^2}$ at the breaking vacuum. 
\mtrem{説明}
In order to decouple the sphaleron process, $\Gamma_{sph} \ll H$ is required.
This condition is equivalent to $\Delta \phi/T\gtrsim 0.9$~\cite{Menon:2004wv}, which is derived at $T \sim \mathcal{O}(100) \GeV$.
Since this condition has only a logarithmic dependence on the temperature, we adopt the condition $\Delta\phi/T>0.9$ as the strongly first-order phase transition
\footnote{With higher temperature, the condition value $0.9$ becomes smaller~\cite{Funakubo:2009eg}.
Thus, the condition $\Delta\phi/T>0.9$ is conservative.}.

\begin{figure}[t]
\begin{center}
\includegraphics[width =13cm]{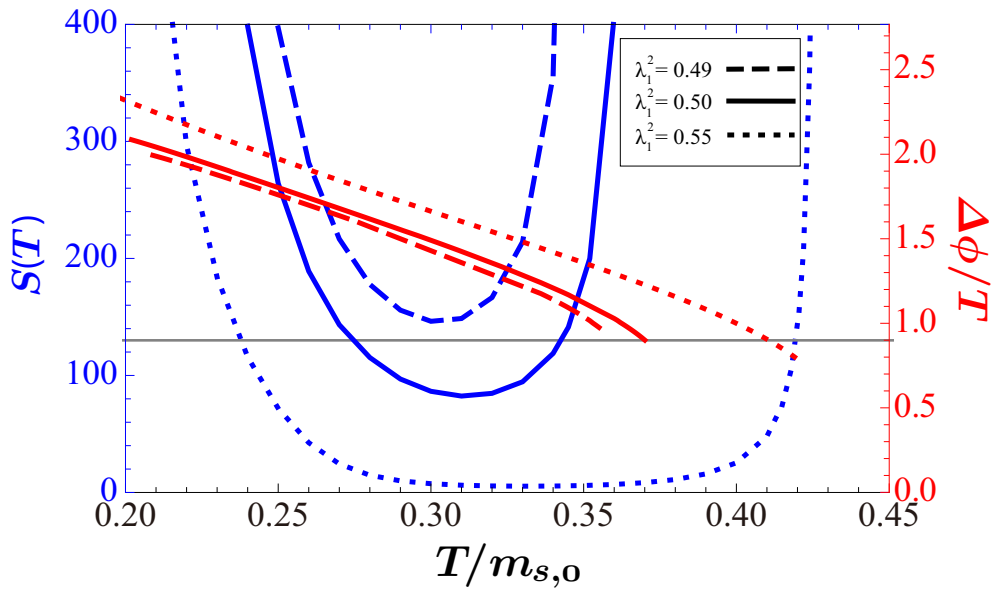}
 \vspace{-.2cm}
\caption[The classical action $S(T)$ for the three-dimensional ($\phi_1,\phi_2,\phi_s$) bounce solution and $\Delta\phi/ T$ as a function of T.]{The classical action $S(T)$ for the three-dimensional ($\phi_1,\phi_2,\phi_s$) bounce solution ({\it Blue}) and $\Delta\phi/ T$ ({\it Red}) as a function of T.
We take $\lambda_1^2 =$ 0.49 ({\it dashed}), 0.50 ({\it thick}) and 0.55 ({\it dotted}). 
The gray line represents $S(T) = 130$ and $\Delta\phi/ T = 0.9$.}
\label{zuzuzu}
 \vspace{-.7cm}
\end{center}
\end{figure}

Figure~\ref{zuzuzu} shows $S(T)$ and $\Delta\phi/T$ as a function of $T$.
The three-dimensional ($\phi_1,\phi_2,\phi_s$) bounce solution $S(T)$ is analyzed numerically by {\tt CosmoTransitions} software package~\cite{Wainwright:2011kj}.
The thick lines correspond to the benchmark point.
The condition for the strongly first-order phase transition is $\Delta\phi/T>0.9$ when $S(T)$ decreases to $130$ at the first time.
Note that the temperature $T$ decreases as the time goes.
From the Figure~\ref{zuzuzu}, we can see that the action $S(T)$ becomes smaller than $130$ at the first time with $\Delta\phi/T\sim 1.1$ when $T/m_{s,0}\sim0.34$ at the benchmark point.
Therefore, the strongly first-order phase transition occurs at this time.
Then the $B+L$ number is generated by the EWBG process and the BAU is generated thanks to the lepton number violating process (see Sec.~\ref{sec_BAU}).
The other lines are drawn with the same parameters at the benchmark point except $\lambda_1$.
Note that the action value $S(T)$ is sensitive to the parameter $\lambda_1$.
With larger $\lambda_1$, the thermal effects on the $\phi_s$ become stronger.
Then, the potential gets more deformed.
As a result, the action value $S(T)$ and $\Delta\phi/ T|_{S = 130}$ become smaller.
With $\lambda_1^2=0.55$, $\Delta\phi/T$ is not larger than $0.9$ when $S(T)$ becomes 130 at the first time.
Thus, the phase transition is not strong.
On the other hand, with smaller $\lambda_1$, $S(T)$ does not decrease to 130.
The strongly first-order phase transition occurs with $0.50 \lesssim \lambda_1^2 \lesssim 0.55$ for the benchmark point.
Figure~\ref{propro} indicates the profile of  the bounce solution at the benchmark point.
From this figure, we find that the typical wall width is $L_w T\sim 30$, and $\Delta\beta \sim 0.1$.
\mtrem{これに関するコメント}
\mtrem{Cosmotransitionの誤差を調べよう}

\begin{figure}[t]
\begin{center}
\includegraphics[width =13cm]{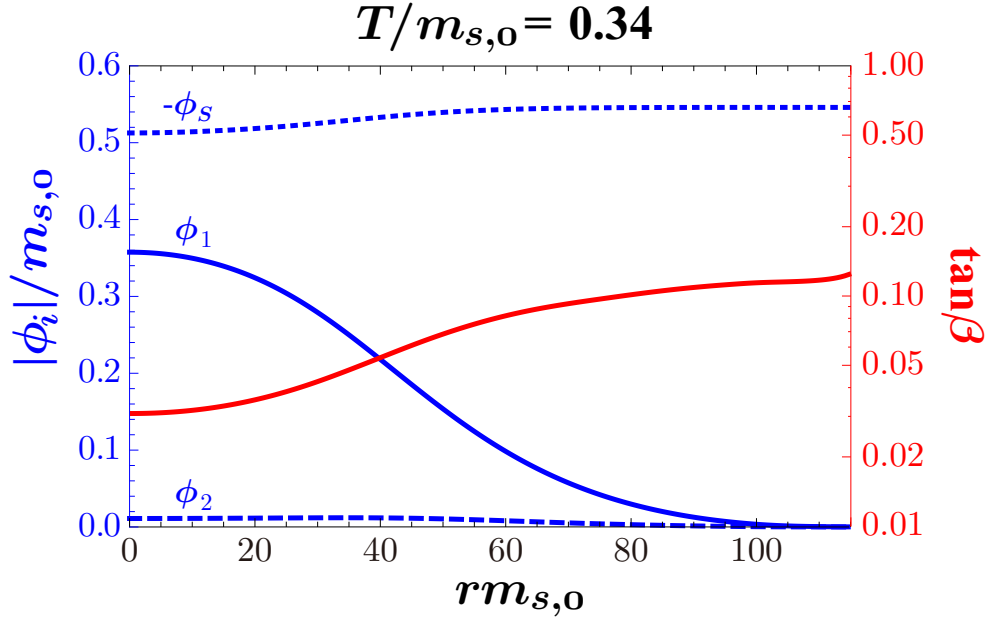}
 \vspace{-.2cm}
\caption[The bounce solution profile for the first-order phase transition at the benchmark point.
]{The bounce solution profile for the first-order phase transition at the benchmark point.
The horizontal axis is the space coordinate $r$ normalized by $m_{s,0}$.
$r=0$ corresponds to the center of the bubble.
The blue lines indicate the field values of $\phi_1$, $\phi_2$ and $-\phi_s$.
The red line corresponds to $\tan\beta$.}
\label{propro}
 \vspace{-.7cm}
\end{center}
\end{figure}

\subsubsection*{Scatter plot}
In order to show the favored region in our scenario, we present a scatter plot in the plane of $\tan\beta_{\rm vac}$ and the charged Higgs boson mass $M_{{\rm charged}}$ (Figure~\ref{scatter}).
We have scanned the following parameter ranges, 
\beq
200\GeV < &M_{\rm charged}& < 2\TeV\,,\nonumber \\
1.5< &\tan\beta_{\rm vac}&<10\,,\nonumber \\
0.3< &\lambda_1^2&<1.0\,,\nonumber \\
0.5 < &t_S/m_{s,0}^3& < 0.7\,,
\eeq
with fixed values $k=1.0, ~\lambda^2=0.5, ~m^2_{\tilde{t}}=m^2_{L'}=m^2_{N'}=m^2_{s,0}$.
We have estimated the bounce action by a simplified way in which we  use the one-dimensional  scalar potential $V_{\rm min}(\phi,T)$. The fitting formula of the Euclidean action is given in Appendix~\ref{FitAction}.
We have checked the error of  the fitting formula of the Euclidean action is at most $\sim 20 \%$ compared with the results of by the three-dimensional scalar potential.
Here we do not consider the mass of the standard model Higgs boson.
It depends on $A_\lambda$ (for low $\tan\beta$ region) and $m^2_{\tilde{t}}$ (for large $\tan\beta$ region) which do not change our result so much.
Thus when $M_{\rm SUSY}=\mathcal{O}(10)~{\rm TeV}$, we can obtain the standard model Higgs boson mass 125 GeV easily with varying $A_\lambda$ or $m^2_{\tilde{t}}$~\cite{Giudice:2011cg}.
Therefore, we set $m_{s,0}=10~{\rm TeV}$ here and do not consider their effects.
At all points in Figure~\ref{scatter} the first-order phase transition ($^\exists S(T) < 130$) occurs.
At the green points the strongly first-order phase transition ($\Delta \phi/T > 0.9$) occurs.
We find that the region with low $\tan \beta$ and the light charged Higgs boson is favored in our scenario.
This is consistent with the intuitive understanding in the previous subsection.

\begin{figure}[t]
\begin{center}
\includegraphics[width =13cm]{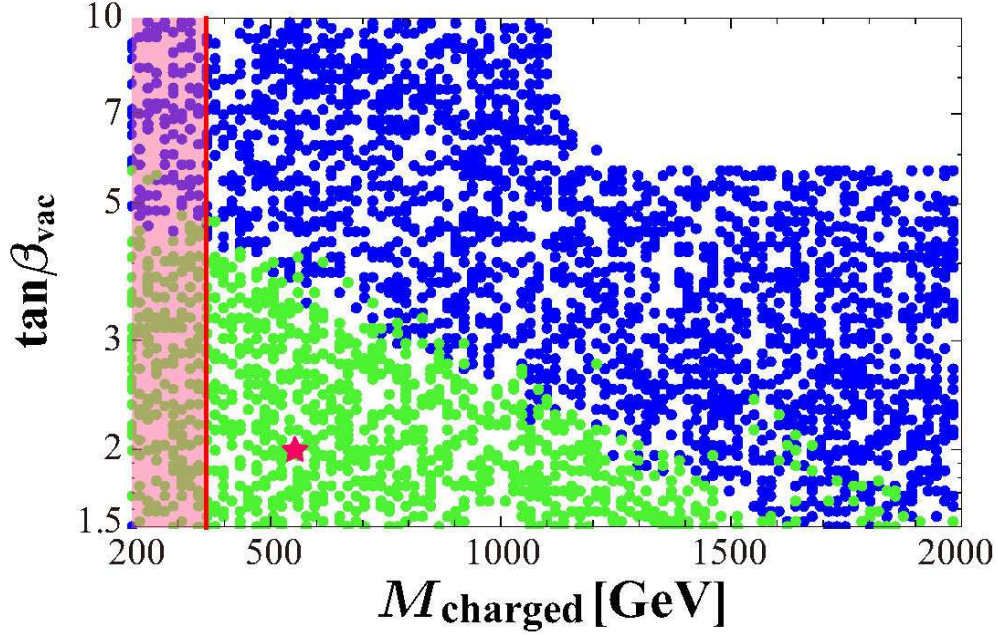}
 \vspace{-.2cm}
\caption[The scatter plot in $\tan\beta_{\rm vac} - M_{\rm charged}$ plane.]{The scatter plot in $\tan\beta_{\rm vac} - M_{\rm charged}$ plane.
The red shaded region shows the exclusion region by $\bar{B} \to X_s \gamma$ search.
At all points the first-order phase transition ($^\exists S(T) < 130$) occurs.
At the green points the strongly first-order phase transition~($\Delta \phi/T > 0.9$) occurs.
The star corresponds to the benchmark point.}
\label{scatter}
 \vspace{-.7cm}
\end{center}
\end{figure}

  \section{Baryon Asymmetry of the Universe}
 \label{sec_BAU}

So far, we have seen that the strongly first-order phase transition can occur at a high temperature in our scenario.
In this section, we show that the proper amount of the BAU can be generated.
Here, we consider the lepton number violating process caused by $\epsilon_N\hat{\bar{N}}'^3$.
This process is needed for the BAU to exist until today since the generated $B+L$ number by EWBG should be converted to the $B-L$ number.

The lepton number ($L$) and the baryon number ($B$) of all multiplets are defined in Sec.~\ref{sec_the_model}.
It is important that only the term $\epsilon_N\hat{\bar{N}}'^3$ violates the lepton number.
To make the discussion clear, we define $N'$ number by the approximate U(1) symmetry $\hat{N}':1,\hat{\bar{N}}':-1$.
This $N'$ number is contained in the $L$ number via the mixing $\epsilon$ (see Eq.~(\ref{sp:z2b})).
Note that the masses of the fermion  and  the scalar components of $\hat{N}',\hat{\bar{N}}'$ are $\mathcal{O}(M_{\rm SUSY})$.

Here, we see the $L$ number decreasing process and the thermal history of our model to show the generation of the BAU.
\subsubsection*{Lepton number decreasing process}
Let us see the details of the $L$ number decreasing process initially. 
There are two steps in this process.
At first, the $L$ number in the standard model sector is converted to the $N'$ number via the mixing terms $\epsilon$.
Then, the $N'$ number decreases due to the term $\epsilon_N\hat{\bar{N}}'^3$.
For simplicity, we consider the situation that the rate of the former process is larger than that of the latter one with assuming $\epsilon>\epsilon_N$.
Thus, the bottleneck process of the $L$ number violation is the process caused by $\epsilon_N\hat{\bar{N}}'^3$.
Therefore, we consider this process only and denote the rate of this process as $\Gamma_{\not{N'}}$.

At first, let us see $\Gamma_{\not{N'}}$ to estimate the effective $L$ number decreasing rate $\Gamma_{\not{L}}$.
We assume that the scalar component of $\hat{\bar{N}}'$ does not decay to two fermion components of $\hat{\bar{N}}'$ kinematically for simplicity
\footnote{At the benchmark point, the scalar component of $\hat{\bar{N}}'$ can decay to two fermion components of $\hat{\bar{N}}'$.
This is avoided by choosing the coupling $\lambda_1\hat{S}\hat{\bar{N}}'\hat{N}'$ larger and the SUSY breaking mass term for the scalar component of $\hat{\bar{N}}'$ smaller.
This choice does not change the result of the previous section.}.
Then, the $N'$ number is violated only by the scattering processes. 
For $T\lesssim \mathcal{O}(M_{\rm SUSY})$, $\Gamma_{\not{N'}}$ can be estimated as
\begin{align}
         \Gamma_{\not{N'}}(T)\sim
         \frac{\epsilon_N^2}{16\pi}
          \frac{{(M_{\rm SUSY}T)}^{3/2}}{M_{\rm SUSY}^2}\exp\left({-\frac{M_{\rm SUSY}}{T}}\right)\,,
\end{align}
and the equilibrium $N'$ number density $n_{N'}$ obeys 
 \begin{align}
         \frac{d n_{N'}}{dt}=-3H n_{N'}-\Gamma_{\not{N'}}n_{N'}+\text{(lepton number conserving processes)}\,.
\end{align}
Then the effective $L$ number decreasing rate $\Gamma_{\not{L}}$ can be written as
\begin{align}
	\Gamma_{\not{L}}(T)=\frac{n_{N'}}{n_L}\Gamma_{\not{N'}}(T)\sim\frac{ \epsilon_{N}^2}{16\pi}  \frac{M_{\rm SUSY}}{N_{\ell}}\exp\left( - \frac{2M_{\rm SUSY}}{T}\right)\,.
	\label{notL}
\end{align}
The $n_L$ is the equilibrium  $L$ number density and $N_{\ell}\sim 10$ is the number of the light components of the leptons\footnote{Strictly speaking, a linear combination of the $B$ number and the $L$ number decreases by the $N'$ number violating process. 
However, the amount of the BAU is changed only by a factor of a few with this effect.
Thus, we do not consider this effect for simplicity.}.
Here we use  $n_{N'} = (1/6) (\susy T)^{3/2} \textrm{exp}(- \susy/T) (\mu_L/T) $ and $n_L =(N_{\ell}/6) T^3 (\mu_L/T) $ \cite{Harvey:1990qw}. 

\subsubsection*{Thermal history}
Now, let us consider the thermal history of our scenario (for the overview see Sec.~\ref{sec:the_scenario}).
At first, the strongly first-order phase transition of the Higgs field occurs at $T_{1st}$.
The Higgs field is trapped at the breaking vacuum of the potential until $T=T_{\rm roll}$.
Then the Higgs field returns to the origin again.
At the benchmark point, $T_{1st}\simeq 0.34~m_{s,0}$ and $T_{\rm roll}\simeq 0.15~m_{s,0}$ hold.
If $\Gamma_{\not{L}} \gtrsim H$ holds during $T > T_{\rm roll} $ and $\Gamma_{\not{L}} \lesssim H$ holds during $T_{\rm roll}>T$, the BAU exists as explained below.
 
At the time $T = T_{1st} $, the strongly first-order phase transition occurs.
In our model, we assume that the $B+L$ number is generated at this time.
This EWBG process is supposed to occur in the time span $\tau_{EWBG}$ and typically $\tau_{EWBG}\ll 1/H$ holds.
Since we consider the situation $\Gamma_{\not{L}}(T_{1st})$ is the same scale of $H(T_{1st})$, the effects of $\Gamma_{\not{L}}$ can be negligible.
Then the $B+L$ number is generated with the $B-L$ number unchanged
\begin{align}
         Y_B(T_{1st})+Y_L({T_{1st}})&>0\,,\\
                  Y_B(T_{1st})-Y_L({T_{1st}})&=0\,,
\end{align}
where $Y_{B/L}$ is defined as the baryon/lepton number density divided by the entropy density. 

After EWBG, the Higgs field is trapped at the breaking vacuum during $T_{1st} > T > T_{\rm roll} $.
The sphaleron process decouples because the field value of the Higgs field is larger than the temperature.
As a result, the $B$ number conserves.
On the other hand, the $L$ number gradually decreases due to the $L$ number violating proccess.
The $L$ number decreasing factor $N_{\rm dec}$ can be estimated as
\begin{align}
         N_{\rm dec}\equiv \int_{t(T_{1st})}^{t(T_{\rm roll})} \Gamma_{\not{L}}dt=\int_{T_{\rm roll}}^{T_{1st}}
         \frac{\Gamma_{\not{L}}(T)}{HT}dT\,.
\end{align}
Thus, just before $T=T_{\rm roll}$, the $L$ number and the $B$ number become
\begin{align}
         Y_{L}(T_{\rm roll})&\simeq e^{-N_{\rm dec}}Y_L(T_{1st})\,,\\
         Y_B(T_{\rm roll})&=Y_B(T_{1st})\,.
\end{align}

After the Higgs field returns to the origin at $T_{\rm roll} > T$, the sphaleron process becomes active again.
Note that the sphaleron process makes the $B+L$ number wash-out towards the thermal equilibrium with conserving the $B-L$ number.
On the other hand, the $B-L$ number decreases by the $L$ number violation process $\epsilon_N$. 
The decreasing factor $N_{w}$ can be estimated as
\begin{align}
         N_w\equiv \int_{t(T_{\rm roll})}^{t(T=0)} \Gamma_{\not{L}} dt
         =\int^{T_{\rm roll}}_{0} \frac{\Gamma_{\not{L}}(T)}{HT}dT\,.
\end{align}
Then the $B$ number and the $L$ number follow
\begin{align}
	Y_B(T_{\rm f})+Y_L(T_{\rm f})&\propto Y_B(T_{\rm f})-Y_L(T_{\rm f})\,, \\
         Y_B(T_{\rm f})-Y_L(T_{\rm f})&\simeq e^{-N_w/c}\left( Y_B(T_{\rm roll})-Y_L(T_{\rm roll})\right)\,,
\end{align}
where $T_{\rm f}$ is the temperature at the sufficiently late time $ M_{\rm SUSY}\gg T_{\rm f}$ and $c\equiv{(n_L-n_B)/n_L}$ is an $\mathcal{O}(1)$ factor.

At the end, $T=T_{\rm f}$, the $B$ number and the $L$ number are estimated as
\begin{align}
         Y_B(T_{\rm f})&\simeq d^{-1}\cdot e^{-N_w/c}\left( 1-e^{-N_{\rm dec}}\right){Y_B(T_{1st})}\,, \label{52} \\
         Y_L(T_{\rm f})&\simeq-c^{-1}\cdot e^{-N_w/c}\left( 1-e^{-N_{\rm dec}}\right){Y_B(T_{1st})}\,, \label{53}
\end{align}
with $d\equiv(n_B-n_L)/n_B$.
If all particles except the standard model particles are heavy enough, $c=79/51$ and $d=79/28$ hold~\cite{Harvey:1990qw}.
In order to obtain the sizable BAU, $N_{\rm dec}\gg 1$ and $N_w\ll 1$ are favored (see Eqs.~(\ref{52},~\ref{53})).
This corresponds to $\Gamma_{\not{L}} \gtrsim H$ during $T > T_{\rm roll} $ and $\Gamma_{\not{L}} \lesssim H$ during $T_{\rm roll}>T$.
Note that both $N_{\rm dec}$ and $N_w$ are proportional to $\epsilon_N^2$ and the quantity $N_{\rm dec}/N_w$ is a function of $T_{1st}$, $T_{\rm roll}$ and $M_{\rm SUSY}$.
Thus, if $N_{\rm dec}/N_w\gg 1$ holds, the sizable BAU can exist until today since we can find the suitable value of $\epsilon_N^2$ which makes large $N_{\rm dec}$ ($N_{\rm dec}\gg 1$) and small $N_w$ ($N_w\ll 1$).
At the benchmark point, we obtain $N_{\rm dec}/N_w\sim 30$
 \footnote{This value depends on the value of the exponential factor in Eq.~(\ref{notL}).
Here, we set this exponential factor as the typical masses of the vector-like fermions $M_{\rm SUSY} = t_S/m_{s,0}^2 \simeq | \phi_s |$ .
However this exponential factor also depends on the masses of the vector-like scalar bosons since at least one boson particle participates in the scattering process.
Typically these masses are heavier than $| \phi_s |$ and this exponential factor becomes larger.
Thus, we have chosen the conservative value here since the ratio $N_{\rm dec}/N_w$ becomes larger with larger exponential factor.}.
To ensure $N_{\rm dec}\gtrsim 1$, we can choose  $\epsilon_N\sim 10^{-5}$.
With this choice of $\epsilon_N$, the generated baryon asymmetry at the EWGB exists until today.
In general, $N_{\rm dec}/N_w\gg 1$ can hold since there is a hierarchy $M_{\rm SUSY} > T_{1st} >T_{\rm roll} $.
Thus, the BAU can exist by this mechanism in our scenario.

\section{Singlino Dark Matter }
\label{sec_DM}

In this section, we show that the singlino dark matter scenario is compatible with our new baryogenesis scenario.
At first, we briefly review  the properties of the singlino dark matter as we have shown thoroughly in the  previous chapter.
Then, we estimate the lifetime of the singlino dark matter with the lepton number violating term.
We show that it does not suffer from experimental constraints.

Let us review the singlino dark matter scenario (the detail is written in the previous chapter).
In our model,  after integrating out the particles with masses above the electroweak scale, the low energy effective Lagrangian becomes
\begin{align}
	\mathcal{L}_{\rm eff}= \mathcal{L}_{\rm SM}-\frac{m_{\tilde{s}}}{2}\bar{\tilde{s}}\tilde{s}-\frac{\lambda_{\rm eff}}{2}h\bar{\tilde{s}}\tilde{s}\,,
\end{align}
where $h$ is the standard model Higgs boson and $\mathcal{L}_{\rm SM}$ is the standard model Lagrangian.
Here, $\tilde{s}$ is the singlino, the lightest neutralino mainly composed by the fermionic component of the singlet superfield $\hat{S}$.
We denote the singlino as the Majorana spinor.
The effective coupling $\lambda_{\rm eff}$ can be estimated as
\begin{align}
	\lambda_{\rm eff}\sim \lambda \frac{v_{EW}}{M_{\rm SUSY}}\sin 2\beta\,.
	\label{lambdaeff}
\end{align}
The singlino mass $m_{\tilde{s}}$ is dominated by the one-loop corrections when $M_{\rm SUSY}$ is large.
In our model, the singlino can get sizable corrections from vector-like multiplets sector.
The singlino mass can be evaluated as
\footnote{Strictly speaking, the singlino mass is promotional to the $A$-terms ($A_{\lambda_1}S\bar{X}'X'$) which are dropped off in the previous discussions.
However, the effects of such $A$-terms $\sim\susy$ to the thermal dynamics are supposed to be small and do not change the previous results.}
\begin{align}
	m_{\tilde{s}}\sim \frac{\lambda_1^2}{(4\pi)^2}M_{\rm SUSY}\,.
\end{align}
In this model, the singlino can be a good candidate of the dark matter.
If $m_{\tilde{s}}\simeq 60~{\rm GeV}$ and $\lambda_{\rm eff}\sim \mathcal{O}(0.01)$,
the singlino dark matter scenario is successful with resonant annihilation via the exchange of the standard model Higgs boson.
Such a situation can be realized when $M_{\rm SUSY}\sim \mathcal{O}(10)~{\rm TeV}$, $\tan\beta\sim \mathcal{O}(1)$ and $\lambda,\lambda_1\sim \mathcal{O}(1)$.
Note that the low $\tan\beta$ and $\mathcal{O}(1)$ couplings are realized with our baryogenesis scenario.
The soft SUSY breaking scale $M_{\rm SUSY}$ is determined by the requirement of the singlino dark matter scenario, especially by the effective coupling $\lambda_{\rm eff}$ Eq.~(\ref{lambdaeff}).

In our model, there are the lepton number violating term ($\epsilon_N$) and the SM-extraparticles mixing terms ($\epsilon$).
Thus, this model does not conserve the $R$-parity and the singlino can decay to the standard model particles.
So, let us estimate the decay rate of the singlino.
Note that the term $\epsilon_N \hat{\bar{N}}'^3$ breaks the lepton number by three $\Delta L=3$.
In addition, the decay process breaks the vector-like multiplet parity $\mathbb{Z}_2$ at least three times.

Let us consider the dominant decay channel $\tilde{s} \to \nu \nu \nu$.
The other channels are more suppressed since the number of final state particles increases if the decay products include charged leptons.
To see the coupling of the $\tilde{s} \nu \nu \nu$, we consider the following fermion four point operator which arises from integrating out the particles whose masses are $\mathcal{O}(M_{\rm SUSY})$
\begin{align}
	\mathcal{O}_{\tilde{s}\nu\nu\nu}=f_{\tilde{s}\nu\nu\nu}
	\epsilon_N \epsilon^3
	\frac{\psi_{\tilde{s}}\psi_\nu\psi_\nu\psi_\nu}{M_{\rm SUSY}^2}\,.
\end{align}
Here, $f_{\tilde{s}\nu\nu\nu}$ is a numerical factor and $\epsilon$ denotes $\epsilon^i$ or $\epsilon_S^i$ defined in Eq.~(\ref{sp:z2b}).
We denote $\psi$'s as the Weyl spinors.
The decay rate of the singlino due to this operator can be evaluated as
\begin{align}
	\Gamma(\tilde{s} \to \nu \nu \nu) &\sim \frac{\epsilon_N^2 \epsilon^6 f_{\tilde{s}\nu\nu\nu}^2}{3072 \pi^3} \frac{m_{\tilde{s}}^5}{M_{\rm SUSY}^4}\,.
\end{align}
The mass of the singlino is favored to be $m_{\tilde{s}}\simeq 60$ GeV in order to realize resonant annihilation via the exchange of the standard model Higgs boson.
On the other hand, the typical value of the $\mathbb{Z}_2$ breaking couplings $\epsilon$ is $\mathcal{O}(10^{-5})$ (see Sec.~\ref{sec_BAU}).
Thus the lifetime of the singlino $\tau_{\tilde{s}}$ can be estimated as
\begin{align}
\label{eq:DM_life}
     \tau_{\tilde{s}}\simeq 0.8\times 10^{36}	
      \left(\frac{10^{-5}}{\epsilon_N}\right)^2
\left(\frac{10^{-5}}{\epsilon}\right)^6
 \left(\frac{10^{-4}}{f_{\tilde{s}\nu\nu\nu}}\right)^2
 \left(\frac{M_{\rm SUSY}}{10~{\rm TeV}}\right)^4 \left(\frac{60~{\rm GeV}}{m_{\tilde{s}}}\right)^5~\text{[sec]}\,.
\end{align}

Now, we estimate the upper bound on the factor $f_{\tilde{s}\nu\nu\nu}$ by a diagrammatic way.
Let us consider the diagrams for the operator $\mathcal{O}_{\tilde{s}\nu\nu\nu}$.
To draw the diagram, we need the vertex $\epsilon_N \hat{\bar{N}}'^3$.
Thus, each diagram includes the vertex $\epsilon_N$ and three propagators of $\hat{\bar{N}}'$.
Since the final state contains three neutrinos, these three propagators of $\hat{\bar{N}}'$ should be converted to them.
Therefore, there are three lines which start from $\hat{\bar{N}}'$ to the neutrino.
We call these lines as lepton lines.
For each lepton line, at least one propagator of a Higgs multiplet or one vacuum expectation value of the Higgs field $v_{EW}$ should be attached~\footnote{
There are also the diagrams in which some lepton lines have no propagator of a Higgs multiplet and no vacuum expectation value of the Higgs field.
However, such a diagram is highly suppressed since a lot of vertices are needed.
Therefore, we ignore such diagrams here.}.
If $v_{EW}$'s are attached to all three lepton lines, the diagram may have no loops and $f_{\tilde{s}\nu\nu\nu}$ is suppressed by $({v_{EW}}/{M_{\rm SUSY}})^3$.
If $v_{EW}$'s are attached to two lepton lines of three, the diagram has at least one loop and $f_{\tilde{s}\nu\nu\nu}$ is suppressed by $(1/16\pi^2)({v_{EW}}/{M_{\rm SUSY}})^2$.
If $v_{EW}$ is attached to one lepton line of three, the diagram has at least one loop and $f_{\tilde{s}\nu\nu\nu}$ is suppressed by $(1/16\pi^2)({v_{EW}}/{M_{\rm SUSY}})$.
If $v_{EW}$'s are not attached to any lepton lines, the diagram has at least two loops and $f_{\tilde{s}\nu\nu\nu}$ is suppressed by $(1/16\pi^2)^2$.
In any cases, the following inequality holds
\begin{align}
	f_{\tilde{s}\nu\nu\nu}\lesssim 10^{-4}\,,\label{eq:snnn}
\end{align}
if $M_{\rm SUSY}=\mathcal{O}(10)\text{ TeV}$.
Note that this estimate of the upper bound on $f_{\tilde{s}\nu\nu\nu}$ is conservative.

From Eq.~(\ref{eq:DM_life}) and Eq.~(\ref{eq:snnn}), the lifetime of the singlino becomes long $\tau_{\tilde{s}}\gtrsim 10^{36}$ sec.~.
On the other hand, there are experimental bounds on the lifetime of the dark matter.
First, the lifetime of the dark matter should be much longer than the lifetime of the universe $\sim10^{17}$ sec.~.
Second, there are constraints from the cosmic ray searches, $\tau_{DM} \gtrsim 10^{29}~{\rm sec.}$~\cite{Fornengo:2013xda}.
Obviously, the lifetime of  the singlino is much longer than the experimental bounds
\footnote{
The experimental bounds by the cosmic ray searches come from the various decay channels of the dark matter.
Especially, the decays to the charged leptons are important.
However, in our model, the decays of the singlino to the charged leptons are more suppressed than the decay to three neutrinos since the number of the final states increases.
Therefore, the bounds can be evaded more easily. }.
Thus there is no problem in the decay of the singlino and the singlino can be a good candidate of the dark matter in our scenario.

\section{Discussions}
\label{5no7}

We comment on the experimental constraints for the light charged Higgs boson and the SM-extraparticles mixings.
First, let us consider the constraints for the light charged Higgs boson.
The relatively light charged Higgs boson and heavy SUSY particles are favored in our scenario.
It means that this model can be regarded as the two-Higgs doublet model at low energy regions.
Even if SUSY particles are heavy, the existence of the light extra scalars is constrained by the flavor and the CP violation physics.
In the viable parameter region of our scenario, the process $\bar{B} \to X_s \gamma$ is the only relevant constraint from the flavor physics.
The red shaded region in Figure~\ref{scatter} is excluded at $95~\%$ C.L. by a current bound~\cite{Hermann:2012fc}.
On the other hand, the electron EDM is one of the severe constraints on a new CP-violating phase \cite{Baron:2013eja}. 
In our scenario, a new CP-violating phase may enter into the potential for the Higgs field through only $A_{\lambda} $ term.
The electron EDM is induced by the mixing between the CP-even and the CP-odd Higgs bosons which is estimated as $\sim \lambda^2 (t_S/m_{s,0}^3)(\lambda A_\lambda/m_{s,0})$.\mtrem{説明}
If $A_\lambda/m_{s,0} \lesssim 0.1$, our scenario is compatible with the current bound of the electron EDM experiments~\cite{Abe:2013qla}.

Second, the flavor changing neutral current appears through the SM-extraparticles mixings.
One of the severe constraints comes from the branching ratio of $\mu \to e \gamma$, Br($\mu \to e \gamma$) $ < 5.7 \times 10^{-13}  ~(90~\%$~CL)~\cite{Adam:2013mnn}.
We have estimated this value at our benchmark point,
\beq
 \textrm{Br}(\mu \to e \gamma) \sim \epsilon^4 \times 10^{-8}.
\eeq
Thus if we take  $\epsilon \lesssim 10^{-2}$, the bound from Br($\mu \to e \gamma$) can be escaped easily.

Finally, we comment on the neutrino masses.
Although this model includes the matters which couple to the neutrinos, neutrino masses are protected to be zero. 
In order to generate the nonzero neutrino masses, we have to extend our model or change the imposed symmetry. 
\coll{For example, let us introduce three right-handed neutrino superfields $\hat{\bar{N}}''$ which have $\mathbb{Z}^R_5$ $R$-symmetry charge $3$, $\mathbb{Z}_3$ symmetry charge $1$ and $\mathbb{Z}_2$ parity even.
We also introduce an extra local U(1) symmetry, and $\hat{\bar{N}}''$ has charge $2$ and other particles do not have charge under this symmetry.
If there is an additional Planck suppressed term $W = ( \frac{\hat{\phi}}{M_{\rm Pl}})^2 \hat{H}_2 \hat{L} \hat{\bar{N}}''$ where $\hat{\phi}$ is the singlet under  $\mathbb{Z}^R_5$ and $\mathbb{Z}_3$,  an even parity under $\mathbb{Z}_2$, and has charge $-1$ under the extra symmetry, 
the appropriate Dirac neutrino masses are generated when the extra symmetry breaking scale is $\langle \phi \rangle \sim 10^{13} $ GeV. }

As this chapter is a first study of a scale free electroweak baryogenesis scenario, much work is left to be done.
First, we have to check whether the proper amount of the baryon number can be generated within our scenario including the explicit CP-violating phases.
Second, the vacuum stability against the charged Higgs field direction has to be checked in detail.

We have to comment on the stability for the charged Higgs field direction of the potential.
At the zero temperature, there is a charge breaking global minimum in the charged Higgs direction if we consider the tree-level potential at the benchmark point (see Appendix \ref{chargeddire}).
This is because the charged Higgs boson mass can become negative in the relatively large $\phi$ region since the field value $|\phi_s|$ becomes small (see \eq{singletvev}).
To see that there is no problem with this minimum, we have checked two conditions.
First, we have checked that the charged Higgs boson mass including the thermal self energy is positive for all time of the universe\footnote{For simplicity, we do not include the mass corrections from the Coleman-Weinberg potential.
We have numerically checked that the Coleman-Weinberg potential  for charged Higgs boson gives typically positive contributions to the mass of the charged Higgs boson.).}.
Second, we have calculated a tunneling rate from the electroweak breaking vacuum to the charge breaking global minimum  at the zero temperature  with tree-level potential.
Then it turned out that the lifetime of the electroweak breaking vacuum is much longer than the one of the universe ($S \sim \mathcal{O}(1000)$).
Thus, we consider that this minimum gives no problem.
The full analysis of the stability against the charged Higgs field direction is complicated and will be done in the future.

\section{Conclusion of Scale Free Electroweak Baryogenesis}
\label{sec_conclusion}

In this chapter, we proposed a new electroweak baryogenesis scenario with the high-scale nMSSM including vector-like multiplets.
We have shown that the strongly first-order phase transition can occur in a high temperature comparable to $M_{\rm SUSY}$.
The proper amount of the BAU can be generated via the lepton number violating process.
Furthermore, the singlino dark matter scenario~\cite{Ishikawa:2014owa} is also compatible with our scenario.
The key points are as follows: 
(i) the thermal mass term for the singlet scalar field generates the global minimum of the potential for the Higgs field far from the origin,
(ii) the lepton number violating process converts the $B+L$ number to the $B-L$ number.
Even though there is the lepton number violating process, the lifetime of the singlino is long enough.
In this baryogenesis process, $M_{\rm SUSY}$ can be an arbitrary value and it is almost a free parameter.
Thus, we call this scenario as a scale free electroweak baryogenesis.
The scale $M_{\rm SUSY}$ will be determined by other requirements.
If $M_{\rm SUSY}\sim \mathcal{O}(10) {\rm ~TeV}$, this scenario is compatible with the proper Higgs boson mass and the right amount of the singlino dark matter without SUSY flavor/CP problem~\cite{Ishikawa:2014owa}.
In addition, this singlino dark matter scenario can be testable by future experiments of the search of the Higgs invisible decay and the direct direction of the dark matter (see previous section). 

Consequently, we have shown the possibility of the high scale baryogenesis scenario.
We hope that this study becomes a first step of scale free electroweak baryogenesis scenarios.


\chapter{Conclusion}
\label{chap6}
\thispagestyle{empty}

The standard model of the particle physics has worked very well for a long time.
Nevertheless, there are many unsolved problems within the SM, for example, the observed dark matter particles and baryon asymmetry of the universe. 
From theoretical viewpoint, the gauge hierarchy problem is still in question.
The supersymmetric models are good candidates of the physics beyond the standard model since  they can solve the hierarchy problem naturally.

The minimal SUSY model contains a supersymmetric dimensionful parameter $\mu$, and  this parameter causes $\mu$ problem, which is also one of the hierarchy problem.
In order to realize nature, namely the $Z$ boson mass is to be at the electroweak scale, 
the $\mu$ parameter  has to know the soft SUSY breaking scale.

The nearly Minimal Supersymmetric Standard Model (nMSSM) is one of the promising models of the new physics: this model can avoid the $\mu$ problem, the domain wall problem, and the tadpole problem simultaneously.
 In addition, this model has natural candidate of the dark matter, namely singlino,  and can generate the baryon asymmetry of the universe.  
 
 \paragraph*{}
In this thesis, we consider the phenomenology of the nMSSM. 
Especially, we focus on the phenomenology of the dark matter and  the baryon asymmetry in the universe by the electroweak baryogenesis mechanism.

First, 
we have studied the phenomenology of the singlino resonant dark matter scenario. 
We find that with high-scale supersymmetry breaking the singlino can obtain a sizable radiative correction to the singlino mass, 
which opens a window for the singlet dark matter scenario with resonant annihilation via the exchange of the Higgs boson.
We have also shown that with high-scale SUSY breaking $\sim 10$ TeV and low $\tan\beta$, the  dark matter relic abundance and the SM Higgs boson mass can be explained simultaneously in this scenario.

Next, 
we have also proposed a new electroweak baryogenesis scenario with the high-scale nMSSM including vector-like multiplets.
We have shown that the strongly first-order phase transition can occur in a high temperature comparable to the soft SUSY breaking scale.
The proper amount of the baryon asymmetry in the universe can be generated via the lepton number violating process.
Furthermore, we have calculated explicitly the lifetime of the singlino and we find that the singlino dark matter scenario is also compatible with our scenario.
The key points are as follows: 
(i) the thermal mass term for the singlet scalar field generates the global minimum of the potential for the Higgs field far from the origin,
(ii) the lepton number violating process converts the $B+L$ number to the $B-L$ number.
Even though there is the lepton number violating process, the lifetime of the singlino is long enough.
In this baryogenesis process, the soft SUSY breaking scale can be an arbitrary value and it is almost a free parameter.
Thus, we call this scenario as a scale free electroweak baryogenesis.
The soft SUSY breaking scale will be determined by other requirements.
If it is $\mathcal{O}(10) {\rm ~TeV}$, this scenario is compatible with the proper Higgs boson mass  without SUSY flavor/CP problem.

\paragraph*{}
Therefore, we find that when the soft SUSY breaking scale is $\mathcal{O}(10)\TeV$, this electroweak baryogenesis scenario is compatible with the singlino resonant scenario.
In addition, these scenarios are also compatible with
the observed mass of the Higgs boson and the constraints by the electric dipole moments measurements and the flavor experiments.

Even for the high-scale SUSY, we have also shown that the parameter region where the singlino dark matter is consistent with the current dark matter relic abundance can be probed by the future experiments.
Hence, the singlino dark matter signal can be a first sign of  the high-scale supersymmetry.

 As a result of these two studies, 
 we conclude that the nMSSM with a high-scale SUSY breaking is valid and can be  probed by the direct direction of the  singlino dark matter.


\newpage
\section*{\Huge{Acknowlegement}}

~~~~~~I would like to express my sincere gratitude to my supervisor Takeo Moroi for 
various instructive discussions, stimulating suggestions and collaborations, especially his insightful comments were invaluable.
 ﻿I would also like to thank Masahiro Takimoto and Kazuya Ishikawa, who are
collaborators for the main parts of this thesis, for fruitful discussions and collaborations, which are  very exciting for me.
 I am very grateful  to Koichi Hamaguchi, Yuji Tachikawa, Motoi Endo and Kazunori Nakayama for 
fruitful discussions, invaluable comments and warm encouragements.
I would like to thank Masaki Yamada for crucial discussions on Chapter 5. 
I am also grateful to Masahiro Ibe, Hitoshi Murayama, Taizan Watari, Yutaka Matsuo and Hiroaki Aihara for careful reading of this manuscript and useful comments and discussions.
Finally, I would like to express my gratitude to all the members of particle physics theory group at University of Tokyo and to my family for their hospitality and nontrivial supports.

\appendix{}
\chapter{Notations and Conventions}
\label{AppA}
\thispagestyle{empty}

\section{Notations}

We use the following metric tensor
\beq
g_{\mu \nu}=g^{\mu \nu }=\begin{pmatrix}1&0&0&0 \\ 0&-1&0&0 \\ 0& 0& -1 &0 \\ 0&0&0&-1\end{pmatrix},
\eeq
with the Greek indices are $\mu = 0,1,2,3,4$, and
the totally antisymmetric tensor is 
\beq 
\epsilon^{0123}=-\epsilon_{0123}=1.
\eeq

The Pauli matrix $\sigma^{a}$ is defined as,
\beq
\sigma^1 = \begin{pmatrix}0&1\\ 1&0 \end{pmatrix}, \  \ \  \sigma^2 = \begin{pmatrix}0&-i\\ i&0 \end{pmatrix}, \ \ \ \sigma^3 = \begin{pmatrix}1&0\\ 0&-1 \end{pmatrix},
\eeq
and $\sigma^{\mu}$ and $\bar{\sigma}^{\mu}$ are defined as follows,
\beq
\sigma^{\mu} = ({\bf 1},~\overrightarrow{\sigma}),\ \ \ \bar{\sigma}^{\mu}=({\bf 1},-\overrightarrow{\sigma}).
\eeq
The antisymmetric tensor for two components is 
\beq
\epsilon_{12} = - \epsilon_{21} = 1.
\eeq

The gamma matrices satisfy the following anti commutation relations,
\beq
\left\{ \gamma^{\mu},\gamma^{\nu} \right\} =2 g^{\mu \nu}.
\eeq
We use the chiral basis gamma matrices, 
\beq
\gamma^{\mu}&=& \begin{pmatrix}0&\sigma^{\mu }\\ \bar{\sigma}^{\mu } &0 \end{pmatrix},\\  \gamma^5=-\frac{i}{4!}\epsilon_{\mu \nu \rho \sigma}\gamma^{\mu}\gamma^{\nu}\gamma^{\rho}\gamma^{\sigma}&=&i\gamma^{0}\gamma^{1}\gamma^{2}\gamma^{3}=\begin{pmatrix}-{\bf 1}&0\\ 0&{\bf 1} \end{pmatrix}.
\eeq

The Gell-Mann matrix  $\lambda^{a}$ is defined as,
\beq
\lambda^1 = \begin{pmatrix}0&1&0\\ 1&0&0 \\ 0&0&0\end{pmatrix}, \  \ \  \lambda^2 = \begin{pmatrix}0&-i&0\\ i&0&0 \\ 0&0&0\end{pmatrix}, \ \ \ \lambda^3 = \begin{pmatrix}1&0&0\\ 0&-1&0\\ 0&0&0 \end{pmatrix},  \ \ \ \lambda^4 = \begin{pmatrix}0&0&1\\ 0&0&0\\ 1&0&0 \end{pmatrix},\non
\lambda^5 = \begin{pmatrix}0&0&-i\\ 0&0&0 \\ i&0&0\end{pmatrix}, \  \ \  \lambda^6 = \begin{pmatrix}0&0&0\\ 0&0&1 \\ 0&1&0\end{pmatrix}, \ \ \ \lambda^7 = \begin{pmatrix}0&0&0\\ 0&0&-i\\ 0&i&0 \end{pmatrix},  \ \ \ \lambda^8 =  \fr{1}{\sqrt{3}}\begin{pmatrix}1&0&0\\ 0&1&0\\ 0&0&-2 \end{pmatrix}.
\eeq

\section{Group theoretical constants}

The generators of a simple Lie group SU$(N)$ are represented by $N \times N$ Hermitian matrix $t^a$, which satisfies ${\rm Tr}[t^a]=0$, and 
$t^a  = \sigma^a /2 ~(a = 1,2,3)$ for SU$(2)$,  $t^a  = \lambda^a /2 ~(a = 1,2,\dots,8)$ for SU$(3)$.
While, in the following  the $t$ for U$(1)_Y$ means the hyper charge $Y_i$ of operated field $\Phi_i$.

The generators are normalized by
\beq
\text{Tr}\left[  t^a t^b \right]  =  T(i) \delta^{ab},\ 
 \ \ T(i)=\left\{ \begin{array}{l}Y_i^2\ \  \text{for} \ {\rm U}(1)_Y,\\
\frac{1}{2}\ \  \text{~~for} \ {\rm SU}(2), \rm{ SU}(3).
\end{array}\right.
\eeq

The structure constants  for  a Lie group $f^{a b c} $ is defined by
\beq
\left[ t^a , t^b \right] = i f^{a b c} t^c.
\eeq

The quadratic Casimir operator for the fundamental representation is defined as,
\beq
\left(\sum_a t^a t^a\right)^i_j = C(i)\delta^i_j,
\eeq
wehre
\beq
C_1(i)&=&Y_i^2\ \ \ \  \textrm{~for~}\textrm{U(1)~and~}\Phi_i,\\
C_2(i)&=&\left\{
\begin{array}{l}
\frac{3}{4}\ \ \ \  \textrm{for~}\textrm{SU(2)~and}\Phi_i = Q,L,H_1,H_2,\\
0\ \ \ \  \textrm{for~}\textrm{SU(2)~and}\Phi_i = \bar{U},\bar{D},\bar{E},S,
\end{array}
\right. \\
C_3(i)&=&\left\{
\begin{array}{l}
\frac{4}{3}\ \ \  \ \textrm{for~}\textrm{SU(3)~and}\Phi_i = Q,\bar{U},\bar{D},\\
0\ \ \ \  \textrm{for~}\textrm{SU(3)~and}\Phi_i = L,\bar{E},H_1,H_2,S.
\end{array}
\right.
\eeq
The quadratic Casimir operator for the adjoint representation is
\beq
\sum_a f_{bde}^a f_{cde}^a &=&  C(G) \delta_{bc},
\eeq
with
\beq
  C(G)&=&\left\{ \begin{array}{ll}0\ \  \text{for} \ \textrm{U(}1),\\
2\ \  \text{for} \ \textrm{SU(}2),\\
3\ \  \text{for} \ \textrm{SU(}3).
\end{array}\right.
\eeq

\chapter{Quantum Corrections}
\thispagestyle{empty}
\abstchapter{In this appendix, we collect the functions of the quantum correction, which are needed  in this thesis. We first present the full set of two-loop RGEs for the coupling constants of the SM and the singlet extension SUSY model. Second the one-loop corrections to the mass of the neutralino are exhibited.
Finally, we present the loop functions which we use in the text.}

\section{Renormalization Group Equations}
\label{RGEqpp}

In this appendix, we assume  the high-scale SUSY mass spectrum, $v_{\rm EW} \ll M_{\rm SUSY}$ and $M_{\rm gaugino} \sim \mu \sim \sqrt{m^2_0} =\mathcal{O}(M_{\rm SUSY})$, where $m^2_0$ represents dimension two soft SUSY breaking mass term  of Higgs and sfermion.
Note that in the split case, $M_{\rm gaugino} \sim \mu \ll M_{\rm SUSY}$, we should consider the RGEs of the Yukawa-like gaugino couplings \eq{yukawagaugino} because the Higgs-Higgsino-gaugino coupling is still active at $v_{EW} < Q < \susy$. 
Therefore, we should take into account the RGE of not only the SM couplings but also the Yukawa-like gaugino couplings at $v_{EW} < Q < M_{\rm SUSY}$.
The RGEs of the split mass spectrum case is written in Refs.~\cite{Giudice:2011cg, Bagnaschi:2014rsa}.

We write the RGEs in the following notation, 
\beq
\frac{d g_i}{d \ln{Q}} = \frac{\beta_{1,i}}{(4 \pi)^2} +  \frac{\beta_{2,i}}{(4 \pi)^4}, 
\eeq
where  $Q $ is the renormalization scale.

\subsection{RGEs below SUSY breaking scale}
\label{RGESM}

First, we present the RGEs up to two-loop order for the SM couplings, $g', g, g_s, y_t, y_b, y_{\tau}$ and $\lambda_{\rm quartic}$ in the $\overline{MS}$ scheme \cite{Machacek:1983tz, Machacek:1983fi, Machacek:1984zw, Bagnaschi:2014rsa}.
In the SM, one-loop level $\beta$ function for the gauge couplings is given as 
\beq
\beta_{{1,g_i}} = - {g_i}^3 \left( \frac{11}{3} C_a(G) - \frac{2}{3} n_f T_a(f) - \frac{1}{3} n_b T_a(b)\right),
\eeq
where $n_f ~(n_b)$ is a number of the gauge multiplet of the Weyl spinor (complex scalar).

\paragraph*{}
The one-loop $\beta$ functions for the SM couplings are 

\beq
\beta_{1,g'} &=& \frac{41}{6} g'^3,\\
\beta_{1,g} &=& -\frac{19}{6} g^3,\\
\beta_{1,g_s} &=& -7 g_s^3,\\
\beta_{1,y_t} &=& y_t\left( -\frac{17}{12} g'^2 - \frac{9}{4} g^2 - 8 g_s^3 + \frac{9}{2} y_t^2 + \frac{3}{2} y_b^2 + y_{\tau}^2\right),\\
\beta_{1,y_b}&=& y_b  \left( -\frac{5}{12}g'^2 -\frac{9}{4}g^2 -8 g_s^2   +\frac{3}{2}  y_t^2  + \frac{9}{2}  y_b^2 +  y_{\tau}^2 \right),\\
\beta_{1,y_{\tau}} &=& y_{\tau} \left( -\frac{15}{4}g'^2 -\frac{9}{4}g^2    +3  y_t^2  + 3 y_b^2 +\frac{5}{2}  y_{\tau}^2 \right).\\
\beta_{1,\lambda_q} &=& 2\lambda_{\rm quartic} \left( 6  \lambda_{\rm quartic} + 6 y_t^2 + 6 y_b^2 + 2 y_{\tau}^2 - \frac{3}{2} g'^2 - \frac{9}{2}g^2 \right) \non
& &- 4(3 y_t^4 + 3 y_b^4 + y_{\tau}^4) + \frac{3}{4} g'^4 + \frac{9}{4} g^4 + \frac{3}{2} g'^2 g^2.
\eeq

\paragraph*{}
The two-loop $\beta$ function for SM couplings are
\beq
\beta_{2,g'} &=& g'^3 \left( \frac{199}{18} g'^2 + \frac{9}{2} g^2 + \frac{44}{3} g_s^2 - \frac{17}{6} y_t^2 - \frac{5}{6}y_b^2- \frac{5}{2} y_{\tau}^2\right),\\
\beta_{2,g} &=& g^3 \left( \frac{3}{2} g'^2 + \frac{35}{6} g^2 + 12 g_s^2 - \frac{3}{2} y_t^2 - \frac{3}{2}y_b^2- \frac{1}{2} y_{\tau}^2\right),\\
\beta_{2,g_s} &=& g_s^3 \left( \frac{11}{6} g'^2 + \frac{9}{2} g^2 - 26 g_s^2 - 2 y_t^2 - 2 y_b^2 \right),\\
\beta_{2,y_t} &=&  y_t \left[ y_t^2\left(  \frac{131}{16} g'^2 + \frac{255}{16} g^2 + 36 g_s^2 - 12 y_t^2 - \frac{11}{4}y_b^2 -\frac{9}{4}y_{\tau}^2 - 6 \lambda_{\rm quartic}\right)\right. \non
& & + y_b^2\left(\frac{7}{48}g'^2 + \frac{99}{16} g^2 + 4 g_s^2 - \frac{1}{4} y_b^2 + \frac{5}{4} y_{\tau}^2 \right) +y_{\tau}^2\left( \frac{25}{8}g'^2 + \frac{15}{8} g^2 - \frac{9}{4} g_{\tau}^2\right)\non
& & + \left. \frac{3}{2} \lambda_{\rm quartic}^2 +\frac{1187}{ 216} g'^4 -\frac{23}{4} g^4 - 108 g_s^4 - \frac{3}{4} g'^2 g^2 + \frac{19}{9} g'^2 g_s^2 + 9 g^2 g_s^2 \right],\\
\beta_{2,y_b} &=&  y_b \left[ y_t^2\left(  \frac{91}{48} g'^2 + \frac{99}{16} g^2 + 4 g_s^2 - \frac{1}{4} y_t^2 - \frac{11}{4} y_b^2 + \frac{5}{4}y_{\tau}^2 \right)\right. \non
& & + y_b^2 \left(\frac{79}{16}g'^2 + \frac{225}{16} g^2 + 36 g_s^2 - 12  y_b^2 - \frac{9}{4} y_{\tau}^2 - 6 \lambda_{\rm quartic}\right) +y_{\tau}^2\left( \frac{25}{8}g'^2 + \frac{15}{8} g^2 - \frac{9}{4} g_{\tau}^2\right)\non
& & + \left. \frac{3}{2} \lambda_{\rm quartic}^2 - \frac{127}{ 216} g'^4 -\frac{23}{4} g^4 - 108 g_s^4 - \frac{9}{4} g'^2 g^2 + \frac{31}{9} g'^2 g_s^2 + 9 g^2 g_s^2 \right],\\
\beta_{2,y_{\tau}} &=&  y_{\tau} \left[ y_t^2\left(  \frac{85}{24} g'^2 + \frac{45}{8} g^2 + 20 g_s^2 - \frac{27}{4} y_t^2 + \frac{3}{2} y_b^2 - \frac{27}{4}y_{\tau}^2 \right)\right. \non
& & + y_b^2 \left(\frac{25}{24}g'^2 + \frac{45}{8} g^2 + 20 g_s^2 - \frac{27}{4}  y_b^2 - \frac{27}{4} y_{\tau}^2\right) +y_{\tau}^2\left( \frac{179}{16}g'^2 + \frac{165}{16} g^2 - 3 g_{\tau}^2 - 6 \lambda_{\rm quartic}\right)\non
& & + \left. \frac{3}{2} \lambda_{\rm quartic}^2 + \frac{457}{24} g'^4 -\frac{23}{4} g^4  + \frac{9}{4} g'^2 g^2  \right],
\eeq

\beq
\beta_{2,\lambda_q} &=& \lambda_{\rm quartic}^2  \left( - 78  \lambda_{\rm quartic} -72  y_t^2 - 72 y_b^2 - 24  y_{\tau}^2 + 18  g'^2 + 54 g^2 \right) \non
& & + \lambda_{\rm quartic} y_t^2 \left( -3 y_t^2 - 42 y_b^2 +  \frac{85}{6} g'^2 + \frac{45}{2}g^2 + 80 g_s^2 \right) \non
& & + \lambda_{\rm quartic} y_b^2 \left(  -3  y_b^2 +  \frac{25}{6} g'^2 + \frac{45}{2}g^2 + 80 g_s^2 \right) + \lambda_{\rm quartic} y_{\tau}^2 \left(  - y_{\tau}^2 +  \frac{25}{2} g'^2 + \frac{15}{2}g^2 \right)\non
& & +  \lambda_{\rm quartic}\left( \frac{629}{24} g'^4 - \frac{73}{8}g^4 + \frac{39}{4} g'^2 g^2 \right) + 4 y_t^4 \left( 15 y_t^2 -3 y_b^2 - \frac{4}{3} g'^2 - 16 g_s^2 \right)\non
& & + y_t^2 \left( - \frac{19}{2} g'^4 - \frac{9}{2} g^4 + 21g'^2 g^2 \right)
 + 4 y_b^4 \left( -3 y_t^2 + 15 y_b^2 + \frac{2}{3} g'^2 - 16 g_s^2 \right)\non
 & & + y_b^2 \left( \frac{5}{2} g'^4 - \frac{9}{2} g^4 + 9 g'^2 g^2 \right)
  + 4 y_{\tau}^4 \left( 5 y_{\tau}^2 - 2 g'^2 \right) + y_{\tau}^2 \left( - \frac{25}{2} g'^4 - \frac{3}{2} g^4 + 11 g'^2 g^2 \right)\non
  & & -\frac{379}{24} g'^6 + \frac{305}{8} g^6 - \frac{559}{24} g'^4 g^2 -  \frac{289}{24} g'^2 g^4.
  \label{lq2loop}
\eeq

\paragraph*{}

\subsection{RGEs above SUSY breaking scale}
\label{RGESUSY}

Next, we have calculated the RGEs up to two-loop order for couplings of the  singlet extension of MSSM (NMSSM), $g', g, g_s, y_t,y_b, y_{\tau},\lambda$ and $\kappa$ in the $\overline{DR}$ scheme. The following results are consistent with Ref.~\cite{Ellwanger:2009dp}.
Note that in the limit of  $\kappa = 0$, the following RGEs reproduce the one in the nMSSM.
It is because the tadpole term does not affect the RGEs of the dimensionless coupling constants.
Furthermore, in the limit of  $\kappa = 0$ and $\lambda = 0$, the following RGEs reproduce the one in the MSSM.

\paragraph*{}
Before we go forward the derivation of the SUSY RGEs, we comment on the matching condition on the Yukawa couplings.
At  SUSY breaking scale, the following relationships are imposed,
\beq
y_{u,SM}(\susy) &=& y_{u,SUSY}(\susy) \sb,\non 
y_{d,SM}(\susy) &=& y_{d,SUSY}(\susy) \cb,
\eeq
where $y_{i,SUSY} $ is the Yukawa coupling in the superpotential, and $y_{i,SM} $ is the one in the SM.
The higher-order corrections to this matching condition (conversion factor from $\overline{MS}$ to $\overline{DR}$ regularization scheme and threshold corrections of the heavy sparticles) are given in Ref.~\cite{Bagnaschi:2014rsa}.
For simplicity, we have omitted its subscript $(SUSY/SM)$ in the text.
For example, in previous section we use the SM Yukawa couplings $y_{i, SM}$, while in this section we use the SUSY Yukawa couplings $y_{i,SUSY}$.

\paragraph*{}
Let us deviate the SUSY RGE of the Yukawa couplings.
First we define the superpotential as follows,

\beq
W=\frac{1}{6}y_{ijk}\Phi^i_0 \Phi^j_0 \Phi^k_0 +\frac{1}{2} M_{ij}\Phi^i_0 \Phi^j_0 +L_{i} \Phi^i_0,
\label{appsuperpote}
\eeq
where $\Phi^i_0$ are  the {\it bare} Chiral superfields.
Thanks to the non-renormalization theorem, a logarithmic dependence of the coupling constant can be related to a logarithmic dependence of the wave function renormalization constant \cite{Seiberg:1993vc, Weinberg3}. 
\beq
\frac{\beta_{n, y_{ijk}}}{(4\pi)^{2n}}&= & y_{ljk} \gamma^{l(n)}_i +y_{ilk} \gamma^{l(n)}_j+y_{ijl} \gamma^{l(n)}_k,\label{appB2} \\
\frac{\beta_{n, M_{ij}}}{(4\pi)^{2n}} &=& M_{l j} \gamma^{l(n)}_i +M_{il} \gamma^{l(n)}_j,\\
\frac{\beta_{n, L_{i}}}{(4\pi)^{2n}}  &= &  L_{l} \gamma^{l(n)}_i \label{appB3},
\eeq
where subscript $n$ represents the loop order, and $\gamma^i_j  $ is an anomalous dimension matrix, 
\beq
\gamma^i_j &\equiv& \frac{1}{2}\frac{d}{d\ln Q}\ln{Z^i_j},\\
 Z^i_j \Phi^{\dag}_{i,R} \Phi^j_R &=& \Phi^{\dag}_{i,0} \Phi^j_0 ,\label{appB1}
\eeq
where $\Phi^i_R$ are  the {\it renormalized} Chiral superfields.

As using the superpotential \eq{appsuperpote}, the anomalous dimension matrix $\gamma^i_j  $ is given as follows \cite{Jones:1983vk,West:1984dg,Bjorkman:1985mi},
\beq
\gamma^{i(1)}_j& =&\frac{1}{(4 \pi)^2 }\left(\frac{1}{2} y^{iml} y_{jml}^{\ast} - 2 g_a^2 C_a(i) \delta ^i_j \right),\label{appB4}\\
\gamma^{i(2)}_j& =&\frac{1}{(4 \pi)^2} \left[ 2 \beta^{(1)}_{g_a} g_a C_a(i) \delta ^i_j - \gamma^{k (1)}_l \left(y^{iml}y_{kmj}^{\ast} + 2 g_a^2 \sum_b (T_a^b)^i_k (T_a^b)^l_j \right) \right]. \label{appB12}
\eeq
where we assume the Chiral superfield $\Phi^i$ is belong to the  fundamental representation of the gauge group $G_a$.
Here,  we use the well-known one-loop level $\beta$ function for the gauge couplings in the SUSY model,
\beq
\frac{\beta_{1,g_a}}{(4\pi)^2} = \frac{1}{(4\pi)^2} g_a^3\left( \sum_i T_a(i) -3 C_a(G)\right).
\eeq
The two-loop level $\beta$ function for the gauge couplings is also related to these functions \cite{Jones:1983vk,West:1984dg,Bjorkman:1985mi}, 
\beq
\frac{\beta_{2, g_a}}{(4 \pi)^4} = \frac{1}{(4\pi)^2}  \left(  2 \beta_{1,g_a} g_a^2 C_a(G) - 2 \sum_i \gamma^{i(1)}_i g_a^3  \frac{C_a(i)}{r} \right), \label{appB13}
\eeq
where $r$ is the number of the generator of the gauge group $G_a$.

\paragraph*{}

In the NMSSM, neglecting all Yukawa coupling except the third generation, the explicit formulae of the one-loop  anomalous dimension ($i=j$) \eq{appB4} are 
\beq
(4 \pi)^2 \gamma_{Q}^{(1)}&=&y_t^2 +y_b^2 -\frac{1}{18}g'^2 -\frac{3}{2}g^2 - \frac{8}{3} g_s^2,\label{appB10} \\
(4 \pi)^2 \gamma_{U}^{(1)}&=&2 y_t^2 -\frac{8}{9}g'^2 - \frac{8}{3} g_s^2,\\
(4 \pi)^2  \gamma_{D}^{(1)}&=&2 y_b^2 -\frac{2}{9}g'^2 - \frac{8}{3} g_s^2,\\
(4 \pi)^2  \gamma_{L}^{(1)}&=&y_{\tau}^2 -\frac{1}{2}g'^2 - \frac{3}{2} g^2,\\
(4 \pi)^2  \gamma_{E}^{(1)}&=&2 y_{\tau }^2 -2 g'^2,\\
(4 \pi)^2  \gamma_{H_1}^{(1)}&=&3y_b^2 +y_{\tau}^2 -\frac{1}{2}g'^2 -\frac{3}{2}g^2 + \lambda^2, \\
(4 \pi)^2 \gamma_{H_2}^{(1)}&=&3y_t^2 -\frac{1}{2}g'^2 -\frac{3}{2}g^2 + \lambda^2, \\
(4 \pi)^2 \gamma_{S}^{(1)}&=& 2 \lambda^2 +2 \kappa^2,\label{appB11} 
\eeq
and off-diagonal one-loop anomalous dimensions ($i \neq j$) are zero.
Using the one-loop anomalous dimensions Eqs.~(\ref{appB10}-\ref{appB11}),
one can derive the two-loop level  RGEs  of the NMSSM in the $\overline{DR}$ scheme.

\paragraph*{}

The one-loop $\beta$ functions for the NMSSM couplings are
\begin{eqnarray}
\beta_{1,g'} & = &11 g'^3, \\
\beta_{1,g} &=& g^3,\\
\beta_{1,g_s} &= &-3 g_s^3,\\
\beta_{1,y_t} & = & y_t \left( -\frac{13}{9} g'^2 -3 g^2 -\frac{16}{3}g_s^2 +\lambda^2   +6 y_t^2  +y_b^2 \right),\\ 
\beta_{1, y_b} &= & y_b \left(  -\frac{7}{9}g'^2  -3 g^2 -\frac{16}{3}g_s^2 +\lambda^2   +y_t^2  +6 y_b^2  +y_{\tau}^2 \right),\\
\beta_{1,y_{\tau}} &= & y_{\tau} \left( -3 g'^2  -3 g^2 + \lambda^2  +3 y_b^2  +4 y_{\tau}^2 \right), \\
\beta_{1, \lambda} &= &   \lambda \left( - g'^2  -3 g^2 +4 {\lambda}^2  +2 {\kappa}^2  +3 y_t^2   +3 y_b^2  + y_{\tau}^2 \right), \\
\beta_{1,\kappa} &= &  \kappa \left( 6 \lambda^2  +6 \kappa^2  \right).
\end{eqnarray}

\paragraph*{}

The two-loop $\beta$ functions for the NMSSM couplings are
\beq
\beta_{2,g'} & = & g'^3  \left(\frac{199}{9} g'^2 + 9 g^2 +\frac{88}{3} g_s^2 - \frac{26}{3}y_t^2 -\frac{14}{3}y_b^2 - 6 y_{\tau }^2 -2 \lambda^2 \right),\label{appB6}  \\
\beta_{2,g} & = &  g^3 \left( 3 g'^2 + 25 g^2 +24 g_s^2 - 6y_t^2 -6 y_b^2 - 2 y_{\tau }^2 -2 \lambda^2 \right),\\
\beta_{2,g_s}&= & g_s^3 \left( \frac{11}{3} g'^2 + 9 g^2 +14 g_s^2 -4 y_t^2 -4 y_b^2 \right),\label{appB62} \\
\beta_{2, y_t} &= & y_t  \biggl( -22 y_t^4 -5 y_b^4  -3  \lambda^4  - 5 y_t^2  y_b^2  - 3y_t^2 \lambda^2  - y_b^2 y_{\tau}^2 - 4  y_b^2 \lambda^2 - y_{\tau }^2 \lambda^2  -2 \lambda^2 \kappa^2  \non
& & + 2 g'^2  y_t^2 + \frac{2}{3} g'^2  y_b^2  +6 g^2 y_t^2  + 16 g_s^2  y_t^2 +\frac{2743}{162} g'^4  +\frac{15}{2} g^4 - \frac{16}{9}g_s^4 \non
& & +  \frac{5}{3}g'^2 g^2 + \frac{136}{27} g'^2  g_s^2  + 8 g^2 g_s^2 \biggl),     \label{appBt}\\
\beta_{2, y_b} &= & y_b  \biggl(  -22 y_b^4  -5 y_t^4 -3 y_{\tau}^4  -3  \lambda^4   -5 y_b^2 y_t^2  - 3y_b^2 y_{\tau}^2-3 y_b^2 y_{\lambda}^2 -4  y_{t }^2  \lambda^2  -2 \lambda^2 \kappa^2 \non
& &+ \frac{2}{3} g'^2  y_b^2  +\frac{4}{3} g'^2  y_t^2  +2 g'^2  y_{\tau}^2  +6 g^2  y_b +16 g_s^2  y_b^2 +\frac{1435}{162} g'^4  +\frac{15}{2} g^4-\frac{16}{9}g_s^4 \non
& & + \frac{5}{3} g'^2 g^2 +\frac{40}{27} g'^2  g_s^2  +8 g^2  g_s^2 \biggl),\\
\beta_{2,y_{\tau}} &=& y_{\tau} \biggl( -10 y_{\tau}^4 -9 y_b^4 -3 {\lambda}^4  -9 y_{\tau}^2 y_b^2 -3 y_{\tau}^2 {\lambda}^2 -3 y_t^2 y_b^2  -3 y_t^2  {\lambda}^2 -2 {\lambda}^2 {\kappa}^2 +2g'^2 y_{\tau}^2 \non
& & -\frac{2}{3}g'^2 y_b^2  + 6g^2 y_{\tau}^2 +16g_s^2  y_b^2 +\frac{75}{2} g'^4  +\frac{15}{2}g^4 +3 g'^2  g^2 \biggl),\label{appBtau} \\
\beta_{2,\lambda} & =  & {\lambda} \biggl( - 10 {\lambda}^4  -9 y_t^4 -9 y_b^4  -3 y_{\tau}^4   - 8 {\kappa}^4  - 9\lambda^2  y_t^2  -9 {\lambda}^2 y_b^2  -3 {\lambda}^2 y_{\tau}^2  -12 {\lambda}^2 {\kappa}^2  \non
& &- 6  y_t^2  y_b^2  +2 g'^2  {\lambda}^2  +\frac{4}{3} g'^2  y_t^2  -\frac{2}{3} g'^2 y_b^2  +2g'^2  y_{\tau}^2  +6g^2 {\lambda}^2  +16 g_s^2  y_t^2  + 16 g_s^2  y_b^2  \non
& &+ \frac{23}{2} g'^4  +\frac{15}{2}g^4 +3 g'^2 g^2  \biggl),\label{appBl} \\
\beta_{2,\kappa} &= &  \kappa \biggl( -24 {\kappa}^4  -12{\lambda}^4 -24 {\kappa}^2 \lambda^2  -18y_t^2 {\lambda}^2  -18 y_b^2  {\lambda}^2  - 6 y_{\tau}^2 {\lambda}^2 + 6 g'^2  {\lambda}^2  \non
 & &+18 g^2  {\lambda}^2       \biggl) .\label{appB7}
\eeq
These formulae are consistent with the RGEs of Ref.~\cite{Ellwanger:2009dp}.

\section{One-loop Corrections to the Mass of the Neutralino}
\label{staubloop}

In this section, we collect the one-loop radiative corrections to the mass of the neutralino \cite{Staub:2010ty}.
Note that, we have found that Ref.~\cite{Staub:2010ty} includes some typos in the equations of the one-loop corrections, and  were provided with fixed one-loop corrections by the author~\cite{private}.

The self-energy matrix for neutralinos is given as follows,
\beq
\Sigma_{i, j}^S(p)&=& 2 \sum_{a=1}^2  B_0(p,m_{\chi_a^-},m_{H^-})m_{\chi_a^-} \Gamma^{L\ast}_{\bar{\chi}_j^0, H^+,\chi_a^-} \Gamma^{R}_{\bar{\chi}_i^0, H^+,\chi_a^-}\non
& & +2 \sum_{a=1}^2  B_0(p,m_{\chi_a^-},M_W )m_{\chi_a^-} \Gamma^{L\ast}_{\bar{\chi}_j^0, G^+,\chi_a^-} \Gamma^{R}_{\bar{\chi}_i^0, G^+,\chi_a^-}\non
& & +  \sum_{a=1}^5\sum_{b=1}^3 B_0(p,m_{\chi_a^0},m_{h_b})m_{\chi_a^0} \Gamma^{L\ast}_{\bar{\chi}_j^0, h_b,\chi_a^0} \Gamma^{R}_{\bar{\chi}_i^0, h_b,\chi_a^0}\non
& & + \sum_{a=1}^5\sum_{b=2}^3 B_0(p,m_{\chi_a^0},m_{A_b})m_{\chi_a^0} \Gamma^{L\ast}_{\bar{\chi}_j^0, A_b,\chi_a^0} \Gamma^{R}_{\bar{\chi}_i^0, A_b,\chi_a^0}\non
& & + \sum_{a=1}^5  B_0(p,m_{\chi_a^0},M_Z)m_{\chi_a^0} \Gamma^{L\ast}_{\bar{\chi}_j^0, G^0,\chi_a^0} \Gamma^{R}_{\bar{\chi}_i^0, G^0,\chi_a^0}\non
& &+6  \sum_{a=1}^2  B_0(p,m_t ,m_{\tilde{t}_a})m_t \Gamma^{L\ast}_{\bar{\chi}_j^0, \tilde{t}_a^{\ast}, t } \Gamma^{R}_{\bar{\chi}_i^0, \tilde{t}_a^{\ast}, t}\non
& &- 8 \sum_{a=1}^2  B_0(p,m_{\chi_a^-}, M_W)m_{\chi_a^-}  \Gamma^{R\ast}_{\bar{\chi}_j^0, W, \chi_a^- } \Gamma^{L}_{\bar{\chi}_i^0,W, \chi_a^-}\non
& &- 4 \sum_{a=1}^5  B_0(p,m_{\chi_a^0}, M_Z)m_{\chi_a^0}  \Gamma^{R\ast}_{\bar{\chi}_j^0, Z, \chi_a^0 } \Gamma^{L}_{\bar{\chi}_i^0,Z, \chi_a^0},
\eeq
where we have neglected the terms which are proportional to the SM fermion mass except top quark.
$\bar{\chi}_i^0$ represents the neutralino of not a mass eigenstate but a gauge eigenstate.
$p$ is the momentum of the external line.

\paragraph*{}

The one-loop corrections from the redefinition of the  neutralino field via the wave function renormalization are given as follows,
\beq
\Sigma_{i, j}^R(p)&=& - \sum_{a =1}^2  B_1(p, m_{\chi_a^-}, m_{H^-}) \Gamma^{R\ast}_{\bar{\chi}_j^0, H^+, \chi_a^- }\Gamma^{R}_{\bar{\chi}_i^0, H^+, \chi_a^- } - \sum_{a =1}^2  B_1(p, m_{\chi_a^-}, M_W) \Gamma^{R\ast}_{\bar{\chi}_j^0, G^+, \chi_a^- }\Gamma^{R}_{\bar{\chi}_i^0, G^+, \chi_a^- }\non
& &  - \sum_{a=1}^3 \sum_{b=1}^3 B_1 (p, m_{\nu_b},m_{\tilde{\nu}_a}) \Gamma^{R\ast}_{\bar{\chi}_j^0, \tilde{\nu}_a^{\ast}, \nu_b }\Gamma^{R}_{\bar{\chi}_i^0, \tilde{\nu}_a^{\ast}, \nu_b }-\frac{1}{2}\sum_{a=1}^5 \sum_{b=1}^3 B_1 (p, m_{\chi_a^0},m_{h_b}) \Gamma^{R\ast}_{\bar{\chi}_j^0, h_b, \chi_a^0 }\Gamma^{R}_{\bar{\chi}_i^0, h_b, \chi_a^0 }\non
&& -\frac{1}{2} \sum_{a=1}^5 \sum_{b=2}^3 B_1 (p, m_{\chi_a^0},m_{A_b}) \Gamma^{R\ast}_{\bar{\chi}_j^0, A_b, \chi_a^0 }\Gamma^{R}_{\bar{\chi}_i^0, A_b, \chi_a^0 } -\frac{1}{2} \sum_{a=1}^5  B_1 (p, m_{\chi_a^0},M_Z) \Gamma^{R\ast}_{\bar{\chi}_j^0, G^0, \chi_a^0 }\Gamma^{R}_{\bar{\chi}_i^0, G^0, \chi_a^0 }\non
&& -3\sum_{a = 1}^6 \sum_{b=1}^3 B_1(p, m_{d_b},m_{\tilde{d}_a}) \Gamma^{R\ast}_{\bar{\chi}_j^0, \tilde{d}_a^{\ast}, d_b }\Gamma^{R}_{\bar{\chi}_i^0, \tilde{d}_a^{\ast}, d_b }
- \sum_{a = 1}^6 \sum_{b=1}^3 B_1(p, m_{e_b},m_{\tilde{e}_a}) \Gamma^{R\ast}_{\bar{\chi}_j^0, \tilde{e}_a^{\ast}, e_b }\Gamma^{R}_{\bar{\chi}_i^0, \tilde{e}_a^{\ast}, e_b }\non
&& -3\sum_{a = 1}^6 \sum_{b=1}^3 B_1(p, m_{u_b},m_{\tilde{u}_a}) \Gamma^{R\ast}_{\bar{\chi}_j^0, \tilde{u}_a^{\ast}, u_b }\Gamma^{R}_{\bar{\chi}_i^0, \tilde{u}_a^{\ast}, u_b }
 -2 \sum_{a=1}^2 B_1(p, m_{\chi_a^-}, M_W) \Gamma^{L\ast}_{\bar{\chi}_j^0, W, \chi_a^- }
\Gamma^{L}_{\bar{\chi}_i^0, W, \chi_a^- }\non
&& -  \sum_{a=1}^5 B_1(p, m_{\chi_a^0}, M_Z) \Gamma^{L\ast}_{\bar{\chi}_j^0, Z, \chi_a^0 }
\Gamma^{L}_{\bar{\chi}_i^0, Z, \chi_a^0 },\\
\Sigma_{i, j}^L(p)&=& - \sum_{a =1}^2  B_1(p, m_{\chi_a^-}, m_{H^-}) \Gamma^{L\ast}_{\bar{\chi}_j^0, H^+, \chi_a^- }\Gamma^{L}_{\bar{\chi}_i^0, H^+, \chi_a^- } - \sum_{a =1}^2  B_1(p, m_{\chi_a^-}, M_W) \Gamma^{L\ast}_{\bar{\chi}_j^0, G^+, \chi_a^- }\Gamma^{L}_{\bar{\chi}_i^0, G^+, \chi_a^- }\non
& &  - \sum_{a=1}^3 \sum_{b=1}^3 B_1 (p, m_{\nu_b},m_{\tilde{\nu}_a}) \Gamma^{L\ast}_{\bar{\chi}_j^0, \tilde{\nu}_a^{\ast}, \nu_b }\Gamma^{L}_{\bar{\chi}_i^0, \tilde{\nu}_a^{\ast}, \nu_b }-\frac{1}{2}\sum_{a=1}^5 \sum_{b=1}^3 B_1 (p, m_{\chi_a^0},m_{h_b}) \Gamma^{L\ast}_{\bar{\chi}_j^0, h_b, \chi_a^0 }\Gamma^{L}_{\bar{\chi}_i^0, h_b, \chi_a^0 }\non
&& -\frac{1}{2} \sum_{a=1}^5 \sum_{b=2}^3 B_1 (p, m_{\chi_a^0},m_{A_b}) \Gamma^{L\ast}_{\bar{\chi}_j^0, A_b, \chi_a^0 }\Gamma^{L}_{\bar{\chi}_i^0, A_b, \chi_a^0 } -\frac{1}{2} \sum_{a=1}^5  B_1 (p, m_{\chi_a^0},M_Z) \Gamma^{L\ast}_{\bar{\chi}_j^0, G^0, \chi_a^0 }\Gamma^{L}_{\bar{\chi}_i^0, G^0, \chi_a^0 }\non
&& -3\sum_{a = 1}^6 \sum_{b=1}^3 B_1(p, m_{d_b},m_{\tilde{d}_a}) \Gamma^{L\ast}_{\bar{\chi}_j^0, \tilde{d}_a^{\ast}, d_b }\Gamma^{L}_{\bar{\chi}_i^0, \tilde{d}_a^{\ast}, d_b }
- \sum_{a = 1}^6 \sum_{b=1}^3 B_1(p, m_{e_b},m_{\tilde{e}_a}) \Gamma^{L\ast}_{\bar{\chi}_j^0, \tilde{e}_a^{\ast}, e_b }\Gamma^{L}_{\bar{\chi}_i^0, \tilde{e}_a^{\ast}, e_b }\non
&& -3\sum_{a = 1}^6 \sum_{b=1}^3 B_1(p, m_{u_b},m_{\tilde{u}_a}) \Gamma^{L\ast}_{\bar{\chi}_j^0, \tilde{u}_a^{\ast}, u_b }\Gamma^{L}_{\bar{\chi}_i^0, \tilde{u}_a^{\ast}, u_b }
 -2 \sum_{a=1}^2 B_1(p, m_{\chi_a^-}, M_W) \Gamma^{R\ast}_{\bar{\chi}_j^0, W, \chi_a^- }
\Gamma^{R}_{\bar{\chi}_i^0, W, \chi_a^- }\non
&& -  \sum_{a=1}^5 B_1(p, m_{\chi_a^0}, M_Z) \Gamma^{R\ast}_{\bar{\chi}_j^0, Z, \chi_a^0 }
\Gamma^{R}_{\bar{\chi}_i^0, Z, \chi_a^0 }.
\eeq
The vertices with  the gauge eigenstate neutralino $\bar{\chi}_i^0$, $X$ and $Y$ ($ \Gamma^{L/R}_{\bar{\chi}_i^0, X, Y }$) are defined in the Ref.~\cite{Staub:2010ty}.

\paragraph*{}

\section{Loop Functions}
\label{appfunctions}

In the calculation of the radiative corrections to the Higgs boson mass from the Coleman-Weinberg potential, we have used the following loop functions, 
 \beq
f(Q^2,x,y)& =  &\frac{1}{x-y}\left( x\ln {\frac{x}{Q^2}}-y\ln{\frac{y}{Q^2}} \right)-1 \nonumber \\
&=&\frac{1}{2}\frac{1}{x-y}(f_1(Q^2,x)-f_1(Q^2,y)),\\
g(x,y)& =  &\frac{1}{(x-y)^3}\left((x+y)\ln{\frac{y}{x}}\right)+\frac{2}{(x-y)^2} \nonumber \\
&=&\frac{1}{4(x-y)^2}\left( f_2(Q^2, x)+f_2(Q^2,y) \right) \non
&& - \frac{1}{(x-y)^3}\left( f_1(Q^2, x)-f_1(Q^2, y)\right),
\eeq
with
\beq
f_1(Q^2, x)&= & 2 x \left( \ln{\frac{x}{Q^2} } -1\right),\\
f_2(Q^2, x)&= &  4\ln{\frac{x}{Q^2}}.
\eeq

\paragraph*{}

In the one-loop threshold corrections of the Higgs quartic coupling at high scale, we have used the following loop functions, 
\beq
\tilde{F}(x) &=& \frac{2 x \ln x}{x^2 -1},\\
\tilde{G}(x) &=& \frac{12 x^2 (1 - x^2 + (1 + x^2) \ln x)}{(x^2 -1)^3},\\
\tilde{H}(x) & = & \frac{3 x (1 - x^4 + 2 x^2 \ln x^2)}{(1 - x^2)^3},\\
\tilde{H}_1(x) & = & \frac{2 x (5 (1 - x^2) + 2 (1 + 4 x^2) \ln x)}{3 (x^2 -1)^2},\\
\tilde{H}_2 (x) &=& \frac{2 x (x^2 -1- 2 \ln x)}{(x^2 -1)^2}, \\
\tilde{f}(x) & =& \frac{3 x (x^2 +1)}{(x^2-1)^2} - \frac{12 x^3 \ln x}{(x^2 - 1 )^3},\\
\tilde{g}(x) &=& - \frac{3(x^4 - 6 x^2 +1)}{2 (x^2 -1)^2} + \frac{6 x^4 (x^2 -3)\ln x}{(x^2 -1 )^3},\\
\tilde{f}_1(x) & =& \frac{6(x^2 +3) x^2}{7 (x^2 -1)^2} + \frac{12 (x^2 -5) x^4 \ln x}{7 (x^2 -1)^3},\\
\tilde{f}_2 (x) & =& \frac{2 (x^2 + 11)x^2}{9 (x^2 -1)^2}+\frac{4 (5 x^2 -17)x^4 \ln x}{9 (x^2 -1)^3},\\
\tilde{f}_3(x) &=& \frac{2 (x^4 + 9x^2 +2)}{3 (x^2 -1)^2} + \frac{4 (x^4 - 7 x^2 -6)x^2 \ln x}{3 (x^2 -1)^3},\\
\tilde{f}_4 (x) &=& \frac{2 (5 x^4 + 25 x^2 + 6)}{7 (x^2 -1)^2}+ \frac{4 (x^4 - 19 x^2 -18)x^2 \ln x}{7 (x^2 -1)^3}, \\
\frac{4}{3}\tilde{f}_5(x,y)&=& \frac{1 + (x + y)^2 - x^2 y^2}{(x^2 -1 )(y^2 -1)} + \frac{2 x^3 (x^2 +1)\ln x}{(x^2 -1)^2(x-y)} - \frac{2 y^3 (y^2 +1) \ln y}{(x - y)(y^2 -1)^2},\\
\frac{7}{6} \tilde{f}_6(x, y)&=& \frac{x^2 + y^2 + x y - x^2 y^2}{(x^2 -1 )(y^2 -1)} + \frac{2 x^5 \ln x}{(x^2 -1)^2(x-y)} - \frac{2 y^5 \ln y}{(x - y)(y^2 -1)^2},\\
\frac{1}{6} \tilde{f}_7(x,y)&=& \frac{1 +x y}{(x^2 -1 )(y^2 -1)} +  \frac{2 x^3 \ln x}{(x^2 -1)^2(x-y)} - \frac{2 y^3 \ln y}{(x - y)(y^2 -1)^2},\\
\frac{2}{3}\tilde{f}_8(x,y) &=&  \frac{x + y}{(x^2 -1 )(y^2 -1)} +  \frac{2 x^4 \ln x}{(x^2 -1)^2(x-y)} - \frac{2 y^4 \ln y}{(x - y)(y^2 -1)^2},
\eeq
and these functions are normalized such that $\tilde{F}(1) = \tilde{G}(1) =\tilde{H}(1) = \tilde{H}_1(1)= \tilde{H}_2(1)=\tilde{f}(1)=\tilde{g}(1)=\tilde{f}_{1/2/3/4}(1)=\tilde{f}_{5/6/7/8}(1,1)=1$.

\paragraph*{}

On the other hand, in the one-loop threshold corrections of the Higgs quartic coupling at weak scale, we have used the following loop functions, 
\beq
F_1(Q) &=& 6 \ln\frac{Q^2}{m_h^2} + \frac{3}{3} \ln \xi -\frac{1}{2} Z\left[\frac{1}{\xi}\right] - Z \left[ \frac{ c_W^2}{\xi}\right] - \ln c_W^2 + \frac{9}{2}\left( \frac{25}{9} - \frac{\pi}{\sqrt{3}} \right), \\
F_0(Q) &=& - 6 \ln \frac{Q^2}{M_Z^2} \left( 1 + 2 c_W^2 - 2 \frac{m_t^2}{M_Z^2}\right) + \frac{3 c_W^2 \xi}{\xi -c_W^2} \ln \frac{\xi}{c_W^2} + 2 Z \left[ \frac{1}{\xi}\right] + 4 c_W^2 Z\left[ \frac{c_W^2}{\xi}\right] \non
& & + \frac{3 c_W^2}{s_W^2}\ln c_W^2 + 12 c_W^2 \ln c_W^2 - \frac{15}{2} (1 + 2 c_W^2)\non
& & -4 \frac{m_t^2}{M_Z^2}\left( 2 Z\left[ \frac{m_t^2}{M_Z^2 \xi} \right] + 4 \ln \frac{m_t^2}{M_Z^2} - 5 \right), \\
F_{-1}(Q) &= &  6\ln\frac{Q^2}{M_Z^2} \left( 1 + 2 c_W^4 - 4 \frac{m_t^4}{M_Z^4} \right) - 6 Z\left[ \frac{1}{\xi}\right] - 12 c_W^4 Z\left[ \frac{c_W^2}{\xi}\right] \non
& & - 12 c_W^4 \ln c_W^2 + 8 (1 + 2 c_W^4) + 24 \frac{m_t^4}{M_Z^4}\left( \ln \frac{m_t^2}{M_Z^2} - 2 + Z\left[ \frac{m_t^2}{M_Z^2\xi} \right] \right),
\eeq
where $\xi = m_h^2 / M_Z^2$, $c_W = \cos \theta_W$, $s_W = \sin \theta_W$ and
\beq
Z[x] &=& \left\{
\begin{array}{l}
2 \zeta \arctan (1 / \zeta) \ \ \ \  \textrm{for }x > 1/4\\
\zeta \ln[(1 + \zeta)/(1 - \zeta)] \ \ \ \   \textrm{for }x < 1/4,\\
\end{array}
\right. \\
\zeta(x) & =& \sqrt{|1 - 4 x|}.
\eeq

\paragraph*{}

In the calculation of the branching ratio of the $\mu \to e \gamma$, we have used the following loop functions, 
\begin{align}
&f_C(x,y)= xy
\left[
\frac{5-3(x+y)+xy}{(x-1)^2(y-1)^2}
-\frac{2\log x}{(x-y)(x-1)^3}
+\frac{2\log y}{(x-y)(y-1)^3}
\right]\,,
\\
&f_N(x,y)= xy
\left[
\frac{-3+x+y+xy}{(x-1)^2(y-1)^2}
+\frac{2x\log x}{(x-y)(x-1)^3}
-\frac{2y\log y}{(x-y)(y-1)^3}
\right]\,,
\end{align}
and these functions are normalized such that  $f_C(1,1)=1/2$ and $f_N(1,1)=1/6$.

\paragraph*{}

In the calculation of the one-loop self energy,
we have used  the following functions\footnote{Our notation of $A$ and $B$ functions are the same  as Ref.~\cite{Pierce:1996zz}.}, which are called Passarino-Veltman function \cite{tHooft:1978xw, Passarino:1978jh, Pierce:1996zz}, 
\beq
A_0(m) &=& (4 \pi)^2 Q^{4-n}\int \frac{d^n q}{i (2 \pi)^n }\frac{1}{q^2 - m^2 + i \epsilon},\\
B_0(p,m_1,m_2) &= & (4 \pi)^2 Q^{4-n}\int \frac{d^n q}{i (2 \pi)^n }\frac{1}{\left[ q^2 - m_1^2 + i \epsilon\right]\left[ (q-p)^2 - m_2^2 + i\epsilon\ \right]},\\
p_{\mu }B_1(p,m_1,m_2) &= & (4 \pi)^2 Q^{4-n}\int \frac{d^n q}{i (2 \pi)^n }\frac{q_{\mu}}{\left[ q^2 - m_1^2 + i \epsilon\right]\left[ (q-p)^2 - m_2^2 + i\epsilon\ \right]},\\
p_{\mu} p_{\nu} B_{21} (p,m_1,m_2)  &+& g_{\mu \nu} B_{22} (p,m_1,m_2) \non
&=&   (4 \pi)^2 Q^{4-n}\int \frac{d^n q}{i (2 \pi)^n }\frac{q_{\mu} q_{\nu}}{\left[ q^2 - m_1^2 + i \epsilon\right]\left[ (q-p)^2 - m_2^2 + i\epsilon\ \right]},
\eeq
where we use the dimensional regulation, $n = 4 - 2 \epsilon$, and $p$ is a momentum of the external line. 

After the integration of the loop momentum,  $A_0$ function becomes 
\beq
A_0(m) = m^2 \left( \frac{1}{\bar{\epsilon}} + 1 - \ln \frac{m^2}{Q^2}\right).
\eeq
Here $1/ \bar{\epsilon}  \equiv 1/ {\epsilon} - \gamma_E + \ln 4 \pi $ and $\gamma_E$ is Euler's constant (0.57721\dots).

$B_0$ function becomes
\beq
B_0(p,m_1,m_2) &= & \frac{1}{\bar{\epsilon}}  - \ln \left( \frac{p^2 }{Q^2}\right) - f_B(x_{+}) - f_B(x_{-}),
\eeq
where
\beq
x_{\pm} &=& \frac{s(p,m_1,m_2)  \pm \sqrt{ s(p,m_1,m_2) ^2 - 4 p^2 (m_1^2 - i \epsilon)}}{2 p^2},\\
s(p,m_1,m_2)  &= & p^2 - m_2^2 + m_1^2,\\ 
f_B(x) &=& \ln (1- x) - x \ln\left(1 - \frac{1}{x}\right) -1.
\eeq

The function $B_1$ can be expressed by $A_0$ and $B_0$ as follows,
\beq
B_1(p,m_1,m_2) = \frac{1}{2 p^2} \left[ A_0(m_2) - A_0(m_1) + s(p,m_1,m_2) B_0(p, m_1,m_2)\right].
\eeq

The zero momentum of the external line  limit, $B_0$ and $B_1$ functions can be expressed as follows,
\beq
B_0(0,m_1,m_2)&=& \frac{1}{\bar{\epsilon}  } - \ln\left( \frac{M^2}{Q^2}\right) + 1+ \frac{m^2}{m^2 - M^2} \ln \left(\frac{M^2}{m^2} \right), \label{B0app}\\
B_1(0,m_1,m_2) &=& \frac{1}{2} \left[ \frac{1}{\bar{\epsilon}  } -\ln \left(\frac{M^2}{Q^2}\right) + \frac{1}{2} + \frac{1}{1-x} +\frac{\ln x}{(1 - x)^2} - \theta (1 - x) \ln x \right],\label{B1app}
\eeq
where $M = \textrm{max}(m_1,m_2)$, $m=\textrm{min}(m_1,m_2)$ and $x = m_2^2 / m_1^2$.

\chapter{Vacuum Transition}
\thispagestyle{empty}

\abstchapter{In this appendix, we  briefly review  the vacuum decay  and the vacuum transition rate per unit space-time volume at zero/finite temperature case. 
Then, we give the fitting formula for Euclidean action in four and three dimensions. }

\section{Vacuum Decay}

When the scalar expectation values on the vacuum are at a global minimum of the scalar potential, this vacuum is stable.
On the other hand, when its scalar expectation values are at a local minimum of one, this vacuum
 becomes unstable.
 Eventually, the unstable vacuum (false vacuum) decay into the global minimum (true vacuum) by a quantum fluctuation (quantum tunneling) and by a
thermal fluctuation (thermal tunneling) for a scalar field located at the local minimum.
 When the vacuum transition rate per unit space-time volume from the false vacuum to the true vacuum is larger than the Hubble parameter, 
a bubble nucleate and  all false vacuum decay quickly.
Namely the universe is filled with the true vacuum.
On the other hand, if  the vacuum transition rate per unit space-time volume is smaller  than the Hubble parameter, the lifetime of the false vacuum is longer than the age of the universe and
it becomes meta-stable vacuum.

 The scalar potential of the SUSY models often has a global minimum which is not an ordinary electroweak symmetry breaking vacuum, and the electroweak symmetry breaking vacuum becomes  unstable \cite{Camargo-Molina:2013sta,Chowdhury:2013dka,Blinov:2013fta,Bobrowski:2014dla}.
 One of the reasons  is that the large $\mu$ term and large $\tb$ lead to the large scalar trilinear couplings like $y_{\ell}\mu \tb  H_2 \tilde{L} \tilde{\bar{E}}$, which can generate the charged breaking global minimum.
 Then, the condition that the electroweak breaking vacuum is not unstable  gives the upper bound on  $\mu$ and  $\tb$.
These vacuum {\it meta-stability} conditions give an allowed region of the deviation from the standard model prediction in SUSY models (e.g.~\cite{Kitahara:2012pb,Carena:2012mw,Kitahara:2013lfa,Endo:2013lva,Endo:2014pja}).

\subsection{Quantum Tunneling at Zero Temperature}

A possibility of the quantum tunneling of the false vacuum had been first suggested by Kobzarev, Okun, and Voloshin \cite{Kobzarev:1974cp}.
Then, Callan and Coleman had established a calculation method  \cite{Coleman:1977py, Callan:1977pt,Coleman}.
The vacuum transition rate from the false vacuum to the true vacuum can be evaluated by semiclassical technique. 
At this time, the imaginary part of the energy of the false vacuum corresponds to the vacuum   transition rate to the true vacuum at zero temperature.
In the semiclassical technique, one  can evaluate  the energy of the false vacuum state using the path integral method in Euclidean space-time.
The vacuum transition rate per unit volume at zero temperature is evaluated as,
\beq
\frac{\Gamma_{\rm trans.}}{V} = A e^{-S_4}.
\eeq
 A precise value of the prefactor $A$ is difficult to evaluate.
However, it does not depend dramatically  on the parameters of the theory, and one can roughly estimate it at the fourth power of the typical electroweak scale in the potential,
\beq
A\simeq (100 \textrm{ GeV})^4.
\eeq 
In contrast, the power index $S_4$ is a sensitive parameter of the vacuum transition rate. 
It can be evaluated by an O(4) symmetric solution as follows,
\beq
S_4 = S_{E4} [\bar{\phi}(\rho )]- S_{E4}[\phi ^f],
\eeq
where  $\rho$ is a radial coordinate in four-dimensional space-time, 
\beq
\rho = \sqrt{(t - t_0)^2+ ({\bf x} - {\bf x}_0)^2 },
\eeq
here the bubble nucleate on $(t_0,~{\bf x}_0)$.
 The Euclidean action in four dimensions $S_{E4}[\phi]$ is as follows,
\beq
 S_{E4} [\phi(\rho )]= 2 \pi^2 \int_{0}^{\infty}  \rho^3 d \rho \left[ \sum_{i} \frac{1}{2} \left( \frac{d \phi_i}{d \rho} \right)^2 + V (\phi_i ) \right],
\eeq
where $\phi_i$ are the real scalar field which construct the scalar potential.
Note that if $\phi_i$ is complex scalar field, the factor $1/2$ is removed.
 $\phi ^f$ represents the value of the fields at false vacuum. The bounce configuration $\bar{\phi}$ is a stationary point of the action, namely $\bar{\phi}$ satisfies the field equations,
 \beq
 \frac{d V(\bar{\phi})}{d \bar{\phi}} = \frac{d^2 \bar{\phi}}{d \rho^2} + \frac{3}{\rho} \frac{d  \bar{\phi}}{d \rho }.
 \eeq
 In addition, the bounce configuration also satisfies the following boundary condition,
\beq
\lim_{\rho \to  \infty} \bar{\phi}(\rho) = \phi ^f,\\
\left. \frac{d \bar{\phi}(\rho)}{d \rho} \right|_{\rho =0} =0.
\eeq

On the other hand, the present Hubble parameter is observed as $H_0 \simeq 1.5 \times 10^{-42}$ GeV. 
When the vacuum transition rate per unit volume $\Gamma_{\rm trans.} / V$ is larger than the fourth power of the current Hubble parameter, 
\beq
A e^{- S_4} >  H_0^4,
\eeq
the lifetime of false vacuum is shorter than the age of the universe and the false vacuum decay quickly.
This inequality leads to 
\beq
S_4 < 4 \ln\left(\fr{A^{1/4}}{H_0}\right)\sim 400.
\eeq
Therefore, when the $S_4 \lesssim  400 $ the false vacuum must decay  into the true vacuum.
This is the vacuum meta-stability condition at the zero tempreture.

\subsection{Thermal Tunneling at Finite Temperature}
\label{finiteTVacuumdecay}

Linde had pointed out that the argument of Coleman is valid only at zero temperature \cite{Linde:1980tt}.
It is because that when the temperature is as large as the typically particle scale, 
the potential changes drastically due to the high-temperature effects.

At the finite temperature, vacuum decay (thermal tunneling) rate can be evaluated by \cite{Linde:1981zj, Quiros:1999jp},
\beq
\fr{\Gamma_{\rm trans.}}{V} &=& A_T e^{-S(T)}\non
& \equiv &A_T e^{-S_3/T},
\eeq
where the prefactor $A_T$ depends on the temperature. However, similarly to the zero temperature case, one can estimate 
\beq
A_T \sim T^4.
\eeq
The power index $S_3$ is the Euclidean action in three dimensions which is evaluated by an O(3) symmetric solution, 
\beq
S_3 = S_{E3} [\bar{\phi}(\rho )]- S_{E3}[\phi ^f],
\eeq
with
\beq
 S_{E3} [\phi(\rho )]= 4 \pi \int_{0}^{\infty}  r^2 d r \left[ \sum_{i} \frac{1}{2} \left( \frac{d \phi_i}{d r} \right)^2 + V (\phi_i ) \right],
\eeq
At the finite temperature, the bounce configuration $\bar{\phi}$  satisfies the field equations,
 \beq
 \frac{d V(\bar{\phi})}{d \bar{\phi}} = \frac{d^2 \bar{\phi}}{d r^2} + \frac{2}{r} \frac{d  \bar{\phi}}{d r},
 \eeq
and
\beq
\lim_{r \to  \infty} \bar{\phi}(r) = \phi^f,\\
\left. \frac{d \bar{\phi}(r)}{d r} \right|_{r=0} =0.
\eeq

\paragraph*{}
The thermal tunneling condition is given as follows,
\beq
\int^{t_{\rm today}}_0 d t \fr{1}{H(t)^3} \fr{\Gamma_{\rm trans.}}{V}  > 1.\label{oomoto}
\eeq
The Hubble parameter during the radiation-dominated era is (the same as \eq{Hubbledef}),
\beq
H^2 &\equiv& \left(\fr{\dot{a}}{a}\right)^2 
= \fr{4 \pi^3}{45 M_{\rm Pl}^2} g_{\ast} T^4\non
&\equiv& \fr{T^4}{4 M_{\rm Pl}^2 \xi^2}. \label{Hubblesono1}
\eeq
The equation of motion \eq{Hubblesono1} and the adiabatic expansion condition, 
\beq
\fr{d (aT)}{d t} =\dot{a}T + a \dot{T} =  0,
\eeq 
lead to
\beq
\fr{d t}{d T} &=& (\dot{T})^{-1} = \left( - \fr{\dot{a}}{a} T \right)^{-1}\non
&=& - \fr{2 M_{\rm Pl} \xi}{T^3}.\label{Hubblesono2}
\eeq 
Substituting Eqs.~(\ref{Hubblesono1}),~(\ref{Hubblesono2}) into \eq{oomoto}, 
the thermal tunneling condition becomes
\beq
- \int^{T_{\rm today}}_{\infty} dT \fr{16 \xi^4 M_{\rm Pl}^4}{T^5} e^{- S_3/T}  > 1,
\eeq
namely 
\beq
\int^{T_c }_0 d T \fr{(2 \xi M_{Pl} )^4}{T^5} e^{- S_3/T}  > 1,
\eeq
where $T_c$ is the critical temperature that the false vacuum and the true vacuum degenerate.
If we take $\xi \sim 3 \times 10^{-2}$ that is the typical value at $T \gtrsim 1 \GeV$ \cite{Wantz:2009it}, the  thermal tunneling condition leads to 
\beq
\fr{S_3}{T} \lesssim \begin{cases}  140 \textrm{~~~~~for~} T_c = 1~\TeV,\\
130 \textrm{~~~~~for~} T_c = 10~\TeV.\\
\end{cases} 
\eeq
Therefore, we have used  $S_3/T \lesssim  130 $  as the thermal tunneling condition in the text.

\section{Fitting Formula for Euclidean Action}
\label{FitAction}

According to Ref.~\cite{Adams:1993zs}, we   give the fitting formula for Euclidean action in four and three dimensions.
Note that in Ref.~\cite{Adams:1993zs}, $\varphi$ denotes real scalar field. On the other hand, we use $\phi$ as a complex scalar field. Thus, a factor $\sqrt{2}$ or $2$ is different from the literature.

\paragraph*{}
Let us consider the following potential for complex scalar $\phi$,
\begin{align}
         V(\phi)=\lambda \phi^2(\phi-\phi_{0,1})(\phi-\phi_{0,2}),
\end{align}
where $\lambda > 0 $ and the dimensional  coefficients $\phi_{0,1}, \phi_{0,2} >0$.
Obviously, this scalar potential has two minimum: the origin $\phi = 0 $ and another point $\phi = \phi_m ~( \sim (\phi_{0,1} + \phi_{0,2})/2)$.
Namely, the origin is a local minimum (false vacuum), and $\phi = \phi_m $ is a global minimum (true vacuum).
If the scalar field expectation value is at a the false vacuum, then this vacuum will be metastable and will decay into the stable true vacuum.

For such the one-dimensional scalar potential, the Euclidean action in four dimensions from the false vacuum  to the true vacuum can be obtained by the following fitting formula \cite{Adams:1993zs},
\beq
S_4  &=&\frac{4\pi^2 }{3\lambda}\frac{1}{(2-\delta)^3}\left[
         13.832\delta-10.819\delta^2+2.0765\delta^3\right],
         \eeq
         with
         \beq
         \delta&\equiv&\frac{8\phi_{0,1}\phi_{0,2}}{(\phi_{0,1}+\phi_{0,2})^2}.
\eeq

Similarly, the Euclidean action in three dimensions  from the false vacuum  to the true vacuum can be obtained by the following fitting formula \cite{Adams:1993zs},
\begin{align}
         S_3=\frac{32\pi }{81\sqrt{\lambda}}(\phi_{0,1}+\phi_{0,2})\frac{\sqrt{\delta/2}}{(2-\delta)^2}\left[
         8.2938\delta-5.5330\delta^2+0.8180\delta^3\right].
\end{align}

\paragraph*{}
In our numerical analysis Figure~\ref{scatter}, we have  reduced the three-dimensional to the one-dimensional scalar potential in the direction of the phase transition.
Then we use the following polynomial fitting, 
\begin{align}
         V(\phi)=(a\phi^2+b\phi+c)\phi^2, 
\end{align}
with
\begin{align}
         a&=\frac{(-3A^2+2A)V_m+A^4V_M}{\left[\phi_m^2(1-A))\right]^2},\\
          b&=-2\frac{(1-2A^2)V_m+A^4V_M}{\phi_m^3(1-A)^2},\\
         c&=\frac{(-4A+3)V_m+A^4V_M}{\left[\phi_m(1-A))\right]^2},\\ 
         A&=\frac{\phi_m}{\phi_M}.
\end{align}
This scalar potential also has two minimum.
The the origin $(\phi = 0,~V(\phi)=0)$ is a local minimum (false vacuum) and  $(\phi_m,~V_m)$ is a global minimum (true vacuum).
Note that we have used this fitting function only when $V_m < 0$, since we need the fitting formula of the scalar potential that  the vacuum on the origin can decay into the vacuum on the global minimum.
This polynomial satisfy the following conditions,
\begin{itemize}
\item The origin is a multiple root.
\item The polynomial passes through $(\phi_M,~V_M)$ and $(\phi_m,~V_m)$.
\item The derivative of $V$ with respect $\phi$ at  $\phi = \phi_m$ vanishes: $V^{\prime}(\phi_m) = 0$.
\end{itemize}
Here we have taken $V_M$ as a local maximum value of the one-dimensional scalar potential.
Finally, $\lambda$, $\phi_{0,1}$ and $\phi_{0,2}$ can be written by the parameter $a,~b,~c$ as follows,
\begin{align}
\lambda &= a,\\
         \phi_{{0,1}},~\phi_{{0,2}}&=
         \frac{-b\pm\sqrt{b^2-4ac}}{2a}\nonumber \\
         &=\phi_M \frac{
         1-2A^2+A^4B\pm(A-1) \sqrt{(2A-1)^2-A^4B}
         }{
       A^3B  -3A+2
         },
\end{align}
         where
\begin{align}
         B&=\frac{V_M}{V_m}.
\end{align}

\chapter{Detail Calculations for Chapter 5}
\thispagestyle{empty}

\abstchapter{In this appendix, we present detail calculations for Chapter \ref{EWBGchap}: the masses of the vector-like matters, the Coleman-Weinberg potential for the vector-like matters and for the top/stop multiplets, and  the conditions of existence of a charge breaking minimum in the charged Higgs direction.}

\section{Coleman-Weinberg Potential}
\label{CWApp}
In this section, we collect the Coleman-Weinberg potentials which are used in Section~\ref{EWBGchap}. Here, we show  explicitly  tadpole and  quadratic terms of the Coleman-Weinberg potential. It is because  that these terms can be  absorbed by the redefinition of the tree parameters (see Section~\ref{subsec:numerical}).

\subsection{Masses of  Vector-like Matters }
\label{vecMASS}
First, the vector-like multiplet superpotential is (see \eq{eq:W_pot}),
\beq
W_{\rm vec.} = \lambda_1 \hat{S} (\hat{\bar{L'}} \hat{L}' + \hat{\bar{E'}} \hat{E}' + \hat{\bar{N'}}\hat{{N'}}) + k \hat{H}_1 (\hat{L}'  \hat{\bar{E'}} + \hat{\bar{L'}} \hat{N}'),
\eeq
where we assume that $\lambda_1$ and $k$ are real for simplicity.
Note that the 
strongly first-order phase transition occurs in $\tb \sim 0$ direction in our model.
So, we neglect the colored vector-like mattes since we assume that they do not have the $H_1^0$ dependence. 
In the following, the doublet matters $\hat{L'}$ and $\hat{\bar{L}}'$ are denoted by
\beq
\hat{L}'&=&\begin{pmatrix}
\hat{L}'_1 \\
\hat{L}'_2
\end{pmatrix},\\
\hat{\bar{L}}'&=&\begin{pmatrix}
\hat{\bar{L}}'_1 \\
\hat{\bar{L}}'_2
\end{pmatrix}.
\eeq

The vector-like fermion mass terms are given as,
\beq
 -\mathcal{L}_{\rm vec.ferm.} &=& (L_1^{'\ast}, \bar{L}'_2, N'^{\ast}, \bar{N}') 
   \begin{pmatrix} 
0 & - \lambda_1 S^{\ast} & 0 & 0\\
- \lambda_1 S &0 & k H_1^0  & 0\\
0& k (H_1^0)^{\ast} & 0 & \lambda_1 S^{\ast} \\
0 & 0 &\lambda_1 S   &0
\end{pmatrix} \begin{pmatrix}
L'_1\\
 \bar{L}_2^{'\ast}\\
  N'\\
    \bar{N}'^{\ast}
    \end{pmatrix}
    \non
 & & + (L_2^{'\ast},\bar{L}'_1, E'^{\ast},\bar{E}') 
  \begin{pmatrix} 
0 &\lambda_1 S^{\ast}  &0 & k (H_1^0)^{\ast}\\
 \lambda_1 S  & 0& 0& 0\\
0 & 0 &0& \lambda_1 S^{\ast} \\
 k H_1^0 &0& \lambda_1 S&0  \\
\end{pmatrix} \begin{pmatrix}
 L'_2\\
\bar{L}_1^{'\ast}\\
   E'\\
   \bar{E}'^{\ast}\\
    \end{pmatrix}.
\eeq
The eight mass eigenvalues are obtained by diagonalizing the vector-like fermion mass  matrix.
The lighter four mass eigenvalues are
\beq
M_{f-}^2&=&\frac{1}{2} \left( 2 \lambda_1^2 |S|^2 + k^2 |H_1^0|^2  -  \sqrt{ k^4 |H_1^0|^4 + 4 \lambda_1^2 k^2 |S|^2 |H_1^0|^2} \right),
\eeq
and the heavier four mass eigenvalues are
\beq
M_{f+}^2 & = & \frac{1}{2} \left( 2 \lambda_1^2 |S|^2 + k^2 |H_1^0|^2 +  \sqrt{ k^4 |H_1^0|^4 + 4 \lambda_1^2 k^2 |S|^2 |H_1^0|^2} \right).
\eeq

\paragraph*{}
\paragraph*{}
The vector-like scalar mass terms are given as,
\begin{eqnarray}
  -\mathcal{L}_{\rm vec.scal.} & = & (\tilde{L}_1^{'\ast} \tilde{\bar{L}}'_2, \tilde{N}^{'\ast}, \tilde{\bar{N}}') \mathcal{M}^2_{\rm vec.scal.neutral}
\begin{pmatrix}
\tilde{L}'_1\\
 \tilde{\bar{L}}_2^{'\ast}\\
  \tilde{N}'\\
    \tilde{\bar{N}}^{'\ast}
    \end{pmatrix}\non
    && 
     + (\tilde{L}_2^{'\ast}, \tilde{\bar{E}}',\tilde{\bar{L}}'_1,\tilde{E}^{'\ast} ) \mathcal{M}^2_{\rm vec.scal.charged} \begin{pmatrix}
 \tilde{L}'_2\\
\tilde{\bar{L}}_1^{'\ast}\\
   \tilde{E}'\\
   \tilde{\bar{E}}^{'\ast}
    \end{pmatrix},
\end{eqnarray}
with
\beq
 \mathcal{M}^2_{\rm vec.scal.neutral}&=&\\
 && {\footnotesize \begin{pmatrix} 
m_{L'}^2 + \lambda_1^2 |S|^2  & 0 & - \lambda_1 k H_1^0 S^{\ast}  & 0\\
0 & m_{\bar{L}'}^2 + k^2 |H_1^0|^2 + \lambda_1^2|S|^2 & 0 & \lambda_1 k H_1^0 S^{\ast}\\
-\lambda_1 k (H_1^0)^{\ast} S & 0 & m_{N'}^2 + k^2 |H_1^0|^2 + \lambda_1^2|S|^2 & 0\\
0 & \lambda_1 k (H_1^0)^{\ast} S &0 & m_{\bar{N}'}^2  + \lambda_1^2|S|^2 
\end{pmatrix}}, \non
  \mathcal{M}^2_{\rm vec.scal.charged} &=& \\
  && {\footnotesize   \begin{pmatrix} 
 m_{L'}^2 + k^2 |H_1^0|^2 + \lambda_1^2|S|^2 &0 & \lambda_1 k (H_1^0)^{\ast}S& 0  \\
 0 &m_{\bar{L}'}^2 + \lambda_1^2 |S|^2  & 0& \lambda_1 k S (H_1^0)^{\ast} \\
 \lambda_1 k H_1^0 S^{\ast}  & 0& m_{E'}^2 + \lambda_1^2|S|^2 & 0\\
 0  & \lambda_1 k S^{\ast} H_1^0 & 0& m_{\bar{E}'}^2  + k^2 |H_1^0|^2 + \lambda_1^2|S|^2\nonumber 
 \end{pmatrix}},
\eeq
where we neglect the $D$ term contributions and the $H_2^0$ dependence because the strongly first-order phase transition occurs in $\tb \sim 0$ direction in our model.

The eight mass eigenvalues are obtained by diagonalizing the vector-like scalar mass  matrix.
When we take the $m_{L'}^2 = m_{\bar{L}'}^2$,   $m_{N'}^2 = m_{\bar{N}'}^2$ and   
 $m_{E'}^2 = m_{\bar{E}'}^2$,  
 the eight mass eigenvalues are given as follows,

\beq
M_{s,{\rm ~neu}1\mp}^2 &=& \frac{1}{2} \left( m_{L'}^2 + m_{N'}^2 + 2 \lambda_1^2 |S|^2 + k^2 |H_1^0|^2 \mp \sqrt{  (m_{L'}^2 - m_{N'}^2 + k^2 |H_1^0|^2)^2 + 4 \lambda_1^2 k^2 |S|^2 |H_1^0|^2} \right), 
\non
M_{s,{\rm ~neu}2\mp}^2 &=&  \frac{1}{2}\left( m_{L'}^2 + m_{N'}^2 + 2 \lambda_1^2 |S|^2 + k^2 |H_1^0|^2 \mp
\sqrt{ (m_{L'}^2 - m_{N'}^2 - k^2 |H_1^0|^2)^2 + 4 \lambda_1^2 k^2 |S|^2 |H_1^0|^2} \right), \non
M_{s,{\rm ~cha}1\mp}^2 &=& \frac{1}{2} \left( m_{L'}^2 + m_{{E'}}^2 + 2 \lambda_1^2 |S|^2 + k^2 |H_1^0|^2 \mp \sqrt{  (m_{L'}^2 - m_{{E'}}^2 + k^2 |H_1^0|^2)^2 + 4 \lambda_1^2 k^2 |S|^2 |H_1^0|^2} \right), 
\non
M_{s,{\rm ~cha}2\mp}^2 &=&\frac{1}{2}\left( m_{L'}^2 + m_{{E'}}^2 + 2 \lambda_1^2 |S|^2 + k^2 |H_1^0|^2 \mp
\sqrt{ (m_{L'}^2 - m_{{E'}}^2 - k^2 |H_1^0|^2)^2 + 4 \lambda_1^2 k^2 |S|^2 |H_1^0|^2} \right).\non
\eeq

\subsection{For Vector-like Matters}

When one takes $m_{L'}^2 = m_{\bar{L}'}^2 = m_{N'}^2 = m_{\bar{N}'}^2 = m_{E'}^2 = m_{\bar{E}'}^2$,  the Coleman-Weinberg potential for the vector-like matters is given as
\beq
V_{\rm CW}^{\rm vec} (H_1^0, S) &= & \frac{1}{32 \pi^2} \biggl[ 4 M_{s-}^4\left( \ln \left( {\frac{M_{s-}^2}{Q^2}} \right) - \frac{3}{2}\right) + 4 M_{s+}^4\left( \ln \left( {\frac{M_{s+}^2}{Q^2}}\right) - \frac{3}{2}\right) \biggl] \non
& &- \frac{1}{32 \pi^2} \biggl[ 4 M_{f-}^4\left( \ln \left({\frac{M_{f-}^2}{Q^2}}\right) - \frac{3}{2}\right) + 4 M_{f+}^4\left( \ln \left( {\frac{M_{f+}^2}{Q^2}} \right)- \frac{3}{2}\right) \biggl],
\eeq
with
\beq
M_{s\mp}^2 = \frac{1}{2} \left( 2 m_{L'}^2 + 2 \lambda_1^2 |S|^2 + k^2 |H_1^0|^2 \mp \sqrt{ k^4 |H_1^0|^4 + 4 \lambda_1^2 k^2 |S|^2 |H_1^0|^2} \right),
\eeq
where $Q$ is the renormalization scale.

If this Coleman-Weinberg potential is expanded around the zero temperature vacuum:
\beq
H_1^0 = 0, ~~~~~~~~~~ S = s_0 =  - \frac{t_S}{m_{s,0}^2},
\eeq
this potential becomes as follows, 
\beq
V_{\rm CW}^{\rm vec} (H_1^0, S) &=& V_{\rm CW,~0}^{\rm vec}
+V_{\rm CW,~s1}^{\rm vec} +(V_{\rm CW,~s1}^{\rm vec})^{\ast} 
+V_{\rm CW,~s2}^{\rm vec}+
V_{\rm CW,~h2}^{\rm vec} \non
&& +V_{\rm CW,~h4}^{\rm vec}  +\dots 
\eeq
with
\beq
V_{\rm CW,~0}^{\rm vec}&=& \frac{8}{32 \pi^2} \left[ \left(m_{L'}^2 + \lambda_1^2 s_0^2\right)^2 \left( \ln\left(\frac{m_{L'}^2  + \lambda_1^2 s_0^2}{Q^2}\right)-\frac{3}{2}\right) - \lambda_1^4 s_0^4 \left( \ln\left(\frac{\lambda_1^2 s_0^2}{Q^2}\right) - \frac{3}{2}\right)\right],\non
\\
V_{\rm CW,~s1}^{\rm vec}& = & \frac{\lambda_1^2 s_0}{2 \pi^2
} \left[ - m_{L'}^2 - \lambda_1^2 s_0^2 \ln \left( \frac{\lambda_1^2 s_0^2}{Q^2}\right)  +\left(  m_{L'}^2 +\lambda_1^2 s_0^2  \right)\ln \left(\frac{m_{L'}^2 + \lambda_1^2 s_0^2}{Q^2} \right) \right] (S - s_0),\non
\\
V_{\rm CW,~s2}^{\rm vec}& = & \frac{\lambda_1^2}{2 \pi^2} \left[- m_{L'}^2- 3 \lambda_1^2 s_0^2 \ln \left( \frac{\lambda_1^2 s_0^2}{Q^2}\right)  + \left(m_{L'}^2+ 3 \lambda_1^2 s_0^2 \right) \ln\left( \frac{m_{L'}^2 + \lambda_1^2 s_0^2}{Q^2}\right) \right] \left|S - s_0\right|^2,\non
\\
\nonumber
\eeq

\beq
V_{\rm CW,~h2}^{\rm vec} &=& \frac{k^2}{4 \pi^2} \left[  - m_{L'}^2  + m_{L'}^2 \ln \left(\frac{m_{L'}^2 + \lambda_1^2 s_0^2}{Q^2}\right) +2 \lambda_1^2 s_0^2 \ln \left(\frac{m_{L'}^2+ \lambda_1^2 s_0^2}{\lambda_1^2 s_0^2}\right)\right] |H_1^0|^2,\non
\\
V_{\rm CW,~h4}^{\rm vec} &=& \frac{k^4}{32 \pi^2} \left[  \frac{2}{3(m_{L'}^2 + \lambda_1^2 s_0^2)^2} \left( 6 \lambda_1^2 m_{L'}^2 s_0^2 + 5 \lambda_1^4 s_0^4 + 6 m_{L'}^4 \ln\left(\frac{m_{L'}^2 + \lambda_1^2 s_0^2}{Q^2}\right) \right. \right. \non
& & ~~~~~~ \left.  + 12 \lambda_1^2 m_{L'}^2 s_0^2 \ln \left(\frac{m_{L'}^2 + \lambda_1^2 s_0^2}{Q^2}\right) 
+ 6 \lambda_1^4 s_0^4 \ln\left( \frac{m_{L'}^2 + \lambda_1^2 s_0^2}{Q^2}\right)\right)
\non
& &  ~~~~~~ \left.- 4 \ln\left(\frac{\lambda_1^2 s_0^2}{Q^2}\right)  -\frac{10 }{3}  \right] |H_1^0|^4.
\eeq
The generated tadpole term $V_{\rm CW,~s1}^{\rm vec}$ and the generated quadratic mass terms  $V_{\rm CW,~s2}^{\rm vec}$ and $V_{\rm CW,~h2}^{\rm vec}$ can be absorbed  by the redefinition of the tree parameters, $t_S,~m_{s,0}^2,~m_1^2$.

\subsection{For Top/stop}
The Coleman-Weinberg potential for the top/stop multiplet is given as
\beq
         V_{\rm CW}^{\rm t} (H_1^0,H_2^0,  S) =\frac{3}{32\pi^2}\left[
         \sum_{\pm}M^4_{t,\pm}\left(\ln\frac{M_{t,\pm}^2}{Q^2}-\frac{3}{2}\right)
         -2M_t^4\left(\ln\frac{M_{t}^2}{Q^2}-\frac{3}{2}\right)
         \right],
\eeq
         with
        \beq
         M_{t,\pm}^2&=&m_{\tilde{t}}^2+ y_t^2|H_2^0|^2\pm y_t\lambda|S||H_1^0|,\\
         M_t^2 &= &y_t^2 |H_2^0|^2,
         \eeq
         where we take $m_{Q}^2 = m_{U}^2 = m_{\tilde{t}}^2$, and neglect the $A$ term and $D $ term contributions.
         If this Coleman-Weinberg potential is expanded around the zero temperature vacuum,
this potential becomes as follows, 
         \beq
V_{\rm CW}^{\rm t} (H_1^0,H_2^0,  S) &=& V_{\rm CW,~0}^{\rm t}
+V_{\rm CW,~2}^{\rm t}  +\dots 
\eeq
with
\beq      
      V_{\rm CW,~0}^{\rm t}&=&\frac{3}{16\pi^2}
         m_{\tilde{t}}^4 \left[ \ln\left(\frac{m_{\tilde{t}}^2}{Q^2}\right) -\frac{3}{2}\right]
         ,\\
        V_{\rm CW,~2}^{\rm t}&=&\frac{3}{16\pi^2}
         \left[
         y_t^2 \lambda^2 s_0^2\ln \left(\frac{m_{\tilde{t}}^2}{Q^2}\right) |H_1^0|^2 
         +2y_t^2 m_{\tilde{t}}^2\left(2\ln\left( \frac{m_{\tilde{t}}^2}{Q^2} \right)-1\right)|H_2^0|^2
         \right].
\eeq
The generated quadratic mass terms $ V_{\rm CW,~2}^{\rm t}$ can be absorbed  by the redefinition of the tree parameters, $m_1^2$ and $m_2^2$.

\section{Charge Breaking Minimum}
\label{chargeddire}
In the nMSSM, there is often a charge breaking minimum in the charged Higgs direction.
Especially, the benchmark point in the text has a charged breaking global minimum. Thus the electroweak symmetry breaking vacuum becomes metastable vacuum.
It is because that when the $\phi$ is relatively large, $|\phi_s|$ becomes small  (see \eq{vevs}).
At this time, small $|\phi_s|$ leads to the light charged Higgs boson mass,
\beq
{M}_{\rm charged}^2=m_1^2+m_2^2+2\lambda^2\phi_s^2+\frac{{g}^2}{2}\phi^2\,.
\eeq
Therefore, the nMSSM tends to have the charge breaking  minimum in the charged Higgs direction.
Note that, when the coupling $\lambda$ is $\mathcal{O}(1)$, the charged Higgs boson is always lighter than the typical mass of the Heavy Higgs boson (see \eq{ch1}),
\beq
M_{\rm charged}^2= M_A^2 - \left(\lambda^2 -\frac{{g}^2}{2} \right) \phi^2  < M_A^2\,.
\eeq

\paragraph*{}
\paragraph*{}
In the following, we consider the condition that the charge breaking minimum occurs.
We can take $\langle H_1^{-} \rangle = 0$ at elsewhere using the SU(2) rotation, 
Thus, the Higgs potential of the nMSSM can be expanded as follows (the same equation as  \eq{Higgspotekuwashi}),
\beq
V_0 &=& m_1^2 |H_1^0|^2 + m_2^2 (|H_2^0|^2+ |H_2^{+}|^2) + m_S^2 |S|^2 + \lambda^2  |H_1^0|^2 |H_2^0|^2    + \lambda^2 |S|^2 (|H_1^0|^2 + |H_2^0|^2 +  |H_2^{+}|^2) \non
 &+& \frac{\bar{g}^2}{8} (  | H_1^0|^4 +  |H_2^0|^4 +  |H_2^{+}|^4 - 2 |H_1^0|^2 |H_2^0|^2 -2 |H_1^0|^2   |H_2^{+}|^2     +2  |H_2^0|^2 |H_2^{+}|^2 )\non
 & +& \frac{g^2}{2} |H_1^0|^2   |H_2^{+}|^2    + ( - \lambda A_{\lambda} S H_1^{0} H_2^0  + t_S S - m_{12}^2 H_1^0 H_2^0 + H.c. ).
\eeq  
Now,
let us consider the minimization condition for $H_1^0, H_2^0, S$ and ``$ H_2^+$".
These conditions  are given as follows,
\beq
\left. \frac{\partial V_0}{\partial H_1^0}\right|_{\rm vev} &=& v_1 \left( m_1^2 + \lambda^2 (v_2^2 + s^2) + \frac{\bar{g}^2}{4}(v_1^2 - v_2^2 -  |v_2^+|^2) + \frac{g^2}{2} |v_2^+|^2 - \lambda A_{\lambda} s \frac{v_2}{v_1} - m_{12}^2\frac{v_2}{v_1} \right) \non
&=& 0, \\
\left. \frac{\partial V_0}{\partial H_2^0}\right|_{\rm vev} &=& v_2 \left( m_2^2 + \lambda^2 (v_1^2 + s^2) + \frac{\bar{g}^2}{4} (v_2^2 - v_1^2 + |v_2^+|^2) - \lambda A_{\lambda } s \frac{v_1}{v_2} - m_{12}^2 \frac{v_1}{v_2} \right) \non
&=& 0,\\
\left. \frac{\partial V_0}{\partial S}\right|_{\rm vev} &=& s \left( m_S^2 + \lambda^2  (v_1^2 + v_2^2 + |v_2^+|^2) + (t_S - \lambda A_{\lambda}v_1 v_2)\frac{1}{s}\right) \non
&=&0,\\
\left. \frac{\partial V_0}{\partial H_2^+}\right|_{\rm vev} &=& (v_2^{+})^{\ast} \left( m_2^2 + \lambda^2 s^2 - \frac{\bar{g}^2}{4}(v_1^2 - v_2^2 -  |v_2^+|^2) + \frac{g^2}{2} v_1^2 \right)\non
&=& 0,
\eeq
namely,
\beq
m_1^2 &=& (m_{12}^2 + \lambda A_{\lambda} s )\frac{v_2}{v_1} -\frac{\bar{g}^2}{4}(v_1^2 - v_2^2 -  |v_2^+|^2) - \frac{g^2}{2} |v_2^+|^2-  \lambda^2 (v_2^2 + s^2),\label{eq3}\\
 m_2^2 &=&  (  m_{12}^2+\lambda A_{\lambda } s) \frac{v_1}{v_2} - \lambda^2 (v_1^2 + s^2) + \frac{\bar{g}^2}{4} (v_1^2 - v_2^2 - |v_2^+|^2), \label{eq4}\\
 s &=& \frac{ - t_S + \lambda A_{\lambda} v_1 v_2} {m_S^2 + \lambda^2 (v_1^2 + v_2^2 + |v_2^+|^2) },\label{eq2}
\eeq
and when we assume $v_2^{+} \neq 0$, then
\beq
m_2^2 = - \lambda^2 s^2 + \frac{\bar{g}^2}{4}(v_1^2 - v_2^2 -  |v_2^+|^2) - \frac{g^2}{2} v_1^2, \label{eq1}
\eeq
where $v_2^+$ represents the vev of the $H_2^+$ scalar field.

Using Eq.~(\ref{eq4}) and  Eq.~(\ref{eq1}), one find the following relationship,  
\beq
 (  m_{12}^2+\lambda A_{\lambda } s) \frac{v_1}{v_2} - \lambda^2 v_1^2 =   - \frac{g^2}{2} v_1^2,
 \eeq
 then, we can get a $\sin 2 \beta$ as a function of $\phi$,
 \beq
\sin 2 \beta = \frac{2 (m_{12}^2+\lambda A_{\lambda } s)}{(\lambda^2 - \frac{g^2}{2}) \phi^2},
\label{sin2beta}
\eeq
where we use 
\beq
v_1&=& \phi \cos \beta, \label{v1}\\
 v_2 &=& \phi \sin \beta. \label{v2}
\eeq
Next, using  Eq.~(\ref{eq3}) and  Eq.~(\ref{eq1}), we can get 
\beq
|v_2^+|^2 = \left ( \frac{\bar{g}^2}{2} - \frac{g^2}{2} \right)^{-1} \left( m_1^2 - m_2^2 + \frac{\bar{g}^2}{2} (v_1^2 - v_2^2) - \frac{g^2}{2} v_1^2 + \lambda^2 v_2^2 - (m_{12}^2+\lambda A_{\lambda } s) \frac{v_2}{v_1} \right). \label{v2plus}
\eeq

It is non-trivial to show a existence of the solution of these conditions Eqs.~(\ref{eq2}),~(\ref{eq1}),~(\ref{sin2beta}) and (\ref{v2plus}).
However, when $A_{\lambda}\sim 0$, these conditions become a bit simple.
Then, \eq{sin2beta} becomes 
 \beq
\sin 2 \beta = \frac{2 m_{12}^2}{(\lambda^2 - \frac{g^2}{2}) \phi^2}.
\label{sin2beta}
\eeq
As one can see, the angle $\beta$ can be expressed as a function of $\phi$ for given input parameters, $\lambda,~m_1^2,~m_2^2,~m_S^2,~t_S$ and $m_{12}^2$.
In addition, $v_1, ~v_2$ and $v_2^+$  can also be expressed as a function of $\phi$, using the fact that $\beta$ can be expressed as a function of $\phi$.
Finally, $s$ can also be expressed as a function of $\phi$ (see \eq{eq2}).

Thus, we can obtain the vevs ($v_1,~v_2,~s,~v_2^+$) that may be the charged breaking vacuum solution as a function  of $\phi$ for  given input parameters.
At this time, if these vevs can satisfy the following three necessary  conditions, 
the solution surely exists, and the charged breaking  minimum exists  for  given input parameters.

Since $\sin 2 \beta \leq 1$,
\beq
\frac{2 m_{12}^2}{\lambda^2 - \frac{g^2}{2}} \leq \phi^2.
\eeq

Since an absolute value of $v_2^+$ is real,
\beq
|v_2^+|^2(\phi)  >0.
\eeq

Since we do not use Eq.~(\ref{eq1}) itself, we should impose it,
\beq
m_2^2  + \lambda^2 s^2 (\phi) - \frac{\bar{g}^2}{4}\left(v_1^2 (\phi) - v_2^2 (\phi)-  |v_2^+|^2 (\phi)\right) + \frac{g^2}{2} v_1^2 (\phi) = 0.
\eeq

In fact, the sample point in the text satisfies the above conditions, and there is a charge breaking global minimum in the charged Higgs direction at the zero temperature.
However, we show that the electroweak symmetry breaking vacuum is actually meta-stable vacuum, and its lifetime is much longer than the one of the universe (see Section~\ref{5no7}).

\bibliography{nMSSMRef.bib}

\begin{thebibliography}{100}
\bibitem{CMS:2014ega}
{\bfseries CMS Collaboration} Collaboration, C.~Collaboration,
``{Precise determination of the mass of the Higgs boson and studies of the
  compatibility of its couplings with the standard model},''.
\end{thebibliography}
\end{document}